%% file: thes-all.tex
\documentclass[english, 12pt, oneside]{mnthesis} 
\pdfoutput=1
\usepackage{amsmath}
\usepackage{amssymb}
\usepackage{graphicx}
\usepackage{babel}

\input{texdefs}

\newcommand{\pfrac}[2]{\left(\frac{#1}{#2}\right)}
\newcommand{\myref}[1]{(\ref{#1})}
\newcommand{\mycite}[1]{\cite{#1}}
\newcommand{\myemph}[1]{#1}

\newcommand{\gev}{\mathrm{GeV}}
\newcommand{\tev}{\mathrm{TeV}}
\newcommand{\mathsubscript}[1]{\mathrm{#1}}
\newcommand{\mbyM}{\frac{m}{M}}
\newcommand{\mssm}{\mathsubscript{MSSM}}
\newcommand{\msusy}{m_{\mathsubscript{susy}}}
\usepackage[pdftex]{hyperref}
\begin{document}
\sffamily 
\phd

\title{\bf The Evolution and Decay of Supersymmetric Flat Directions
  in the Early Universe and Their Role in Thermalizing the Universe}
\author{Matthew G. Sexton} \campus{University of Minnesota}
\program{Physics} \director{Prof. Marco Peloso}

\submissionmonth{December} 
\submissionyear{2008} 

\abstract{I study the post-inflation oscillation and decay of light coherent scalar field condensates that may develop during an inflationary phase of the universe.  In particular, the light scalars studied are a composition of the scalar particles of a supersymmetric theory which correspond to the flat directions of the theory's scalar potential.  Some toy models that possess supersymmetric flat directions are presented and numerical solutions for the evolution of the scalar fields are obtained.  Both analytic and numeric results suggest that such condensates, if they existed in the early universe, can decay through a rapid and nonperturbative process long before these condensates could significantly affect the thermalization of the universe.}
\copyrightpage
\acknowledgements{I would like to thank my Adviser Marco Peloso who I worked closely with on much of the material presented here and who provided invaluable suggestions and comments.  I would also like to thank A.~Emir~G\"umr\"uk\c{c}\"uo\u{g}lu and Professor Keith Olive who also worked on these problems and who contributed helpful comments and suggestions.  Finally, I would like to thank the other members of my Dissertation committee; Professors John Wygant, Joseph Kapusta, and Dan Cronin-Hennessey for their time and helpful comments.}
\dedication{This work is dedicated to my wife Dohi, to our boys Owen and Ryan and to our parents whose encouragement made this work possible.}
\beforepreface
\figurespage
\tablespage
\afterpreface            

\newpage 
\pagenumbering{arabic} \setcounter{page}{1}
\newpage \chapter{Introduction} \label{sec-introduction}
          
Inflation and Supersymmetry are central topics in modern cosmology and
modern particle physics.  There is indirect evidence in support of
inflation \cite{inflation-evidence}, and there is theoretical and
phenomenological motivation for supersymmetry \cite{susy-evidence}.
This thesis is devoted to some problems in early universe cosmology
assuming both inflation and low energy supersymmetry are a part of
nature.  The discussion will be of the dynamics of coherent scalar
fields in the early universe and their possible effects on Cosmology.
One coherent scalar field to be reviewed is the inflaton which is the
proposed field responsible for inflation.  The other scalars, which
are the focus of the work, are those of a supersymmetric theory such
as the Minimal Supersymmetric Standard Model (MSSM). These fields may
have been prepared in a coherent state through the process of
inflation.  Our goal is to obtain an order of magnitude estimate for
the decay (or decoherence) time of these coherent scalar fields after
the conclusion of the inflation.

It is worthwhile to briefly consider the larger context of fundamental
scalar fields in nature.  Scalar fields are associated with many
strange effects in particle physics and cosmology.  For example, the
Higgs Mechanism, which is believed to be responsible for Electroweak
symmetry breaking, is mediated by a fundamental scalar Higgs field(s).
Also, inflation is believed to be due to the evolution of a coherent
scalar field called the inflaton.
Similarly, the present observed acceleration of the universe
\cite{supernova} may be caused by the dynamics of a scalar field
called quintessence \cite{caldwell-etall-1998}.  While composite
scalar fields (bound states) have been observed, not one fundamental
scalar field has ever been observed.  All the observed fields of the
standard model of particle physics are spin 1/2 fermions or vector
gauge bosons.  The yet to be observed Higgs boson is in fact the only
fundamental scalar field in the standard model. One possibility is
that fundamental scalars simply do not exist in nature.  For instance,
there exist well motivated models of particle physics which allow for
Electroweak symmetry breaking without a fundamental scalar Higgs field
\cite{technicolor}.  However, there appears to be no a~priori reason
to disallow fundamental scalars, and on the contrary, nature may be
signaling their existence in the above examples.

Scalar fields also possess peculiar classical and quantum mechanical
properties.  For instance, without some protective mechanism, the
effective masses of scalar fields are not stable to quantum
corrections and they will run up to the fundamental scale of ones
theory which is typically the Planck scale $M_P\sim 10^{19}\,\gev$ or
the Grand Unification Scale $M_{GUT}\sim 10^{16}\,\gev$.  This
sensitivity to the ultraviolet scale is one aspect of what is commonly
known as the Hierarchy problem.  The second aspect to the Hierarchy
problem is why these scales should be so much larger than the scale of
Electroweak symmetry breaking which is of order $\tev$.
We do not attempt to explain the Hierarchy of scales here, but this
Hierarchy, which amounts to thirteen or sixteen orders of magnitude,
will in fact play a role in our models and analyses.

A second peculiar property of scalar fields is that they may obtain a
uniform classical Vacuum Expectation Value (VEV) over all the
observable universe.  The above three examples of the Higgs mechanism,
inflation, and quintessence are all essentially consequences of this
property that scalar fields may obtain a VEV.
Other fields such as the gravitational field and gauge vector fields
may obtain Vacuum Expectation Values, but for these to be uniform over
our observable universe implies a preferred direction to space, or
could be simply reinterpreted as a scalar component to these fields
(in the case of cosmological perturbations for instance).
Vector and tensor fields can be classical on local scales of course,
as in a laser or in the gravitational fields of stars and planets.
Only a scalar field can possess a Lorentz invariant vacuum expectation
value which would be essentially unobservable except for its
modification of fundamental properties of nature.  Specifically, in the
Higgs mechanism, the Lorentz invariant scalar Higgs VEV couples to the
leptons and quarks of the standard model in such a way to give them
the effective masses that we measure in the laboratory.  The proposed
inflaton evolved in time and thus broke Lorentz invariance; however,
it was a uniform VEV over a large patch of space which resulted in a
largely uniform energy density and temperature in the early universe.
The remnants of this uniform bath are observable today as the nearly
perfect thermal spectrum of the cosmic microwave background
\cite{blackbody}.  The concept of a vacuum expectation value of a
scalar field is so fundamental to modern particle theories, that when
it arises in a model, it is simply referred to as ``the'' vacuum of
the model.

The first property mentioned above of the instability of the scalar
field's mass to quantum corrections can be corrected by making ones
theory supersymmetric which cancels the instability and allows for a
hierarchy of scales to be phenomenologically viable.  Note that
supersymmetry does not explain why the hierarchy is there however.
The second property mentioned is a generic property of bosonic fields
that they may coalesce into a Bose-Einstein condensate.\footnote{Yet a
  third peculiar property of scalar fields discussed in
  Section~\ref{sec-inflation} is that scalar fields can have an
  equation of state with negative pressure.}  To understand how such
VEVs can develop in nature, it is unavoidable to look to cosmology for
guidance.

The theory of inflation proposes that there was a stage during the
early universe in which space stretched at an accelerated rate.  If
the inflation was exponential, the line element would be written
$ds^2=dt^2 - e^{2Ht}d{\vec x}^2$.  This inflation ended when the agent
which was responsible for the acceleration had spent enough of its
energy into the expansion that the exponential expansion could no
longer continue.  If one neglects quantum effects during
inflation, then all other species of particles are diluted to
negligible amounts by the inflation, with the only remaining
constituent to the energy density being the inflaton.  The model of
inflation solved a few outstanding problems of cosmology at the time
it was conceived \cite{guth-1981}, and other useful models of
inflation were soon developed \cite{new-inflation,chaotic-inflation}.
The model of chaotic inflation \cite{chaotic-inflation} is the one
which will be applied here.

A first ingredient to our study is the fact that inflation can magnify
the quantum fluctuations of a field into an observable classical wave
with non-negligible energy density.  It may be shown in particular
that this happens to scalar fields whose mass is sufficiently small
\cite{VEVgrowth}.  This effect on the quantum fluctuations of a field
is actually a generic property of all inflation models because it is
an effect of the accelerated expansion itself.  Specifically, all that
is required is a small enough event horizon during inflation for the
fluctuations to propagate past and become frozen
\cite{mukhanov-cosmology-2005}.  The event horizon during inflation is
approximately the inverse Hubble parameter $H^{-1}$ which can be as
small as $10^{-30}$ meters in many inflation models.  It is not so
unbelievable to imagine a quantum fluctuation propagating over such a
short distance.
Finally, it is noted that this effect of inflation on the quantum
fluctuations of a field is believed to be the source of the
cosmological perturbations \cite{cosmological-perturbations} which has
left a signature in the large scale structure of the universe
\cite{sdss} and in the Cosmic Microwave Background \cite{wmap}.
While the study of the cosmological perturbations is a central theme
in cosmology, it will not be the focus of this work.

The second ingredient to our study is that supersymmetry and the MSSM
in particular provides many such light scalar fields.  If inflation
lasted a long enough time, then some set of these scalar fields would
have obtained non-negligible VEVs by the conclusion of inflation
through the above effect on their quantum fluctuations.  Such scalar
field VEVs may then have implications for cosmology.  For instance,
one beneficial feature of these scalar field VEVs is their potential
to contribute to baryogenesis \cite{affleck-dine-1985}.  There may
also be adverse effects from light scalar field VEVs in the early
universe, and these will be discussed shortly.

The scalar fields we are interested in are the scalar particles of the
MSSM.  In general, these fields are coupled to one another, and if
during inflation any one of these begin to obtain a VEV, it will
induce an effective mass in the other fields it couples to.  It may
even induce an effective mass to itself through quartic
self-interactions of the MSSM.  To see this, one may expand the field
as $\phi=\Phi+\gd\phi$ where $\Phi$ is the VEV (to be treated
classically) and $\gd\phi$ are the quantum fluctuations.  Considering
a potential such as $V=m^2|\phi|^2 + \gl|\phi|^4$ and expanding in
$\gd\phi$, the mass term of the perturbations may be extracted.
Schematically, one obtains,
\begin{eqnarray}
m^2 |\phi|^2 + \gl|\phi|^4 \rightarrow (m^2+\gl\Phi^2)\gd\phi^2
\end{eqnarray}
Then assuming the bare mass $m$ is small (specifically $m<H$), and
assuming inflation lasts a sufficiently long time, the VEV $\Phi$ will
accumulate during inflation, and the above effective mass of the
perturbations will grow.  In some patches of the universe, the
effective mass may grow to exceed $m$ and become comparable to $H$.
After this instant, the growth of the VEV in these patches will stop.
In this way, the interaction terms of the scalar potential in the MSSM
can suppress the growth of VEVs during inflation to be orders of
magnitude below the Planck mass.  However, in the MSSM and other
supersymmetric theories, there are special combinations of the scalar
fields for which the renormalizable potential is exactly zero!  These
are known as supersymmetric flat directions.  The flat directions are
tilted slightly by supersymmetry breaking masses $m\sim \tev$, but
these masses are small compared to the Hubble parameter during
inflation which can be $10^{13}\,\gev$ in many models.  In this case,
assuming there are no other higher order Planck suppressed
interactions in the scalar potential, and assuming inflation lasts a
relatively long time, the scalar field VEV can become larger than
$M_P$.  Because the mass is small, the energy density of these VEVs
should still remain below that of the inflaton during inflation.
However, after inflation has ended and after the inflaton condensate
has decayed, there are scenarios in which the energy density of the
light scalar fields can in fact overtake that of the inflaton's decay
products.  One potential adverse effect of such scenarios is that if
the light scalars possess a baryonic or leptonic charge, as in the
Affleck-Dine mechanism \cite{affleck-dine-1985}, they could yield an
overly large baryon asymmetry \cite{affleck-dine-1985, linde-1985}.
Additionally if the scalars dominate the inflaton, their fluctuations
will be the cosmological perturbations, and then ones modeling must be
sensitive to the associated phenomenological constraints from the CMB
(see \cite{enqvist-mazumdar-2003} and references there in).

The above problem of a flat direction field coming to dominate the
energy density of the universe is one example of a generic problem in
reconciling many extensions of the standard model with inflationary
cosmology.  It is known as the moduli problem.  The problem is that
weakly interacting light fields with large enough energy density may
persist in the universe for an extended period of time and may upset
the successful predictions of the big bang cosmology.  For instance if
the fields decay late, they may generate too much entropy and result
in an unacceptably small baryon asymmetry \cite{kolb-etall-1983,
  ellis-etall-1986} or they may upset the
the predictions of nucleosynthesis \cite{ellis-etall-1986}.
Alternatively, if these fields are stable, they may provide too much
dark matter to be consistent with observation such as in the case of
the gravitino problem to be discussed shortly.

However, there can be benefits of having flat direction VEVs.  One
benefit is the mechanism of baryogenesis mentioned earlier
\cite{affleck-dine-1985}.  Another possible benefit is a lower
reheating temperature \cite{allahverdi-mazumdar-2005}, but this
benefit may be difficult to realize \cite{olive-peloso-2006,
  sexton-etall-2008}, and why this is the case will be reviewed
shortly.
In general, the cosmological effects of flat directions depend
strongly on the evolution of the inflaton and flat direction fields
during the post-inflation phase of the universe but before the thermal
phase.  This is the phase of thermalization or ``reheating'' as it is
commonly known and we shall briefly discuss this.  After inflation has
ended, the inflaton dominates the energy density of the universe, and
it will be mostly uniform over large patches of space, but it will
oscillate in time and otherwise behave as a condensate of massive zero
momentum particles in any particular patch.  It will decay by some
unknown mechanism into the quanta of the standard model or of MSSM or
of some other extension of the standard model.  The reheating
temperature mentioned above is simply the temperature achieved by the
inflaton's decay products once they have thermalized.  The reheating
temperature can also be loosely thought of as the highest temperature
that the universe ever reached.\footnote{The thermalization stage is
  called reheating because in the first models of inflation
  \cite{new-inflation} the universe was initially in a thermal phase
  which was followed by the inflationary phase.  The ``reheating'' was
  then a return to thermal equilibrium after the inflationary phase.}

An upper bound for the reheating temperature is determined by the
energy density of the inflaton at the conclusion of inflation if one
assumes that decay and thermalization happen instantly at this time.
The energy density at this time is related to the Hubble parameter
through the Friedmann equation, and measurements of the CMB put a
bound on the Hubble parameter at the end of inflation of about $H_I <
10^{15}\,\gev$.  From this, a bound on the reheating temperature
$T_{RH}< 10^{16}\,\gev$ is obtained (see Section~\ref{sec-reheating}).
At the opposite extreme, the reheating temperature must be greater
than a MeV because nucleosythesis happens at this temperature and
requires a thermal bath.  Thus, there is approximately 19 orders of
magnitude of uncertainty in the reheating temperature.  However, we note
that in most models, the reheating temperature is significantly lower
than the above upper bound to account for a finite decay time of the
inflaton.  To summarize, both $H_I$ and $T_{RH}$ are two important benchmarks for a
cosmological model as both quantities set the energy scale for the
physical processes which may occur.  For instance,
one successful feature of inflation is that it can dispose of
dangerous heavy relic particles or topological defects such as
magnetic monopoles which are prevalent in Grand Unified theories and
also generated in large quantities at GUT scale temperatures
\cite{guth-1981}.  Such particles are disposed of by the effect of
dilution.  Inflation thus solves the monopole problem, but there are
other heavy relics which can be generated thermally and perhaps
non-thermally after inflation has ended.  If the reheating temperature
is higher than about $10^6-10^9\,\gev$ then one such hypothetical
heavy particle, the gravitino, can be produced in large enough
quantities to either upset the predictions of nucleosynthesis, or if
the gravitino is stable, provide too much dark matter
\cite{gravitinos-weinberg, gravitinos-moroi, gravitinos-olive}.  Thus,
there is incentive for cosmologists to construct models with naturally
lower reheating temperatures.  A simple and well motivated mechanism
to lower the reheating temperature is if the inflaton decays through
Planck suppressed interactions, here referred to as a gravitational
decay.  The reheating temperature is then restricted to be below
approximately $10^9\,\gev$ which is still not quite enough to prevent
the above mentioned gravitino problem \cite{gravitinos-weinberg,
  gravitinos-moroi, gravitinos-olive}.  The reheating temperature may
be further reduced if the interactions which thermalize the inflaton
decay products are somehow suppressed.  These interactions would
presumably be gauge interactions, and a simple mechanism for
suppressing them is if the gauge bosons which mediate the interactions
have acquired large masses
\cite{affleck-dine-1985,allahverdi-mazumdar-2005}.  In fact, the flat
direction VEVs just discussed can provide this mechanism because the
VEVs typically break many or all of the standard model gauge
symmetries
\cite{gherghetta-kolda-martin-1995,dine-randall-thomas-1995b}.  The
effective vacuum established will then consist of heavy gauge vector
bosons among other heavy states. This would indeed suppress the
thermalizing interactions and it could lower the reheating temperature
as mentioned above.  However, to obtain a significant suppression, it
requires the flat directions to remain coherent over a long time and
in such cases the flat direction's energy density may come to dominate
over that of the inflaton \cite{olive-peloso-2006,ellis-etall-1987}.
Flat direction domination can be problematic as discussed above.

%
The suppression of thermalization depends on the flat direction VEVs
remaining coherent over a relatively long time, and this has spurred a
recent discussion over the time scale for the decay of flat direction
VEVs \cite{allahverdi-mazumdar-2005, olive-peloso-2006,
  sexton-etall-2008, allahverdi-shaw-campbell-2000,
  postma-mazumdar-2004, allahverdi-mazumdar-2007,
  allahverdi-mazumdar-2008,basboll-etall-2007,basboll-2008} of which
this thesis is a part.  A perturbative estimate of the decay rate of
the flat direction VEVs support the hypothesis of a long lived flat
direction \cite{allahverdi-mazumdar-2005}, but this estimate clearly
depends on the VEVs remaining coherent. The VEVs will decay
eventually, putting the universe back into the gauge symmetric state,
but until the decay, the vacuum is in the broken state discussed
above.  However the vacuum is also evolving -- so strictly speaking it
is not a vacuum -- and it has been pointed out in
\cite{olive-peloso-2006} that the evolution may in fact be
non-adiabatic.  The time scale of the evolution is long compared to
the fast (heavy) scales such as the oscillation of the
inflaton.\footnote{For a uniform field such as the inflaton or the
  flat direction condensates, the mass of the field is also the
  natural frequency of classical oscillation of the field.  Thus, a
  heavy mass means a fast oscillation.} In this sense it is a good
approximation to treat the background VEVs as a vacuum. However, the
flat direction VEVs typically possess more than one complex degree of
freedom, and the relative evolution of these light degrees of freedom
can be a source of non-adiabatic evolution.  The corresponding light
quanta to the flat direction degrees of freedom are thus not
necessarily adiabatically invariant.  The argument along the above
lines is proposed and developed in
\cite{olive-peloso-2006,sexton-etall-2008}, and the results indicate
that non-adiabatic evolution may contribute to flat direction decay.
Much of the new material presented in this thesis is based on the work
of \cite{sexton-etall-2008} and also extends and clarifies some of
these results.  In the remainder of the introduction, this work is
described.

The decay of the flat directions can equivalently be described as the
onset of spatial instability to the flat direction condensates.  Like
the inflaton, the flat direction VEVs can be treated as condensates of
massive zero momentum scalar particles, but with masses approximately
$10^{10}$ times smaller than that of the inflaton.  The decay of the
condensates proceeds by ``kicking'' quanta from the zero momentum modes
into higher momentum states through the effect of the
non-adiabatically evolving background.  In position space, this simply
corresponds to the development of spatial instability of the
condensate.

If the condensate is perfectly uniform over space and if all fields
are treated classically, a spatial instability will not develop.
However, one expects there to be small noise fluctuations on top of
the uniform field.  Even assuming the absence of classical noise,
there is quantum noise, and the dynamics of the uniform field is not
necessarily stable to either source of noise.\footnote{Alternatively,
  if the condensate is coupled to other fields which can have
  classical or quantum noise, spatial instability can develop in the
  these other fields.  An example is provided in
  Section~\ref{sec-inflation}} Specifically, both classical and
quantum noise can be amplified by the effect of the evolving
background which couples to the noise perturbations through effective
and time dependent mass terms.  The instability is thus produced by
parametric resonance.

The formalism here used to quantify the above instability is that of
quantum fields in external backgrounds, and it is commonly used in
quantum field theory and cosmology \cite{mukhanov-winitzki-2007}.  It
is typically applied to handle the case of a single field with a time
dependent frequency.  The formalism was extended to handle multiple
scalar and fermionic fields in \cite{nilles-peloso-sorbo-2001} and it
is further developed for scalars here.  As an example of these new
developments, consider in the single field analysis, the quantity
$\dot\go / \go^2$.  Specifically, one indication of nonadiabatic
evolution is when the frequency of a state changes in a time
comparable to or faster than one period of oscillation for the state,
or $\dot\go /\go^2 \simg 1$.  This measure of non-adiabatic evolution
is here generalized to the case of multiple scalar fields where one
must not only keep track of the time evolution of the
eigenfrequencies, but also the time evolution of the eigenvectors.  In
particular, if any eigenvector has changed during a time comparable to
or greater than the oscillation period of any state of the system, the
evolution can be non-adiabatic.  These quantities can thus be arranged
in a matrix. The matrix expression is relatively simple and is
presented here and in \cite{sexton-etall-2008}, but to our knowledge
has not been reported elsewhere.  As in the single field case, the
elements of this matrix provide a convenient preliminary measure of
non-adiabatic evolution.  They are applied in this thesis for this
specific purpose.

An alternative method for studying the spatial instability of the flat
directions, which contrasts to the above formalism, is the lattice
simulation of the classical field theory \cite{lattice-easy}.  Such
classical simulations are possible when the condensates have large
occupation numbers. For instance, lattice methods have been performed
successfully to study nonperturbative decay of the inflaton condensate
\cite{lattice-examples}.  Lattice methods may in principle also be
applied to study flat directions but there are no published examples
to date.  However as these simulations are very demanding with respect
to computer memory, the types of models that can be studied are thus
limitted.
The methods used here are more amenable to analytic arguments, are
less computationally demanding than lattice methods, and these methods
also incorporate quantum effects by construction.  Lattice methods are
not applied in this work, but have not been ruled out for future work.

One limitation to our formalism is that it is performed at quadratic
order in the perturbations.  Interaction terms are not included, and
also the production of quanta is not reflected by a corresponding
depletion of the evolving flat direction condensate.  Our models
strictly do not conserve energy in this respect and thus become
unreliable when the energy density of the produced quanta becomes
comparable to the energy density in the flat direction VEVs.  However,
the formalism is known to adequately model the initial phase of
exponential particle production.  For instance, a comparison of the
two methods for a toy inflation model has been found in qualitative
agreement \cite{lattice-compare}.  Our goal is to obtain an order of
magnitude estimate of the VEV's decoherence (decay) time, and by
determining the initial growth exponent, this goal may be achieved.
The effects of interactions will be discussed at the conclusion of
Section~\ref{sec-reheating}.

Another important aspect to studying the decay of flat directions
which is developed here is the handling of the gauge symmetry.  Due to
the complexity involved in studying flat directions with the full
$SU(3)\times SU(2)\times U(1)$ standard model gauge group, only $U(1)$
gauge models are here presented.\footnote{Note however, that in
  \cite{sexton-etall-2008} the method is generalized to $SU(N)$ models
  which behave as $N$ copies of the $U(1)$ model.}  Despite this
limitation, our analysis correctly handles the gauge fixing of the
perturbations and in particular the vector longitudinal modes
(Goldstone mode) which do not necessarily decouple from the other
scalar perturbations.  Choosing the unitary gauge is the most
convenient choice and once this gauge is fixed, the constraints of
gauge current conservation become transparent.  We show how the
constraints of current conservation lead to non-trivial consequences
for the evolution of the flat direction VEVs.

We apply the above gauge fixing to two simple $U(1)$ models involving
both a single flat direction and also multiple flat directions
\cite{olive-peloso-2006}.  The multiple flat direction model will be
shown to yield non-adiabatic evolution.  One expects the perturbations
to the flat direction VEVs to also be the states involved in the
non-adiabatic evolution because these are the light states of the
vacuum and so evolve on the same time scale as the vacuum.  To
understand why the model can lead to a parametric resonance, one first
tabulates the degrees of freedom.  The multiple flat direction model
consists of the gauge vector and four charged complex scalars.  The
time-like component of the vector is non-dynamical so this removes one
degree of freedom, and the gauge fixing removes a second degree of
freedom.  There remains ten propagating physical degrees of freedom.
Three of these degrees of freedom will belong to the massive vector
and one will belong to a heavy Higgs state of the same mass as the
vector. There remains six light degrees of freedom, and these
necessarily correspond to three flat direction degrees of freedom of
the model.  However not all of the flat directions necessarily obtain
VEVs, and in the solutions we consider, the gauge current constraint
mentioned above allows only two of the three flat directions to obtain
a VEV.  The remaining flat direction which did not obtain a VEV mix
with the Higgs perturbation and this system of three real fields mixes
strongly and results in parametric resonance.

It will be shown that all three states including the heavy Higgs will
be produced in the resonance.
It will also be found that the production of the heavy states compared
to that of the two light quanta is consistent with the principle of
equipartition of the energy.  The equipartition here will not be that
of a thermal bath, but rather a property of the strong coupling
between the modes, and more akin to the equipartition seen in some
classical systems of strongly coupled oscillators
\cite{equipartition,equipartition2}.  The equipartition observed in
the numeric results is also obtained by direct analysis of the
Heisenberg equations of motion from which a scaling property of these
equations is obtained.
In particular, this scaling allows us to rescale our numerical
solutions, which are limited by machine precision to a small hierarchy
of approximately nine orders of magnitude to the phenomenologically
relevant case in which the hierarchy is up to sixteen orders of
magnitude.

The numeric results of this thesis will suggest that a system of
multiple flat directions can decay nonperturbatively for a broad range
of parameters via parametric resonance.  Numerical estimates of the
decay time will be part of these results.  In addition to the numeric
results, some formal results stand on their own. In particular the
scaling arguments, the adiabaticity parameters, the reformulation of
the Heisenberg equations of motion (not mentioned above), and the
gauge fixing procedure are all new results, and potentially useful
tools in further study.

The plan of the thesis is as follows:
In Section~\ref{sec-susyflat}, we lay out the formalism for
identifying and cataloguing of MSSM flat directions.  This includes a
motivation of the scalar potential (the D and F terms) of the MSSM,
the formalism of gauge invariant monomials for identifying flat
directions, and the lifting of flat directions by the
nonrenormalizable superpotential.  This section concludes with a brief
discussion of supersymmetry breaking, and the spectrum of soft
sparticle masses.

In Section~\ref{sec-heiseqs} we discuss the formalism for describing
the quantum evolution of a system of coupled scalar fields in a time
dependent background, or the Heisenberg equations of motion. In
particular we continue the program begun in
\cite{nilles-peloso-sorbo-2001} by presenting some new formal tools
and discussing the solution of the Heisenberg equations when there is
a hierarchy of mass scales. We also draw the connection between the
evolution and an equipartition of energy.

In Section~\ref{sec-inflation} a brief introduction to classical
inflation is presented followed by a description of the mechanism by
which quantum fluctuations of light scalar field such as the SUSY flat
directions will be converted into a VEV through inflation.  We discuss
the post-inflation evolution of the VEVs and the likelihood of
obtaining large VEVs of order the Planck Mass.

Section~\ref{sec-reheating} contains discussion of different scenarios
for reheating but is focused on a standard case in which the bulk of
the produced entropy is from the inflaton.  In our discussion of
reheating, we emphasize nonperturbative dynamics or ``preheating''
scenarios as they are known in the literature. The calculation of the
reheating temperature and the baryon asymmetry is discussed.  The
potential effects of flat directions on the reheating process is also
discussed

In Section~\ref{sec-gaugefixing} we explain the procedure for gauge
fixing a generic model of U(1) flat directions to the unitary gauge.
We then apply the procedure to two simple U(1) symmetric models of
flat directions, and we calculate the formal expressions which
describe the non-perturbative decay of one of the models.

In Section~\ref{sec-results} the multiple flat direction model is
solved numerically and the results for the production of quanta is
presented.  A useful scaling relation is described which allows us to
rescale our results from an energy scale dictated by limitations of
the computer to the much smaller Electroweak scale masses that are
phenomenologically relavant.

We conclude in Section~\ref{sec-conclusions} with some discussion of
the implications of our results, and in particular which features may
be generically applicable to the MSSM.  We also indicate future
directions of study.

In the Appendix, a brief discussion of the notation of the thesis is
provided, as well as some calculations too lengthy to show in the main
text.

\newpage \chapter{Supersymmetry and Flat Directions of the Scalar Potential} \label{sec-susyflat}
In this section, we present a brief introduction to supersymmetric
field theory in order to motivate the scalar potential in these
theories, and the ``flat directions'' which are the minima of the
scalar potential.  We describe the methods for identifying flat
directions of a supersymmetric theory which involve gauge invariant
monomials \cite{buccella-etall-1982}.  We then discuss the terms of the
superpotential which are expected to lift the flat directions
\cite{gherghetta-kolda-martin-1995,dine-randall-thomas-1995b},
providing a natural cutoff for the VEV a flat direction may obtain
during inflation.  We conclude with a brief discussion of
supersymmetry breaking and the spectrum of soft sparticle masses which
also appear in the scalar potential and which characterize the
classical evolution of a flat direction during and after inflation.
Much of the background material related to supersymmetry was gathered
from the review~\cite{martin-susyprimer}.

         \section{The Lagrangian of a Supersymmetric Field Theory and its Scalar Potential} \label{sec-susyflat-scalarpotential} 
Below we introduce supersymmetric field theory and the scalar
potential in these theories.

A supersymmetry transformation will transform a bosonic field into a
fermionic one and vice-versa while preserving the form of the terms
in the Lagrangian.  For instance, the canonical kinetic terms for a
complex spin-0 scalar $\phi$, and that of a two component spin-1/2
fermion $\psi$ in the Lagrangian are \footnote{The notation here is
  ${\bar\sigma}^\mu=(\sigma_0, -\vec \sigma)$ and
  ${\sigma}^\mu=(\sigma_0,\vec \sigma)$ where
  $\sigma_0=\left(\begin{array}{ll}1 &0 \\ 0 & 1\end{array} \right)$
  and $\vec\sigma$ are the Pauli spin matrices
  $\sigma_1=\left(\begin{array}{ll}0 &1 \\ 1 & 0\end{array} \right)$,
  $\sigma_2=\left(\begin{array}{ll}0 &-i \\ i & 0\end{array} \right)$,
  $\sigma_3=\left(\begin{array}{ll}1 &0 \\ 0 & -1\end{array} \right)$
  .} 
\begin{equation}
\der_\mu \phi^* \der^\mu \phi + i\psi^\dagger {\bar\sigma}^\mu\der_\mu \psi
\end{equation}
A supersymmetry transformation should have the effect $\{\phi,\psi\}
\rightarrow \{\tilde\phi,\tilde\psi\}$ such that the form of the above
kinetic terms will have not changed after the transformation.  A
simpler symmetry which one may compare to supersymmetry, is the
symmetry of the following Lagrangian involving two complex scalar
fields $\phi_1$ and $\phi_2$,
\begin{equation}
\der_\mu \phi_1^* \der^\mu \phi_1 + \der_\mu \phi_2^* \der^\mu \phi_2 - m^2(|\phi_1|^2 +|\phi_2|^2)
\end{equation}
This Lagrangian is invariant under SU(2) transformations of the
doublet $(\phi_1,\phi_2)$ which rotates the scalar fields $\phi_1$ and
$\phi_2$ into each-other.  The transformation matrix can be written
$e^{i\vec\epsilon\cdot\vec\tau}$ where
$\vec\epsilon=(\epsilon_1,\epsilon_2,\epsilon_3)$ is a vector
specifying the rotation and $\tau = \frac12 (\gs_1,\gs_2,\gs_3)$ are
the group generators. The infinitesimal form of this transformation is,
\begin{equation}
\left(\begin{array}{l}\phi_1\\ \phi_2\end{array}\right)   \rightarrow 
(1+i\vec\epsilon\cdot\vec\sigma)\left(\begin{array}{l}\phi_1\\ \phi_2\end{array}\right).  
\label{susyflat-su2transformation}
\end{equation}
For example, the standard model Higgs field transforms in this way.  A
supersymmetry transformation is slightly more involved than an SU(2)
transformation however as the generators of the supersymmetry
transformation are anti-commuting spinors.  
Under an infinitesimal supersymmetry transformation a scalar
$\phi$ and a fermion $\psi$ will transform via an anti-commuting two
component spinor $\epsilon$ as follows,
\begin{equation}
\gd \phi \rightarrow \phi+ \epsilon^\alpha \psi_\alpha
\;\;\;,\;\;\; 
\psi_\alpha \rightarrow \psi_\alpha+ i(\sigma^\mu \epsilon^\dagger)_\alpha\der_\mu \phi \label{susytransformation1}
\end{equation}
where $\alpha=1,2$ indexes the two components of the spinor. This
infinitesimal SUSY transformation is to be compared with the
infinitesimal $SU(2)$ transformation~\myref{susyflat-su2transformation}
shown above.  There is a further subtlety in the above SUSY
transformation that for the transformation to be consistent, it is
necessary that the number of bosonic degrees of freedom in $\phi$
match the number of fermionic degrees of freedom in $\psi$.  At a
cursory level there is a problem since the spinor is described by two
complex values while the scalar is described by one complex value.  Of
course, there are only two on-shell degrees of freedom in both the
scalar and spinor, but the other two spinor degrees of freedom can go
off-shell when calculating a quantum amplitude,\footnote{by off-shell,
  we mean that the Heisenberg uncertainty relation allows the
  relation $p_\mu p^\mu = m^2$ to be violated for finite periods of
  time (ie for a virtual particle).} and we will want our
transformation to be applicable both on-shell and off-shell.  In order
to match both the on-shell and off-shell degrees of freedom, a complex
auxiliary field $F$ is introduced.  This field will not be dynamical.
It will be rather a Lagrange multiplier in the classical sense, but it
will make up the difference of off-shell degrees of freedom for the
scalar so that the fermion may consistently ``rotate'' into the spin-0
boson.  The auxiliary $F$ field is necessary for the supersymmetry
transformation to be consistent.  The combination of a complex spin-0
scalar $\phi$, a two component spinor $\psi$, and the auxiliary field
$F$ is known in the literature as a ``chiral multiplet'', and it
transforms consistently into itself under supersymmetry
transformations.  The chiral multiplets have the following form in the
Lagrangian,
\begin{equation}
\mathcal{L}_{\mathsubscript{chiral}} = \der_\mu \phi_i^* \der^\mu \phi_i + i\psi_i^\dagger {\bar\sigma}^\mu\der_\mu \psi_i + F_i^* F_i + \mathcal{L}_{\mathsubscript{chiral\;int}} 
\label{Lchiral}
\end{equation}
which is simply the canonical kinetic terms shown previously, plus a
quadratic part in the the auxiliary field $F$, plus interactions
between multiplets encoded in $\cL_{\mathsubscript{chiral\;int}}$ to be specified
shortly.  Every standard model fermion (a quark or lepton) is expected
to belong to a ``chiral multiplet'' such as above, and so is expected
to have a spin-0 counterpart (a squark or slepton).  Additionally, the
as yet to be discovered Higgs sector fields are also expected to
belong to chiral multiplets, and have their respective spin-1/2
counterparts (Note the MSSM requires two Higgs doublets versus one for
the Standard Model).

The interactions, $\mathcal{L}_{\mathsubscript{chiral\;int}}$ shown above must
preserve supersymmetry as well as gauge invariance, and these two
constraints are enough to restrict its form.  Without going into the
details of the derivation, $\mathcal{L}_{\mathsubscript{chiral\;int}}$ may be encoded
into a complex quantity referred to as the superpotential $W$.  The
interaction Lagrangian and the superpotential are then defined,
\begin{eqnarray}
\mathcal{L}_{\mathsubscript{chiral\;int}} &=& -\frac12 \left(W^{ij}\psi_i\psi_j + W^i F_i\right) + \mbox{c.c.} \label{Lchiralint}\\
W &=& b_a \Phi^a + m_{ab}\Phi^a\Phi^b + \gl_{abc}\Phi^a\Phi^b\Phi^c \label{susyflat-superpotential} \\
W^i &=& \frac{\delta W}{\delta \Phi_i} \;\;\;,\;\;\;
W^{ij} = \frac{\delta^2 W}{\delta\Phi_i \delta\Phi_j}  \nonumber 
\end{eqnarray}
where the standard convention of writing $W$ in superfield notation is
used.  The superfield $\Phi_a$ is a compact notation to represent both
the scalar $\phi_a$ as well as its fermionic superpartner $\psi_a$.
Note also that $W$ is not in truth a potential because it is in
general complex valued.  To summarize, $W$ is invariant under both
supersymmetry and the gauge symmetries, and it encodes the information
required to construct $\cL_{\mathsubscript{chiral\;int}}$ as shown above.

The above defined scalar fields and their superpartners must be
charged under the standard model gauge group.  Our SUSY models then
require gauge vector particles $A_\mu^a$, and each of these will
possess a fermionic superpartner, $\lambda^a$ (a gaugino) for
supersymmetry to be preserved.  For the same reasons as we required an
auxiliary field $F$ for the chiral multiplet, we will require a real
scalar auxiliary field $D$ for the gauge vector multiplet.  The
counting of the off-shell and on-shell degrees of freedom in this case
are as follows. The vector and spin-1/2 particles both have two
on-shell degrees of freedom.  Off-shell, the vector has three degrees
of freedom as the gauge symmetry always allows us to eliminate one of
the degrees of freedom.  We must then add one degree of freedom to
make up the difference, and this is provided by a real scalar
auxiliary field $D$.  The vector gauge multiplet thus contains a real
vector $A_\mu^a$, a spin-1/2 fermion $\lambda$ and real scalars $D^a$
which appear in the Lagrangian as follows
\begin{eqnarray}
\cL_{\mathsubscript{total}} &=& \cL_{\mathsubscript{chiral}} + \cL_{\mathsubscript{gauge}} + \cL_{\mathsubscript{additional\;int}} \label{susyflat-Ltot}\\
\mathcal{L}_{\mathsubscript{gauge}} &=& -\frac14 F_{\mu\nu}^a F^{\mu\nu a} - i \gl^{\dagger a}{\bar\sigma}^\mu \mathcal{D}_\mu \gl^a + \frac12 D^a D^a 
\label{Lgauge} \\
 F_{\mu\nu}^a &=& \der_\mu A_\nu^a - \der_\nu A_{\mu}^a + g f^{abc}A_\mu^b A_\nu^c \nonumber \\
\mathcal{D}_\mu \lambda^a &=& \der_\mu \lambda^a + g f^{abc} A_\mu^b \lambda^c \nonumber
\end{eqnarray}
where $f^{abc}$ are the gauge structure constants for the specific
vector gauge particle contained in the multiplet.  The interactions of
these gauge particles is partly specified by replacing gauge covariant
derivatives for the regular derivatives in \myref{Lchiral}.  However,
there are additional interactions between the gauge particles and the
chiral multiplets which are allowed by supersymmetry and gauge
symmetry.  These are parametrized by $\cL_{\mathsubscript{additional\;int}}$ above and have the
form,
\begin{equation}
\cL_{\mathsubscript{additional\;int}} = -\sqrt2 g \left[(\phi^* T^a \psi)\lambda^a + \mbox{c.c}\right] + g(\phi^*T^a\phi)D^a \label{Lmixedint}
\end{equation}
where $g$ is the relevant gauge coupling.  To be complete, the
supersymmetry transformation for the fields after inclusion of both
chiral and gauge vector multiplets is generalized
\cite{martin-susyprimer} from \myref{susytransformation1} to,
\begin{eqnarray*}
\delta \phi &=& \epsilon \psi_i\\
\delta \psi_{i\ga} &=& i(\gs^\mu\epsilon^\dagger)_\ga \mathcal{D}_\mu\phi_i + \epsilon_\ga F_i  \\
\delta F_i &=& i\epsilon^\dagger {\bar\gs}^\mu \mathcal{D}_\mu\psi_i +\sqrt2 g (T^a\phi)_i\epsilon^\dagger\lambda^{\dagger a} \\
\delta A^a_\mu &=& \frac{1}{\sqrt2}\left(\ge^\dagger {\bar\gs}^\mu \gl^a - \gl^{\dagger a}{\bar\gs}^\mu \ge \right)\\
\delta \lambda^a_\ga &=& \frac{i}{2\sqrt2}(\sigma^\mu {\bar\gs}^\nu \ge)_\ga F_{\mu\nu}^a + \frac{1}{\sqrt2}\ge_\ga D^\ga\\
\delta D^a &=& \frac{i}{\sqrt2}\left(\ge^\dagger {\bar\gs}^\mu \mathcal{D}_\mu\gl^a - \mathcal{D}_\mu\gl^{\dagger a}{\bar\gs}^\mu \ge \right)
\label{susytransformation2}
\end{eqnarray*}
The transformation is quite complicated, but this should not obscure
the general idea which is very simple.  To summarize, the Lagrangian
of our supersymmetric field theories is defined by the
Eqs.~\ref{Lchiral}~-~\ref{Lmixedint}.
The scalar potential of the candidate theory is determined by solving
the equations of motion for the auxiliary fields $F^\ga$ and $D^a$
(which are Lagrange Multipliers) and substituting these back into the
Lagrangian.  The result of this procedure is the following,
\begin{eqnarray}
V(\phi,\phi^*) &=& \frac12 \sum_\ga D^\ga(\phi,\phi^*)^2 + \sum_a |F^a(\phi)|^2 \label{susyflat-scalarpotential} \\
D^\ga &=& g_\ga \sum_{a,b} \phi^* T^a \phi \label{Dsol} \\
F_a(\phi) &=& \frac{\der W}{\der \phi^a} \label{Fsol}
\end{eqnarray}
where the first line is the scalar potential, and the latter two lines
are the equations of motion for the auxiliary fields.  
Note that the scalar potential ultimately arises as a constraint from
the demands of both supersymmetry and gauge invariance since (i) the
auxiliary fields were required to satisfy the demand of supersymmetry
and (ii) the interaction Lagrangians were required to satisfy both
demands.  Note also that the scalar potential is strictly greater than
or equal to zero.  The above expressions for the scalar potential, and
the auxiliary fields will be referred to frequently in the subsequent
sections.

With the scalar potential now motivated, we return briefly to a larger
scope and to the analogy between supersymmetry and the SU(2) symmetry.
One requirement that both supersymmetry and the SU(2) symmetry demand
is that the fields which are transforming into each-other must have
the same mass.  However, in the same way that the SU(2) symmetry may
be broken spontaneously, the supersymmetry may also be broken
spontaneously so that it is effectively hidden from us.  One
consequence of supersymmetry breaking is that the masses of a particle
and its superpartner generally become different.  We conclude that if
the fields in nature are truly supersymmetric, then supersymmetry is
necessarily broken, because experimenters have never observed a
fundamental boson with the same mass and charge as the electron.  If
such a particle exists, its mass must be larger than the mass scale
probed by existing colliders.  The topic of supersymmetry breaking
will be raised again when we discuss the the MSSM and the
parametrization of supersymmetry breaking in the the MSSM's effective
Lagrangian.  The scalar potential above will then be augmented by soft
SUSY breaking contributions.

In the next section we identify the roots to the scalar
potential~\myref{susyflat-scalarpotential}, which are the minimum
field configurations called flat directions.

         \section{D-Flat Directions and How to Identify Them} \label{sec-susyflat-derivemonomials} 
The Lagrangian of supersymmetric field theories just discussed in
general possesses directions in field space for which the scalar
potential~\myref{susyflat-scalarpotential} is zero.  The fields which
compose these flat directions may thus in principle be excited to
large classical field strengths at no cost to the potential energy.
As we saw in Section~\ref{sec-inflation}, inflation provides a
mechanism by which these large VEVs for flat directions may develop.
Of course, there will be soft supersymmetry breaking additions to the
effective scalar potential as well as non-renormalizable terms which
may lift the flat directions.  Both types of terms will limit the
growth of the flat direction VEVs during inflation with the limiting
effects of the non-renormalizable terms becoming more relevant at
larger field strengths (as these terms possess higher powers of the
VEV amplitude).  Also, there may be supergravity corrections in the
effective potential which, if present could strongly limit the growth
of a VEV \cite{dine-randall-thomas-1995, dine-randall-thomas-1995b}.
The presence of SUGRA corrections are model dependent however, and
there exist well-motivated models of supergravity such as ``no-scale''
supergravity \cite{noscale-sugra} in which such corrections are
absent at tree-level and thus supressed \cite{olive-etall-1999}.

In this section, we temporarily put aside the above mentioned
subtleties of lifting a flat direction, since our first task is to
identify the flat directions.  To solve for all the roots of the
scalar potential~\myref{susyflat-scalarpotential} is a formidable
task, but one may use gauge symmetry to simplify the problem.
Specifically, to solve for these special field configurations, one may
use the theorem of \mycite{buccella-etall-1982} which relates
solutions of $D^\ga(\phi_a)=0$ to the gauge invariant polynomial of
the fields $\phi_a$ composing the flat direction.  Having identified
these solutions, one may then impose the F-flatness constraint
$F_a=0$, to further isolate the flat direction manifold. The theorem
of \cite{buccella-etall-1982} is summarized next.

To begin, define the set of $\Phi_a \ne 0$ to be a particular solution of the
D-flat constraint, $D^\ga=0$ which, using \myref{susyflat-scalarpotential}, is
written,
\begin{equation}
\Phi_a^* T^\ga \Phi_a = 0 \;\;\mbox{ for all } \ga \label{dflat-constraint}
\end{equation}
In the following we refer to $\phi_a$ as fields, although the proof
only requires that they are complex quantities which transform under
the gauge symmetries.  Recall that an infinitesimal gauge
transformation may be written
\begin{equation}
\phi \rightarrow (1 + i T^\ga \ge^\ga) \phi \\
\end{equation}
where $T^\ga$ are the gauge generators of the group, and $\ge^\ga$ are
infinitesimal parameters.  Assuming one can construct a gauge invariant
polynomial $I(\phi_a)$ of the same set of the $\phi_a$, and performing
an infinitesimal gauge transformation on the $\phi_a$, the function $I$
transforms as,
\begin{eqnarray}
I &\rightarrow& I(\phi_a + i T^\ga \ge^\ga \phi_a) \\
&\rightarrow& I(\phi_a) + \frac{\der I}{\der\phi_a} T^\ga \ge^\ga \phi_a
\end{eqnarray}
The gauge invariance of $I$ then requires,
\begin{equation}
\frac{\der I}{\der\phi_a} T^\ga \phi_a = 0 \mbox{ for all }\ga
\end{equation}
This statement for the gauge invariance of $I$ becomes the statement
of D-flatness \myref{dflat-constraint} if one chooses $\phi_a=\Phi_a$
and if one can additionally show that 
\begin{equation}
\left. \frac{\der I}{\der\phi_a}\right|_{\phi=\Phi} = C \Phi_a^*  \;\;\;\mbox{for all }a \label{buccella-constraint}
\end{equation}
where $C$ is a nonzero complex factor independent of $a$.  The authors
of \cite{buccella-etall-1982} point out that this additional
condition~\myref{buccella-constraint} is typically obtained.

As an example of the above correspondence between flat directions and
gauge invariant polynomials, consider a toy model with a $U(1)$
symmetry consisting of two complex scalar fields $\phi_a$ and $\phi_b$
with charges $+1$~and~$-1$.  The D-term in the scalar potential~\myref{susyflat-scalarpotential} is,
\begin{equation}
\frac12 g^2(|\phi_a|^2-|\phi_b|^2)^2
\label{susyflat-toymodel}
\end{equation}
and the flat direction is simply
$\{\Phi_a=fe^{\gs+\gth}\;,\;\Phi_b=fe^{\gs-\gth}\}$ where $f$ is real
and $\gth$ is a complex phase which can be eliminated via gauge
transformation.  The gauge invariant polynomial corresponding to this
flat direction is the monomial, $I_2=\phi_a\phi_b$, and the
condition~\myref{buccella-constraint} may be checked,
\begin{eqnarray}
\left. \frac{\der I_2}{\der\phi_a}\right|_{\phi=\Phi} &=& \Phi_b = \pfrac{\Phi_b\Phi_a}{f^2}\Phi_a^* = C\Phi_a^*\\
\left. \frac{\der I_2}{\der\phi_b}\right|_{\phi=\Phi} &=& \Phi_a = \pfrac{\Phi_a\Phi_b}{f^2}\Phi_b^* = C\Phi_b^*
\end{eqnarray}
where it is apparent that the common complex factor in this case is
$C=\frac{I(\Phi)}{f^2}$.  We note that the MSSM monomials studied in
the following sections will satisfy the constraint
(\ref{buccella-constraint}) in a similar way as above.

To summarize, by constructing gauge invariant polynomials $I(\phi_a)$
with the property \myref{buccella-constraint}, we implicitly find
solutions to the D-flat constraint.  Next, the F-flat constraints may
be imposed.  To be precise, a D-flat direction is lifted, when there
is a nonzero $F^a=\frac{\der W}{\der\phi^a}$ in the scalar potential~\myref{susyflat-scalarpotential} which can happen when there exists a term in
the superpotential~\myref{susyflat-superpotential} with one or less fields that
do not obtain a VEV in the flat direction.  
It is helpful to notice that the terms of the superpotential are
themselves gauge invariant monomials and there are some flat
directions which are lifted just by their corresponding monomial
appearing in the superpotential.

One comment is in order with regard to the flat direction VEVs and to
the larger goal of studying the dynamics of these flat directions
which is perhaps obvious, but we will mention none-the-less.  The
VEVs $\Phi_a$ can take on a continuum of values -- they are not
isolated solutions to the D-flatness constraint -- and the solution
space is typically a $n$-dimensional surface with $n>1$.  Hence, there
is no obstacle to constructing space-time dependent solutions
$\Phi_a(x^\mu)$.  The price to pay of course is an increase in the
energy of the system from the kinetic and gradient terms for $\Phi_a$
which would appear in the Lagrangian.  This suggests that a flat
direction VEV with an initially uniform value such as would be
generated during inflation can decay via spatial instability -- one
need only study the dynamics to determine if this realized in any
particular model.  This is done in later sections, but we now return
to the topic of identifying the flat directions.

The above procedure for determining the F and D-Flat directions is
systematically carried out for the MSSM fields in
\mycite{gherghetta-kolda-martin-1995} and in
\mycite{dine-randall-thomas-1995b}. In the next section, we outline
the procedure and we present a few examples.

         \section{Identifying D-Flat Directions of the MSSM} \label{sec-susyflat-mssmmonomials} 
The correspondence between flat directions and gauge invariant
polynomials can be exploited to catalog the possible flat directions
in the MSSM by computing all the monomials invariant under the
standard model gauge group $SU(3)\times SU(2)_L\times U(1)_Y$. This
has been done in \mycite{gherghetta-kolda-martin-1995} and
\mycite{dine-randall-thomas-1995b}.  The procedure of
\mycite{gherghetta-kolda-martin-1995} for constructing monomials of
the MSSM fields is summarized as follows: First, the color indices of
all the fields are contracted to form $SU(3)$ singlets in a basis of
monomials denoted $B_3$.  The $SU(2)_L$ indices of the fields in $B_3$
are then contracted to form all the possible $SU(3)\times SU(2)_L$
singlets in a basis $B_{32}$.  Finally this basis $B$ of all
$SU(3)\times SU(2)_L\times U(1)$ singlets are formed by combining the
polynomials in $B_{32}$ into hypercharge zero combinations.  It can
then be shown that not all of monomials constructed in this way are
independent.  There are nonlinear relations between some monomials,
and the authors of \mycite{gherghetta-kolda-martin-1995} are able to
reduce the set of monomials with these relations.

We will explain this in slightly more detail.  The notation used is
that of \mycite{gherghetta-kolda-martin-1995} in which Greek letters
$\ga,\gb,\gg$ refer to $SU(2)_L$ indices, latin letters $a,b,c,...$
refer to color indices, and latin letters $i,j,k,...$ refer to family
(generation) indices.  The uppercase latin letters $I,J,K,...$ are
also used below and their meaning should be clear from the context
they are written.

The basis $B_3$ is constructed by contracting the color indices of the
squark fields.  The $SU(3)$ invariant combinations are,
\begin{eqnarray}
(q_I\bq_J) &\equiv& q_I^a \bq_{aJ} \\
(q_Iq_Jq_K) &\equiv& q_I^a q_J^b q_K^c \epsilon_{abc} \\
(\bq_I\bq_J\bq_K) &\equiv& \bq_{aI} \bq_{bJ} \bq_{cK} \epsilon^{abc}
\end{eqnarray}
where $I,J,K,...=1...6$ index the quark, and the symmetry under
interchange of $I,J,K...$ is not specified.  For the squark $SU(2)_L$
doublets, the following two classes of monomials are possible,
\begin{eqnarray}
(Q_i Q_j Q_k)^\ga &\equiv& Q_i^{\gb a}Q_j^{\gg b}Q_k^{\ga c} \epsilon_{abc} \epsilon_{\gb\gg} \;\;\;\mbox{not all family indices equal}\\
(QQQ)_4^{(\ga\gb\gg)} &\equiv& Q_i^{\ga a} Q_j^{\gb b} Q_k^{\gg c} \epsilon_{abc}\epsilon^{ijk}
\end{eqnarray}
where the first of the above transforms as an $SU(2)_L$ doublet and
the second of the above transforms in the spin-3/2 representation.
The $SU(2)_L$ singlets for the basis $B_{32}$ are constructed from the
doublets via,
\begin{equation}
(\gvf_I \gvf_J) \equiv \gvf_I^\ga \gvf_J^\gb \epsilon_{\ga\gb}
\end{equation}
Finally, the basis $B$ is constructed by combining the monomials of
$B_{32}$ which appear with the hypercharges $Y=\{-2,-1,0,1,2\}$ into
monomials of the form
\begin{equation}
\xi_0 \;\;,\;\; 
\xi_1\xi_{-1}\;\;,\;\; 
\xi_2\xi_{-2}\;\;,\;\; 
\xi_2\xi_{-1}\xi_{-1}\;\;,\;\; 
\xi_{-2}\xi_1\xi_1
\end{equation}
The following relations between monomials are also applicable to
reducing the above bases $B_3$, $B_{32}$ and $B$ into smaller bases,
\begin{eqnarray*}
q_I^a(q_Kq_Lq_M) &=& q_K^a(q_Iq_Lq_M) + q_L^a(q_Kq_Iq_M) + q_M^a(q_Kq_Lq_I) \\
(q_Iq_Jq_K)(\bq_L\bq_M\bq_N) &=& (q_I\bq_L)(q_J\bq_M)(q_K\bq_N) \pm (\mbox{permutations}) \\
(\gvf_I\gvf_J)(\gvf_K\gvf_L) &=& (\gvf_I\gvf_K)(\gvf_J\gvf_L) + (\gvf_I\gvf_L)(\gvf_K\gvf_J) \\
(\xi^I_1\xi^J_{-1})(\xi^K_1\xi^L_{-1}) &=& (\xi^I_1\xi^L_{-1})(\xi^K_1\xi^J_{-1}) \\
(\xi_2\xi^I_1\xi^J_{-1})(\xi_2\xi^K_1\xi^L_{-1}) &=& (\xi_2\xi^I_1\xi^L_{-1})(\xi_2\xi^K_1\xi^J_{-1}) \\
&...&
\end{eqnarray*}
where the first three above have been obtained from the properties of
Levi-cevita symbol.  It is instructive to look at a few simple
examples of monomials (with the gauge index structure also shown),
\begin{eqnarray}
LL \bee &=& L^\ga L^\gb \epsilon_{\ga\gb} \bee \\
Q \bd L &=& Q^\ga_a \bd^a L^\gb \epsilon_{\ga\gb}\\
Q \bar u H_u &=& Q^\ga_a {\bar u}^a H_u^\gb \epsilon_{\ga\gb}\\
\bar u \bar d \bar d &=& \bq_{a} \bq_{b} \bq_{c} \epsilon^{abc}
\label{monomialexamples}
\end{eqnarray}
where the family indices have been suppressed.
Also note these particular monomials will appear again in the next
section when R-parity is discussed.  Tables of the numerous other
monomials can be found in
\cite{gherghetta-kolda-martin-1995,dine-randall-thomas-1995b}.  In the
next section, we look at the lifting of the D-flat directions from the
F-terms and from the soft supersymmetry breaking terms of the MSSM.

         \section{The F-terms of the Scalar Potential and Lifting of D-Flat Directions} \label{sec-susyflat-lifting} 
The D-flat directions are lifted by three main contributions; the
renormalizable F-terms, the non-renormalizable F-terms, and soft
supersymmetry breaking terms of the Lagrangian
\cite{gherghetta-kolda-martin-1995,dine-randall-thomas-1995b}.  We
will discuss the renormalizable F-terms first.  The renormalizable
F-terms are determined from the
superpotential~\myref{susyflat-superpotential} via \myref{Fsol}. The
MSSM superpotential has the form,
\begin{equation}
W_{\mssm} = \bar u y_u Q H_u - \bar d y_d Q H_d - \bar e y_e L H_d + \mu H_u H_d \label{susyflat-mssmsuperpotential}
\end{equation}
where the $y$'s above are $3\times 3$ dimensionless yukawa matrices
which operate in the family space and which are assumed to have
magnitude of order unity.  The gauge invariant contraction of indices
is done in the same way as in the previous section, and in fact the
above terms in the superpotential are nothing else but gauge invariant
monomials.  The fields $H_u$ and $H_d$ are the two Higgs $SU(2)_L$
doublets of the MSSM, and the $\mu$ parameter has dimensions of mass
and it is taken to be order of magnitude TeV
\mycite{martin-susyprimer}.  There are some terms that could have been
included into the MSSM superpotential, but are disallowed because they
violate baryon number conservation or lepton number conservation.
These terms are,
\begin{equation}
W_{\mathsubscript{omitted}}=
\frac12 \gl^{ijk} L_iL_j \bee_k + 
\gl^{\prime ijk} L_iQ_j\bd_k + 
\mu^{\prime i}L_i H_u + 
\frac12 \gl^{\prime\prime ijk} {\bar u}_i {\bar u}_j {\bar d}_k
\label{Womitted}
\end{equation}
The first three terms above will lead to lepton number violation and
the last term will lead to baryon number violation.  One disastrous
consequence of such terms is a rapid decay of the proton.
Rather than imposing $B$ and $L$ conservation separately, all of these
dangerous terms above can be eliminated with the assumption of a
discrete symmetry called R-parity,
\begin{equation}
P_R = (-1)^{3(B-L)+2s} \label{Rparity}
\end{equation}
where $s=\pm\frac12$ is the spin quantum number of the field involved.
From the above definition, it can then be shown that all of the
standard model particles (including the Higgs Bosons) have R-parity
$P_R=+1$, while all of the superpartners have R-parity $P_R=-1$.  That
R-parity is conserved by the MSSM means that all the terms in the MSSM
Lagrangian should have an R-parity of $P_R=+1$.  One notable consequence of
assuming R-parity is that
the lightest supersymmetric particle is required to be absolutely
stable, and thus a promising dark matter candidate.\footnote{For a
  more comprehensive discussion of R-parity, see
  \cite{martin-susyprimer}}

To see how the D-flat directions may become lifted by the F-terms,
consider the four flat directions specified previously in
\myref{monomialexamples}.  The $LLe$, $QdL$, and $udd$ monomials of
\myref{monomialexamples} would clearly be lifted by the corresponding
terms in (\ref{Womitted}).  However R-parity prevents these terms, and
thus allows these three directions to be both D-flat and F-flat.
R-parity does not protect the $QuH_u$ direction however, so this
direction is lifted.

Note that for those flat directions that are not lifted by the
F-constraints such as $LLe$, $QdL$, and $udd$, their flatness will be
preserved at higher loop corrections in the renormalization group by
the SUSY non-renormalization theorems.  One must then look at
non-renormalizable terms in the superpotential to determine when a
particular D-flat direction is lifted.  The non-renormalizable
superpotential is composed of gauge invariant monomials of order $n=4$
and higher, and the lifting is determined in the same manner as is
outlined in Section~\ref{sec-susyflat-derivemonomials}.  The
dimensionful coefficients to these terms are assumed to be in inverse
powers of the GUT scale or some other ultraviolet cutoff $M$.  The
potential can thus be written as a series expansion
\cite{dine-randall-thomas-1995b},
\begin{equation}
V(\Phi) = m^2 |\Phi|^2 + V_{n>3}(|\Phi|) \;\;\;,\;\;\;V_{n>3}(|\Phi|) \equiv \sum_{n=4}^\infty |\lambda_n|^2 M^4 \left|\frac{\Phi}{M}\right|^{2(n-1)}  
\label{susyflat-potentialseries}
\end{equation}
where $m$ is a soft supersymmetry breaking mass to be discussed in the
following section, $\Phi$ is the VEV of the flat direction, and $n$
indexes the corresponding nonzero term in the superpotential.  The
lifting of all the MSSM flat directions by terms in the superpotential
is discussed in detail in \cite{gherghetta-kolda-martin-1995,
  dine-randall-thomas-1995b}, and
in~\cite{gherghetta-kolda-martin-1995}, the corresponding complex
dimensionality of the flat direction subspace is also listed.  The
flat direction VEV obtained during inflation is partly controlled by
the leading nonzero term in the superpotential which is specified here
and in \cite{gherghetta-kolda-martin-1995} by the index $n$.  The main
result of this section which will be applied later is,
\begin{equation}
  \begin{array}{l}\mbox{The majority of
    the MSSM flat direction degrees of freedom}\\\mbox{remain flat at least until } n= 3,4,5,6 \end{array}
\label{susyflat-liftingindex}
\end{equation}
There are only two complex degrees of freedom remaining which could
remain flat until $n=7,9$.  We note that while $n$ is a dummy index
running over integers $n>3$ above, for the remaining chapters we will
typically use $n$ to mean the index corresponding to the lowest
non-vanishing term in the series~\myref{susyflat-potentialseries}.  We
emphasize that the existence of these terms which can lift the flat
direction are model dependent.  Based on known symmetries these terms
can be present, but this does not mean that they actually are present.

         \section{SUSY Breaking and the Spectrum of Soft Sparticle Masses} \label{sec-susyflat-susybreaking} 
The effects of SUSY breaking on a model of flat directions appear in
the parameters of the effective scalar potential and if one further
assumes supergravity, the effects may also appear in the Kahler
potential, or equivalently the kinetic terms
\cite{dine-randall-thomas-1995b,dine-randall-thomas-1995}.  In our
models, we only consider effects on the scalar potential, and in
particular the mass terms.  However, we briefly discuss the
supergravity case as well as the implications of the spontaneous
supersymmetry breaking dynamics occurring during inflation or
reheating.

In general, global supersymmetry is broken when the scalar potential
obtains a VEV, and this can only happen when the equations $F=0$ and
$D=0$ cannot be simultaneously satisfied for any value of the fields.
The F and D terms of the scalar potential then become supplemented by
supersymmetry breaking corrections with a characteristic energy scale
$\msusy\sim \tev$ which is the energy scale suggested by particle
phenomenology and cosmology.  Models in which the MSSM fields are
directly responsible for SUSY breaking are difficult to reconcile with
phenomenology.  Instead the belief is that supersymmetry is broken by
``hidden'' sector fields \cite{martin-susyprimer} which communicate
the breaking to the MSSM ``visible'' sector indirectly through a
flavor blind interaction such as gravity (Planck-scale Mediated
Supersymmetry Breaking) or through the standard model gauge
interactions (Gauge Mediated Supersymmetry Breaking).  The most
obvious effect on the ``visible'' sector is the lifting of the
sparticle mass spectrum to the TeV scale.  The focus here will in fact
be on the mass spectrum of the squarks, sleptons and scalar Higgs
fields since these particles comprise the possible MSSM flat
directions.

A generic superpotential~\myref{susyflat-superpotential} can have mass
parameters, and the MSSM~\myref{susyflat-mssmsuperpotential} contains
one such term which is the $\mu H_u H_d $ term ($\mu$ is also assumed
to be of the $\tev$ scale).  A new breaking potential $\cL_{\mathsubscript{soft}}$ is added to the model in which
only one particle from any supersymmetric boson-fermion pair need
appear.\footnote{The standard model partners do not appear in
  $\cL^{\mathsubscript{scalar}}_{\mathsubscript{soft}}$ since these masses come only from the Higgs
  VEV after Electroweak symmetry breaking.}  In this potential, there
will be mass terms for the squarks, sleptons, Higgs fields and
gauginos, as well as cubic interaction terms \cite{martin-susyprimer},
but we restrict our attention to the scalar mass terms which appear
with the following parametrization,
\begin{eqnarray*}
\cL^{\mathsubscript{scalar}}_{\mathsubscript{soft}}
&=& \tilde{\bar u} \mathbf{m^2_{\bar u}}\tilde{\bar u}^\dagger + 
\tilde{\bar d} \mathbf{m^2_{\bar d}}\tilde{\bar d}^\dagger + 
\tilde{\bar e} \mathbf{m^2_{\bar e}}\tilde{\bar e}^\dagger + 
\tilde{L} \mathbf{m^2_{L}}\tilde{L}^\dagger + 
\tilde{Q} \mathbf{m^2_{Q}}\tilde{Q}^\dagger  \\
&\mbox{}& + m^2_{H_u}H_u^* H_u + m^2_{H_d}H_d^*H_d + (b H_u H_d + c.c)
\end{eqnarray*}
where $\mathbf{m^2_i}$ are $3\times 3$ Hermitian matrices in the
family space.
When one speaks of the spectrum of masses, it is the above parameters
which specify this.  There are some mass degeneracies for the particle
states which are immediate consequences of gauge symmetry.  Namely
that the components of any SU(2) doublet are degenerate in mass and
similarly for any SU(3) triplet.
This fact is implied by the above notation.  Additionally, it is
expected that the overall mass scale for the above terms should be the
$\tev$ scale.  Also there is phenomenological evidence related to
$CP$ violation, Lepton number conservation, and flavor changing
neutral currents which further restricts the values of the above mass
terms \cite{martin-susyprimer}.  Without going into detail, we note a
simple paradigm for satisfying these phenomenological constraints
which is to assume each of the above matrices $\mathbf{m^2_{i}}$ are
proportional to the identity matrix.  This leaves 6 parameters to
specify the matrices completely.  This paradigm is known in the
literature as ``soft supersymmetry-breaking universality''.  However,
we note that in stronger versions of universality all the mass
matrices are degenerate to a single mass scale $m_0$.  The
universality paradigm is connected to the expectation discussed above
that supersymmetry is broken by a hidden sector and communicated to
the MSSM sector by a flavour independent interaction.  The flavour
independence of the interaction would naturally lead to degeneracies
imposed by universality \cite{martin-susyprimer}.

This is not the whole story however because the mass parameters are
energy scale dependent, and they ``run'' with renormalization group
equations.  The degeneracy is only exact at some universal scale which
is usually taken to be the GUT scale.  The renormalization group
equations for the MSSM are then used to solve for the mass parameters
at the typical interaction energy scale $Q$ of ones experiment.  In
our case, the energy scale of the flat direction dynamics is $Q\sim
\tev$ as the flat directions mass and decay products would be of this
scale.
The renormalization group equations result in corrections to the
masses which are on the order $\ln(Q/Q_0)$ where the input scale
$Q_0\sim M_{GUT}$ is assumed.  In this sense the corrections are
``soft'' compared to a theory without supersymmetry which typically
has quadratic corrections with the energy scale $Q$, and thus enormous
mass corrections \cite{martin-susyprimer}.  The notation ``soft'' for
our susy breaking potential $\cL_{\mathsubscript{soft}}$ thus refers
to this property.  If the renormalization group equations are solved
assuming some form of universality, typically one finds the first two
flavors of any family in the squarks and sleptons remain nearly
degenerate, but the third flavor of a family receives corrections
within a factor of ten or so of this degeneracy
\cite{martin-susyprimer}.

To summarize, the assumptions we make on the spectrum of scalar masses
in our later models is modest, that they are all near the TeV scale
but not all degenerate.  It is noted however that universality is not
necessary in the MSSM, and the spectrum may in fact have less
degeneracy but probably not more.

There can be other new effective interactions to the scalar potential
which appear due to supersymmetry breaking and which can be mediated
by new heavy quanta of mass at or slightly below the Grand Unification
Scale $M_{GUT}\sim 10^{16}\,\gev$.  The class of terms we are
interested in are those which violate the difference of Baryon and
Lepton number $B-L$. Such interactions are not part of the MSSM, but
they are present in some Grand Unified models such as SO(10) models
(for example, see \cite{morgan-1991}).  We do not attempt a discussion
of Grand Unified models which are far beyond the scope of this thesis.
However we will motivate the nature and origin of these effective
interactions by again considering our toy model.  Our toy model will
in fact not contain all the desired elements we wish (it will not
allow for $CP$ violation), but it will be possible to proceed.
Specifically, we will determine the effective interactions applied in
our later numerical simulations of Section~\ref{sec-results}.

The toy model introduced in Section~\ref{sec-susyflat-derivemonomials}
is composed of two complex scalar fields $\phi_1$ and $\phi_2$ charged
oppositely under a U(1) symmetry and with masses of the scale
$\msusy$.  In addition to the local U(1) gauge symmetry, the
Lagrangian and in particular the D-term, is invariant under two
separate global gauge transformations $\phi_1\rightarrow
e^{i\gL_1}\phi_1$ and $\phi_2\rightarrow e^{i\gL_2}\phi_2$.  The model
will thus conserve separately the numbers of $\phi_1$ and $\phi_2$
quanta.
One can loosely think of these as individual lepton flavor numbers for
instance.  Now consider adding to this theory a new neutral heavy
vector $X$ which violates the number conservation of $\phi_1$ and
$\phi_2$ with a decay such as $X\rightarrow \phi_1\phi_2$ or
$X\rightarrow \phi_1^*\phi_2^*$.  The heavy particle would also
mediate a process such as $\phi_1\phi_2\rightarrow\phi_1^*\phi_2^*$
which violates the $\phi_1$ number and $\phi_2$ number but conserves
the gauge charge.  Since these $X$ particles are very heavy, they
would be created as virtual quanta to mediate scattering such as
above, but never appear on-shell as an initial or final state
quanta.\footnote{This effective four scalar vertex may be compared to
  the 4-fermion vertex one obtains after integrating out the $W$ boson
  from the electro-weak Lagrangian.}  The quanta would be unobservable
and the above interaction would appear in all respects as due to a
quartic interaction of the form $\gl\phi_1^2\phi_2^2 + c.c$.  In
particular the new quartic interaction would be interpretted as a
correction to the classical Lagrangian and thus a correction to the
classical equation of motion and dynamics.

Our next task is to estimate the magnitude of the coupling $\gl$ based
on the tree-level interaction. The scattering
$\phi_1\phi_2\rightarrow\phi_1^*\phi_2^*$ is modified by the presence
of the condensate $\langle \phi_1\rangle = \langle \phi_2\rangle
\equiv \Phi$ and in particular the propagating heavy particle will
acquire an effective mass $m^2_{X_\mathsubscript{eff}}\sim
\mbox{max}\left[|\Phi|^2,m_X^2\right]$ where $m_X\siml M_{GUT}$ is the
mass in the absence of the flat direction VEV.  It is assumed that the
VEV is large, $|\Phi| > m_X$ so $m^2_{X_\mathsubscript{eff}}\sim
|\Phi|^2$.  Thus the effective coupling acquires a factor $1/
m^2_{X_\mathsubscript{eff}}$ from the X propagator and a factor
$\msusy^2$ which is the typical momentum transfer at the energy scale
we are interested in.  To summarize, our effective couplings have the
following form and magnitude,
\begin{equation}
  \gl \phi_1^2\phi_2^2 + \gl^* {\phi_1^*}^2{\phi_2^*}^2 
  \hspace{3em},\hspace{3em}
  \gl \sim \frac{\msusy^2}{|\Phi|^2}
\label{susyflat-lambdacouplings}
\end{equation}
Again, this interaction is U(1) gauge invariant but violates the
$\phi_1$ and $\phi_2$ number conservation.  We emphasize that in
realistic models, the leading order $CP$ violating effects appear at
one loop order or higher \cite{affleck-dine-1985} and the magnitude of
the coupling can be model dependent.  However, as a first
estimate~\myref{susyflat-lambdacouplings} is sufficient.

The above interactions can be used to model the classical generation
of a baryon asymmetry or what is commonly known as Affleck-Dine
Baryogenesis \cite{affleck-dine-1985}.  Recall the three criterion for
Baryogenesis are known to be (i) baryon number violating interactions,
(ii) $C$ and $CP$ violation and (iii) out of thermal-equilibrium
evolution \cite{sakharov-1967}.  The first criterion is present in the
above model, and the third criterion is also present since the fields
will be in coherent states and thus far from thermal equilibrium.  The
$C$ and $CP$ violation is not as obvious.  One must consider the phase
of the flat direction VEV $\Phi\equiv |\Phi|e^{\gS}$ as well as the
phase of the complex coupling $\gl=|\gl|e^{\Theta}$.  Either one of
these phases may be removed at some fixed time by a field redefinition
$\Phi=e^{i\gL_0}\Phi_*$, but the relative phase may not be removed,
and this is the source of $C$ and $CP$ violation
\cite{affleck-dine-1985}.  Our tree-level analysis is insufficient to
establish this relative phase, and in fact, our toy model which only
possesses two fields and a U(1) symmetry is too restrictive to allow
for $C$ and $CP$ violation at all.  Hence we simply insert this phase
by hand.  A careful discussion here of of the origin of the $C$ and
$CP$ violating phase would require a long digression, but it is
sufficient to note that these effects are present in the standard
model, and can also be present in the MSSM and Grand Unified models.
Our goal was to motivate the origin and magnitude of the above quartic
coupling~\myref{susyflat-lambdacouplings}, and this has been done.

To conclude, we return to the topic of the source of supersymmetry
breaking, and its dynamics.  The potential
$\cL_{\mathsubscript{soft}}$ may be time dependent in the context of
cosmology, and one should be sensitive to this fact
\cite{lyth-riotto-review-1998}.  For instance if supersymmetry
breaking in the hidden sector is somehow tied to inflation, it is not
guaranteed that $\msusy\sim\tev$ is the correct mass scale for the
scalar particles during inflation or the early stages of reheating.
Similarly, there may be competing mechanisms which break
supersymmetry.  An example of a competing susy breaking mechanism is
provided in \cite{dine-randall-thomas-1995b, dine-randall-thomas-1995}
in the context of supergravity.  These authors point out that the
inflaton itself breaks supersymmetry by virtue of it having a VEV, and
this breaking can be communicated to other fields through the Kahler
potential.  In this situation, there may be corrections to the masses
of the fields of order the Hubble parameter $H$ with the sign of these
corrections being a free parameter.  Such corrections, if present
would presumably be the dominant contribution to the flat directions
potential, and thus a dominant contribution to the evolution of the
flat direction during inflation and reheating
\cite{dine-randall-thomas-1995b, dine-randall-thomas-1995}.  However
there exist well-motivated models of supergravity such as ``no-scale''
supergravity \cite{noscale-sugra} in which such mass corrections are
absent at tree-level and thus supressed \cite{olive-etall-1999}.  In
the following we simply assume that the flat directions are lifted
only by $\cL_{\mathsubscript{soft}}$ and by possible
non-renormalizable terms of the superpotential mentioned in the
previous subsection.  Similarly, we assume
$\cL_{\mathsubscript{soft}}$ is time independent during the later
stages of inflation and during reheating.

\newpage \chapter{Quantization of Scalar Fields in External Backgrounds} \label{sec-heiseqs} 
Here we introduce the formalism to describe quantum fields in external
backgrounds.  Some well known examples in which this formalism has
been successfully applied are Hawking radiation \cite{hawking-1975},
the cosmological perturbations, \cite{cosmological-perturbations}
Schwinger pair production \cite{schwinger-pairproduction}, and
resonant preheating of the inflaton field after inflation
\cite{preheating}.  In all these examples, there is a bosonic field
which has acquired a vacuum expectation value.  The VEV evolves
classically and acts as a source for the other fields to which it is
coupled.  The coupling then may lead
to a time evolution of the parameters of the Lagrangian for the other
fields which is non-adiabatic and which may result in the production
of quanta.

This physical picture and the accompanying formalism are the main
tools we apply to study the evolution of the flat direction field(s)
during and after inflation.  The cases we study in fact involve
multiple scalar fields which mix and the mixing can provide more
possibilities for non-adiabatic evolution
\cite{nilles-peloso-sorbo-2001, nilles-peloso-sorbo-2001b}.  In this
section, we determine the differential equations which describe the
quantum evolution of these multiple fields
\cite{nilles-peloso-sorbo-2001}, and we explain how to extract the
particle production from the solutions.  We discuss how to manipulate
a model's Lagrangian to extract the driving terms which appear in
these differential equations.  It is then possible to arrange the
driving terms into a simple quantitative measure of non-adiabaticity
which generalizes the measure of non-adiabaticity commonly used in the
single field case.  To conclude, we develop a series solution to the
equations, and determine the scaling properties of these solutions
under changes in the energy scales of the model.  The material here is
thus focused on the differential equations and the techniques for
determining, analyzing and interpretting the equations and solutions.
The techniques presented are applied in later sections.

         \section{Diagonalizing a System of Multiple Scalar fields, Bogolyubov Transformations, and the Heisenberg Equations of Motion} \label{sec-heiseqs-formalism} 
We are interested in studying actions consisting of multiple scalar
fields such as the following,
\begin{equation}
S =  \int d^4x \frac12 \left(\phi_i,_\mu \phi_i,_\mu - \phi_iM^2_{ij}\phi_j \right)
\end{equation}
where the mass matrix $M^2$ is time dependent and non-diagonal.  Note
that the actions we consider in later sections acquire the above time
dependent form only after splitting the action into a classical time
dependent background and quantum fluctuations away from this
background.  The fields $\phi_i$ will then represent the quantum
fluctuations.  For convenience, the above action is transformed to
momentum space and also written in matrix notation with the fields
arranged into $N$-vectors $X=\left\{\phi_1,\phi_2,,,\phi_N \right\}$,
\begin{equation}
S = \int d\eta \;d^3k \frac12 \left({X'}^\dagger X' - X^\dagger \gO^2 X \right) \;\;\;,\;\;\; \gO_{ij}^2= M_{ij}^2 + k^2 \gd_{ij}
\label{heiseqs-ac}
\end{equation}
Additionally, the time coordinate is $\eta$ which will later be
identified as the conformal time, and time derivatives are denoted
with a prime.
The system \myref{heiseqs-ac} is studied using the canonical
quantization and the Bogolyubov transformations to relate initial time
Schroedinger operators to the Heisenberg operators.  With the
solutions to the Heisenberg equations of motion, expectation values of
observables may then be determined.  Specifically, we calculate the
production of quanta, and we determine properties of the derived
equations which permit a large production of quanta.  The reader is
referred to appendix~\ref{calc-bogtransform-general} or to
reference~\cite{nilles-peloso-sorbo-2001} for further details of this
formalism.

To motivate the discussion of the multi-field case~\myref{heiseqs-ac},
consider first a simpler one-dimensional system, a harmonic oscillator
with a time dependent frequency.  The action is,
\begin{equation}
S_{\mathsubscript{harmonic oscillator}} = \int d\eta \frac12\left({x'}^2 -  \go(\eta)^2 x^2 \right)
\end{equation}
In the Heisenberg picture, operators evolve with the same equations as
their corresponding classical equations of motion.  Thus the operators
$x$ and $p$, satisfy,
\begin{eqnarray}
x' = p \nonumber \\
p' = -\go^2 x \label{heiseqs-shoeqs}
\end{eqnarray}
The position and momentum operators $x$ and $p$ may be converted into
the complex basis $\hat a$ and ${\hat a}^\dagger$ ,
\begin{eqnarray}
x &=& \frac{1}{\sqrt{2\go}}\left(\hat a + {\hat a}^\dagger\right) \nonumber \\
p &=& i\sqrt{\frac{\go}{2}}\left(\hat a - {\hat a}^\dagger\right) \label{heiseqs-xp}
\end{eqnarray}
which are the familiar raising and lowering operators that satisfy the
commutation relations $[\hat a,{\hat a}^\dagger]=1$.  In this basis,
the Hamiltonian $H=\frac12\left(p^2+ \go^2x^2\right)$ is written,
\begin{equation}
H = \go\left({\hat a}^\dagger \hat a + \frac12\right) \equiv \go\left( \hat n + \frac12\right)
\label{heiseqs-hamiltonian1d}
\end{equation}
where $\hat n={\hat a}^\dagger \hat a$ is a number operator ${\hat
  n}|n\rangle=n|n\rangle$ which returns the principle quantum number
$n$ with $n=0$ being the ground state.  Now, if the frequency $\go$ is
constant, then the state of the system will exhibit a trivial time
dependence -- if it begins in any particular energy eigenstate, it
remains in this eigenstate.  Transitions may occur when time
dependence is introduced to the Hamiltonian, either by addition of new
terms or by giving the frequency $\go$ time dependence.  The latter
case of a time dependent frequency is the one considered.
The time evolution of the operators must be determined from the
Heisenberg equations of motion~\myref{heiseqs-shoeqs}.  To solve for the
evolution, first parameterize the operators with a Bogolyubov
transformation which will relate the raising and lowering operators
$\hat a(\eta)$ and ${\hat a}^\dagger(\eta)$ at some time $\eta$ to the initial
Schroedinger operators at time $\eta_0$.\footnote{Since our problems
  involve a time dependent Hamiltonian, the Bogolyubov transformations
  are time dependent.  A space dependent Hamiltonian would have called
  for a space dependent transformation.  In this sense, the Bogolyubov
  transformation is a general tool (see
  appendix~\ref{calc-bogtransform-general}).  } The Schroedinger
operators are defined, $a \equiv {\hat a}(\eta_0)$ and $a^\dagger \equiv
{\hat a}^\dagger(\eta_0)$, and the Bogolyubov transformation is then,
\begin{eqnarray}
\hat a(\eta) &=& \alpha(\eta) a + \beta(\eta) a^\dagger \nonumber \\
\hat{a}^\dagger(\eta) &=& \alpha^*(\eta) a^\dagger + \beta^*(\eta) a \;.
\label{heiseqs-bogsimple}
\end{eqnarray}
The reason for this transformation is that the Heisenberg operators
$\hat a$ and ${\hat a}^\dagger$ lead to the canonical
form~\myref{heiseqs-hamiltonian1d} in which $\hat a$ and ${\hat
  a}^\dagger$ are the annihilation/creation operators of the physical
eigenstates at any given time.  The Bogolyubov transformation thus
allows us to compute expectation values of observables at a given time
once a solution for $\ga$ and $\gb$ has been determined.

Recalling that the canonical commutation relations must be ensured at
all times, one substitutes \myref{heiseqs-bogsimple} into $[\hat a,{\hat
  a}^\dagger]=1$ and then requires $[a, a^\dagger]=1$ to determine the
following constraint on the parameters $\ga$ and $\gb$,
\begin{equation}
|\ga|^2-|\gb|^2 = 1\;. \label{heiseqs-2}
\end{equation}
However, performing the same substitution on $[\hat a,\hat a]=0$ does
not lead to a constraint because the relation is satisfied trivially.
The evolution of the $\ga$ and $\gb$ are subsequently determined by
substituting \myref{heiseqs-xp} into the Heisenberg equations of
motion \myref{heiseqs-shoeqs} which yields,
\begin{eqnarray}
\ga' &=& -i\go \ga + \frac{\go'}{2\go}\gb \nonumber \\
\gb' &=& i\go \gb  +\frac{\go'}{2\go}\ga \label{heiseqs-bogeqs}
\end{eqnarray}
It can be verified that these differential equations will preserve the
constraint~\myref{heiseqs-2}.
It is thus only necessary to enforce the constraint on the initial
values $\ga(\eta_0)$ and $\gb(\eta_0)$.  The ground state of the
system is defined by $\hat a (\eta_0)|0\rangle = 0$, where $\hat
a(\eta_0) = a$ which means $\gb(\eta_0)=0$ and $\ga(\eta_0)=1$.  The
equations~\myref{heiseqs-bogeqs} can then be solved analytically or
numerically, and with a solution for $\ga(\eta)$ and $\gb(\eta)$, the
time evolution of the operator is known.  For example, putting the
initial state as the ground state $|0\rangle$, the expection value of
the number operator, ${\hat n}=\hat{a}^\dagger \hat{a}$ at some time is
determined,
\begin{eqnarray}
\langle \hat n(\eta) \rangle &=& \langle 0|(\alpha^* a^\dagger + \beta^* a) (\alpha a + \beta a^\dagger) |0\rangle \nonumber \\
&=& \beta^*\beta \label{heiseqs-shonum}
\end{eqnarray}
Note finally from \myref{heiseqs-bogeqs}~and~\myref{heiseqs-shonum},
that if $\go'$ is zero, then $\langle n \rangle=const.$ and the state
of the system will not change, which is the expected result.  However,
if $\frac{\go'}{2\go} \gtrsim \go$, then the time dependence becomes
nontrivial, and the system can make transitions.  In other words,
energy can be pumped into and out of the system by the effect of the
rapidly changing frequency $\go$.  The quantity $\frac{\go'}{2\go}$
will be referred to as an adiabatic parameter since it quantifies the
likelihood of transitions between the eigenstates of $\hat n$.

We proceed with the multi-field system \myref{heiseqs-ac} with the goal
of reproducing the results of \cite{nilles-peloso-sorbo-2001}, and we
refer to the harmonic oscillator example where appropriate.  The major
difference between this system and the harmonic oscillator is the
presence of multiple frequencies and thus multiple energy eigenstates
for the system to populate.  Additionally, the mass matrix is time
dependent and not necesarily diagonal.  One may diagonalize $\gO^2$
with a time dependent rotation matrix $C(\eta)$,
\begin{equation}
C^T(\eta)\gO^2(\eta)C(\eta) = \go^2(\eta)
\label{heiseqs-cmatrix}
\end{equation}
where $\go$ is the diagonal eigenfrequency matrix. Note the
transformation $C$ is time dependent, so we cannot generally perform
this change of basis on the Lagrangian without introducing new kinetic
structures, but if the mass matrix is evolving slowly enough, the
column vectors of $C$ will define the physical eigenstates of the
system.  Now define conjugate momenta in the nondiagonal basis $\Pi_i
\equiv \frac{\der\cL}{\der{X'}_i} = X_i'$, and define the fields and
conjugage momenta in the diagonal basis $\hat X$ and $\hat \Pi$ as
follows \cite{nilles-peloso-sorbo-2001},
\begin{equation}
X = C \hat X \nonumber \;\;\;,\;\;\; \Pi = C \hat \Pi 
\label{heiseqs-conjugatemomenta}
\end{equation}
with
\begin{eqnarray}
{\hat X}_i &=& \int \frac{d^3k}{(2\pi)^{3/2}} \left[ e^{ikx}h_{ij}(\eta) a_j + e^{-ikx}h_{ij}^*(\eta) a_j^\dagger \right] \nonumber \\
\hat{\Pi}_i &=&    \int \frac{d^3k}{(2\pi)^{3/2}} \left[ e^{ikx} {\tilde h}_{ij}(\eta) a_j + e^{-ikx}{\tilde h}_{ij}^*(\eta)a_j^\dagger \right]
\label{heiseqs-modefunctions}
\end{eqnarray}
which, in the case of positive eigenvalues $\go_i>0$, are equivalently
written,
\begin{eqnarray}
{\hat X}_i &=& \int \frac{d^3k}{(2\pi)^{3/2}} \frac{1}{\sqrt{2\go_i}}\left[ e^{ikx}{\hat a}_i(k,\eta) + e^{-ikx} {\hat a}_i^\dagger(k,\eta) \right] \nonumber \\
\hat{\Pi}_i &=& \int \frac{d^3k}{(2\pi)^{3/2}} \; i\sqrt{\frac{\go_i}{2}} \left[ e^{ikx}{\hat a}_i(k,\eta) - e^{-ikx} {\hat a}_i^\dagger(k,\eta) \right]
\label{heiseqs-diagonalbasis}
\end{eqnarray}
where the Bogolyubov transformation is now,
\begin{eqnarray}
{\hat a}_{i}(k,\eta) &=& \ga_{ij}(k,\eta) a_{j}(k) + \gb_{ij}^*(k,\eta) a_{j}^\dagger(k) \nonumber \\
{\hat a}_{i}^\dagger(k,\eta) &=&  \ga^*_{ji}(k,\eta) a_{j}^\dagger(k)  +  \gb_{ji}(k,\eta) a_{j}(k) 
\label{heiseqs-bogt} 
\end{eqnarray}
and for consistency between
representations~\myref{heiseqs-modefunctions}~and~\myref{heiseqs-diagonalbasis},
the following relations will hold 
\begin{eqnarray*}
h &=& \frac{1}{\sqrt{2\go}}(\ga+\gb) \\
{\tilde h} &=& \frac{i\go}{\sqrt{2\go}}(-\ga +\gb)
\end{eqnarray*}
Note $\hat X$ and $\hat\Pi$ satisfy canonical commutation relations
$[{\hat X}_i(x),{\hat\Pi}_j(y)]=\gd(x-y)\gd_{ij}$, and one may check
using the orthogonality of $C$, that the nondiagonal fields satisfy
the same commutation relations, $[X_i(x),\Pi_j(y)]=[C_{ik}{\hat
  X}_k,C_{jm}{\hat\Pi}_m] =
\gd(x-y)\gd_{km}C_{ik}C_{jm}=\gd(x-y)\gd_{ij}$.  The constraints on
$\ga$ and $\gb$ coming from the commutation relations
$[a_i,a^\dagger_j]=\gd_{ij}$ and $[a_i,a_j]=0$ are respectively,
\begin{eqnarray} 
\ga \ga^\dagger - \gb^*\gb^T &=& 1 \label{heiseqs-mc1}\\
\ga \gb^\dagger - \gb^*\ga^T &=& 0 \label{heiseqs-mc2}
\end{eqnarray}
These constraints reduce to the constraint~\myref{heiseqs-2} when $N=1$.
The Hamiltonian has the same form in either basis,
\begin{eqnarray}
H &=& \int d^3k \frac12 \left(\Pi^\dagger \Pi + X^\dagger \gO^2X\right) \\
&=& \int d^3k \frac12 \left({\hat \Pi}^\dagger \hat \Pi +\hat{X}^\dagger \go^2\hat{X}\right)
\end{eqnarray}
though in the diagonal basis, after normal ordering, has the well
known form,
\begin{equation}
H = \sum_i\int d^3k\; \go_i \hat{n}_i
\;\;\;\mbox{with}\;\;\;
\hat{n}_i \equiv \hat{a}^\dagger_i(k) \hat{a}_i(k)
\end{equation}
where the occupation number operator ${\hat n_i(k)}$ has been defined.
The diagonal basis ${\hat X}_i$ thus corresponds to the physical
eigenstates of the system at any time during the evolution,
ie.~eigenstates of ${\hat n}$. The Heisenberg equations of motion for
the fields are,
\begin{eqnarray}
X' &=& \Pi = \frac{\der\mathcal{L}}{\der X'} \nonumber \\
\Pi' &=& -\gO^2 X
\end{eqnarray}
and when expressed in terms of $\alpha$ and $\beta$ take the form,
\begin{eqnarray}
 \alpha' &=& \left(-i \omega -I\right)\alpha + \left(\frac{\omega'}{2\omega}-J\right)\beta \nonumber \\
 \beta' &=& \left(i \omega-I\right) \beta + \left(\frac{\omega'}{2\omega} -J\right)\alpha 
\label{heiseqs-bogequations} \\
I,J &=& \frac12 \left(\sqrt{\omega} \Gamma \frac{1}{\sqrt{\omega}} \pm \frac{1}{\sqrt{\omega}}\Gamma \sqrt{\omega} \right) \;\;\;,\;\;\; \Gamma = C^T C' 
\label{heiseqs-defineIJGamma}
\end{eqnarray}
where the antisymmetric $\Gamma$ and $I$ matrices and the symmetric
$J$ matrix have also been defined
\cite{nilles-peloso-sorbo-2001}. Written using index notation, these
equations are equivalently written,
\begin{eqnarray*} \begin{array}{l}
\ga_{ij}' = -i\go_i\ga_{ij} - I_{ik}\ga_{kj} + \frac{\go_i'}{2\go_i}\gb_{ij} -J_{ik}\gb_{kj}\\[0.3em]
\gb_{ij}' = i\go_i\gb_{ij} - I_{ik}\gb_{kj} + \frac{\go_i'}{2\go_i}\ga_{ij} -J_{ik}\ga_{kj} \\[0.3em]
I_{ij},J_{ij} = \frac12 \Gamma_{ij}\left(\sqrt{\frac{\go_i}{\go_j}} \pm \sqrt{\frac{\go_j}{\go_i}}\right)
\end{array} 
\hspace{2em}\mbox{no summation on }i,j
\end{eqnarray*}
The Gamma matrix also may be written in terms of the eigenvectors
$C_i$ as follows,
\begin{equation}
\Gamma_{ij} = C_i \cdot C_j'
\end{equation} 
and thus the elements $\gG_{ij}$ contain the information for the rate of
change of the eigenvectors,
\begin{equation}
C_i' = \sum_j -\Gamma_{ij} C_j
\end{equation}
As in the SHO example~\myref{heiseqs-shoeqs}, the
equations~\myref{heiseqs-bogequations} preserve the constraints
\myref{heiseqs-mc1} and~\myref{heiseqs-mc2} at all times if the
constraints are satisfied at some arbitrary time.
Using~\myref{heiseqs-bogt}, the occupation number of the i'th state is
determined,
\begin{eqnarray} 
\langle \hat{n}_i(\eta) \rangle &=&
\langle \gO_0| \hat{a}^\dagger_i(k,\eta) \hat{a}_i(k,\eta) |\gO_0\rangle \\
&=& \left(\beta^* \beta^T\right)_{ii}   \;\;\;,\;\;\;\mbox{no summation on }i
\label{heiseqs-occnumber}
\end{eqnarray}
where $|\gO_0\rangle$ is the initial vacuum, and where the
Schroedinger operators $a_i(k)$ annihilate this initial vacuum.  Note
that the occupation number is not a good quantum number when $\go^2$
is negative.  This is most easily seen by trying to compute the
expectation value of the Hamiltonian which becomes imaginary.  In this
case, the modes are tachyonic, and a particle interpretation of the
states is not appropriate.  However the field operators are still well
defined when written in terms of the mode functions $h$ and ${\tilde
  h}$ \cite{mukhanov-winitzki-2007} introduced
in~\myref{heiseqs-modefunctions}.
In the absence of a particle interpretation, observables such as a
correlation function $\langle\phi(x)\phi(y)\rangle$ may still be
evaluated in the mode function
decomposition~\myref{heiseqs-modefunctions}.  A calculation of this
sort is performed in Section~\ref{sec-inflation} for just such a
circumstance.  We will use the representation $\ga$ and $\gb$ whenever
the particle interpretation is available.  

Finally, we mention that the Heisenberg
equations~\myref{heiseqs-bogequations} may be reduced to a smaller
apparently equivalent set of equations by applying the algebraic
relations~(\ref{heiseqs-mc1}-\ref{heiseqs-mc2}). This reduction is
shown in~\myref{sec-heiseqs-reduce}.

         \section{Expressing Lagrangians in Canonical Form} \label{sec-heiseqs-canonicalL}  
Typically the Lagrangians one wishes to study are not initially in the
form~\myref{heiseqs-ac}, but must be put into this form.  Our models
are no exception, so we here present some useful techniques for
manipulating a Lagrangian to extract the matrix $\Gamma$ and the
eigenvalues $\go_i$ which specify the Heisenberg equations of
motion~\myref{heiseqs-bogequations}.  Working in momentum space, the
form of Lagrangian we are typically faced with contains additional
mixed terms specified by a known antisymmetric matrix $K$~\footnote{if
  $K$ is not anti-symmetric, it can be made so by addition of suitable
  total derivative terms.  Similarly, if the kinetic term posesses a
  time dependent matrix factor this may be diagonalized and made of
  the form~\myref{heiseqs-precanonicalL}},
\begin{equation}
\mathcal{L} = \frac12 \left({X^\dagger}' X'  +  {X^\dagger}' K X + X^\dagger K X' - X^\dagger \tgO^2 X \right)
\label{heiseqs-precanonicalL}
\end{equation}
where $X$ is a vector composed of an arbitrary number of real scalar
fields, $\tgO^2$ is a known symmetric matrix, and both $\tgO^2$ and
$K$ depend on time. Note that the matrix $\tgO^2$ is not the frequency
matrix due to the presence of a nonzero $K$.  To remove the mix terms
proportional to $K$, one considers a vector $Y$ which is related to
$X$ by a time dependent rotation, $Y \equiv \cR X$ with the condition
that the representation $Y$ puts the Lagrangian in the canonical form.
An analytic solution for $\cR$ is not always possible, but since $\cR$
is a basis dependent quantity, one expects it should not appear in
physical results.  We thus do not attempt to solve for $\cR$.  By
expanding out the kinetic term $\frac12 {Y'}^TY$ and comparing
to~\myref{heiseqs-precanonicalL}, one may
write~\myref{heiseqs-precanonicalL} in the canonical form,
\footnote{ A
relation between the kinetic term for $Y$ and that of $X$ is
obtained by expanding,
\begin{eqnarray*}
\frac12 {Y'}^T Y' &=&\frac12 {X'}^TX' + {X'}^T \cR^T \cR' X + \frac12 X^T {\cR'}^T \cR' X \\
  &=& \left(\frac12 {X'}^TX' + {X'}^T  K X\right) + \frac12 Y^T \cR K^T K \cR^T Y 
\end{eqnarray*}
where we have identified $\cR^T \cR' \equiv K$, from which it also
follows that ${\cR'}^T \cR' = K^T K$.} 
\begin{equation}
\mathcal{L} = \frac12 {Y'}^T Y' - \frac12 Y^T \cR(\tgO^2 + K^T K )\cR^T Y
\label{heiseqs-canonicalL}
\end{equation}
where one reads from above that the frequency matrix is
$\gO^2=\cR(\tgO^2 + K^T K )\cR^T$.  Since the eigenvalues of $\gO^2$
are invariant to a rotation of $\gO^2$, one can determine the the
diagonal frequency matrix $\omega$ by diagonalizing the known matrix
$(\tgO^2 + K^T K)$ with a rotation $U$.  Specifically,
\begin{eqnarray}
\Omega^2 = \cR(\tgO^2+ K^TK)\cR^T = \cR U\go^2U^T \cR^T
\nonumber 
\end{eqnarray}
It is also clear from the above that the previously defined $C$
matrix~\myref{heiseqs-cmatrix} will have the specific form $C=\cR U$. The
matrix $\Gamma$ is then computed,
\begin{eqnarray}
\Gamma \equiv C^T C' = U^T \cR^T (\cR U)' &=& U^T \cR^T \cR' U + U^T U' \nonumber \\
&=& U^T K U + U^T U'
\label{heiseqs-gammaform2}
\end{eqnarray}
and thus both $\go$ and $\Gamma$ matrices are determined without an
explicit form for the rotation $\cR$.  To summarize, given the
Lagrangian in a form~\myref{heiseqs-precanonicalL} with the two
matrices $K$ and $\tgO^2$ specified, one may extract the relevant
quantities for the evolution~(\ref{heiseqs-bogequations}), namely the
eigenfrequency matix $\go$ and the matrix $\Gamma$ describing the rate
of change of the eigenvectors.

After having put the Lagrangian into one of the above
forms~\myref{heiseqs-ac}~or~\myref{heiseqs-canonicalL}, it may be that
some of the fields which compose the vector $X$ and which have nonzero
entries in $K$ and $\tgO^2$ may in fact decouple.  A decoupling may
not always be obvious from examination of $K$ or $\tgO^2$, but a
computation of the $I$ and $J$ matrices can often indicate a hidden
supression or decoupling.  It may then be possible to perform
additional rotations on $X$ to decouple the fields and thus reduce the
Lagrangian further.  If the rotation is time-dependent, the
transformation will induce terms in $K$ and $\tgO^2$ as discussed
above, but the induced terms may be chosen to yield cancellations,
decoupling, or other simplifications of these matrices.

One specific case of interest is when the Lagrangian can be expanded
in a small parameter.  In our models the small parameter will be the
ratio of two mass scales $\mbyM$.  We will see later that the strong
mixing (unsupressed in the expansion $\mbyM$) in the Lagrangian
dominates the Heisenberg Equations, and the weak mixing (mixing
supressed by powers of $\mbyM$) in fact contributes neglibly to the
dynamics and may therefore be neglected.  The fields which mix only
weakly should then be decoupled with a small rotation, a rotation
which deviates from the identity by small terms in positive powers of
$\mbyM$.  The strongly mixed parts of the Lagrangian will remain
unchanged by such a rotation, but the transformation may be chosen to
decouple the weakly mixed fields in the Lagrangian to a corresponding
order in $\mbyM$.  The decoupled fields may then be discarded.  By
this procedure, the remaining Lagrangian should only contain the
minimal field content, those fields which mix strongly and thus may
lead to a non-adiabatic evolution.
The reader is referred to Section~\ref{sec-gaugefixing} and
Appendix~\ref{calc-u1-lagrangian} for specific examples employing this
procedure.  We now move on to further discuss the form of the
Heisenberg equations and its interpretation and solution.

         \section{Quantifying Non-adiabatic Evolution} \label{sec-heiseqs-nonadiabatic}  
To gain intuition for the Heisenberg
Equations~\myref{heiseqs-bogequations}, notice that there are two
qualitatively different parts to the equations; a "unitary" part
represented by the anti-Hermitian matrices $(\pm i\omega - I)$ and a
"nonunitary" part represented by the symmetric matrix
$\left(\frac{\omega'}{2\omega}-J\right)$.  One can check that the
unitary part preserves the total occupation number
$\mbox{Tr}[\beta^*\beta^T]$, but the matrix $\beta^*\beta^T$ is not
preserved in general which indicates a possible conversion from one
species of particles to another from this evolution.  The non-unitary
part will boost (or contract) both $\alpha\alpha^\dagger$ and
$\beta^*\beta^T$ in~\myref{heiseqs-mc1} while keeping the difference
invariant -- preserving \myref{heiseqs-mc1}-\myref{heiseqs-mc2}. This
part of the evolution will change the total occupation number in the
system.  Note however, that the unitary part can not be neglected as
the mixing between species effects the evolution and so has physical
implications.  We make one final rewriting of the Heisenberg
equations~\myref{heiseqs-bogequations}.  By extracting factors of
$\sqrt{\omega}$, the equations are written
\begin{eqnarray}
\alpha' &=& \sqrt{\omega}\left(-i  - \frac{1}{\sqrt{\omega}}I\frac{1}{\sqrt{\omega}}\right)\sqrt{\omega}\alpha 
+ \sqrt{\omega}\left(\frac{\omega'}{2\omega^2}-\frac{1}{\sqrt{\omega}}J\frac{1}{\sqrt{\omega}}\right)\sqrt{\omega}\beta 
\nonumber \\
\beta' &=& \sqrt{\omega}\left(i  - \frac{1}{\sqrt{\omega}}I\frac{1}{\sqrt{\omega}} \right) \sqrt{\omega}\beta 
+ \sqrt{\omega}\left(\frac{\omega'}{2\omega^2} -\frac{1}{\sqrt{\omega}}J\frac{1}{\sqrt{\omega}}\right)\sqrt{\omega}\alpha
\label{bogequations2}
\end{eqnarray}
The equations are now written entirely in terms of the following matrices,
\begin{equation}
\sqrt{\omega}\;\;\;,\;\;\;
\left(1  \pm i \frac{1}{\sqrt{\omega}}I\frac{1}{\sqrt{\omega}}\right) \;\;\;,\;\;\;
A = \left(\frac{\omega'}{\omega^2}-\frac{1}{\sqrt{\omega}}2J\frac{1}{\sqrt{\omega}}\right) \nonumber
\end{equation}
where the matrix $A$ has been defined.  One recognizes the diagonal
part of $A$ as adiabatic parameters from the single-field analysis.
The off-diagonal parts of this matrix may also be interpreted as
adiabatic parameters as follows.  One expects nonadiabatic evolution
when either the oscillation frequency of the physical state or the
field composition of the state (its eigenvector) has changed in a time
comparable to or faster than the period of one oscillation of the
state $\frac{2\pi}{\omega}$.  The latter condition on the field
composition is clearly relevant only if there are two or more fields
from which the state is composed.  Specifically, one may guess
nonadiabatic evolution when $|\hat C_i'| > \omega_i$ or specifically
when,
\begin{equation}
\sqrt{\sum_j \left(\frac{\Gamma_{ij}}{\omega_i}\right)^2} > 1
\label{adiabaticguess}
\end{equation}
This condition is not quite correct however, since in the case of
degenerate eigenstates, the eigenvectors have no preferred directions
in their subspace, and so the rate of change of these eigenvectors in
this subspace is not physically meaningful.  This feature is taken
into account by the $A$ matrix (as well as the $J$ matrix) which has
the form,
\begin{eqnarray}
A &=& \frac{\omega'}{\omega^2} - \left( \Gamma \frac{1}{\omega} - \frac{1}{\omega}\Gamma \right) \nonumber \\
A_{ij}  &=& \left\{
\begin{array}{ll}
\frac{\omega'_i}{\omega_i^2} & i=j \\
\Gamma_{ij}\left(\frac{1}{\omega_i}-\frac{1}{\omega_j} \right) &i\ne j
\end{array}\right\}
\;\; \mbox{no summation}
\label{heiseqs-adiabaticitymatrix}
\end{eqnarray}
and it handles both limiting cases,
\begin{eqnarray}
\lim_{\omega_i\rightarrow\omega_j} A_{ij} =0 \;\;\;,\;\;\;i\ne j  
\nonumber \\
\lim_{\omega_j >> \omega_i} A_{ij} = \left(\frac{\Gamma_{ij}}{\omega_i}\right)  \;\;\;,\;\;\;i\ne j 
\nonumber
\end{eqnarray}
so for nondegenerate frequencies the condition \myref{adiabaticguess}
is asymptotically achieved.  We thus take the elements of $A$ as our
adiabatic parameters.  If any one element of the matrix is greater
than unity $|A_{ij}|>1$, we expect nonadiabatic evolution.  We note
finally that large $A_{ij}$ are a necessary condition for net
production of quanta, but not a sufficient condition.  Thus the matrix
elements $A_{ij}$ may be treated as useful indicators of particle
production but not conclusive indicators.

         \section{Mass Hierarchy, Equipartition, Averaging,  A Leading Order Solution, and Scaling} \label{sec-heiseqs-scaling}  
In this subsection, we discuss the solutions of the Heisenberg
equations of motion when the states of the theory have a hierarchy of
two mass scales, the scale $m\sim TeV$ of the flat direction, and the
scale of the flat direction VEV $\langle \phi_i\rangle \siml M_P$.
For notational convenience we will denote the heavy scale simply by
$M$ and the relevant perturbation parameter is then $\frac{m}{M}$.
The quanta of the theory will also have masses of these two scales,
but the nonperturbative dynamics, and thus the particle production
will depend on the slow oscillations of frequency $\sim m$.  If one
wished, it is possible to consistently remove the heavy fields of the
theory leaving behind an effective theory of the light
states. However, this assumes the heavy states are always produced as
virtual particles -- there never being enough energy available to
produce the heavy particle.  This is not a good assumption.  Our
system of flat direction VEVs are extremely far from thermal
equilibrium, so there is no notion of temperature to clearly indicate
whether the heavy degrees of freedom will be ``frozen'' or
``unfrozen.''  Additionally if nonperturbative effects are important,
there can be processes with many light particles annihilating to
create a heavy particle.  We thus include the heavy states in the
theory, though we use the hierarchy of scales as a perturbation
parameter to approximate the dynamics of the full theory.
Specifically, the approximation will allow us to numerically solve our
system at a small value of the ratio $\frac{m}{M}$ and to consistently
rescale the solutions to correspond to a smaller value of this ratio.
This is useful because when obtaining numerical solutions, there are
limitations of machine precision which prevent one from taking the
ratio $\frac{m}{M}$ at the value of physical interest which is
$\frac{m}{M}\siml \frac{TeV}{10^{-2}M_P} \sim 10^{-14}$.

In the following, we present a leading order solution to the
Heisenberg equations of motion~\myref{heiseqs-bogequations} valid
during a parametric resonance of the heavy and light states, and valid
after an averaging has been applied to the fast
oscillations.\footnote{By fast oscillations we mean those with
  frequency $\sim M$. The slow oscillations have frequency $\sim m$.}
The above mentioned scaling relations will follow once the leading
order solution to these equations are identified.  The first step is
to determine the conditions on the
equations~\myref{heiseqs-bogequations} to yield a parametric
resonance, and specifically conditions which take into account the
expansion in $\frac{m}{M}$.  Recall, that after fixing the initial
conditions $\ga=\mathbf{1}$ and $\gb=0$, the equations are uniquely
specified by the driving matrices $I,J,\go$ and $\frac{\go'}{\go}$
which are in turn dependent on the matrices $\Gamma$ and $\go$.  The
driving terms can be organized into a series in half integer powers of
$\pfrac{m}{M}$.  We will only require the leading order terms of these
expansions to determine the leading order solutions of the equations.
For instance, the heavy states have frequencies $\omega_A\sim M$ which
are zeroth order, and the light states have frequencies $\omega_a\sim
m$ which are first order.  Also, for the models we consider, both
$\Gamma$ and $\go$ will evolve only on the slow time scale
$m$.\footnote{In principle, $\Gamma$ and $\go$ could also evolve on
  the fast scale $M$, but such cases would correspond to a coherent
  classical motion of the heavy Higgs-like excitations which are
  orthogonal to the flat directions.  This possibility is discussed
  again in Section~\ref{sec-reheating-nonperturbativeflat}} For
instance, the rotations $C$ which bring the system to diagonal form
will evolve on this time scale.  Thus $C'\sim m$ assuming the $C_{ij}$
are not further suppressed in any way, and so $\Gamma_{ij}\sim m$.
Now denoting the heavy fields with upper case indices $A,B,C$ and
light fields with lower case indeces $a,b,c$, the leading order
driving terms which appear in the equations are tabulated as follows,
\begin{equation}
\begin{array}{lll}
\omega_a \sim \sqrt{m^2 + k^2} &  \omega_A \sim M  &\\[0.4em]
\frac{\omega_a'}{\omega_a} \sim \frac{m^3}{m^2 + k^2} & \frac{\omega_A'}{\omega_A} \sim m & \\[0.4em]
I_{ab},J_{ab} \sim m  & I_{AB},J_{AB} \sim m & I_{Ab},J_{Ab} \sim \sqrt{m M}
\end{array}
\label{heiseqs-leadingdrivingterms}
\end{equation}
where the dependence on $k$ is also shown assuming $k\siml m$.  We
also note by examination of the definition for $J_{ij}$
in~\myref{heiseqs-defineIJGamma}, that a degeneracy in eigenvalues,
$\go_i = \go_j$ will lead to further suppression in $J_{ij}$, greater
than the order listed above.

We may also make a heuristic guess of the leading order solutions
$\{\ga, \gb\}$.  An important consideration is that we are seeking a
series solution in the specific case of non-adiabatic evolution when
the occupation numbers of the states $n_i \equiv(\gb^*_{ij}\gb_{ij})$
are expected to be large.  Assuming there is no suppression in the
driving terms, all the light states should have comparable occupation
numbers $n_a\sim n_b$ at any given time.  We do not expect the heavy
states to have comparable occupation numbers to the light states.
Assuming there is sufficient mixing between the heavy and light
states, we do however expect the energy density of any heavy degree of
freedom to be comparable to that of a light degree of freedom, and
thus $\go_A n_A \sim \go_a n_a$ which then implies
\begin{equation}
\left\{\frac{\ga_a}{\ga_A},\frac{\gb_a}{\gb_A}\right\} \sim \sqrt{\frac{\go_a}{\go_A}} \sim \sqrt{\frac{m}{M}}
\label{heiseqs-equipartition}
\end{equation}
This is essentially an assumption of ``equipartition'' though it does
not depend on the presence of a thermal bath.  Rather the
``equipartition'' is due to strong coupling between the states
combined with non-adiabatic evolution for at least one of the states
\cite{equipartition}.
The heuristic guess of equipartition also allows one to make some
statements about how the solutions rescale under a change in the
parameters $m$ and $M$.  Specifically, so long as the non-adiabatic
evolution does not vanish as one varies these parameters, we still
expect the relation~\myref{heiseqs-equipartition} to hold for the
solutions.  However, the argument assumes there is a large production
of quanta for the light states, and says nothing about this amount.
To determine this one must directly examine the Heisenberg equations
of motion which is done next.

The analysis of the Heisenberg equations is facilitated by separating
the two extreme time scales set by $m$ and $M$.  As shown above, the
driving terms of the Heisenberg equations all evolve on the slow time
scale, though there are fast oscillations of frequency $M$ introduced
to the dynamics due to the presence of terms $\go_A \ga_A$ and
$\go_A\gb_A$ in the equations.  Note that the slow oscillations will
appear as constants on the time scale of the fast oscillations.
Similarly, we expect the slow oscillations to be only sensitive to the
average value of the fast evolving quantities.  We should be able to
formally average the fast oscillations from the Heisenberg equations
leaving behind the slow dynamics which is relevant to the
non-adiabatic evolution.  The fast oscillations are averaged over by
integrating the equations over exactly one or a few periods
$\frac{2\pi}{M}$ to obtain a difference equation.  The driving terms
can be treated as constants over this very small integration region
and they may thus be extracted from the integration.  Writing the
Heisenberg equations in matrix notation,
\begin{equation}
\frac{d}{d\eta}\left(\begin{array}{l}\ga \\\gb \end{array} \right) 
=
\left(\begin{array}{ll}-i\go - I & \frac{\omega'}{2\omega}-J \\ \frac{\omega'}{2\omega}-J & +i\go -I \end{array} \right) 
\left(\begin{array}{l}\ga \\\gb \end{array} \right)
\end{equation}
this equation is of the form $X' = A X$ where $A$ is an approximately
constant matrix over short time intervals and $X$ is a vector.  The
general solution is $X(\eta_f) = e^{A(\eta_f-\eta_i)}X(\eta_i)$.  From
this one constructs a difference equation over a very small time
interval of time $\gD\eta$ which corresponds one fast oscillation
period.  The exponential may be approximated over this small time
$e^{A\gD\eta}\approx\mathbf{1}+A\gD \eta$.  The difference equation is
$X\left(\eta+\gD\eta\right)-X(\eta) = A X(\eta)\gD\eta$
which for small time $\gD\eta$ will simply be a differential, except
the functions $X$ have been averaged over a small window of time.  By
this argument, we re-write our differential equations,
\begin{equation}
\frac{d}{d\eta}\left(\begin{array}{l}\bar\ga \\\bar\gb \end{array} \right) 
=
\left(\begin{array}{ll}-i\go - I & \frac{\omega'}{2\omega}-J \\ \frac{\omega'}{2\omega}-J & +i\go -I \end{array} \right) 
\left(\begin{array}{l}\bar\ga \\\bar\gb \end{array} \right)
\label{heiseqs-averagedeqs}
\end{equation}
where we have introduced the time averaged quantities $\bar \ga$ and
$\bar \gb$.  The equations are identical in form to our starting
equations because the driving terms were treated as constants over the
short time interval.  Apparently nothing has changed except the
notation, but the reason this averaging step has any meaning is that
now ${\bar\ga}'\sim m\bar\ga$ and ${\bar\gb}'\sim m\bar\gb$ where
beforehand we had for some components $\ga_{Ai}'\sim M\ga_{Ai}$ and
$\gb_{Ai}'\sim M \gb_{Ai}$.  This subtle difference allows us to
construct the leading order behavior of the solutions to
equations~\myref{heiseqs-averagedeqs} in the small parameter
$\frac{m}{M}$. The equations are first written using index notation
again with upper-case $A,B,C$ corresponding to the heavy modes,
\begin{eqnarray*}
\bga_{AB}' &=& -i\go_A\bga_{AB} - I_{AC}\bga_{CB} - I_{Ac}\bga_{cB} + \frac{\go_A'}{2\go_A}\bgb_{AB} - J_{AC}\bgb_{CB} - J_{Ac}\bgb_{cB}\\
\bgb_{AB}' &=&  i\go_A\bgb_{AB} - I_{AC}\bgb_{CB} - I_{Ac}\bgb_{cB} + \frac{\go_A'}{2\go_A}\bga_{AB} - J_{AC}\bga_{CB} - J_{Ac}\bga_{cB}\\
\bga_{Ab}' &=& -i\go_A\bga_{Ab} - I_{AC}\bga_{Cb} - I_{Ac}\bga_{cb} + \frac{\go_A'}{2\go_A}\bgb_{Ab} - J_{AC}\bgb_{Cb} - J_{Ac}\bgb_{cb}\\
\bgb_{Ab}' &=&  i\go_A\bgb_{Ab} - I_{AC}\bgb_{Cb} - I_{Ac}\bgb_{cb} + \frac{\go_A'}{2\go_A}\bga_{Ab} - J_{AC}\bga_{Cb} - J_{Ac}\bga_{cb}\\
\bga_{aB}' &=& -i\go_a\bga_{aB} - I_{aC}\bga_{CB} - I_{ac}\bga_{cB} + \frac{\go_a'}{2\go_a}\bgb_{aB} - J_{aC}\bgb_{CB} - J_{ac}\bgb_{cB}\\
\bgb_{aB}' &=&  i\go_a\bgb_{aB} - I_{aC}\bgb_{CB} - I_{ac}\bgb_{cB} + \frac{\go_a'}{2\go_a}\bga_{aB} - J_{aC}\bga_{CB} - J_{ac}\bga_{cB}\\
\bga_{ab}' &=& -i\go_a\bga_{ab} - I_{aC}\bga_{Cb} - I_{ac}\bga_{cb} + \frac{\go_a'}{2\go_a}\bgb_{ab} - J_{aC}\bgb_{Cb} - J_{ac}\bgb_{cb}\\
\bgb_{ab}' &=&  i\go_a\bgb_{ab} - I_{aC}\bgb_{Cb} - I_{ac}\bgb_{cb} + \frac{\go_a'}{2\go_a}\bga_{ab} - J_{aC}\bga_{Cb} - J_{ac}\bga_{cb}
\end{eqnarray*}
In the first four equations above we notice
using~\myref{heiseqs-leadingdrivingterms} that for instance
$\frac{\bga_{Ai}'}{\go_A\bga_{Ai}}\sim \frac{m}{M}$ and so the
left-hand-sides may in fact be neglected in the series approximation.
From the second pair of equations, we may then infer there exists a
solution with relative order $\{\ga_A,\gb_A\} \propto
\sqrt{\frac{m}{M}}\{\ga_a,\gb_a\}$ which was guessed by our heuristic
argument of equipartition~\myref{heiseqs-equipartition}.  There are
also many terms above which are subleading and may be discarded.  The
resulting leading order equations are obtained to be,
\begin{eqnarray}
       0 &=& -i\go_A\bga_{AB}\nonumber\\
       0 &=&  i\go_A\bgb_{AB} \nonumber\\
       0 &=& -i\go_A\bga_{Ab}                 - I_{Ac}\bga_{cb}                                                - J_{Ac}\bgb_{cb}\nonumber\\
       0 &=&  i\go_A\bgb_{Ab}                 - I_{Ac}\bgb_{cb}                                                - J_{Ac}\bga_{cb}\nonumber\\
\bga_{aB}' &=&                                - I_{ac}\bga_{cB} + \frac{\go_a'}{2\go_a}\bgb_{aB}                 - J_{ac}\bgb_{cB}\nonumber\\
\bgb_{aB}' &=&                                - I_{ac}\bgb_{cB} + \frac{\go_a'}{2\go_a}\bga_{aB}                 - J_{ac}\bga_{cB}\nonumber\\
\bga_{ab}' &=& -i\go_a\bga_{ab} - I_{aC}\bga_{Cb} - I_{ac}\bga_{cb} + \frac{\go_a'}{2\go_a}\bgb_{ab} - J_{aC}\bgb_{Cb} - J_{ac}\bgb_{cb}\nonumber\\
\bgb_{ab}' &=&  i\go_a\bgb_{ab} - I_{aC}\bgb_{Cb} - I_{ac}\bgb_{cb} + \frac{\go_a'}{2\go_a}\bga_{ab} - J_{aC}\bga_{Cb} - J_{ac}\bga_{cb}
\label{heiseqs-leadingeqs}
\end{eqnarray}
We should check that the leading order solutions to the equations do
not exhibit fast oscillations, and in fact, this is guaranteed through
the heavy modes $\{\ga_{Ai},\gb_{Ai}\}$ being algebraically solved in
terms of the light modes in the second pair of equations.  In this
sense, the heavy modes have been removed from the dynamics, though the
heavy quanta are still allowed to be produced. We also learn from the
above that $\{\ga_{AB},\gb_{AB}\}$ are zero to leading order.

Finally, we may show the scaling of the leading order solution to the
averaged equations $\bga,\bgb$ under changes in the parameters which
was our original purpose.  Supposing a solution has been determined
for the Heisenberg Equations at some small value of the parameter
$\frac{m}{M}$.  Then the scales are changed as follows $m\rightarrow
\mu m$ and $M\rightarrow \gamma M$ for which the ratio
$\frac{m}{M}\rightarrow\frac{\mu m}{\gamma M}$ is still sufficiently
small.  Using the scaling properties of the driving
terms~\myref{heiseqs-leadingdrivingterms}, one may infer rescaled
solutions to the leading order equations~\myref{heiseqs-leadingeqs} in
terms of the known solutions.  Table~\ref{table-rescaling} illustrates
the relation.  Notice in particular that because the light mode
solutions $\ga_{aj},\gb_{bj}$ are invariant under the rescaling, their
initial conditions are preserved, but this is not true for the heavy
modes $\ga_{Aj},\gb_{Aj}$.  Having averaged over the fast
oscillations, the history of the heavy states has been erased.  This
is consistent with the dynamics of the modes being attracted to an
equipartition of energy.
\begin{table}[!htbp]
\begin{center}
\begin{tabular}{|c|l|l|}
\hline 
& \textbf{known sol.} & \textbf{rescaled sol.} \\
\hline 
&&\\
\textbf{Scales} & $m$ , $M$ & $\mu m$ ,  $\gg M$ \\
&&\\
\hline
&&\\
\textbf{Solutions} & $\bga_{ij}$ , $\bgb_{ij}$, $n_i$ & $\begin{array}{rrr}\sqrt{\frac{\mu}{\gg}}\bga_{Aj}  &, \sqrt{\frac{\mu}{\gg}}\bgb_{Aj} &, \frac{\mu}{\gg}n_A \\ \bga_{aj} &, \bgb_{aj} &, n_a\end{array}$ \\
&&\\
\hline
&&\\
\textbf{Functional dependence} &$\eta$ , $k$ & $\eta_{new}=\frac{\eta}{\mu}$ , $k_{new}=\frac{k}{\mu} $ \\[0.2em]
e.g. $\bga_k(\eta)\rightarrow \bga_{k_{new}}(\eta_{new})$ &&\\
&&\\
\hline
\end{tabular}
\caption[Rescaling of Solutions to the Heisenberg Equations of
Motion]{\label{table-rescaling} With known solutions to the leading
  order Heisenberg Equations~\myref{heiseqs-leadingeqs} at the scales
  $m$ and $M$, one may infer rescaled solutions at the scales $\mu m$
  and $\gamma M$ assuming both $\frac{m}{M}$ and $\frac{\mu m}{\gg M}$
  are sufficiently small.  The table shows how the solutions and their
  functional dependence change under the rescaling.  The upper case
  index $"A"$ corresponds to a heavy mode, the lower case $"a"$
  correspond to a light mode, and $i,j$ correspond to either heavy or
  light.}
\end{center}
\end{table}

To summarize, the solutions to Heisenberg equations have been shown to
exhibit a clear scaling at leading order in the expansion
$\frac{m}{M}$.  The scaling of the solutions is shown in
Table~\ref{table-rescaling}.  The scaling has been determined from a
heuristic analysis applying the equipartition principle and from a
direct analysis of the differential equations.  The scaling will
additionally be verified numerically in later sections over a range of
ratios $\frac{m}{M}$ and finally the scaling will be applied to
extrapolate numerical solutions for small values of $\frac{m}{M}$ to
the very small values of $\frac{m}{M}\sim 10^{-14}$ of physical
interest.  A byproduct of obtaining the scaling relations was the
physical prediction of equipartition of the energy during the
non-adiabatic evolution.

         \section{Reformulating the Heisenberg Equations} \label{sec-heiseqs-reduce} 
In this section we present a nontrivial re-characterization of the
Heisenberg Equations of motion as written
in~\myref{heiseqs-bogequations} in which the
constraints~(\ref{heiseqs-mc1}-\ref{heiseqs-mc2}) are built-in.  The
equations and the constraints again are,
\begin{eqnarray}
& \alpha' = \left(-i \omega -I\right)\alpha + \left(\frac{\omega'}{2\omega}-J\right)\beta \nonumber \\
& \beta' = \left(i \omega-I\right) \beta + \left(\frac{\omega'}{2\omega} -J\right)\alpha \nonumber \\
& \ga \ga^\dagger - \gb^*\gb^T = 1 \label{heiseqs-reduce3}\\
& \ga \gb^\dagger - \gb^*\ga^T = 0 \label{heiseqs-reduce4}
\end{eqnarray}
where the dependent variables $\ga$ and $\gb$ are square matrices,
$\omega$ is the diagonal eigenfrequency matrix, $I$ is antisymmetric
and $J$ is symmetric.  We attempt to re-write these equations in
quadratic variables $n=\gb^*\gb^T$.  Recall this is a hermitian matrix
which yields the occupation numbers on the
diagonal~\myref{heiseqs-occnumber}.  We compute,
\begin{eqnarray}
n' &=& \gb^{*\prime}\gb^T + \gb^* \gb^{T\prime} \nonumber \\
&=&\left[(-i\go - I)\gb^* + \left(\frac{\go'}{2\go} - J \right)\ga^* \right]\gb^T + \gb^*\left[\gb^T(i\go+I) + \ga^T\left(\frac{\go'}{2\go} - J \right) \right] \nonumber \\
&=& (-i\go-I)n + n(i\go+I) + \left(\frac{\go'}{2\go} - J \right)\ga^*\gb^T + \gb^*\ga^T\left(\frac{\go'}{2\go} - J \right) \nonumber \\
\mbox{} \label{heiseqs-reduce5}
\end{eqnarray}
Examining this equation, it seems appropriate to also define the
elements of the matrix $m\equiv\ga^*\gb^T = \gb\ga^\dagger$ as
dependent variables (notice the constraint~\myref{heiseqs-reduce3} is
applied in this definition). We compute the time derivative of $m$ to
check if the system of equations closes,
\begin{eqnarray}
m' &=& \ga^{*\prime}\gb^T + \ga^* \gb^{T\prime} \nonumber \\
&=& \left[(i\go -I)\ga^*+ \left(\frac{\go'}{2\go} - J \right)\gb^*\right]\gb^T +\ga^* \left[\beta^T\left(i \omega + I\right) + \alpha^T\left(\frac{\omega'}{2\omega} -J\right) \right] \nonumber \\
&=& (i\go -I)\ga^*\gb^T + \ga^*\gb^T(i\go+I) + \left(\frac{\go'}{2\go} - J \right) n + (n^*+1)\left(\frac{\go'}{2\go} - J \right)  \nonumber \\
\mbox{} \label{heiseqs-reduce6}
\end{eqnarray}
where the constraint~\myref{heiseqs-reduce3} was applied in the last
step to substitute $\ga^*\ga^T=(n^*+1)$.  The system of
equations~(\ref{heiseqs-reduce5}-\ref{heiseqs-reduce6}) does in fact
close with the new variables $n$ and $m$.  The original variables
$\ga$ and $\gb$ are $N\times N$ complex matrices subject to two
algebraic constraints~(\ref{heiseqs-reduce3}-\ref{heiseqs-reduce4}).
Analogously, the new variables $n$ and $m$ are $N\times N$ complex
matrices subject to structurally equivalent algebraic constraints
$n^\dagger=n$ and $m^T =m$ which makes $n$ hermitian and $m$
symmetric.  The number of independent variables of the old and new
representation are thus seen to match, and
the two systems of equations are apparently equivalent.  However, one
must be careful because our new variables are quadratic expressions,
and it is still not obvious whether some physical information has not
been lost because of the nonlinear transformation.  It may be possible
to write $\alpha$ and $\beta$ explicitly in terms of $m$ and $n$ and
thus show a strict equivalence, but this computation is not performed
here.

It is worthwhile to express the equations in a Real representation
using Real symmetric matrices $n_R\equiv Re(n)$ and $m_R\equiv Re(m)$
and the Real anti-symmetric matrices $n_I\equiv Im(n)$ and $m_I \equiv
Im(m)$, and the system is then written in the form,
\begin{eqnarray}
n_R'&=& n_RI-In_R +\go n_I - n_I\go + \left(\frac{\go'}{2\go} - J \right)m_R + m_R \left(\frac{\go'}{2\go} - J \right) \nonumber \\
n_I'&=& n_II-In_I -\go n_R + n_R\go + \left(\frac{\go'}{2\go} - J \right)m_I - m_I \left(\frac{\go'}{2\go} - J \right) \nonumber \\
m_R'&=& - \go m_I- m_I \go - I m_R + m_R I + \left(\frac{\go'}{2\go} - J \right)n_R + n_R\left(\frac{\go'}{2\go} - J \right) \nonumber \\
&\mbox{}& +\;  \left(\frac{\go'}{2\go} - J \right) \nonumber \\
m_I'&=&  \go m_R + m_R \go - I m_I + m_I I + \left(\frac{\go'}{2\go} - J \right)n_I + n_I\left(\frac{\go'}{2\go} - J \right) 
\label{heiseqs-reducefinal}
\end{eqnarray}
One notable feature of these equations is that the initial conditions
one usually considers of vanishing occupation number are extremely
simple -- they are $n=m=0$.  There is only one term above in the third
set of equations which provides a seed to drive the evolution of the
system $\left(\frac{\go'}{2\go} - J \right)$, and this quantity has
already been identified in the previous sections as being responsible
for particle production.

We make a few final notes regarding the above reformulation of the
Heisenberg Equations of motion.  One point to be made is that the
constraints~(\ref{heiseqs-reduce3}-\ref{heiseqs-reduce4}) are
essentially enforcing canonical commutation relations (which was
discussed in the previous section), and apparently the reformulated
equations now have the commutation relations automatically enforced!
To build upon this formulation, one might attempt to construct
analogous equations for fermions in which the anti-commutation
relations are incorporated.  This is in the same spirit of the program
begun in \cite{nilles-peloso-sorbo-2001} in which the equations
\myref{heiseqs-bogequations} were originally presented in a
multi-field context.  Also, the dimension of the system was never
specified, so the equations are generally applicable.  We note that
the single field case of the above equations has been noted and
applied already in \cite{zeldovich-starobinsky-1971}.

Finally, while this new formulation may in fact be more efficient for
numerical computation than the original
formulation~\myref{heiseqs-bogequations}, it was derived only after
the numerical results had been obtained and verified
via~\myref{heiseqs-bogequations}.  Conseqently, these extra
computations were not required and so not performed.  The material of
this section may thus be considered tangential to the thrust of the
thesis.

\newpage \chapter{Inflation and the Evolution of Scalar Fields in the Early Universe} \label{sec-inflation} 
The goal of this section is to introduce inflationary cosmology, and
to explain the process by which inflation converts the quantum
fluctuations of a light scalar field $\phi$ such as a flat direction
into observable classical perturbations of the field \cite{VEVgrowth}.
To an observer present at some time after the conclusion of inflation,
the spectrum of perturbations for the field $\phi$ which are larger
than the observer's Hubble radius will sum up to appear as a
homogeneous classical field or a vacuum expectation value (VEV)
defined $\langle \phi\rangle_{patch} \equiv \Phi$.  Because a quantum
process has generated the field, different observers in different
patches of the universe will observe different realizations of $\Phi$.
We will consider patch sizes which correspond to the present
observable universe, and we determine the variance of the distribution
of $\Phi$ for an ensemble of such patches.  The computed variance will
then be a measure for the typical amplitude $\Phi$ obtained in any
arbitrary patch, and thus a best estimate for the value in our
observable universe.

In computing the variance, we also take into account limitations to
its growth due to a finite duration of inflation as well as due to the
effects of higher order terms in the fields potential which will
cutoff the growth of the VEV \cite{ellis-etall-1987}.  The main result
of this section is a characterization of the post-inflation initial
conditions for the flat direction fields $\langle\phi_i\rangle$ prior
to their oscillation and subsequent decay during reheating which will
be studied in detail in
Sections~\ref{sec-reheating}-\ref{sec-results}.
A critical input to the generation of the above perturbations is the
Hubble parameter during inflation $H_I$.  We conclude by briefly
discussing the COBE normalization of the inflaton's potential, and a
bound on $H_I$ set by non-observation of a gravitational wave
signature in the Cosmic Microwave Background.

         \section{The Classical Theory of Inflation} \label{sec-inflation-classical} 
We first review the basics of inflation, beginning with a list of the
problems in the big bang cosmology that inflation
resolves.\footnote{The list of problems that inflation resolves can be
  found in many references on cosmology.  The list presented here was
  compiled from the reviews \cite{rubakov-cosmology-2005,
    liddle-review-1999, riotto-inflation-2002,linde-inflation-2007}}

\begin{description}
\item[Horizon problem:] The Cosmological horizon defines the
  maximum distance from any observer from which particles travelling
  to the observer are reaching the observer at some time $t$.  It is
  thus the size of a causally connected patch in the universe at this
  time.  The horizon problem is that approximately $\sim 30,000$
  patches that were causally disconnected at the time of the last
  scattering of the CMB photons, appear to have the same temperature
  to better than one part in $10^{-4}$ as measured by us today. There
  is no reason apriori that these regions should have the same
  temperature.  \textbf{Note:} The Cosmological horizon is not to be
  confused with the event horizon of an observer in an accelerated
  reference frame.  The event horizon marks a region of space about
  an observer from which an emitted light signal at a time $t$ will
  \myemph{never} reach the observer.

\item[Flatness problem:] The spatial curvature contribution to
  the Friedmann equation is measured by us today to be $< 0.02$,
  compared to the contribution from matter, radiation and dark energy
  \cite{wmap}.  This is consistent with a flat or nearly flat
  universe.  A nearly flat solution is an unstable solution of the
  non-inflationary cosmology.  This means that even assuming the
  curvature contribution is as large as $10^{-2}$ today, it had to be
  smaller in the past.  For instance at the Electroweak epoch, the
  contribution would need to be $<10^{-26}$.  Why is the curvature so
  small?  One might postulate that the curvature is exactly zero by
  some symmetry, but a dynamic mechanism which drives this term to
  small values might also be a solution.

\item[Entropy problem:] The entropy of our current observable
  patch of the universe can be estimated to be proportional to the
  number of CMB photons in our patch $\sim 10^{88}$.  The entropy in
  this volume should remain essentially constant as one evolves the
  universe to early times.  A legitimate question is ``where did the
  entropy come from?''  Did it come from one (or a few) explosive
  event(s) or is it coming in a more continuous process?  In either
  case, one is challenged to provide a workable mechanism to create
  this entropy.  One notes that both the horizon and flatness problems
  may be cast into the form of the entropy problem, and in this sense
  the entropy problem is not necessarily a different
  problem.\footnote{In Guth's original paper \cite{guth-1981} the
    words ``entropy problem'' never appear but his argument very
    clearly draws the connection between entropy and the horizon and
    flatness problems.}

\item[Isotropy, homogeneity problems:] Why is it that for each
  direction one looks, the universe appears mostly the same, and for
  different points in the universe the conditions also appear mostly
  to be the same?

\item[Problem of primordial inhomogeneity's:] The CMB and
  large scale structures tell us that the early universe had density
  perturbations $\frac{\gd \rho}{\rho}\sim 10^{-5}$ at the time of
  recombination and earlier.  Where did these small perturbations come
  from?

\item[Gravitino Problem:] The MSSM is a globally supersymmetric
  theory, but supersymmetry may also be promoted to a local symmetry
  in which case the SUSY algebra includes general coordinate
  transformations.  Local supersymmetric theories thus include the
  graviton as well as the graviton's fermionic superpartner, the
  gravitino.  The gravitino acquires mass during the breaking of local
  supersymmetry through a super-Higgs mechanism.  The gravitino should
  also only couple to matter gravitationally, so its lifetime may be
  very long \footnote{Though if the Gravitino is the LSP its lifetime
    would be necessarily infinite}.  Additionally, gravitinos will be
  produced thermally in the early universe.  The gravitino problem is
  that an overproduction of gravitinos in the early universe leads to
  severe phenomenological consequences such as upsetting
  nucleosynthesis, diluting the baryon number of the universe, or in
  the case of a stable gravitino, leading to an overly large dark
  matter component.  In order to avoid the gravitino problem, the
  reheating temperature must be less then approximately
  $10^6-10^9\,\gev$ to suppress the production of these particles where
  this range depends on the gravitino mass and the dominant decay
  channel of the gravitino \cite{gravitinos-moroi, gravitinos-olive}.

\item[Monopole (topological defect) problem:] Topological defects are
  stable field configurations of many grand unified field theories
  (GUTs), and defects such as magnetic monopoles are expected to form
  if the universe cools through a temperature of about $10^{16}\,\gev$
  which is the approximate unification scale of these theories.  For
  Grand Unified theories to be viable, one needs to explain why these
  topological defects have not been observed.\footnote{For an
    introduction to magnetic monopoles, see \cite{iliopoulos-2008}.}

\end{description}

The entropy problem is related to the assumption of evolving
adiabatically from the present state of the universe to very early
times.  In Guth's paper \cite{guth-1981} in which he introduces
inflation, he connects the assumption of adiabatic expansion to the
flatness and horizon problems, and he quantifies these problems as
fine tuning problems of the initial conditions.  He then points out,
"...both problems could disappear if the assumption of adiabaticity
were grossly incorrect" \cite{guth-1981}.  He then proposes a period
of accelerated expansion as the solution.  Going beyond the context of
Guth's paper, here is how the above problems in cosmology are
resolved: The acceleration, if it lasts long enough, has one effect of
transforming a small causally connected region into a larger region of
numerous apparently causally disconnected regions after the
acceleration has ended.
This solves the horizon problem.  the acceleration also has the effect
of driving any initial curvature contribution in the Friedmann
equation to very small values by the end of the acceleration period,
thus solving the flatness problem.  The acceleration will additionally
dilute the particle density of all particle species, including
topological defects such as monopoles or exotic species such as
gravitinos.  This solves the monopole problem and partly solves the
gravitino problem.\footnote{The gravitino problem is only partly
  resolved because the gravitinos may then be produced after inflation
  during reheating or afterwards if the reheating temperature is high
  enough.}  Then, if the acceleration is isotropic, it will lead to a
homogeneous and isotropic universe.  However, the acceleration also
has the effect of red shifting all fluctuations of the fields in the
universe well outside any observers Cosmological horizon after the
acceleration phase, so that the fields effectively reside in their
ground states -- the entropy per comoving volume has thus been driven
to nearly zero.  \footnote{Note the current acceleration of the
  universe if it continues will have the same effect leading to a
  nearly zero entropy density for radiation for any individual patch.
  The radiation entropy will never be exactly zero as there is a
  finite entropy associated with the de~Sitter event horizon
  \cite{gibbons-hawking-1977,bunch-davies-1978}} We clearly do not
want to wind up with an empty universe.  Additionally, the consistency
of the Friedmann equation requires some agent to be responsible for
the acceleration.  There are problems with identifying the agent of
the acceleration with Einstein's cosmological constant; one cannot get
the acceleration phase to finish, nor will there be any large entropy
production.  In Guth's paper, he models the agent as a Lorentz
invariant false vacuum state.  Other researchers later consider the
agent as a scalar field \cite{new-inflation,chaotic-inflation}.  In
either case, the entropy problem may be resolved via the large entropy
generated by the decay of the false vacuum or by the decay of scalar
field once the inflation has ended.  Guth's contribution was to
recognize that the monopole, flatness, entropy and horizon problems
could in principle be resolved by this concept of inflation
\cite{guth-1981}.
It was soon after recognized that inflation may additionally resolve
the problem of primordial inhomogeneities.

We would like to see explicitly how the accelerated expansion is
obtained.  A first step is to write down Einstein's equations with the
Friedmann Robertson Walker metric which is a good description of our
universe during every stage of its evolution, including the inflation
stage.  The FRW metric has the form,
\begin{equation}
ds^2 = dt^2 - R(t)^2 \left(\frac{dr^2}{1-\kappa r^2}+r^2d\gO^2 \right)
\end{equation}
where the spatial part of the metric is both homogeneous and isotropic
(or maximally symmetric), and where the $\kappa$ term specifies the
spatial curvature of the universe.  Substituting the FRW metric into
the left-hand-side of Einstein's equations $G_{\mu\nu}=8\pi
GT_{\mu\nu}$, and substituting the energy momentum tensor of a perfect
fluid into the right-hand-side, one obtains the Friedmann Equation,
\begin{equation*}
H^2 = \pfrac{\dot R}{R}^2 = \frac{8\pi }{3}G \rho - \frac{\kappa}{R^2}
\end{equation*}
where $\rho$ is the energy density of the fluid, and $p$ would be its
pressure.  As mentioned above, the curvature contribution to the
Friedmann equation, which is the term proportional to $\kappa$ above,
is negligible today and even more negligible at earlier epochs under the
assumption of adiabatic evolution, so we set $\kappa=0$, and do not
mention the curvature term again.  The FRW metric and the Friedmann
equation we consider for the universe are then,
\begin{eqnarray}
ds^2 &=& dt^2-R^2 d{\vec x}^2 \\
H^2 &=& \pfrac{\dot R}{R}^2 = \frac{8\pi }{3}G \rho
\label{inflation-friedmann}
\end{eqnarray}
The Einstein Equations additionally result in an equation for the
acceleration of the scale factor,
\begin{equation}
\frac{\ddot R}{R} = \frac{-4\pi G}{3} (\rho +3 p)
\label{inflation-accelerationeq}
\end{equation}
and there is also the statement of the conservation of energy and
momentum for a perfect fluid in an expanding universe,
\begin{equation}
d (\rho R^3) = - p d\left( R^3 \right)
\label{inflation-firstlaw}
\end{equation}
which is equivalently the first law of thermodynamics for the
fluid. It can be shown that only two of the three
equations~(\ref{inflation-friedmann},\ref{inflation-accelerationeq},\ref{inflation-firstlaw})
are independent.  From~(\ref{inflation-firstlaw}), if we assume an
equation of state $p=w\rho$, then one may solve for the energy density
$\rho$ in terms of the scale factor,
\begin{equation}
\rho = \rho_0 \pfrac{R}{R_0}^{-3(1+w)}
\label{inflation-energydensity}
\end{equation}
where $\rho_0$ and $R_0$ are the energy density and scale factor at
some epoch (usually the current epoch).  Now we explore the dynamics
for different equations of state for the dominating fluid.  The
equation of state for a matter fluid is $w=0$, as it may be treated as
non-interacting pressure-less dust.  The
equation~(\ref{inflation-energydensity}) then tells us
$\rho_{matter}\propto R^{-3}$ which is interpreted rather simply as
the expanding universe simply dilutes the particle number density by
an inverse factor of the volume $R^{-3}$.  For a relativistic particle
species whose particles satisfy $E\approx k$, the equation of state is
$w=1/3$.  The particle number density is diluted as it is for matter,
and the wave-number $k$ is also red-shifted by $R^{-1}$ to give the
final result $\rho_{radiation} \propto R^{-4}$.  

For both matter and radiation, the acceleration
equation~(\ref{inflation-accelerationeq}), implies a decelerating
universe.  We then notice the curious, yet remarkable fact that for a
fluid with an equation of state $w<-1/3$, the acceleration equation
gives a positive acceleration.  Additionally, in the specific case
$w=-1$, Eq.~(\ref{inflation-energydensity}) tells us the energy
density is constant.  For a constant energy density, Friedmann's
equation~(\ref{inflation-friedmann}) requires the Hubble parameter to
be constant, and the solution is an exponential expansion $R\propto
e^{Ht}$.

One simple way to achieve the equation of state $w=-1$ is with
Einsteins Cosmological constant which gives an exactly constant vacuum
energy $\rho_{vac}$ in Friedmann's equations.  This mechanism is not
viable for inflation since it is not clear how one would end the
inflationary stage and begin a radiation dominated era necessary for
the hot big bang.  \footnote{Additionally, there are no cosmological
  density perturbations generated in a pure de~Sitter background.}
The choice that Guth made was a Lorentz invariant false vacuum state
\cite{guth-1981} which would mimic a cosmological constant, and yield
the equation of state $w=-1$.  However, he recognized there were
problems with this picture with regard to the conclusion of the
inflationary stage.  The true vacuum would develop through bubble
nucleation from the false vacuum, and most of the latent heat for the
decay would be concentrated in the bubble walls.  The entropy of the
universe would thus need to be created through collisions of the
bubble walls.  One problem was that assuming enough collisions occurs,
they produce too large density perturbations, and a second larger
problem is that in order to obtain enough inflation to solve the
horizon and flatness problems, one requires the false vacuum decay to
be so slow that the collisions would in fact rarely occur
\cite{guth-1981}.  Shortly after Guth's paper, these problems were
resolved in the general context of a scalar field evolving on a
potential and achieving $w\simeq-1$ for the initial stage of the
motion \cite{new-inflation,chaotic-inflation}.  We explain this latter
case next as it is simpler and also the commonly accepted paradigm.
Specifically, the inflation model we assume is the chaotic inflation
model of Linde's~\cite{chaotic-inflation}.
We first write the action for a single real scalar field, the proposed
inflaton, which is minimally coupled to gravity,
\begin{equation}
S = \int d^4x \sqrt{-g}\left[\frac12 g^{\mu\nu}\der_\mu\chi \der_\nu\chi + V(\chi)\right]
\end{equation}
Note the inflaton field will be denoted by $\chi$ here and in the
remaining sections.  The classical equations of motion for the
inflaton are obtained to be,
\begin{equation}
\ddot\chi -\nabla^2\chi+ 3 H \dot\chi = -\frac{\der V}{\der\chi} \hspace{5em} H=\frac{\dot R}{R}
\label{inflation-scalarfieldeqs}
\end{equation}
One may also show that the energy and momentum tensor for the scalar
field is,
\begin{equation}
T_\mu^\nu = \chi,_\mu\chi,^\nu - \left(\frac12 \chi,_\ga\chi^\ga - V(\chi) \right)\gd_\mu^\nu
\end{equation}
which may be put in the form of the energy momentum tensor for a perfect fluid
\begin{eqnarray}
T_\mu^\nu = (\rho + p)U_\mu U^\nu - p \gd_\mu^\nu 
\;\;\;\mbox{with}\;\;\;
\left\{\begin{array}{l}
\rho=\frac12 \chi,_\ga\chi,^\ga + V(\chi) \\
p = \frac12 \chi,_\ga\chi,^\ga - V(\chi) \\
U^\mu = \frac{\chi,^\mu}{\sqrt{\chi,^\ga\chi,_\ga}} \;\;\mbox{valid when }\chi,^\ga\chi,_\ga > 0 \\
\end{array}\right.
\label{inflation-emtensorscalar}
\end{eqnarray}
with energy density $\rho$ and pressure $p$ and where $U^\mu$ defines
the local four-velocity of the fluid.  In the case where the scalar
field has only time dependence, $\chi,_\ga\chi,^\ga={\dot\chi}^2$, we
notice from the expressions of the pressure and energy density, that
the equation of state $p=-\rho$ is approximately obtained if the
condition $\frac12\dot\chi^2 \ll V(\chi)$ is satisfied.  In this case,
the Friedmann equation approximates,
\begin{equation}
H_I^2\approx \frac{8\pi G}{3}V(\chi_I) \;\;\;\mbox{when}\;\;\; \left.\dot\chi^2\right|_{\chi=\chi_I} \ll V(\chi_I)
\end{equation}
We want the equation of state $w\simeq-1$ to be satisfied for as long a
time as possible to solve the above mentioned problems in cosmology,
so we impose another condition that the second derivative of the field
also be small, or looking at the equation of
motion~(\ref{inflation-scalarfieldeqs}), the condition is $\ddot\chi
\ll H\dot\chi$ which then implies $3H\dot\chi=-V'$.  Both conditions
on $\dot\chi$ and on $\ddot\chi$ can then be compactly written using
two dimensionless parameters known as the slow-roll parameters
$\epsilon$ and $\eta$ defined as follows,
\begin{eqnarray}
{\dot\chi}^2 \ll V
\;\;&\rightarrow& \;\;
\epsilon \equiv \frac12 M_p^2 \pfrac{V'}{V}^2 \\
\ddot\chi  \ll H\dot\chi
\;\;&\rightarrow&
\eta \equiv M_p^2 \pfrac{V''}{V}
\label{inflation-slowroll}
\end{eqnarray}
with inflation being obtained when $\ge\ll 1$ and $|\eta|\ll 1$.  The
convenience of these parameters is that they indicate when inflation
is obtained given only information about the potential, and where the
field is sitting on the potential.  

The above represents a basic picture of inflation.  Before moving on
to discuss quantum fluctuations during inflation, we return to
quantify inflation's solution to the horizon problem and also to make
some comments on the duration of inflation.
In the context of inflation, the horizon problem is solved by putting
the largest observable scales today within the event horizon during
inflation so that these regions are in causal contact.  The
cosmological horizon length today is approximately
$H_{today}^{-1} \sim 10^{26}\;m \approx 10^4\,\mbox{Mpc}$.  Observers
today are thus able to see fluctuations up to this approximate length
but not greater.  This length $L \simg H_{today}^{-1}$ may be
evolved backwards in time in proportion to the scale
factor.\footnote{Fluctuations of this length and larger are frozen
  from the time they exitted the horizon during inflation to the time
  they re-enter the horizon today.  They change only by being
  stretched with the scale factor.}  The minimum number of e-folds of
the scale factor during inflation required to put this length scale
within the event horizon is obtained from the expression,
\begin{equation}
\frac{H_{today}}{H_I^*} = \frac{R_{RH}}{R_{today}}\frac{R^*_I}{R_{RH}}
\end{equation}
We define $R_{RH}$ as the scale factor at the start of a radiation
dominated universe which is here assumed to also be the end of
inflation (though see Section~\ref{sec-reheating}). We define $R_I^*$
as the scale factor during inflation when our scale of interest has
exitted the horizon.  We then parametrize $\frac{R^*_I}{R_{RH}}\equiv
e^{N_{efolds}}$ where $N_{efolds}$ has been defined which is the
number of e-folds to the end of inflation.  For the remaining ratio
above, we can use $\frac{R_{RH}}{R_{today}}\approx
\frac{T_{today}}{T_{RH}}$ where $T_{today}\approx 10^{-13}\,\gev$ is
the measured temperature of the CMB today.  Taking the logarithm of
our equation and reorganizing the fractions to put the unknown
quantities $T_{RH}$ and $H_I^*$ together, one obtains,
\begin{eqnarray}
N_{efolds} &\sim& 66 - \log\pfrac{T_{RH}}{H_I^*} \label{inflation-60efolds}
\end{eqnarray}
where we have used $H_{today}\approx 2.33 \cdot 10^{-42}\,\gev$.  For
the number of e-folds of observable inflation, we will hereafter use
$N_{efolds}\approx 60$.

A final ingredient to our later analysis is the duration of inflation.
One may determine this time by rewriting the slow roll approximation
$3H \dot\chi = -V'$ in a form that may be integrated,
\begin{equation*}
\int_{t_{init}}^{t_{final}} dt = \int_{\chi_{init}}^{\chi_{final}} \frac{-3 H}{V'} d\chi
\end{equation*}
Assuming a quadratic potential, one obtains,
\begin{equation}
\Delta t = 3\frac{(H_{init}-H_{final})}{m_\chi^2}
\label{inflation-duration1}
\end{equation}
One must specify the initial and final Hubble parameters which are
fixed by the initial and final values of the field.  It is reasonable
to choose $\chi_{init}\sim \sqrt{\frac{M_P}{m_\chi}}M_P$ since
beginning with fields larger than this results in a situation where
the quantum fluctuations of the inflaton dominate over the classical
evolution, and inflation is then ``eternal'' \cite{eternal-inflation}.
With this choice, we have $\frac{H_{init}}{H_{final}} \sim
\frac{\chi_i}{\chi_f} \sim \sqrt{\frac{M_p}{m_\chi}}$, and using
$H_{init} \mg H_{final}$ we approximate the duration of inflation to
be,
\begin{equation}
\Delta t 
\sim \frac{H_{final}}{m_\chi^2}\sqrt{\frac{M_p}{m_\chi}}
\label{inflation-duration2}
\end{equation}
This concludes our development of the classical theory of inflation,
and we now consider quantum fluctuations.

         \section{Quantum Fluctuations of Light Scalar Fields During Inflation} \label{sec-inflation-quantum} 
Neglecting quantum fluctuations, the effect of inflation on any other
field is to stretch the profile of the field and dilute its energy
density to negligible amounts.  However, the story is quite different
when one considers the quantum fluctuations of the field.
Specifically, a large quantum fluctuation of a field may exist for a
short period of time in accordance with the uncertainty relations, and
for fluctuations whose wavelength are far inside the event horizon,
the field will see space as approximately Minkowskian.  However the
inflation will stretch this fluctuation, and once the perturbation's
wavelength has stretched past the event horizon, the fluctuation can
no longer propagate -- it is frozen.  Hence, if the horizon scale is
small enough -- which is to say if the Hubble parameter is big enough
-- then the mode will be frozen at a relatively larger value.  This
effect is known to occur for scalar fields with small mass, for the
scalar perturbations of the inflaton field (which have a small
effective mass), and also for the transverse part of the metric tensor
perturbations. \footnote{It is the equation of motion of the field
  which governs whether the field survives with non-negligible value
  at the event horizon.  The equation of motion for the metric
  perturbations can be put in the same form as the equations for a
  scalar field.\cite{rubakov-etall-1982}} The result of the inflaton
perturbations are the cosmological density perturbations which seed
the structure formation of the universe and leave their imprint on the
CMB.  The results of the tensor perturbations are primordial
gravitational waves which may also leave an imprint on the CMB. These
latter two topics are not the focus of this dissertation, but they
will be mentioned again briefly when we discuss phenomenological
bounds placed on inflation from CMB observations.  The case we now
consider is the first case mentioned above of a light test scalar
field during inflation.

We wish to study the evolution of the scalar field $\phi(x,t)$ during
and after inflation for some patch of the universe.  If the patch
width $L$ is taken to be the scale of the cosmological horizon at some
time after inflation, then an observer in this patch will have no
knowledge of long wavelength fluctuations larger than his/her horizon.
The sum of all these long wavelength fluctuations will appear to the
observer as a uniform field which is a Vacuum Expectation Value (VEV)
for the field.  Additionally, the VEV is in general different for
different patches owing to the random quantum fluctuations which
seeded these VEVs.  Sampling over the whole ensemble of patches, one
determines a mean of zero $\langle\phi\rangle=0$, but the variance is
not zero $\langle\phi^2\rangle \ne 0$, and the variance will be the
relevant measure of the amplitude of fluctuations.\footnote{Higher
  correlation functions could be appreciable if their are large cubic
  or higher order terms in the potential $V(\phi)$.  For the moment,
  is assumed these terms are suppressed, so all statistical information
  is contained in the variance -- the statistics are Gaussian.
  Non-Gaussianity will be discussed in
  Section~\ref{sec-inflation-vevestimate}.}  In our problem, the patch
of interest is the observable universe today, but our computation of
the variance will be done at a time during inflation.  Knowing that
any fluctuation is frozen after it exits the event horizon during
inflation and remains frozen until this fluctuation later returns
within the Cosmological horizon after inflation, we may simply perform
our sum at the instant during inflation when the scales of interest
have exitted the event horizon.  The largest observable fluctuations
today were determined to have exitted the horizon during inflation
approximately 60 e-folds before the end of
inflation~\myref{inflation-60efolds}.  Thus, our computation will be
done at this time $t_{60}$, and the value of $L$ at this time will
simply be the inverse Hubble parameter $L = H^{-1}_{60}$.  Finally, as
should be clear from the above, when we determine the variance, the
sub-horizon wavelengths will not be included in the sum.  The variance
will then be denoted $\langle \phi^2\rangle_{kL\siml 1}$ indicating
that only wavelengths $\frac{2\pi}{k} \simg L$ are being summed.

To perform the computation we change to conformal time $dt = R d\eta$,
and we redefine the scalar field $\phi=\frac{\gvf}{R}$ and work with
the rescaled field $\gvf$.  We express $\gvf$ in the momentum space
decomposition outlined in the previous section (and in
Appendix~\ref{calc-bogtransform-general}), which again is,
\begin{equation}
\gvf(x,\eta) = \int \frac{d^3k}{(2\pi)^{3/2}}\left( e^{ik\cdot x} h_k(\eta) a + e^{-ik\cdot x}h^*_k(\eta) a^\dagger\right)
\label{inflation-modeexpansion}
\end{equation}
where in our conformal coordinates, the momentum is related to the
physical momentum by $\mathbf{k}_{phys}=\frac{\mathbf{k}}{R}$.  Using
the above mode expansion, the variance is determined to be (see
Appendix~\ref{calc-variance}),
\begin{equation}
\langle \gvf^2 \rangle_{kL \siml 1} \approx \frac12 \int_0^{L^{-1}} \frac{d^3k}{(2\pi)^3} |h_k|^2
\label{inflation-variance}
\end{equation}
One must now determine the time evolution of the mode functions
$h_k(\eta)$ which follow from the field equations.  The field
equations for $\phi$ are the same as
in~(\ref{inflation-scalarfieldeqs}), though now written in terms of
rescaled fields $\gvf$ and in conformal time units as follows,
\begin{equation}
\gvf'' - \nabla^2 \gvf + \left(m^2 R^2 - \frac{R''}{R} \right)\gvf = 0
\label{inflation-scalarfieldeqs2}
\end{equation}
upon substitution of (\ref{inflation-modeexpansion}), the mode
functions $h_k(\eta)$ will satisfy the equation,
\begin{equation}
h_k'' + \go_k^2(\eta)h_k =0 
\;\;\;,\;\;\; 
\go_k^2(\eta) = \left(k^2+m^2 R^2(\eta) - \frac{R''}{R}\right)
\label{inflation-modeequation1}
\end{equation}
which is the equation for an oscillator with time dependent frequency
as was studied in the previous section.  To model the inflation in the
slow roll regime, the metric is approximately de~Sitter,
\begin{equation}
ds^2 = R^2 (d\eta^2-d{\vec x}^2)
\;\;,\;\;
R = e^{H_I t} =  \frac{-1}{H_I \eta}
\end{equation}
where $H_I$ is the Hubble parameter during inflation which is
approximately constant.  The equations for the
modes~(\ref{inflation-modeequation1}) with the above metric is,
\begin{equation}
h_k'' +\left[k^2 + \left(\frac{m^2}{H_I^2}  - 2\right)\frac1{\eta^2}\right] h_k =0
\label{inflation-modeequation2}
\end{equation}
This may be put in the form of Bessel's equation, and the solutions
$h_k(\eta)$ thus determined (see
Appendix~\ref{calc-variance}).  The solutions must be
canonically normalized
at early times (or large k) where the mode functions are well
approximated by simple plane waves.  Specifically, the
equation~(\ref{inflation-modeequation2}) for the modes at sub-horizon
scales, asymptotes to $h_k''+k^2h_k=0$ with solutions $e^{\pm ik\eta}$.
Referring to~(\ref{inflation-modeexpansion}), we want $h_k(\eta) =
\frac{1}{\sqrt{2k}}e^{-ik\eta}$ at early times for our fields to be
canonically normalized.  The normalization is obtained to be,
\begin{equation}
h_k(\eta) = \sqrt{\frac{-\eta\pi}{4}}e^{i\left(\frac{n\pi}{2}+\frac{\pi}{4} \right)} \left(J_n(-k\eta) + i Y_n(-k\eta)\right) 
\;\;\;,\;\;\; 
n^2= \frac94 - \frac{m^2}{H^2}
\end{equation}
These solutions are considered in the limit of small mass $m<H_I$
where $n^2>0$, and large mass $m>H_I$ where $n^2<0$.  The
variance~\myref{inflation-variance} is converted back in terms of the
physical momenta $k_{phys}=\frac{k}{R}$, and in terms of the original
fields $\phi=\gvf/R$, and the integration is performed over the
dimensionless variable $x=\frac{k_{phys}}{H_I}$ in the long wavelength
limit $x<1$.  The results are,
\begin{equation}
\langle\phi^2 \rangle_{k_{phys}<H_I}
\approx \left\{ \begin{array}{ll}
H_I^2 \pfrac{2^{(2n-5)}\gG(n)^2}{\pi^4}\int_0^1 \frac{dx}{x} x^{\pfrac{2m}{3H_I}^2}  &\mbox{for } m \ll H_I \\[0.5em]
H_I^2\pfrac{1}{6|n|(2\pi)^3} & \mbox{for } m \mg H_I
\end{array}\right.
\label{inflation-varianceresult}
\end{equation}
Notice the result~\myref{inflation-varianceresult} is scaling with the
Hubble parameter as $H_I^2$ which is a measure of the vacuum energy
density during inflation, $H_I=\frac{8\pi G}{3}\rho_\phi$.  Thus, the
higher the energy scale of inflation, the larger the variance in our
test field.
However, in the case of a large mass, the
variance~\myref{inflation-varianceresult} is suppressed by a factor
$\frac1{n}\approx \frac{H_I}{m}$.  The reason for the suppression is
that a massive field can only propagate a distance of order
$\frac1{m}$, so the field will have been damped to small values long
before it ever reaches the de~Sitter event horizon.  In the case of a
small mass, the integration in~\myref{inflation-varianceresult} is
logarithmically divergent in the limit $m\rightarrow 0$, but with
finite and small $m$, yields the result,
\begin{equation}
\Phi_{rms} =\sqrt{\langle\phi^2\rangle_{k_{phys} \siml H_I}} \approx H_I\pfrac{H_I}{4\pi^{3/2}m}
\label{inflation-variancesmallmass}
\end{equation}
To understand the source of the logarithmic divergence as
$m\rightarrow 0$, one must consider the time evolution of the VEV
during inflation.  The results~\myref{inflation-varianceresult} were
obtained assuming a de~Sitter metric which implies an infinite
duration of inflation during which the mode functions $h_k(\eta)$ are
fixed points of the de~Sitter evolution, and thus the variance at a
given length scale is invariant in time.\footnote{See
  Appendix~\ref{calc-variance} and also the discussion in
  reference~\cite{mukhanov-cosmology-2005}} If one takes into account
that there was an initial time $t_0$ in which $\Phi$ is zero over all
the patches in the universe, the time dependence of the variance is
seen to have the form~\cite{VEVgrowth},
\begin{equation}
\pfrac{3 H_I^4}{8\pi^2 m^2}\left(1-\exp\left[-\pfrac{2 m^2}{3 H_I}(t-t_0) \right]\right)
\label{inflation-variancelinde}
\end{equation}
which, at later times $(t-t_0) \mg \frac{H}{m^2}$, asymptotes to the
results~(\ref{inflation-varianceresult}-\ref{inflation-variancesmallmass})
differing only by a factor of order one.  In the case of early times
$t \ml \frac{H}{m^2}$ the variance is proportional to
$\langle\phi^2\rangle \propto H_I^3 (t-t_0)$ which is also valid in
the limit of zero mass.  The time evolution is in fact that of a
random walk for the field VEV. 
In the case inflation lasts a sufficiently long time, the logarithmic
divergence in the variance would emerge as a result of a random walk
over an indefinite number of time steps.  However in this case, the
VEV may instead be cutoff by higher order terms in the fields
potential $V(\Phi)$ as discussed in \cite{ellis-etall-1987}.
Specifically, a reasonable guess for the flat direction potential is
the series expansion defined in \myref{susyflat-potentialseries} which
again is,
\begin{equation}
V(\Phi) = m^2 |\Phi|^2 + V_{n>3}(|\Phi|) \;\;\;,\;\;\;V_{n>3}(|\Phi|) \equiv \sum_{n=4}^\infty |\lambda_n|^2 M_P^4 \left|\frac{\Phi}{M_P}\right|^{2(n-1)}
\label{inflation-potentialseries}
\end{equation}
where the heavy scale in the expansion is assumed to be the Planck
mass and the coefficients are assumed $\lambda_n\sim 1$.  If the
variance becomes large enough that the higher order terms in the
potential are comparable to the mass term, then the variance
results~(\ref{inflation-varianceresult}-\ref{inflation-variancelinde})
become a poor description of the distribution.  Higher order
correlation functions become large, imparting potentially large
deviations from gaussianity.  We do not attempt to calculate these
higher order n-point functions.  However, the behavior of the
distribution's width as a function of time may still be approximated.
Assuming the flat direction VEVs begin at some time during inflation
with negligible Vacuum Expectation Value $\Phi(t_0)=0$, the
distribution will be a Gaussian and its width (variance) will grow
according to the early time behavior
of~\myref{inflation-variancelinde}, and it will continue to grow until
either the asymptotic value is reached or until the lowest
non-vanishing higher order term in the
potential~\myref{inflation-potentialseries} becomes comparable to the
mass term.  In the latter case, the distribution begins to acquire a
nongaussian component due to these higher order terms and the
evolution ceases to follow~\myref{inflation-variancelinde}.  One may
guess the width of the distribution will remain essentially unchanged
after this time, but with the shape deviating further away from the
gaussian.  By this argument, the width of the distribution is
approximately,
\begin{equation}
\frac{\Phi_{\mathsubscript{width}}}{M_P} \sim \pfrac{m}{nM_P}^{\pfrac{1}{n-2}}
\label{inflation-vevwidth}
\end{equation}
which was obtained by equating the lowest non-vanishing higher order
term in the potential to the mass term, and dropping factors of order
one.  An alternative approach for estimating the width of the
distribution is to split the field into a classical background and a
fluctuation, $\phi=\Phi+\gd\phi$ where the fluctuation $\gd\phi$ is
quantized.
In this argument (outlined in the Introduction), one would expect the
growth of $\Phi$ to stop once the effective mass of the perturbations
becomes the same order as the Hubble parameter,
$m_{\mathsubscript{eff}}\sim H_I$ so that the quantum fluctuations
cannot propagate past the event horizon and thus the accumulation of
the VEV will stop.  Specifically, the equation for the fluctuations
$\gd\gvf$ are the same as~\myref{inflation-scalarfieldeqs2}, but with
an effective mass, $m_{\mathsubscript{eff}}\approx\left. \frac{d^2V}{d\phi^2}
\right|_{\gvf=\Phi}$, where the mass $m$ has been neglected.
One then approximates $m_{\mathsubscript{eff}}$ using the leading non-vanishing term
in the series expansion~\myref{inflation-potentialseries}.  This
argument leads to the following bound for $\Phi$,
\begin{equation}
\frac{\Phi^{\mathsubscript{hard}}_{\mathsubscript{bound}}}{M_P} \sim \pfrac{H_I}{nM_P}^{\frac{1}{n-2}}
\label{inflation-vevbound}
\end{equation}
which is also bounded below $M_P$ but larger
than~\myref{inflation-vevwidth} for a give value of $n$ (see
Figure~\ref{fig-vevbound}).  Either of the above
estimates~\myref{inflation-vevwidth}~or~\myref{inflation-vevbound} for
the width are acceptable for our purposes, though the second estimate
is established when the mass of the field can be neglected against the
higher order terms so it implies the distribution has evolved longer
in time compared to the first case.  Thus it is a larger bound.  The
latter bound may also be written in terms of the energy densities of
the flat direction and inflaton as follows,
\begin{eqnarray}
\rho_{\Phi\,\mathsubscript{bound}} &\sim& \rho_\chi \frac{1}{n^2}\pfrac{H_I}{nM_P}^{\frac{2}{n-2}}
\label{inflation-rhophibound}
\end{eqnarray}
where the quadratic term of the potential has been neglected in
obtaining this.  From this expression it is clear that $\rho_\Phi$ is
bounded by $\rho_\chi$ which is expected.  For instance, taking
$H_I\sim 10^{-6}M_P$, the ratio of the energy densities is maximized
when $n\sim 20$ where $\frac{\rho_{\Phi}}{\rho_\chi}\sim
10^{-4}$. \footnote{The approximation in~\myref{inflation-rhophibound}
  gives $\frac{\rho_{\Phi}}{\rho_{\chi}}\rightarrow 0$ in the limit of
  large $n$.  This is expected, since in this limit, one expects the
  quadratic term of the potential to dominate, and thus obtain
  $\rho_\Phi = m^2 M_P^2$.  Note the quadratic term, although
  neglected in
  obtaining~\myref{inflation-vevbound}~and~\myref{inflation-rhophibound},
  is still present.}  Note that since the quantum fluctuations cannot
propagate past the horizon once $\Phi^{\mathsubscript{hard}}_{\mathsubscript{bound}}$ is reached, it
is essentially a hard bound beyond which larger values of the VEV are
impossible to obtain.
Finally, note that neither of the above two bounds on the distribution
says anything about the shape of the distribution which is presumably
very different than a gaussian in both cases.~\footnote{One example of
  a strongly nongaussian distribution is one which develops a minimum
  at the origin and a peak at some nonzero value of $|\Phi|$.  If the
  peak is sufficiently narrow, then in fact the field has been driven
  to an attracting surface (or fixed points if it is a real field)}

There is one final consideration to be made which is that the Hubble
parameter is not a constant during inflation, but decreases slowly as
the inflaton field rolls down the potential.  It is thus inappropriate
to use a single value for $H_I$ in the expressions for the
variance~(\ref{inflation-variance}-\ref{inflation-variancelinde}).
The calculations may be adjusted to take this into account, and if
this is done, the infrared divergence noted above will return as the
spectrum will have obtained a ``red'' tilt
\cite{riotto-inflation-2002}.  One must therefore impose an infrared
cutoff to the corresponding momentum integral which physically can
correspond to a finite duration of inflation.
This calculation is not performed here.  However, let us make some final
crude estimates to take into account the duration of inflation.  From
the result~\myref{inflation-variancelinde}, one sees that a time
$\Delta t_{\mathsubscript{asymptotic}}\simg \frac{H_I}{m^2}$ is necessary for the
asymptotic value~\myref{inflation-variancesmallmass} to develop.
Assuming zero temperature, the duration of inflation was determined
for a quadratic potential in~\myref{inflation-duration1} to be $\Delta
t_{\mathsubscript{inflation}} \sim
\frac{H_I}{m_\chi^2}\sqrt{\frac{M_P}{m_\chi}}$. Taking $H_I$ to be the
Hubble parameter at the conclusion of inflation, and comparing these
two times,
\begin{eqnarray*}
\frac{\Delta t_{\mathsubscript{asymptotic}}}{\Delta t_{\mathsubscript{inflation}}} &\sim& \pfrac{m_\chi}{m}^2\sqrt{\frac{m_\chi}{M_P}} \\
&\sim& 10^{17}
\end{eqnarray*}
where the Hubble parameter during inflation has been fixed by the COBE
normalization to $H_I\sim10^{-6}M_p$ assuming a potential
$V(\chi)=\frac12 m^2\chi^2$ (see Section~\ref{sec-inflation-cobe}).  The
above estimate suggests that inflation may not last long enough for
the asymptotic variance~\myref{inflation-variancesmallmass} to be
reached.  The variance will thus approximately grow as $H_I^3\Delta t$
through the whole evolution reaching the following value at the end of
inflation.
\begin{eqnarray}
\frac{\Phi_{RMS}}{M_P} &\sim& \sqrt{H_I^3\Delta t_{\mathsubscript{inflation}}} \nonumber \\
 &\sim& \pfrac{H_I}{M_P}^2\pfrac{M_P}{m_\chi}^{5/4} \nonumber \\
&\sim& 10^{-4}
\label{inflation-durationbound}
\end{eqnarray}
At the conclusion of Section~\ref{sec-inflation-classical}, we
determined that the Hubble parameter at the start of inflation was a
factor $\sqrt{\frac{M_P}{m_\chi}}$ larger than at the conclusion of
inflation.  If we were to take this into account carefully as
discussed above, one would expect the above estimate of $\Phi_{RMS}$
to increase.  Also note that this approximation assumes that the
variance for $\Phi$ at the start of inflation is fixed at zero, but
this is not guaranteed especially in the context of eternal inflation
\cite{eternal-inflation}.  Vacuum expectation values which have grown
near to the two
bounds~\myref{inflation-vevwidth}~and~\myref{inflation-vevbound} are
thus not impossible.

         \section{Estimating the Distribution of VEVs} \label{sec-inflation-vevestimate} 
The previous discussion was to explain the mechanism of the growth of
flat direction VEVs.  We gathered that two limiting factors could be
(a) the duration of inflation and (b) the higher order terms of the
flat directions potential~\myref{inflation-potentialseries}.  It was
finally noted that our estimates of the
variance~(\ref{inflation-variance}-\ref{inflation-variancelinde}) did
not take into account that the Hubble parameter changes from the
beginning of inflation to the end of inflation making these results
incomplete.  However, our estimates for $\Phi_{width}$ and
$\Phi_{bound}$ are still applicable.

\begin{figure}[th]
\begin{center}
\includegraphics[scale=0.7]{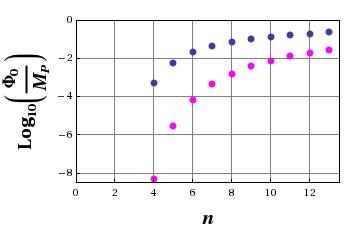}
\caption[Estimating the Distribution of VEVs Obtained During
Inflation]{\label{fig-vevbound} Estimates of the width (lower curve) and
  absolute upper bound (upper curve) for the distribution of flat
  direction VEVs as a function of the lowest nonzero term in the
  superpotential indexed by $n$
  (see~\myref{susyflat-potentialseries}~and~\myref{inflation-potentialseries}).
  The width and upper bound curves are obtained from
  formulas~\myref{inflation-vevwidth}~and~\myref{inflation-vevbound}.
  In the above $m\sim 10^{-16}M_P\sim TeV$ and $H\sim 10^{-6}M_P$
  which is fixed by the COBE normalization for a quadratic inflaton
  potential (See Section~\ref{sec-inflation-cobe})}
\end{center}
\end{figure}

Hereafter, we assume that higher order terms $V_{n>3}$ of the
potential~\myref{inflation-potentialseries} are indeed present and
that the distribution of VEVs has evolved so that the
bounds~\myref{inflation-vevwidth} and~\myref{inflation-vevbound} are
appropriate.  For instance, these bounds may be easily obtained if
inflation is eternal.  Both bounds are shown in
Figure~\ref{fig-vevbound} as a function of the lowest non-vanishing
term in the superpotential indexed by $n$ (see
\myref{susyflat-liftingindex}).

Note that to keep the discussion in the previous section as simple as
possible, the calculations were performed assuming a real scalar
field.  The MSSM flat directions however are complex scalars. The
above
results~\myref{inflation-vevwidth}~and~\myref{inflation-vevbound} are
still valid for a complex field.  However a complex scalars may also
be charged under some symmetry.  In the case of the MSSM flat
directions, this charge is typically the difference in the baryon and
lepton numbers, $B-L$ and the symmetry is the conservation of the
$B-L$ charge.  Our estimates of the variance would apply to the
modulus of the complex field.  One expects the phase of the field to
be randomized as well, though the distribution need not be uniform on
the circle if $C$ and/or $CP$ symmetries are explicitly violated in
the theory.

Another consideration is that there are many MSSM flat directions, and
if some set of them obtains large VEVs during inflation, then this
may exclude another complementary set of flat directions from
obtaining VEVs as a consequence of the large induced mass.  Thus,
when the quantum fluctuations of the flat directions are generated
during inflation, there will be competition between mutually exclusive
flat directions.  If inflation lasts long enough, some set of the flat
directions may have obtained large VEVs at the exclusion of a
complementary set.\footnote{since the effective masses of the
  complementary set have become large enough to suppress the
  propagation of their fluctuations through the horizon} As the
process is random, some patches of the universe may be dominated by
one set of VEVs and another patch may be dominated by a complementary
set, and these may smoothly interpolate with patches with small or no
VEVs at all.  
A comprehensive analysis of the resulting distributions of VEVs would
be a difficult task, and far beyond the scope of this work.  However
for our purposes, it is sufficient for us to have made an estimate for
the approximate width of these distributions as quantified in
Figure~\ref{fig-vevbound}.  

What is relevant to our models is not the distribution of VEVs at the
conclusion of inflation, but the distribution at a later time when
$H\siml m$, and in particular the distribution's tail.  Our models
will describe the evolution just after $R_{H\sim m}$.  The motion from
the end of inflation to this instant may in fact alter the shape of
the distribution's tail.  These issues are discussed next.

         \section{Post-Inflation Classical Evolution of Flat Direction and Inflaton VEVs} \label{sec-inflation-vevevolution} 
In the previous two sections, the focus was on the statistical
distribution of VEVs obtained by a flat direction by the conclusion of
inflation. In particular, both a width and absolute upper bound to the
distribution was estimated
in~\myref{inflation-vevwidth}~and~\myref{inflation-vevbound} and shown
in Figure~\ref{fig-vevbound}.  Because this thesis is focused on the
fate of very large flat direction VEVs, we will consider only the
range of initial VEVs from this distribution falling between these two
bounds and we study their subsequent classical evolution after
inflation has ended.
In general, the distribution of VEVs valid at the end of inflation
will be transformed to a slightly different distribution at the
instant when $H\sim m$.  After this instant, the classical evolution
has a relatively simple form.  This state of the evolution is also the
stage relevant to our models.
We then discuss the early generation and subsequent conservation of
the $B-L$ charge belonging to the flat direction fields.  Discussion
of the quantum effects (decay) on the inflaton and flat directions
after inflation is postponed to Section~\ref{sec-reheating} which
builds upon the description here.  Elements of the following
discussion can also be found in
\cite{affleck-dine-1985,ellis-etall-1987, dine-randall-thomas-1995b,
  dine-kusenko-2003}.

Recall from Section~\ref{sec-inflation} that inflation ends when the
inflaton has evolved to a location on its potential when the slow-roll
conditions~(\ref{inflation-slowroll}) are no longer valid.  Thus, the
inflaton field begins to move in the sense ${\dot\chi}^2\sim V(\chi)$,
and begins oscillations according to the equation of
motion~(\ref{inflation-scalarfieldeqs}) and the Friedmann equation.
The spatial gradient term is negligible compared to the other terms
initially, so the inflaton remains a homogeneous field over the
horizon volume as its oscillations begin.  The solutions are damped
oscillations (see Appendix~\ref{calc-inflatonoscillation}) about
$\chi=0$ with frequency $m_\chi \sim 10^{-6}M_P$,
\begin{equation}
\chi(t) = \frac{M_p}{\sqrt{3\pi}m_\chi t} \sin(m_\chi t) + \cO\left(\frac{M_P}{(m_\chi t)^2} \right)
\label{inflation-inflatonoscillations}
\end{equation}
From this, one calculates the energy density of the inflaton in terms
of the scale factor $R$ and one sees the evolution to be that of a
matter dominated universe,
\begin{equation}
\rho_\chi \approx m_\chi^2M_p^2 \pfrac{R_0}{R}^3 
\label{inflation-rhochi}
\end{equation}
where $R_0$ is the scale factor at the onset of inflaton oscillations
or equivalently, the end of slow roll inflation, and where the
prefactor has been dropped in order to simplify later calculations.
The behavior of the energy density as a function of the scale factor
$\frac{R}{R_0}$ is shown in Figure~\ref{fig-rhoevolution}.
%
\begin{fullpagefigure}
\begin{center}
\mbox{}

\vspace{-3em}
\includegraphics[scale=0.8]{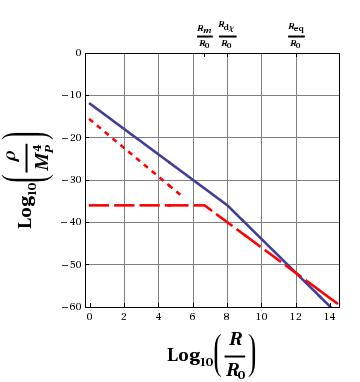}
\caption[Evolution of the Energy Density of the Inflaton and Flat
Direction]{\label{fig-rhoevolution}Evolution of the energy density of
  the inflaton (solid curve) and a flat direction VEV (long dashes) at
  the conclusion of inflation.  The horizontal axis is Logarithmic in
  the scale factor where $R_0$ is the scale factor at the onset of
  inflaton oscillations.  The flat direction (long dashes) has been
  given an initial amplitude $|\Phi_0|\sim 10^{-2}M_P$ which
  corresponds to the the lower bound of Figure~\ref{fig-vevbound} with
  $n=10$.  Also shown is a case assuming the same potential but with
  the VEV having reached the upper bound of Figure~\ref{fig-vevbound}.
  In this latter case, the early motion of the flat direction (short
  dashed curve) is dominated by the higher order terms $V_{n>3}$ in
  the potential~\myref{inflation-potentialseries}, and so the initial
  energy density can be orders of magnitude higher.  However, assuming
  the field is in slow-roll, one may argue that $\rho_{\Phi}$ should
  decrease slightly faster than $\rho_\chi$ as shown for example in
  the short dashed curve. As $\Phi$ decreases, the terms $V_{n>3}$ in
  the potential~\myref{inflation-potentialseries} become comparable to
  the mass term, and the evolution is expected to smoothly transition
  between the two dashed curves (transition not shown).  Note the
  analysis of this last case is somewhat heuristic and not by any
  means definite.  Also shown are $R_{d\chi}$, the scale factor at the
  instant of the inflaton's gravitational decay and $R_{eq}$, the
  scale factor at the instant the flat direction would overtake the
  inflaton~\myref{rhoevolution-Req} assuming $\Phi$ has not already
  decayed. For a comparable analysis on which this figure is based
  see~\cite{ellis-etall-1987}}
\end{center}
\end{fullpagefigure}

Note, the inflaton energy density $\rho_\chi$ will dominate over the
energy density of the flat direction initially and thus drive the
Hubble parameter.  The precise evolution of the flat direction is
determined through the flat direction's equations of motion and the
Friedmann equation which are,
\begin{eqnarray}
& \ddot\Phi + 3H\dot\Phi + \frac{d V}{d\Phi} = 0  \label{inflation-flatdireq} \\
& \left(\frac{\dot R}{R}\right)^2 = \frac{8 \pi G}{3}\left( |\dot\Phi|^2 + V(\Phi) + \rho_\chi\right) \nonumber
\end{eqnarray}
where we have assumed one complex degree of freedom for the flat
direction $\Phi$, and where $V(\Phi)$ has the proposed
form~\myref{inflation-potentialseries}.  To develop an understanding
for how the flat direction evolves, consider two different scenarios
for $\Phi_0$ which lie on the two curves of
Figure~\ref{fig-rhoevolution}.  The scenario with $\Phi_0$ on the
lower curve is simpler, so we begin with this.  Recall from the
previous section that in this scenario the mass term is comparable to
the contribution of the higher order terms of the potential,
$m^2\Phi^2\sim V_{n>3}$.  At the conclusion of inflation, the field
will be frozen, but will begin to move when the condition
$\frac{d^2V}{d\Phi^2} \sim H^2$ is satisfied which is then the
condition $m\sim H$.  Thus the flat direction VEV is frozen until the
scale factor $R_{m}$ as shown in Figure~\ref{fig-rhoevolution}.
Another way to state this is that the field value is nearly frozen
until the age of the universe $H^{-1}$ is comparable to the period of
one oscillation of the field $m^{-1}$.  The inflaton will still
be executing damped
oscillations~\myref{inflation-inflatonoscillations} at this instant.
The subsequent motion of $\Phi$ will be discussed shortly.

Next consider the scenario with $\Phi_0$ on the larger
bound~\myref{inflation-vevbound} and thus on the higher curve of
Figure~\ref{fig-vevbound}.  The energy density of the flat direction
in this scenario can be many orders of magnitude above the first
scenario, but still sub-dominant to the inflaton energy density (see
Figure~\ref{fig-rhoevolution}).  Also in this scenario, the VEV will
begin to move immediately at the conclusion of inflation, since
$\frac{d^2V}{d\Phi^2} \sim H^2$ is satisfied immediately.  The exact
motion of $\Phi$ at the onset of inflaton oscillations depends on the
nonlinearities of the potential.  When the field moves, it is
reasonable to assume that the energy density will decrease as fast or
faster than the inflaton.  For instance, if one assumes the field is
in slow-roll (neglecting $\ddot\Phi$ in the equation of motion), then
this may be shown. A situation such as this is depicted in
Figure~\ref{fig-rhoevolution}.
In this case, the energy density would decrease as fast or fast than
the inflaton until the mass term becomes comparable to the higher
order terms of the potential $m^2|\Phi|^2 \sim V_{n>3}(\Phi)$.  At
this instant, the field would reside on the lower curve of
Figure~\ref{fig-vevbound}.  The evolution would then smoothly make a
transition to the case of a mass dominated potential which was
described above.  During this whole phase of the evolution the
assumption is that the field is in slow roll.  What we draw from this
analysis is that initial field values larger than the
width~\myref{inflation-vevwidth} may evolve to approximately the same
field value and energy density at the instant $H\sim m$.

\begin{figure}[th]
\begin{center}
\includegraphics[scale=0.5]{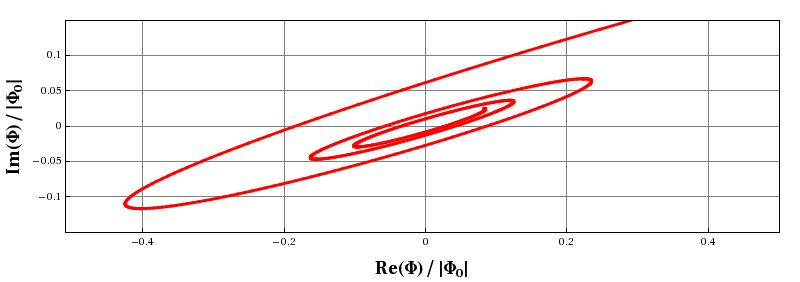}
\caption[Example Evolution of a Flat Direction
Field]{\label{fig-phievolution}Evolution of the complex scalar field
  $\Phi$ for a sample choice of parameters. The parameters chosen are
  $A_i=\frac12$, $A_r=-\sqrt{\frac{15}{8}+\frac{17}{8\sqrt5}}$,
  $\gd_i=\frac{\pi}{10}$, and $\gd_r=\frac{7\pi}{10}$.  The charge
  asymmetry as computed by formula~\myref{inflation-chargeasymmetry}
  is $\ge_{B-L}=\frac18\sqrt{7-\frac{6}{\sqrt{5}}} \approx 0.26$ }
\end{center}
\end{figure}

After this instant $H\sim m$, the equation of
motion~\myref{inflation-flatdireq} is approximately linear in $\Phi$,
and the motion will be that of damped linear oscillator with a time
dependent damping term $H$.\footnote{ The Hubble parameter does in
  fact depend on $\Phi$ as shown in~\myref{inflation-flatdireq}, but
  only very weakly.  Strictly speaking, this is a nonlinear system,
  but it is approximately linear since $H$ is driven mostly by the
  inflaton energy density $\rho_\chi$.}  The solutions are similar to
that of the inflaton,
\begin{equation}
\Phi \approx \pfrac{R_m}{R}^{3/2}\Phi_m\left[A_r\cos(m t + \gd_r) + i A_i\sin(m t + \gd_i) \right] + \cO\pfrac{R_m^3}{R^3}
\label{inflation-flatdiroscillations}
\end{equation}
which evolves with a matter-like energy density,
\begin{equation}
\rho_\phi \sim m_\phi^2 \Phi_m^2 \pfrac{R_m}{R}^3
\end{equation}
and with $A_r$, $A_i$, $\gd_r$, $\gd_i$ being dimensionless
coefficients of the evolution.  Three of these coefficients are fixed
by initial conditions and the fourth is fixed by the normalization,
\begin{equation}
A_r^2 \cos^2\gd_r + A_i^2\sin^2\gd_i = 1
\label{inflation-vevnorm}
\end{equation}
This evolution is shown for a sample set of parameters in
Figure~\ref{fig-phievolution} where one sees the field executes a
elliptic-like trajectory about the origin while spiralling to smaller
values due to Hubble expansion.
%
During this time of damped linear oscillations, the ratio of the flat
directions energy density to that of the inflaton is fixed to the
following ratio (assuming the inflaton and/or flat direction have not
already decayed),
\begin{eqnarray}
\frac{\rho_\Phi}{\rho_\chi} \sim \pfrac{\Phi_m}{M_P}^2
\label{inflation-chiphiratio}
\end{eqnarray}
and we see that so long as $\Phi_m < M_p$, the inflaton will dominate
the total energy density.  If this condition is not satisfied, then
the flat direction will overtake the inflaton even before the flat
direction begins oscillations.  As we have just discussed, the higher
order terms in the potential, if present, will keep the flat direction
VEV bounded below the Planck mass.\footnote{If however the
  potential~\myref{inflation-potentialseries} contains only the soft
  mass parameters and exactly zero higher order terms, and if
  inflation lasts a long enough time, then in principle one may obtain
  $\Phi_m > M_P$, and the flat direction may overtake the inflaton.
  This could result in a second stage of inflation driven by the flat
  direction VEV.  However, this situation cannot lead to large enough
  cosmological perturbations, and is thus not viable.}

One other element to the classical evolution of the flat direction is
the charge associated with the motion of the flat direction.  The
charge of the flat direction is not a standard model gauge charge but
rather it will typically be the difference of lepton number and baryon
number $B-L$.\footnote{Note however that in many Grand Unified models,
  the discrete $B-L$ symmetry may be promoted to a continuous gauge
  symmetry $U(1)_{B-L}$ \cite{martin-susyprimer}.} The field $\Phi$ is
then by definition a gauge singlet field.  As was mentioned in the
previous subsection~\ref{sec-inflation-vevestimate}, the phase of the
flat direction VEV is also randomized during inflation, though the
distribution would presumably depend on the specific breaking of $C$
and $CP$ symmetries of ones model.  Additionally if there exist $B-L$
violating interactions such as those discussed in
Section~\ref{sec-susyflat-susybreaking} which involve the flat
direction fields, then a $B-L$ charge can be generated dynamically
through the flat directions classical equations of motion.  This is
essentially the well known baryogenesis mechanism of Affleck and
Dine~\cite{affleck-dine-1985}.  If inflation has driven the flat
direction VEVs close enough the lower curve of
Figure~\ref{fig-vevbound}, then such nonlinear terms if present can in
principle contribute a phenomenologically relevant $B-L$ charge.
Once the field has reduced in amplitude past $H\sim m$, the charge
will then be conserved.  The charge in this regime is determined from
our solutions~\myref{inflation-flatdiroscillations} as follows,
\begin{eqnarray*}
n_{B-L} &=& (i \Phi^* \dot\Phi - i \dot\Phi^*\Phi) \\
&=& - \pfrac{R_m}{R}^3(m\Phi_m^2)A_i A_r \cos(\gd_i-\gd_r) 
\end{eqnarray*}
and the number charge per co-moving volume $(n_{B-L}R^3)$ is thus
conserved.  Notice, the charge can take positive and negative values
which will correspond to the handedness of the rotation of its
trajectory in the complex plane.  Also notice that given the
constraint~\myref{inflation-vevnorm}, the charge density is extremized
when $(\gd_i-\gd_r)=\pm\pi$ and $A_i=A_r=1$, and in the
case $\gd_i-\gd_r=\pm\pi/2$, the charge is zero.  Defining the ratio
of the $B-L$ number charge density to the total particle density as
the parameter $\ge_{B-L}$, one determines,
\begin{equation}
\ge_{B-L}\equiv \left|\frac{n_{B-L}}{n_{\Phi}}\right| = A_i A_r \cos(\gd_i-\gd_r) 
\label{inflation-chargeasymmetry}
\end{equation}
where this ratio was determined at $R_m$. The quantity $\ge_{B-L}$ is
the $B-L$ number per particle of the condensate, and it is bounded
between $\pm 1$.  In the absence of $B-L$ violation, the net charge
should be zero and the ellipse will have collapsed to a line passing
through the origin.  In the above we have parametrized the motion,
but we have not explained how the charge asymmetry is developed
through the nonlinear terms.  An example of such terms was introduced
in~\ref{susyflat-lambdacouplings}.  To see how a charge is developed,
redefine $\Phi$ such that its initial phase is zero and thus $\Phi$
initially lies on the real axis.  As discussed
after~\myref{susyflat-lambdacouplings} this field redefinition implies $\gl$ is explicitly complex (not on
the real axis).  Also assuming $\dot\Phi=0$ initially, the equation of
motion for the imaginary part of $\Phi$ at the initial time reduces to
\begin{equation}
\ddot\Phi_{\rm{Im}} = - \mbox{Im}[\gl] \Phi_{\rm{Re}}^3
\end{equation}
and thus the field will begin to rotate off the real axis into the
complex plane.  Whether the rotation is clockwise or counter-clockwise
depends on the phase of $\gl$.  In our later numerical simulations we
will instead take $\gl$ to be Real valued and the initial phase to be
nonzero which is an equivalent situation.
The amount of baryon asymmetry (computed later in
Section~\ref{sec-reheating-etab}) is dependent largely on the initial
VEV $\Phi_0$ as well as on the existence of interactions such
as~\myref{susyflat-lambdacouplings}.  At later times when the higher
order interactions such as these become sub-dominant to the mass term,
the built up charge will be conserved to a good approximation
\cite{affleck-dine-1985}.
The $B-L$ asymmetry may also be smaller or larger depending on how the
inflaton and flat directions proceed to evolve.  For instance if the
flat direction energy density can overtake that of the inflaton, the
baryon asymmetry can be order one.  This situation would require very
large sources of later entropy to dilute the $B-L$ charge density to
be phenomenologically viable.  Alternatively, if the inflaton remains
the dominant fluid, the $B-L$ asymmetry obtains a suppression
inversely proportional the amount of entropy released by inflaton
decay.  These issues will be discussed again in
Section~\ref{sec-reheating}.

         \section{COBE Normalization and the Energy Scale of Inflation} \label{sec-inflation-cobe} 
Before continuing with our discussion of the evolution of the flat
direction fields, we make a brief digression to discuss two very
important inputs to our study from observational cosmology.  These are
phenomenological constraints on the Hubble parameter during inflation,
$H_I$ obtained from the Cosmic Microwave background observations.  As
we have seen from the previous section in the
results~(\ref{inflation-variance}-\ref{inflation-variancelinde}) the
Hubble parameter or equivalently the energy scale of
inflation\footnote{since during slow-roll inflation one has
  $H_I^2\approx \pfrac{8\pi}{3M_P^2}\rho_\chi$ with $\rho_\chi \approx
  V(\chi_I)$.} controls the amplitude of the perturbations one obtains
for a light scalar test field present during inflation.  It can be
shown that the amplitudes of the perturbations of the scalar and
tensor modes of the gravitational field (primordial density
perturbations and gravity waves) are also controlled by the Hubble
parameter in a similar way
\cite{cosmological-perturbations,rubakov-etall-1982}.  A large value
for $H_I$ thus implies large cosmological perturbations which may be
in conflict with observations of the CMB.  The cosmological
perturbations are not the focus of this thesis, but we shall briefly
describe a constraint on $H_I$ related to the tensor modes as well as
a constraint between $H_I$ and inflaton's potential known as the COBE
normalization, related to the scalar modes.

The tensor modes are like the scalar test fields just discussed in
that the only physical parameter which appears in their spectrum is
the Hubble Parameter \cite{cosmological-perturbations,
  rubakov-etall-1982}.  In this sense, a signature of primordial
gravitational waves in the CMB is a direct measure of the energy scale
of inflation.  If one supposes the gravitational waves are a sizeable
contribution to the CMB anisotropy, then the measured anisotropy
places an upper bound on $H_I$ which is approximately,
\begin{equation}
H_I \lesssim 10^{-4}M_P
\label{inflation-gravitywavebound}
\end{equation}
where $H_I$ is the Hubble parameter at the end of
inflation.\footnote{This argument can be found in
  \cite{rubakov-cosmology-2005}}  This bound is very conservative as
the primordial gravitational wave contribution to the CMB is expected
to be sub-dominant.

The scalar modes of the gravitational field are proportional to the
perturbations of the inflaton field via Einsteins equation. The scalar
modes may also be written in terms of the density perturbations
$\frac{\gd \rho}{\rho}$ at the time of last scattering of the CMB
photons, and these are related to the measured angular anisotropies of
the CMB.  Without going into these details, one obtains the COBE
normalization of the inflaton's potential which is,\footnote{See
  \cite{lyth-riotto-review-1998,riotto-inflation-2002} or one of the
  many other references on inflation for a derivation of this
  relation.}
\begin{eqnarray}
\frac{H_I^2}{10\pi^{3/2}\dot\chi} &\sim& \frac{\gd\rho}{\rho}  \\
\frac{V^{3/2}}{M_P^3V'} &\sim& 10^{-5}
\label{inflation-cobenorm1}
\end{eqnarray}
where the slow-roll approximation has been applied and where $H_I$ and
$\chi$ are taken at the instant of 60 e-folds from the end of
inflation which are to match the amplitude of the density perturbation
on the largest observable scales today.  The left-hand-side of this
expression is proportional to the spectrum of perturbations, which
again scales with powers of $H_I$, but the spectrum is also dependent
on the shape of the potential which appears in the denominator of the
second expression as $V'$.  The COBE
normalization~\myref{inflation-cobenorm1} is a ``normalization'' of
$V(\chi)$ because, once the shape of the potential has been specified,
then this equation fixes the value of $H_I$ or equivalently fixes the
energy scale.  For example, the inflaton potential we consider is a
quadratic potential $V(\chi)=\frac12 m_\chi^2\chi^2$.  Using the above
formula, one obtains,
\begin{equation}
m_\chi \sim 10^{-6}M_P
\label{inflation-cobenorm}
\end{equation}
which is the value we will apply throughout.  From this one determines
the Hubble parameter at the end of inflation to be $H_I \sim
10^{-6}M_P$.

The implications of the COBE normalization cannot be underestimated
(For a general discussion, see \cite{olive-1999}).  Once one chooses a
form for the inflaton potential, the COBE normalization, sets the
energy scale for all of the field dynamics and particle decays which
follow inflation.  In setting the energy scale it also suggests the
particle theories which one may choose to study.  For example, the
quadratic inflaton potential considered is the simplest one can
imagine in the chaotic inflation paradigm.  This falls into the class
of polynomial potentials of which many models of inflation can be
mapped into \cite{kofman-reheating-2008,lyth-riotto-review-1998}.  The
COBE normalization then yields approximately the same energy scale
$H_I\sim 10^{16}\;GeV$
for all these potentials \cite{lyth-riotto-review-1998} which is also
the energy scale $M_{GUT}$, at which the three gauge couplings appear
to unify (see Figure~\ref{fig-gaugecouplings}).  It is thus natural
to consider supersymmetry and Grand Unified theories in combination
with these potentials because one can then make predictions for the
dynamics of these theories.

It is worth noting that the quadratic potential is consistent with CMB
observations, though other potentials such as a quartic potential are
now disfavored \cite{wmap}.  We note finally that some models of
inflation can allow for a lower energy scale of inflation.  Many of
these are well motivated and consistent with observations, but do not
easily fit into the context of our discussion which assumes
supersymmetry, $B-L$ violating effects from the GUT scale, and large
vacuum expectation values of flat direction fields which are generic
to supersymmetric theories, but may not be present in other theories.
We return to the discussion of the dynamics and decay of the inflaton
and flat direction fields.

\newpage \chapter{Reheating and Preheating} \label{sec-reheating}  
In this section we describe reheating, which is the process by which
the large entropy of the universe is generated through the decay of
the inflaton and through the subsequent thermalization of the inflaton
decay products.  Reheating is thus the particle genesis for the
universe. A target quantity in the study of reheating is the reheat
temperature $T_{RH}$ which is the temperature of the inflaton's decay
products once they have thermalized (assuming the inflaton energy
density dominates).  Another target quantity is the baryon asymmetry
quantified by the difference of the baryon and lepton number densities
divided by the entropy density, $\frac{n_{B-L}}{s}$.  This quantity is
expected to be conserved after the universe has reached a thermal
state, and is related to the observed baryon to photon ratio today
\cite{pdg-2008}.  
There are other possible observables related to
reheating which will be mentioned at the conclusion of this section,
but the focus here will be on $T_{RH}$ and $\eta_{B}$ both of which
are dependent upon the dynamics of the flat directions.\footnote{While
  upper and lower bounds on $T_{RH}$ may be established, a reliable
  determination is not currently feasible due to uncertainties in
  cosmology and particle theory.  In the future, if these
  uncertainties are overcome, a reliable determination of $T_{RH}$
  might be possible}

To begin, a framework for determining $T_{RH}$ and $\eta_B$ is
outlined, based on the Hubble parameter during inflation, $H_I$ the
decay rate of the inflaton $\Gamma_\chi$, and the rate of
thermalization of the inflaton decay products $\Gamma_{therm}$.
We then discuss three perturbative decay scenarios of the inflaton;
decay into fermions, decay into scalars, and a Planck suppressed
decay.  We describe the subsequent thermalization process of the
inflaton decay products in the absence of flat direction VEVs, and
then in their presence, explaining how in the latter case the
thermalization may be delayed due to the effective vacuum established
by the flat directions.  However, this presumes the flat directions
persist in a coherent state for a relatively long time.  We note that
the perturbative decay channels of the flat direction are suppressed
by powers of the VEV \cite{affleck-dine-1985}, so this presumption is
not unwarranted, but indications are that a fast nonperturbative decay
of the flat direction VEV may be possible \cite{olive-peloso-2006}.
To motivate the analysis of nonperturbative flat direction decay, we
discuss the now well studied nonperturbative decay of the inflaton
commonly referred to in the literature as ``preheating''
\cite{preheating}.  Based on this analysis we present a preliminary
discussion of the nonperturbative decay of the flat directions.
Figure~\ref{fig-reheating} shows how the elements in the above
scenarios are related.  The reader is also referred to the following
books and reviews from which the background material here was gathered
\cite{olive-1999, mukhanov-cosmology-2005, rubakov-cosmology-2005,
  kofman-reheating-1996,
  kofman-reheating-2008,linde-inflation-2007}.
\begin{fullpagefigure}
\begin{center}
\includegraphics[scale=0.55]{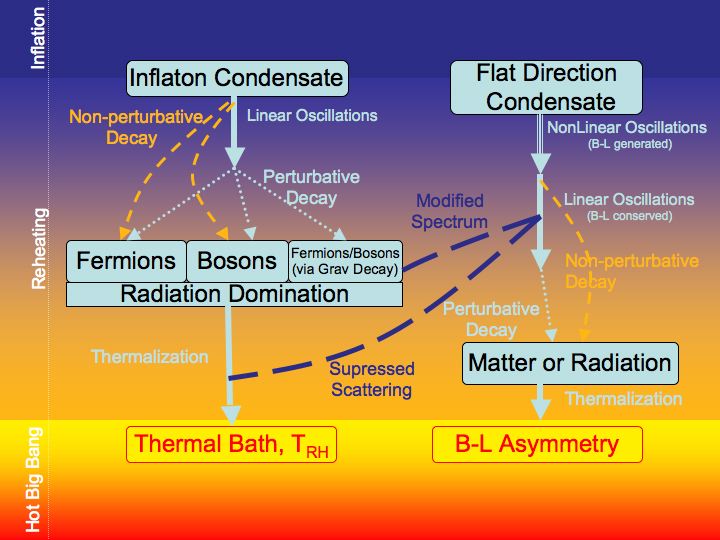}
\caption[Schematic of Reheating Involving Flat
Directions]{\label{fig-reheating} Schematic of the model of reheating
  involving flat directions.  For the inflaton, decay may proceed via
  couplings to fermions and/or bosons through perturbative or
  nonperturbative mechanisms.  The inflaton may also decay through
  gravitationally suppressed coupling (perturbative decay).  Also
  shown is the flat direction evolution. The flat directions establish
  an effective vacuum into which the inflaton will decay and in which
  its products will thermalize, but the vacuum is effective only as
  long as the flat direction VEV remains coherent.  Thermalizing
  interactions can be suppressed due to exchanged particles having
  very large effective masses.  Additionally, the flat direction VEVs
  perturbative decay rate is strongly suppressed (see
  equation~\myref{reheating-gammaflat}) for the same reason that the
  mediating particles are very massive \cite{affleck-dine-1985}.  If
  the VEV is sufficiently long lived, its energy density may come to
  dominate over that of the inflaton.  If this happens, the reheating
  temperature $T_{RH}$ will be determined by flat direction decay and
  thermalization of its decay products instead of that of the
  inflaton. The figure shows the case of inflaton domination because
  our belief is that flat directions will decay quickly by
  nonperturbative mechanisms and thus removing its induced effective
  vacuum.  Such nonperturbative decay mechanisms are introduced in
  this section and further developed in the remaining sections.}
\end{center}
\end{fullpagefigure}

         \section{Determining the Reheating Temperature} \label{sec-reheating-trh} 
Here we continue to develop the picture of the post-inflation
evolution of the inflaton and flat directions which was begun in
Section~\ref{sec-inflation-vevevolution}.  The emphasis in the
previous discussion was on the classical evolution of these
fields. The emphasis here is on quantum effects, namely the decay
rates of these fields $\Gamma_\chi$, and $\Gamma_\Phi$, as well as the
rate of thermalization of the inflaton decay products,
$\Gamma_{therm}$.  While the basic picture is developed, a simple
framework for calculating the reheating temperature $T_{RH}$ and the
baryon asymmetry is also established.  The main difficulty and the
main source of uncertainty is the reliable computation of the decay
rates and thermalization rates and this is postponed to the next
section.

To begin, consider the following simple upper bound for the reheat
temperature which assumes that at the end of inflation, the inflaton
decays
in less than one Hubble time.  Specifically, all of the inflaton
condensate decays and thermalizes with negligible decrease in the
total energy density in less than one Hubble time.  Under these
assumptions, and applying the upper bound on the Hubble parameter
during inflation, $H_I < 10^{-4} M_p$ discussed in
Section~\ref{sec-inflation-cobe}, we equate $H_I$, to the Hubble
parameter during a thermal radiation dominated epoch during which time
the temperature is related to $H$ as follows
(see~\myref{calc-thermodynamics-relateTH}),
\begin{equation}
T=\pfrac{45}{4g_* \pi^3}^{1/4} \sqrt{H M_P} \hspace{4em}
g_*(T) = g_b + \frac78 g_f
\end{equation}
where $g_*(T)$ is the effective number of relativistic (massless)
degrees of freedom at the temperature $T$ with $g_b$ and $g_f$ being
the number of bosonic and fermionic degrees of freedom.  In the
standard model $g_* = 106.75$, and for supersymmetric GUTs $g_* \siml
1000$ from which we may obtain a convenient formula for the reheating
temperature
\begin{equation}
T_{RH}\sim 0.1\sqrt{H_{RH}M_P}
\label{reheating-relateTH}
\end{equation}
where $H_{RH}$ is the Hubble parameter at the instant of
thermalization.  From this expression and the upper bound $H_I <
10^{-4}M_P$, one determines,
\begin{eqnarray}
T_{RH} < 10^{16}\;GeV
\label{reheating-upperbound}
\end{eqnarray}
which is thus an absolute upper bound on $T_{RH}$ independent of any
assumptions on the inflaton's potential or other interactions.  An
absolute lower bound on the reheating temperature is obtained from the
fact that nucleosynthesis occurs in a thermal bath at a temperature of
approximately 1 MeV \cite{hannestad-2004}, so
\begin{equation}
T_{RH} \gtrsim 10^{-3}GeV
\label{reheating-lowerbound}
\end{equation}
It is remarkable that there exists approximately $19$ orders of
magnitude of uncertainty in $T_{RH}$.
However, the above are phenomenological bounds on $T_{RH}$.  If one
makes assumptions on the particle theory applicable during reheating,
this range may be narrowed.  We will make some minimal assumptions on
our particle theory in order to estimate the reheating temperature,
the first of which is an assumption of a quadratic inflaton potential
$\frac12m_\chi^2\chi^2$ whose mass $m_\chi\sim 10^{-6}M_P$ is fixed by
the COBE normalization~\myref{inflation-cobenorm}.  This assumption
also fixes the Hubble Parameter at the end of inflation,
\begin{equation}
H_I \sim 10^{-6}M_p 
\hspace{2em}\mbox{with}\hspace{2em}
V(\chi)=\frac12 m_\chi^2\chi^2
\end{equation}
so if the inflaton were to decay and thermalize instantly (within one
Hubble time) the reheating temperature in this case is $\sim
10^{15}\;GeV$ which is six orders of magnitude above the temperature
on needs to avoid an overproduction of the gravitinos
~\footnote{Recall in the context of the Gravitino problem discussed in
  Section~\ref{sec-inflation-classical}, that the upper bound on the
  reheat temperature is in the range $T_{RH}\lesssim 10^6-10^9\;GeV$
  determined in \cite{gravitinos-moroi,gravitinos-olive}.}  In light
of this, one may look for mechanisms which would suppress either the
inflaton's decay or the thermalization of its products.  In
particular, the role of flat directions in suppressing the
thermalization of the inflaton decay products will be discussed in the
following subsections.  However, before making such assumptions, we
erect a simple framework for determining $T_{RH}$ as well as the
baryon asymmetry $\eta_B$ with model assumptions entering through
decay and reaction rates $\Gamma_i$.  The only further assumptions we
make in this framework are the quadratic inflaton potential, and the
Friedmann equation.

We determined in Section~\ref{sec-inflation-vevevolution} the
classical evolution of both the inflaton and a flat direction after
the inflationary stage.  The solutions are damped oscillations of the
form~\myref{inflation-inflatonoscillations} for the inflaton with
frequency $m_\chi \sim 10^{-6}M_P$
and~\myref{inflation-flatdiroscillations} for a flat direction with
frequency $m\sim 10^{-16}M_P$.  For both of these homogeneous scalar
fields, there is an equivalent quantum interpretation as zero
temperature (and thus zero momentum) condensates of scalar particles
of mass $m_\chi$ and $m$ respectively.  These particles can have decay
mechanism(s) into lighter species which we parametrize with the decay
rates $\Gamma_\chi$ and $\Gamma_\phi$.  The decays are quantum effects
which modify the respective classical equations of motion.  Consider
for instance, the inflaton.  One estimates the number of $\chi$
particles in the inflaton condensate as,
\begin{eqnarray}
n_\chi &=& \frac{\rho_\chi}{m_\chi} = \frac{1}{2m_\chi} \left({\dot\chi}^2 + m_\chi^2\chi^2\right) \\
&\approx& \frac{\dot\chi^2}{m_\chi} \approx m_\chi\chi^2
\end{eqnarray}
where the virial theorem has been applied in the last two steps to
approximate the number density.  The reaction rate equation for the
$\chi$ particles is
\begin{equation}
\frac1{R^3}\frac{d(R^3 n_\chi)}{dt} = - \Gamma_{\chi} n_\chi \\
\label{reheating-ratechi}
\end{equation}
where $n_\chi$ is the number density of the inflaton condensate.
Substituting the expressions for the number density $n_\chi$, we
obtain,
\begin{eqnarray}
\ddot\chi + (3H+\Gamma_\chi) \dot\chi + m_\chi^2 \chi &=& 0 
\label{reheating-eominflaton}
\end{eqnarray}
which is the equation of motion~\myref{inflation-scalarfieldeqs}
including a decay term which appears on the same footing as the Hubble
parameter.~\footnote{There are also quantum corrections to the
  inflaton mass in the equation of motion, but we assume these effects
  are already taken into account.}  A similar equation may be derived
for the complex flat direction field.
For the moment, we consider just the inflaton decay based on the
equation~\myref{reheating-eominflaton}. When the decay rate is small
compared to the Hubble parameter $\Gamma_\chi < H$, the solution to
the equations of motion is the
result~\myref{inflation-inflatonoscillations}.  Though when the
expansion has slowed to a point where $H\siml\Gamma_\chi$, the decays
soon dominate over the coherent classical oscillation,
and the condensate is soon depleted.  This happens at a value of the
scale factor we define as $R_{d\chi}$.  Since the inflaton mass is
orders of magnitude above the mass of most of its expected decay
products, the products will be relativistic.  For $R>R_{d\chi}$, that
is after inflaton decay, the energy density evolves in a radiation
dominated universe (see Figure~\ref{fig-rhoevolution}).
In this regime after the initial decay, the decay products are not
necessarily in a thermal state however \cite{ellis-etall-1987}, and
there will be scatterings, particle creation and particle destruction
to bring the system to a thermal and chemical equilibrium
\cite{ellis-etall-1987,davidson-sarkar-2000}.  This thermalization
rate is parametrized with a second reaction rate $\Gamma_{therm}$.
If, at the time of decay, the thermalization rate is faster than the
Hubble expansion, which implies $\Gamma_{therm} > \Gamma_{\chi}$, then
thermalization can be thought of as happening instantly at
$R_{d\chi}$.  The reheat temperature is then determined
with~\myref{reheating-relateTH} by setting $H\sim\Gamma_\chi$ in this
expression.  If on the other hand the thermalization process is slow,
and $\Gamma_{therm}<\Gamma_\chi$, then the reheating temperature is
determined by setting $H \sim \Gamma_{therm}$ in this expression.
Summarizing,
\begin{equation}
T_{RH}\sim 0.1\sqrt{H_{RH}M_P}
\;\;\;,\;\;\;
H_{RH} \sim \mbox{Min}[\Gamma_\chi,\Gamma_{therm} ] 
\label{reheating-TRH}
\end{equation}
Reheating is thus determined by the slowest reaction rate.  Note that
we nowhere assumed the reaction rates $\Gamma_\chi$ and
$\Gamma_{therm}$ were perturbative results.  The rate $\Gamma_\chi$ in
particular can equivalently parametrize a nonperturbative decay.

The presence of flat directions with large vacuum expectation values
can modify the above history of reheating through the value of inputs
$\Gamma_\chi$ and $\Gamma_{therm}$. Furthermore, if the flat
directions have large enough expectation values, and if they are
sufficiently long lived, the flat directions can come to dominate the
total energy density.  For this to happen, the inflaton must decay
before or during the flat directions oscillations so that the ratio
$\rho_\Phi/\rho_\chi$ is no longer constant but scales as $R$ until a
significant portion of the flat direction particles decay or
up-scatter into a relativistic species (see
Section~\ref{sec-inflation-vevevolution} and
Figure~\ref{fig-rhoevolution}). Referring to
figure~\ref{fig-rhoevolution}, the value of the scale factor at which
the flat direction energy density would overtake that of the inflaton
is obtained in~\myref{rhoevolution-Req} to be
\begin{equation}
\frac{R_{eq}}{R_0} \sim \pfrac{M_P}{\Phi_0}^2 \pfrac{m_\chi}{\Gamma_\chi}^{2/3}
\label{reheating-Req}
\end{equation}
Then assuming that the flat direction decay rate $\Gamma_\Phi$ is a
constant, the condition for the flat direction to dominate is that
$R_{eq} < R_{d\Phi}$ which may be rewritten (see
formula~\myref{rhoevolution-flatdirdominates}),
\begin{equation}
\frac{\Gamma_\Phi}{\Gamma_\chi} < \pfrac{\Phi_0}{M_P}^4
\label{reheating-flatdirdominates}
\end{equation}
The region of flat direction domination is shown in the plane
$(\Phi_0, \Gamma_\Phi)$ in Figure~\ref{fig-flatdirdominates}.

One may also wish to compute the reheating temperature in these cases
when the flat direction dominates, but they imply a baryon asymmetry
of order one, and thus require either an extremely small $\ge_{B-L}$
or a later ``additional'' source of entropy to dilute the B-L charge
to the required amount.  As the first case is not obviously natural
and the second case adds complications to the simple picture of
inflation, flat direction domination is not pursued here any further
than establishing the parameter boundaries in
Figure~\ref{fig-flatdirdominates}.

         \section{Determining the Baryon Asymmetry} \label{sec-reheating-etab} 
The next task is to estimate the baryon asymmetry within the above
context.  It was noted in Section~\ref{sec-susyflat-lifting} that the
MSSM flat directions may carry a $B-L$ charge.  It was also mentioned
that such a charge may be generated by the early motion of the flat
direction through Planck suppressed $B-L$ violating interactions, and
afterwards approximately conserved.  While only a baryonic charge is
strictly necessary for baryogenesis, it may be that a nonzero $B-L$
charge is required \cite{arnold-1987}, though see
\cite{kuzmin-etall-1987}.  The reason is that the sum of the baryon
and lepton numbers $B+L$ is violated by Electroweak processes known as
sphalerons.\footnote{For an introduction to sphalerons, see
  \cite{mukhanov-cosmology-2005}.  The first work on the generation of
  the Baryon asymmetry through sphalerons was
  \cite{kuzmin-etall-1985}} Howver the quantity $B-L$ is conserved by
Electroweak interactions, so any baryon asymmetry generated may be
erased by these sphaleron processes when they are in equilibrium
\cite{arnold-1987}.

In the following we will estimate the quantity $\frac{n_{B-L}}{s}$
which is the ratio of the $B-L$ number density to the entropy density.
This quantity is an adiabatic invariant since the $B-L$ number per
co-moving volume $R^3 n_{B-L}$ is conserved soon after the flat
direction oscillations begin and the entropy per co-moving volume
$R^3s$ is conserved from $T_{RH}$ to the present epoch.  The
conversion of $B-L$ to $B$ by sphalerons is quantified approximately
by the relation \cite{olive-baryogenesis-1994},
\begin{equation}
n_B= \frac{28}{79}n_{B-L}
\label{reheating-sphaleron}
\end{equation}
The baryon-to-entropy ratio today is $\pfrac{n_B}{s}_{obs}\sim
10^{-10}$ which is obtained from the observed baryon-to-photon
ratio.\footnote{What is often reported in the literature is the baryon
  number to photon number ratio today
  $\eta=\frac{n_B}{n_\gamma}\approx 5\times 10^{-10}$. The photon
  number is related to the entropy by $s\approx 7 n_\gamma$ which
  accounts for the neutrinos in the universe.}  A challenge for any
fundamental particle theory is to obtain this baryon-to-entropy ratio
from first principles.  There are many proposed mechanisms of
baryogenesis \cite{baryogenesis-examples}, but the mechanism which is
intimately connected with the study of flat directions is that of
Affleck and Dine \cite{affleck-dine-1985}.  We next compute the baryon
to entropy ratio in the Affleck Dine baryogenesis scenario.

Recall from Section~\ref{sec-inflation-vevevolution}, the number
density $n_{B-L}$ is computed from the classical
trajectory~\myref{inflation-chargeasymmetry}, and once generated the
$B-L$ number per co-moving volume is then nearly conserved.  The
resulting baryon to entropy ratio depends on how much this charge is
diluted by the entropy generated through inflaton decay and
thermalization.  Specifically, the entropy per co-moving volume is not
fixed until after a thermal state is reached at $T_{RH}$, so it should
be computed at or after this instant.  However, the resulting
expression is the same if it were computed at the time of inflaton
decay.  The reason is that we have assumed the inflaton energy density
dominates throughout, and the inflaton decay products behave like
radiation before and after thermalization. The entropy density is
related to the energy density through $s \sim \rho_{rad}^{3/4}$
from~\myref{thermodynamics-relatesrho}, so the entropy per co-moving
volume is largely insensitive to when the inflaton decay products
thermalize.  The main approximation here is that the flat direction
energy density is negligible compared to the inflaton energy density.
As shown in Figure~\ref{fig-rhoevolution}, the dominance of the
inflaton is not guaranteed.  The entropy density and the $B-L$ number
density are both computed at the time of inflaton decay $R_{d\chi}$ to
obtain the desired ratio,
\begin{eqnarray}
\frac{(n_{B-L}R^3)_{d\chi}}{(sR^3)_{d\chi}} &\sim& 
\pfrac{\ge_{B-L}\frac{\rho_\Phi}{m}}{\rho^{3/4}}_{d\chi} \\
&\sim& \ge_{B-L} \pfrac{\rho_\Phi}{\rho_\chi}_{d\chi}\frac{\rho_{d\chi}^{1/4}}{m}  \\
&\sim& \ge_{B-L} \pfrac{|\Phi_0|}{M_P}^2\pfrac{\Gamma_{\chi}}{M_P}^{1/2} \pfrac{M_P}{m}
\end{eqnarray}
where in the last step, the relation~\myref{inflation-chiphiratio} has
been applied as well as the relation $\rho_{d\chi}\sim \Gamma_\chi^2
M_P^2$ which follows from $H_{d\chi}\sim \Gamma_\chi$ and the
Friedmann equation. The baryon-to-entropy ratio is then estimated
using the sphaleron conversion~\myref{reheating-sphaleron} and
assuming a gravitational decay $\Gamma_\chi\sim
\frac{m_\chi^3}{M_P^2}$ with $m_\chi \sim 10^{-6}M_P$, $m\sim
10^{-16}M_P$.  One obtains,
\begin{equation}
\frac{n_B}{s} \sim \ge_{B-L}\pfrac{\Phi_0}{M_P}^2 10^7
\label{reheating-baryontoentropy}
\end{equation}
Recall from~\myref{inflation-chargeasymmetry} that $\ge_{B-L}$
quantifies the $B-L$ number per particle in the flat direction
condensate with $\ge_{B-L}=\pm 1$ corresponding to a maximally charged
condensate.
Also, our constraint that the inflaton energy density dominates
translates to the constraint $\Phi_0 < M_P$ as shown
in~\myref{inflation-chiphiratio}.  Using both considerations, the
above baryon to entropy ratio may still be significantly greater than
one if the condensate is maximally charged and $\Phi_0 \siml M_p$.  It
has been argued by Linde \cite{linde-1985} that generically the baryon
to entropy ratio should not exceed one.  In the example given in
\cite{linde-1985}, it is the thermal corrections to the masses of the
states involved which result in the ratio being bounded below unity,
but the example assumes a case of flat direction domination.  It is
not immediately obvious how one would generalize such arguments to our
situation and assumptions.  We do not attempt such corrections here.
The above ratio~\myref{reheating-baryontoentropy} is sufficient to
establish the leading dependence on the parameters $\Phi_0$ and
$\ge_{B-L}$.  Specifically, one concludes that to have a successful
Affeck-Dine baryogenesis, one requires either $\ge_{B-L} \ll 1$ and/or
$\Phi_0\ll M_P$ given our assumption of a gravitational decay (and
precluding any additional contributions of entropy such as from moduli
decay).  We emphasize that the baryon asymmetry, while sourced by
early motion of the flat direction, appears to be largely independent
of the flat direction's later dynamics, so long as the inflaton
dominates the energy density.  This result for the baryon to entropy
ratio~\myref{reheating-baryontoentropy} and the above brief
qualifications are thus the conclusion of our discussion of
baryogenesis.

         \section{Perturbative Decay and Thermalization of the Inflaton} \label{sec-reheating-perturbative} 
The perturbative decay of the inflaton and thermalization of its decay
products was studied shortly after the first models of inflation were
developed \cite{reheating-perturbative} (see also the text-books
\cite{mukhanov-cosmology-2005, kolb-turner}).  In the following, we
briefly present the generic scenarios as well as some recent
developments regarding the thermalizing interactions which follow
inflaton decay \cite{davidson-sarkar-2000}.

In the absence of a flat direction VEV, the inflaton decays into a
vacuum in which supersymmetry is assumed to be softly broken at a
scale $\msusy\sim \tev$.
We assume that there exist light states $m_\psi, m_\xi\ll m_\chi$ in
this vacuum to which the inflaton can decay, specifically, spin 1/2
fermions via $\chi\rightarrow \psi\psi$ or spin-0 bosons via
$\chi\rightarrow \xi\xi$.
The corresponding terms in the Lagrangian would appear as,
\begin{equation}
\cL_{int} = 
- g \chi\xi^2 
- h\chi\bar\psi \psi
\label{reheating-Lgeneric}
\end{equation}
where the coupling $h$ is dimensionless, and $g$ has dimensions of
mass.  The decay rates for these interactions can be computed, and
assuming that the final states have masses $m_\psi,m_\xi\ll m_\chi$,
the perturbative decay rates are,
\begin{equation}
\Gamma(\chi \rightarrow \xi\xi) = \frac{g^2}{8\pi m_\chi} 
\;\;\;,\;\;\;
\Gamma(\chi \rightarrow \psi\psi) = \frac{h^2m_\chi}{8\pi} 
\label{reheating-gammaxipsi}
\end{equation}
The form of these rates could also be obtained from dimensional
analysis.  In the case these rates are gravitationally suppressed, one
has $g\sim m_\chi \pfrac{m_\chi}{M_P}$ and $h\sim\frac{m_\chi}{M_P}$,
resulting in,
\begin{equation}
\Gamma_{Grav} \sim \frac{m_\chi^3}{M_p^2}
\label{reheating-gammagrav}
\end{equation}
which could also have been obtained by dimensional analysis and
independently of the spin of the final state particles.  The
gravitationally suppressed decay would proceed in approximately
$\frac{m_\chi}{2\pi \Gamma_{Grav}} \sim 10^{13}$ oscillations of the
inflaton.  To determine $T_{RH}$ assuming the gravitational decay, the
thermalization rate $\Gamma_{\mathsubscript{therm}}$ must be computed.  We use the
results of \cite{davidson-sarkar-2000} in which it is shown that a
thermalization process can be expanded in a series involving the
coupling $\ga$ as well as the small quantity $\frac{T_{RH}}{m_\chi}$
which quantifies how effective the process is at transferring the high
momenta inflaton products $k\sim m_\chi$ to the reheat temperature.
It is shown in~\cite{davidson-sarkar-2000} that the dominant
thermalization processes after perturbative inflaton decay are
$2\rightarrow 3$ inelastic scattering with a rate,
\begin{eqnarray}
& \Gamma_{2\rightarrow 3} \sim \ga \pfrac{m_\chi}{T_{RH}}^2 \Gamma_{2\rightarrow 2} \\
& \Gamma_{2\rightarrow 2} \sim n_\chi \gs
\;\;\;,\;\;\;
\gs = \pfrac{\ga}{m_\chi}^2\pfrac{R}{R_{d\chi}}^2
\label{reheating-gamma2to3}
\end{eqnarray}
where $\Gamma_{2\rightarrow 2}$ is the elastic scattering rate, $\gs$
is the corresponding elastic cross-section, and $n_\chi$ is the number
density for the inflaton condensate (which should be of the same order
as the number density of its decay products).

\begin{figure}[th]
\begin{center}
\includegraphics[scale=0.5]{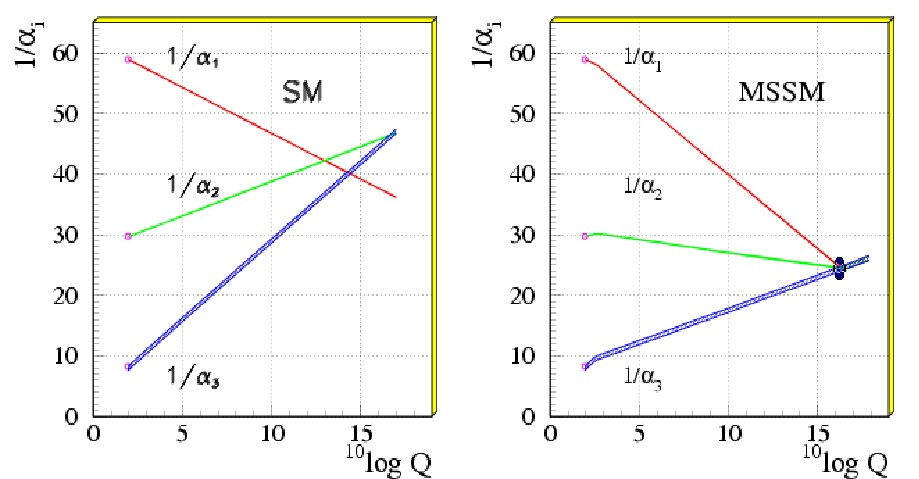}
\caption[Running of the Gauge Couplings in the Standard Model and the
MSSM]{\label{fig-gaugecouplings} Running of the gauge couplings in the
  standard model and the MSSM. The inflaton decay products are
  expected to thermalize through gauge interactions.  With our
  assumptions, one expects the typical momentum transfer of the
  interactions to be $T_{RH}\siml 10^9\,\gev$
  from~\myref{reheating-TRHgravNoflat}.  Thus the relevant gauge
  couplings can be read from the righthand plot at $Q\sim 10^9\,\gev$.
  Note also the apparent gauge coupling unification in the MSSM at the
  scale $M_{GUT}\sim 10^{16}\,\gev$.  Figure obtained
  from~\cite{pdg-2008} }
\end{center}
\end{figure}

Note that the thermalization rate $\Gamma_{2\rightarrow 3}$ is
suppressed by three powers of the gauge coupling $\ga^3$, but then
enhanced by the factor $\pfrac{m_\chi}{T_{RH}}^2$ which quantifies the
processes efficiency in momentum transfer.  One may determine the the
thermalization rate and reheat temperature by a self-consistent method
by first assuming $\Gamma_{\mathsubscript{therm}} > \Gamma_\chi$ so that $T_{RH}\sim
\sqrt{\Gamma_\chi M_P}$, and then determining the conditions on the
parameters for this to be true.
By this method, one requires the gauge coupling to satisfy $\ga^{-1} >
100$ (see~\myref{rhoevolution-alphanoflat} and reference
\cite{olive-peloso-2006}).  Keeping in mind that this is an order of
magnitude approximation, this condition on $\ga$ may be compared
against the expected values of the gauge couplings, shown in
Figure~\ref{fig-gaugecouplings}, where one would presumably choose the
renormalization scale at or near a momentum transfer of order $Q\sim
T_{RH}$.  Assuming the condition on the gauge couplings is satisfied,
the thermalization happens immediately after inflaton decay and the
reheat temperature is
\begin{equation}
T_{RH} \sim 10^9\;\gev \hspace{3em}\mbox{gravitational decay, w/o flat direction VEVs}
\label{reheating-TRHgravNoflat}
\end{equation}
Again, this temperature corresponds to the gravitationally suppressed
decay~\myref{reheating-gammagrav} in the absence of flat directions.
It will be a reference case in the remaining discussion.

%

         \section{Perturbative Decay and Thermalization of the Inflaton With Flat Directions} \label{sec-reheating-perturbativewithflat} 
Next we consider the effect of flat direction VEVs on the inflatons
decay rate $\Gamma_\chi$ and on the thermalization rate of the decay
products $\Gamma_{therm}$.  During inflation, a subset of the MSSM
flat directions have presumably accumulated a large condensate of
particles in their zero momentum state (these are the VEVs).  It is
useful to think of this condensate as an effective vacuum.  From
Section~\ref{sec-susyflat}, it was noted that this vacuum typically
breaks all the standard model gauge symmetries, and thus the spectrum
of particles is quite different than the unbroken vacuum.  In
particular, the induced particle masses are of order of the obtained
VEVs which may be nearly at the Planck Mass.  One notes however that
the VEVs evolve slowly in time with frequencies $\sim \msusy$ as
discussed in Section~\ref{sec-inflation-vevevolution}.  For the
inflaton whose mass is much greater than $\msusy$, the vacuum will
look stationary.  However, for the light states\footnote{The light
  states will be the quanta of the flat directions or their
  superpartners as well as any other MSSM quanta which did not acquire
  masses through the Higgs mechanism} of mass $\siml \msusy$ the
vacuum will be moving relatively fast assuming the particle momenta is
small, $k\siml \msusy$.  There can be nonperturbative effects for
processes involving these states, some of which will be focused on
later.

However for the inflaton and for its relativistic decay products, the
vacuum appears nearly static (adiabatic) but it contains a large
number of heavy MSSM states.  We note two effects of these heavy
states. The first effect is that the perturbative decay of the
inflaton into these heavy states will be kinematically blocked when
$\Phi_i \simg m_\chi$.  However, if there is at least one light state
in the spectrum for the inflaton to decay to, and assuming the
coupling is not suppressed, perturbative inflaton decay should not be
suppressed.  Even in the case with all gauge symmetries broken such
light states must exist.  These will be the flat direction states or
their fermionic superpartners which could be the quarks and leptons of
the standard model or the Higgsinos if the Higgs fields are the flat
direction VEVs.  The presence of these states is guaranteed by
supersymmetry, and the supersymmetry breaking can only lift the masses
to the TeV scale.\footnote{Only the Higgsinos should have mass $\sim
  TeV$.  The quark and lepton masses should be that measured in the
  standard model obtained by the Higgs mechanism.}
The second effect is that the gauge bosons have acquired the same
large masses $\siml M_P$ and this will suppress the thermalization
processes which are assumed to be mediated by gauge bosons.  Of
course, if there are any remaining unbroken gauge symmetries, one
expects the remaining gauge boson(s) to efficiently mediate the
thermalization of the particles they are coupled to.

Hereafter, it is assumed (i) the inflaton decays gravitationally
implying a perturbative decay, (ii) that \myemph{all} gauge symmetries
have been broken and (iii) there do exist light MSSM states into which
the inflaton may decay (namely the flat direction states or their
fermionic superpartners).  Note the $2\rightarrow 3$ thermalization
rate determined by the arguments of \cite{davidson-sarkar-2000} relied
up the existence of massless (or light) states which could be emitted
as the third particle in the reaction.  Assuming the existence of such
states, and under the above three assumptions, $\Gamma_\chi$ is given
by~\myref{reheating-gammagrav}, and the thermalization rate is
obtained by replacing $\gs \sim \frac{\ga^2}{\Phi^2}$
in~\myref{reheating-gamma2to3}, and recomputing.  In particular, for
reheating to occur at the instant of inflation decay, one requires
$\Gamma_{2\rightarrow 3} > \Gamma_\chi$, which translates to the
condition,
\begin{equation*}
\ga^3 > \frac{1}{100}\pfrac{m_\chi}{M_P}^5\pfrac{M_P}{m}^2\pfrac{\Phi_0}{M_P}^2
\end{equation*}
which has been determined in the
Appendix~\myref{rhoevolution-alphawithflat}.  This is rewritten as a
condition on $\Phi_0$,
\begin{equation*}
\Phi_0 < \pfrac{\ga}{10^{-2}}^{3/2} 10^{-3} M_P
\label{reheating-trhatgammachi}
\end{equation*}
We may also determine the reheating temperature in the case this
$\Phi_0$ does not satisfy the above constraint.  The expression for
$T_{RH}$ in this case is determined in~\myref{rhoevolution-trh1} and
shown as a function of $\Phi_0$ in Figure~\ref{fig-trhsupression}.  It
is again emphasized that these results are order of magnitude
estimates, but the supression shown in the figure is apparently a mild
one given our assumptions.
\begin{fullpagefigure}
\begin{center}
\includegraphics[scale=1]{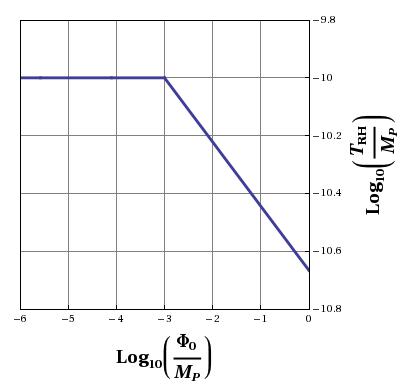}
\caption[Supression of the Reheating
Temperature by a Flat Direction]{\label{fig-trhsupression} Approximate behavior of
  $T_{RH}$ as a function of $\Phi_0$ assuming gravitational decay of
  the inflaton and suppressed perturbative $2\rightarrow 3$
  interactions which control the thermalization. The curve is
  determined from formula~\myref{rhoevolution-trh1} which is obtained
  by the methods of Section~\ref{sec-reheating-perturbative}.  The
  reduction of the reheating temperature is less than an order of
  magnitude.  The temperature $10^{-10}M_P$ corresponds to
  thermalization occurring instantly after inflaton decay.}
\end{center}
\end{fullpagefigure}

         \section{Perturbative Decay of Flat Directions} \label{sec-reheating-perturbativeflat} 
The suppression of $\Gamma_{therm}$ through the presence of the flat
direction condensate as well as the case of flat direction domination
both assume the condensate remains coherent for many inflaton
oscillations.  This is partly justified by making a perturbative
estimate for the decay of the flat direction quanta into the effective
vacuum established by the condensate.  One expects the decay to be
mediated by particles with induced mass of order of the flat direction
VEV, and to decay into light fermions and bosons $m_\psi,m_\xi \ll
\msusy$ of the effective vacuum.  From this, we estimate a tree
level decay rate
\begin{equation}
\Gamma^{pert}_\phi \sim \frac{\ga^2 m^3}{|\Phi|^2}
\label{reheating-gammaflat}
\end{equation}
If one takes into account the time dependence of $\Phi$, one may
determine the decay rate of the flat direction $\Gamma_\Phi^{pert}$
which is the above rate $\Gamma_\phi^{pert}$ at the instant when
$H\sim \Gamma_\phi^{pert}$.  Assuming the decay happens after inflaton
decay, the rate has the form,
\begin{equation}
\Gamma_\Phi^{pert} = \pfrac{\ga m}{|\Phi_0|}^{4/5} \pfrac{m}{\Gamma_\chi}^{1/5}m
\label{reheating-gammaPhipert}
\end{equation}
which has been determining in~\myref{rhoevolution-GammaPhipert}.  The
flat directions perturbative decay rate as specified by this formula
is plotted as a function of $|\Phi_0|$ in
Figure~\ref{fig-flatdirdominates}.  One sees that for $|\Phi_0|\simg
10^{-2}M_P$, the flat direction energy density will overtake the
inflaton.

\begin{fullpagefigure}
\begin{center}
\includegraphics[scale=1]{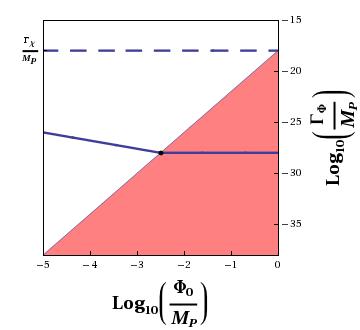}
\caption[Flat Direction Domination]{\label{fig-flatdirdominates} Flat
  Direction Domination: The shaded region above defines the region of
  the parameter space $(\Phi_0,\Gamma_\Phi)$ in which the flat
  direction energy density will eventually dominate over that of the
  inflaton.  The region is determined through
  formula~\myref{reheating-flatdirdominates}. In the case of a
  perturbative decay of the flat direction condensate, the decay rate
  of the condensate $\Gamma_\Phi$ is shown as the solid line from
  formula~\myref{reheating-gammaPhipert}. In the case the flat
  direction dominates, its decay-rate becomes independent of
  $\Phi_0$.}
\end{center}
\end{fullpagefigure}

One reason why the perturbative result should be viewed with
skepticism was mentioned above and this is the vacuum (condensate) is
evolving on the time scale $\msusy$ which is also the mass of the
flat directions.\footnote{A second reason the perturbative result
  may be incomplete is that one expects the inflaton decay products to
  scatter off the flat direction condensate, and this may deplete the
  condensate.  The interactions would be suppressed by powers of
  $\Phi_i$ since the mediating particles are heavy, but it is not
  obvious the suppression would be as severe as
  in~\myref{reheating-gammaflat} since the kinematics are different --
  the inflaton decay products have an energy on the order of $T_{RH}$
  instead of $\msusy$. We do not pursue this decay mechanism here
  beyond noting its presence.}  The vacuum thus does not appear static
to the flat direction quanta.  An adiabatic vacuum is an implicit
assumption in the perturbative computation of the decay
rate~\myref{reheating-gammaflat} for instance.  Hence the perturbative
results involving the flat direction quanta may be completely
wrong. Nonperturbative effects may in fact dominate, and perhaps
facilitate a rapid decay of the flat direction even before the
inflaton decays.  Nonperturbative effects are discussed next; first
for the inflaton, and then for the flat directions.

         \section{Nonperturbative Decay of the Inflaton: Preheating} \label{sec-reheating-nonperturbative} 
The above estimates for the reheat temperature assume that
$\Gamma_\chi$, $\Gamma_{therm}$, and $\Gamma_\phi$ are obtained via
perturbative processes in which the coupling is assumed small and the
number of quanta involved in any reaction is also small.  This
assumption on $\Gamma_\chi$ is indeed good in the case of a
gravitational decay of the inflaton, and so the perturbative estimate
is appropriate.  In fact, except for the content of this subsection
and Section~\ref{sec-reheating-perturbative}, we universally assume a
gravitational coupling for the inflaton and thus a perturbative
inflaton decay.  However, it is worthwhile to consider the
non-gravitational decay mechanisms which may be nonperturbative.  The
nonperturbative decay of the inflaton will prove to be a useful
example for later studying nonperturbative flat direction decay.

The perturbative estimate for the decay of a species of particles is
not appropriate if the relevant coupling is large.
However, even when the coupling is small, if the fields involved have
acquired VEVs, the perturbation expansion in the coupling may still
break down.  Essentially, the probability amplitude for a scattering
with a condensate particle is very high due to the large occupation
number.
In this situation, one splits the field $\chi$ into a classical
background $\langle\chi\rangle$ and quantum fluctuations $\gd\chi$,
\begin{equation}
\chi = \langle\chi\rangle + \gd\chi
\end{equation}
The classical background, which describes the condensate, is assumed
known, and the computations done using the fluctuations $\gd\chi$ will
not suffer from the breakdown of perturbation theory present in the
original field description.  One must then be aware that any
observable computed in the new field description may in principle
correspond to a large number of quanta in the original field
description.  For example, one must be wary of making kinematic
arguments which rely on there being a small number of initial state
particles involved in the decays.  These kinematic arguments can
easily fail when the coupling is large and/or the fields have acquired
VEVs.  The computations should then be done using the above new field
description using the methods of Section~\ref{sec-heiseqs}, and the
results may then be interpreted in both frameworks.

It has been shown that nonperturbative effects can be relevant to
inflaton decay.  These effects are referred to in the literature as
``preheating'' \cite{preheating}.  There are at least two striking
consequence of preheating which are (1) that particles with masses
orders of magnitude larger than the inflaton mass can be produced
through decay, and (2) that the production of quanta can occur through
an exponential growth.  We will work through one example of
nonperturbative inflaton decay following the development of
\cite{mukhanov-cosmology-2005} in which both these features will
emerge.
In the next section, we discuss the possibility of a nonperturbative
decay for the flat directions, as it is the objective of this thesis
to show that nonperturbative effects can also be relevant for the flat
directions.

We consider inflaton decay through the scalar coupling shown
in~\myref{reheating-Lgeneric}, and we perform a perturbative
calculation first from which the exponential growth of quanta is
apparent even in the case of a very small coupling.  The physical
effect is that of Bose-Einstein condensation.  In the decay
$\chi\rightarrow \xi\xi$, the momenta of the product $\xi$'s are each
$|\bk|= \frac{m_\chi}{2}$.  If one neglects the redshift due to Hubble
expansion, one can see the occupation numbers in this momentum mode
will quickly become large, and Bose-Einstein effects should be taken
into account. To quantify the production, we write down the rate
equation describing the growth in the number density of the $\xi$
particles which is,
\begin{equation}
\frac{1}{R^3}\frac{d\left(R^3 n_\xi\right)}{dt} = 2 \Gamma_\chi n_\chi
\label{reheating-ratexi}
\end{equation}
which is to be compared to the corresponding equation describing the
depletion of the inflaton field~\myref{reheating-ratechi}.  We already
presented the decay rate $\Gamma_\chi=\frac{g^2}{8\pi m_\chi}$ in the
absence of Bose-Einstein effects (see~\myref{reheating-gammaxipsi}).
Now we must obtain the correction due to Bose-Einstein effects.  The
matrix element for the process $\chi\rightarrow \xi \xi$ is,
\begin{equation}
\left|\langle n_\chi - 1 , n_{\mathbf{k}} + 1 , n_{-\mathbf{k}} + 1 |
a^\dagger_{\mathbf{k}}a^\dagger_{-\mathbf{k}}a_{\chi}
|n_\chi , n_{\mathbf{k}} , n_{-\mathbf{k}}\rangle \right|^2 = (n_\xi(\mathbf{k})+1)^2 n_\chi\gd^3(\mathbf{k}_\xi)
\end{equation}
and the matrix element for the reverse process $\xi\xi\rightarrow\chi$ is,
\begin{equation}
\left|\langle n_\chi + 1 , n_{\mathbf{k}} - 1 , n_{-\mathbf{k}} - 1 |
a_{\mathbf{k}}a_{-\mathbf{k}}a^\dagger_{\chi}
|n_\chi , n_{\mathbf{k}} , n_{-\mathbf{k}}\rangle \right|^2 = n_\xi(\mathbf{k})^2 (n_\chi\gd^3(\mathbf{k}_\xi)+1)
\end{equation}
Note that the inflaton condensate populates the $k=0$ momentum mode so
the occupation number for this mode is $n_\chi\gd^3(\mathbf{k}_\xi)$
where $n_\chi$ is the inflaton condensates number
density.
For the $\xi$ spectrum, $n_\xi(\mathbf{k})$ is a strongly peaked
function at $|\mathbf{k}|=\frac{m_\chi}{2}$, but not as narrow as the
delta function.  The net rate of decrease of inflaton condensate
particles is proportional to the difference of the above two matrix
elements which is proportional to
$n_\chi(1+2n_{\mathbf{k}})$,
where we have applied $n_\chi\gd^3(\mathbf{k}=0) \mg 1$. Defining the
occupation number of the $\xi$ particles as $n_\xi(k)\equiv
n_\xi(\mathbf{k}) = n_\xi(-\mathbf{k})$,
the decay rate is approximately,
\begin{equation}
\Gamma_\chi\approx \frac{g^2}{8\pi m_\chi}\left(1 + 2 n_\xi\left(k=\frac{m_\chi}{2}\right)\right)
\label{reheating-res1}
\end{equation}
where the prefactor is necessary to be consistent with our previous
result for $\Gamma_\chi$.  To determine the number density $n_\xi$ we
must integrate $n_\xi(k)$ over all momenta. If we can determine the
approximate width of the function $n_\xi(k)$, we should then be able
to approximate this integration.  To do this, express the potential as
follows,
\begin{equation}
-\frac12 m_\xi^2 - g\chi\xi^2 = -\frac12\left[m_\xi^2 +2g\chi_0\sin (m_\chi t)\right] \xi^2 
\end{equation}
where the solution to the equation of motion for $\chi$ has been used,
$\langle\chi\rangle = \chi_0 \sin (m_\chi t)$.  We have also assumed
the scale factor is kept constant and neglected the depletion of the
condensate.  In the above $\chi$ is treated classically, and the
potential is written as the effective mass for the $\xi$ field. The
momenta of the produced $\xi$ particles are then,
\begin{equation}
k = \sqrt{\pfrac{m_\chi}{2}^2 - m_\xi^2 - 2g\chi_0\sin(m_\chi t)}
\end{equation}
Additionally, if both $g\chi_0$ and $m^2_\xi$ are small compared to
$m_\chi^2$, then the particles are produced in narrow band of phase
space, centered at $k=\frac{m}{2}$ with width
\begin{equation} 
\gD k \approx m_\chi \pfrac{4g\chi_0}{m_\chi^2}
\end{equation} 
We can then integrate $n_\xi(k)$ over momenta to approximate the
number density $n_\xi$,
\begin{eqnarray}
n_\xi \approx \frac{4\pi k_0^2 n_k \gD k }{(2\pi)^3}  &\approx& \frac{m g\chi_0 n_k}{2\pi^2} \\
&\approx& \frac{g n_\chi n_k}{2\pi^2\chi_0} 
\end{eqnarray}
where the virial theorem has been applied in the last step.  Using
this last relation, the decay rate~\myref{reheating-res1} may be
written
\begin{equation}
\Gamma_\chi \approx \frac{g^2}{8\pi m_\chi}\left(1 + \frac{2\pi^2\chi_0}{g}\frac{n_\xi}{n_\chi}\right)
\end{equation}
Substituting this into the rate equation~\myref{reheating-ratexi}, and
assuming $n_k > 1$, one obtains,
\begin{equation}
n_\xi \propto \exp \left[\frac{\pi^2 g\chi_0}{m_\chi^2}N\right]
\label{reheating-expgrowth}
\end{equation}
where $N=\frac{m_\chi t}{2\pi}$ is the number of oscillations of the
inflaton field.  The strength of the effect can be read from the
exponent of~\myref{reheating-expgrowth} by determining the number of
oscillations of the inflaton at which the depletion of the inflaton
condensate becomes appreciable.  Note that when approximating the
resonance, we assumed a small coupling $g < \frac{m_\chi^2}{\chi_0}$,
so even in the case of small coupling the exponential growth is
obtained.  The above example of narrow resonance demonstrates that the
perturbative results are not always adequate.  We should note however,
that the narrow resonance is sensitive to the expansion of the
universe. When this is included, the effect is diminished as the
occupation numbers of any particular momentum mode are constantly
shifting to lower momenta and reducing the occupation number at
$k=m_\chi/2$.  Additionally, one should take into account the
depletion of the inflaton condensate and interactions in which
produced $\xi$ particles back-scatter on the inflaton condensate.
Inclusion of such effects, which are also necessary for energy
conservation, can also change the final result.

Note that when the coupling is large, $g > \frac{m_\chi^2}{\chi_0}$,
our above result is invalid as the resonance is no longer narrow.  In
this case, the exponential production may still be obtained but the
resonance is now for a broad range of momenta.  We must use the
methods of Section~\ref{sec-heiseqs} which are generally applicable.
For the case of large coupling consider instead the quartic potential
$\frac12\tg\chi^2\xi^2$ which is similar to the flat direction
potentials later considered.  Here, the coupling $\tg$ is
dimensionless, and the regime of strong coupling is $\tg
>\frac{m}{\chi_0}$.  The perturbative decay corresponds to the
annihilation process $\chi\chi \rightarrow \xi\xi$.  We perform the
nonperturbative calculation for the production of $\xi$ quanta
treating $\chi$ as a classically evolving field $\langle\chi\rangle =
\chi_0 \sin (m_\chi t)$.  Again, we are freezing the scale factor, and
neglecting the depletion of the inflaton condensate.  The effective
potential for the $\xi$ field in this case is,
\begin{equation}
-\frac12 \left(m_\xi^2 + \chi_0^2\sin^2(m_\chi t) \right)\xi^2
\end{equation}
and the equation of motion for the momentum modes $\xi_k$ are,
\begin{equation}
\ddot \xi_k + \go(t)^2 \xi_k =0
\;\;\;,\;\;\;
\go(t)^2 = k^2 + m_\xi^2 + \chi_0^2 \sin^2(m_\chi t)
\end{equation}
which is of the form discussed in Section~\ref{sec-heiseqs}, so the
methods Section~\ref{sec-heiseqs} can thus be applied.  We will not
work through the details of the calculation which may be done either
numerically or analytically (see \cite{mukhanov-cosmology-2005}).  We
know that $\xi$ particle production may occur when the adiabatic
parameter $\frac{\dot\go}{\go^2}$ is greater than one.  The adiabatic
parameter is computed,
\begin{equation}
\frac{|\dot\go|}{\go^2} = \frac{m_\chi\tg^2\chi_0^2|\sin(m_\chi t)\cos(m_\chi t)|}{(k^2 + m_\xi^2 + \tg^2\chi_0^2 \sin^2(m_\chi t))^{3/2}}
\label{reheating-inflatonadiabaticparam}
\end{equation}
and notice that for very small masses $m_\xi$ and momenta, the
adiabatic parameter is greater than one for short durations of time
when $\sin^2(m_\chi t)\ll 1$, or expanding about a time $m_\chi
t_*=n\pi$ in the small time interval $\ge \ll m_\chi^{-1}$, the
adiabatic parameter is approximated,
\begin{equation}
\frac{|\dot\go|}{\go^2} = \frac{m_\chi^2 \tg^2 \chi_0^2 \ge}{(k^2 + \tg^2\chi_0^2 m_\chi^2 \ge^2)^{3/2}}
\end{equation}
The upper bound of the resonance band is then the maximum value of
$k^2$ such that $\frac{|\dot\go|}{\go^2}>1$ is still true, or when 
\begin{eqnarray}
k^2 &<& \left[m_\chi \tg^2\chi_0^2 (m_\chi \ge)\right]^{2/3} - \tg^2\chi_0^2(m_\chi\ge)^2 \\
 &\lesssim& m_\chi \tg \chi_0
\end{eqnarray}
where the right-hand-side of the above expression has been maximized
with respect to variation of $\ge$.  We note immediately, that momenta
greater than $m_\chi$ can be generated which implies that many $\chi$
particles are involved in the decay process.  Specifically, if we take
$\chi_0=M_p$ and $m_\chi=10^{-6}M_p$, then the momenta of the produced
quanta may be as large as $k\sim 10^3 m_\chi$.  Additionally, when the
occupation numbers are computed using the methods of
Section~\ref{sec-heiseqs}, one finds an exponential growth of
particles during these short time intervals, and a quick depletion of
the inflaton condensate (within 25 oscillations) for a range of
couplings $10^{-3}>\tg>10^{-6}$ and with $m_\chi=10^{-6}M_p$ (results
quoted from \cite{mukhanov-cosmology-2005}).

         \section{Nonperturbative Decay of the Flat Direction VEVs: Preliminaries} \label{sec-reheating-nonperturbativeflat} 
This section provides the motivation for our choice of flat direction
models in Section~\ref{sec-gaugefixing}, and is based on the arguments
of \cite{olive-peloso-2006, sexton-etall-2008}.  The focus is on the
nonperturbative decay mechanism and establishing criterion for an
efficient nonperturbative decay.  We begin by noting the above model
of inflaton decay via coupling $\tilde g \chi^2\xi^2$ is not
appropriate to model flat direction decay as the flat directions are
complex fields.  The analogous coupling would be
\begin{equation}
\tilde g |\Phi|^2\xi^2
\label{reheating-trialcoupling}
\end{equation}
where $\Phi$ is the flat direction VEV, and $\xi$ is a real light
scalar field.  The motion of the flat direction is typically an
elliptical motion in the complex plane as shown in
Figure~\ref{fig-phievolution}.  The induced mass $m_{\xi}\sim
|\Phi|^2$ will never pass through zero and thus will never result in a
large adiabatic parameter~\myref{reheating-inflatonadiabaticparam} as
happens for the inflaton.  To shown this, one may estimate the
adiabatic parameter shortly after the onset of flat direction linear
oscillations to be,
\begin{eqnarray}
\frac{\dot\go_{\xi}}{\go_{\xi}^2} &\siml& \frac{\msusy}{\ge_{B-L}|\Phi_0|}
\label{reheating-adiabaticguess}
\end{eqnarray}
where $\go_{\xi} \sim m_{\xi} \sim |\Phi_0|$ and where $\ge_{B-L}$ is
again the $B-L$ number per particle in the flat direction condensate
defined in~\myref{inflation-chargeasymmetry}.  Taking $\Phi_0\sim
10^{-2}M_P$ One then requires $\ge_{B-L}\siml
\frac{\msusy}{|\Phi_0|}\siml 10^{-14}$ in order to obtain an order
one adiabatic parameter.  One might suggest that the early nonlinear
evolution of $\Phi$ prior to $R_{H\sim m}$ may contribute to
non-adiabatic evolution, but this evolution would be model dependent
and also conditional on the higher order terms of the flat direction
potential~\myref{inflation-potentialseries} dominating over the mass
term (see Section~\ref{sec-inflation}).  Our goal is to look for a
model independent decay mechanism so we focus on the behavior during
the linear oscillations of the flat direction when $R>R_{H\sim m}$.

There is another problem with the
coupling~\myref{reheating-trialcoupling}, and it is that in the
context of the MSSM with all gauge symmetries broken, the field $\xi$
should correspond to either the heavy Higgs-like scalars or to the
orthogonal light states which are flat direction quanta $\gd\phi_a$.
For couplings to the heavy scalars, one naively expects the
conclusion~\myref{reheating-adiabaticguess} to
apply.\footnote{Although we will see these heavy states may in fact be
  produced} For the flat direction perturbations, this conclusion does
not necessarily apply.  Additionally, as has been discussed in
Section~\ref{sec-susyflat-lifting}, one rather expects a situation in
which multiple flat directions have acquired VEVs so the effective
mass terms~\myref{reheating-trialcoupling}, should properly be
represented as a matrix expression composed of the heavy and light
states.  The matrix elements generally acquire time dependent terms of
the sort $\dot\Phi_a$.  Because the time dependence is on the same
scale $\msusy$ as the light eigenstates of the mass matrix, one
expects nonperturbative particle production.  Then through analysis
and numerics, as discussed in Section~\ref{sec-heiseqs}, one may
determine whether this is in fact the case.  We note that the
production in the multifield case is sourced by both the rapidly
changing eigenvalues and rapidly changing eigenvectors (see
Section~\ref{sec-heiseqs}.  We also note, that for reasons discussed
in Section~\ref{sec-heiseqs-scaling}, the heavy Higgs-like states of
the theory should not immediately be discarded as they will
generically mix with the light states and may be produced on shell.
Thus the mass matrix above $\cM^2$ should be expanded to include these
heavy states.  Specific models which further illustrate the above
sketch will be presented in detail in Section~\ref{sec-gaugefixing}
which follows.

Before concluding, we note some potential obstacles to the
nonperturbative decay of the flat directions.  The first potential
obstacle is the kinematic availability of states for the flat
direction to decay to.  Specifically, the flat direction quanta have
mass $\msusy$, and there may be limited or zero phase space for decay
of flat direction quanta into other states of the same mass if there
are equal numbers in the initial and final state.  For instance, if
the only allowed reactions were $2\rightarrow 2$ reactions such as an
s-channel process $\phi_1^*\phi_1\rightarrow \phi_2^*\phi_2$, the
resonance band would be limited by the difference of the masses $k^2
\sim m_1^2-m_2^2$.  However, if the process is nonperturbative there
may be more particles in the initial state than in the final state
$(\phi_1^*\phi_1)^N\rightarrow \phi_2^*\phi_2$ which broadens the
resonance by the ratio $N$.  As we saw at the conclusion of the
previous section for the case of the inflaton, the ratio could be as
large as a thousand.  This can make kinematics less of an obstacle.
However, the flat direction condensate may possess a certain amount of
$B-L$ charge at the start of linear oscillations, and one must confirm
that $B-L$ conservation will not constrain the decay channels of the
flat direction.  For instance in the annihilation example just
mentioned, $(\phi_1^*\phi_1)^N\rightarrow \phi_2^*\phi_2$, if the
initial state possesses nonzero $B-L$, then the final charge must
match.  This could restrict the allowed interactions to $N\rightarrow
N$, and potentially restrict the decay rate if $m_1\approx m_2$.
Charge conservation issues such as this are raised for instance in
\cite{allahverdi-mazumdar-2008}. However, a supression of the flat
direction decay as envisaged in \cite{allahverdi-mazumdar-2008} relies
heavily on the condensate being maximally charged, and as we will see
in the Results, a sizeable charge asymmetry is still consistent with
nonperturbative decay.

A second potential obstacle concerns the heavy states should they
happen to be produced in our calculations.  By dimensional analysis,
the subsequent decay rate of our heavy states should be,
\begin{equation}
\Gamma_{heavy} \sim \ga M
\label{reheating-heavydecay}
\end{equation}
where $\ga$ is a gauge coupling and $M$ is the heavy scale.  Since the gauge couplings are all
less than one (see Figure~\ref{fig-gaugecouplings}) at the TeV scale
and above, this suggests that $\Gamma_{heavy} < M$.  This
conclusion is also supported by the analogy to the standard models
Electroweak symmetry breaking and the decay width of the heavy W and Z
bosons or the standard model Higgs.  These states are analogous to the
heavy states in our Lagrangians, and their widths are measured (or
expected in the case of the Higgs) to be less than their mass
\cite{pdg-2008}.  If this inequality is not obtained for our heavy
states, meaning that the decay width of our heavy states is comparable
to or larger than the mass of the states, then these states are
virtual.  This would make the computation of their occupation number
nonsensical as the notion of a particle state is not available.

From the above considerations, our expectations are that once we have
chosen a realistic flat direction model, one may obtain a resonance
band for momenta $k\lesssim \msusy$. One must verify that
interactions exist within the model to mediate the decay, and one must
verify that the classical dynamics of the VEVs can lead to an
adiabaticity matrix with elements greater than order one.  The only
way to verify these two points is to specify a model, solve the
classical evolution, and apply the methods of
Section~\ref{sec-heiseqs}.  These issues are explored in the following
section for some simple models of flat directions.

         \section{Summary of Reheating} \label{sec-reheating-summary} 
The above discussion of reheating is only a sampling of the scenarios
for the reheating process considering very generic decay mechanisms
(see Figure~\ref{fig-reheating}).  The full range of scenarios is vast
and corresponds closely to the vast array of inflaton models
\cite{kofman-reheating-2008}, the large uncertainty in the energy
scale of inflation, and the unknown appropriate particle/field
description during reheating.  These are major open questions.  It is
worth noting that answers to the latter two questions may come in the
not too distant future as observational cosmologists are seeking to
measure gravitational waves (and thus the energy scale of inflation)
and particle physicists apply the LHC to characterize the TeV scale.

In the remaining material, we aim to answer questions related to the
effects of the flat directions on a situation of gravitational decay
of the inflaton.  We saw that the thermalization process can be
potentially delayed by the presence of flat direction VEVs, but more
importantly, we saw that the flat direction can in fact dominate the
energy density of the universe if it does not decohere or decay
quickly.  The following sections are focused on the realistic
modelling and simulation of flat direction dynamics with the goal of
estimating the decay time of the flat directions, and thus resolving
these questions.

\newpage \chapter{Modeling Flat Direction Decay} \label{sec-gaugefixing} 
The previous four sections aimed to present the broad picture of
inflation and reheating in the context of MSSM flat directions, and to
present technical tools for studying the evolution of scalar fields in
the early universe.  In this section, the picture necessarily narrows
in order to make a concrete calculation.  Specifically, the full
$SU(3)\times SU(2)\times U(1)$ gauge symmetry of the standard model is
here truncated to its $U(1)$ part, and the full spectrum of scalar
fields in the MSSM is truncated to a case of two complex scalars and a
case of four complex scalars.  The two cases correspond to one and
multiple flat directions respectively.  To describe the effective
vacuum in these models, the fundamental scalar fields are split into a
classical background (which are the flat direction VEVs) plus
quantized fluctuations away from this background.  It is then
necessary to fix a gauge, and we fix both the background fields and
the fluctuations to the unitary gauge.  The backgrounds will be chosen
to possess only time dependence and the perturbations to the
background will possess the full space-time dependence.  However, by
evolving the perturbations one may determine how quickly spatial
fluctuations develop (flat direction decay).
The goal of this section is thus to express the two Lagrangians of our
models in a canonical form that may be solved numerically in
Section~\ref{sec-results}.


         \section{Gauge Fixing U(1) Symmetric Models} \label{sec-gaugefixing-general} 
The actions we consider consist of $N$ complex scalars $\phi_i$ which
are similar to be the scalar particles of the MSSM with the main
difference being the gauge group here contains only the U(1) part.  We
develop two simple U(1) models which have the generic form,
\begin{equation}
S = \int d^4x \sqrt{-g}\left\{-\frac14 F_{\mu\nu}(B)^2 + \sum_i^N |D_\mu\phi_i|^2 - V(\phi_1,\phi_2,...\phi_n) \right\} 
\label{gaugefixing-u1action}
\end{equation}
where $B_\mu$ is the U(1) gauge field, $q_i$ are the charge assignments
of the scalars and
$D_\mu\phi_i = \left(\partial_\mu - i e q_i B_\mu \right)\phi_i$
is the covariant derivative.  The fields thus transform under the
gauge symmetry as follows,
\begin{eqnarray*}
B_\mu &\rightarrow& B_\mu + \frac{\gl,_\mu}{e} \\
\phi_i &\rightarrow& e^{iq_i\gl} \phi_i
\end{eqnarray*}
We fix the fields to the unitary gauge where the degrees of freedom of
the fields match the physically propagating degrees of
freedom.\footnote{Equivalently, there is no extra gauge-fixing
  constraint to enforce on the classical equations of motion or
  Gupta-Bleuler constraint to enforce on the quantized fluctuations.}
This gauge choice provides the most straightforward and intuitive
field description in our canonical quantization.  A simple way to fix
to the unitary gauge is to combine the fields in the Lagrangian into
gauge invariant combinations (for example see
\cite{mukhanov-cosmology-2005}).  To do this, we first decompose the
complex scalars into the representation $\phi_i = \frac{f_i}{R}
e^{i\gs_i}$ with $f_i>0$ and with the scale factor appearing in the
denominator.  Of course, this decomposition will lead to a coordinate
singularity if any one of the $f_i$ ever becomes zero.  However, our
$\phi_i$ acquire nonzero vacuum expectation values, and by assumption
will not cross the origin except in the specific limiting case where
any of the phases $\gs_i$ are constants of the evolution.  The
decomposition is thus acceptable.  To economize notation we will also
write many expressions in this section setting the scale factor to a
constant, $R=1$.  The time dependence of the scale factor will be
restored in the final results using the prescription derived in
Appendix~\ref{calc-u1-lagrangian}.  Using this representation, $\phi_i
= \frac{f_i}{R} e^{i\gs_i}$, the gauge transformation is,
\begin{eqnarray*}
B_\mu &\rightarrow& B_\mu + \frac{\gl,_\mu}{e}  \\
\gs_i &\rightarrow& \gs_i + q_i\gl \\
f_i &\rightarrow& f_i
\end{eqnarray*}
The gauge fixing procedure is to construct linear combinations of the
phases $\gs_i$ among themselves and with $B_\mu$ which are
individually gauge invariant.  The unique gauge invariant combination
for the vector is,
\begin{equation}
\tB_\mu = B_\mu - \frac{\ga,_\mu}{g}
\;\;\;,\;\;\;
\ga = \frac{\sum_i q_i \gs_i}{\sum_i q_i^2}
\;\;\;\mbox{which transforms}\;\;\;
\ga \rightarrow \ga+\gl
\label{gaugefixing-gaugeinvariantvector}
\end{equation}
It is quick to verify that $\ga$, which will be the goldstone mode,
transforms under the gauge symmetry as specified, and also quick to
verify that the vector $\tB_\mu$ is gauge invariant.  One may then
combine the phases $\gs_i$ into a basis of gauge invariant
combinations $\theta_i$.
This will be done concretely for the two examples below.  We simply
note here that our basis of gauge invariant phases $\theta_i$ will be
$N-1$ dimensional and unique up to a time independent $O(N-1)$
rotation.  We note finally that the above gauge fixing procedure may be
equivalently interpreted as performing a specific gauge
transformation on the fields in which the combination of phases
$\ga$ is transformed explicitly to zero.

With the decomposition $\phi_i=f_ie^{i\gs_i}$, the kinetic terms of
the scalars may be written,
\begin{equation}
\sum_i |D\phi_i|^2 
= \sum_i \left[(\der_\mu f_i)^2 + f_i^2(\der_\mu \gs_i)^2\right] - B_\mu J^{\mu} + \frac12\mathcal{M}^2(f_i)  B_\mu B^{\mu}
\label{gaugefixing-scalarkineticterm}
\end{equation}
where the gauge current $J_\mu$ and the effective mass of the vector
$\cM^2$ are defined,
\begin{eqnarray}
J_\mu &\equiv& \left(e\sum_i i  q_i \phi^\dagger_i \phi_i,_\mu + h.c. \right) 
= -2e \sum_i  q_i f_i^2 \der_\mu\gs_i \nonumber \\
\mathcal{M}^2(|\phi_i|)&\equiv& \frac{{e}^2}{2}\sum_i q_i^2 |\phi_i|^2
\label{gaugefixing-scalarkinetictermb}
\end{eqnarray}
In the unitary gauge, the gauge kinetic
terms~(\ref{gaugefixing-scalarkineticterm}-\ref{gaugefixing-scalarkinetictermb})
retain their form except for the unphysical phase $\ga$ being fixed to
zero.  Similarly, the gauge current term $J_\mu$ is now a ``physical''
quantity and this offers some computational convenience.
Specifically, because our fields are assumed only time dependent, then
gauge charge conservation requires $\der_0\langle J_0\rangle = 0$ or
equivalently $\langle J_0\rangle = J_0^{init}$.  Here, $J_0^{init}$ is
the initial gauge charge density, and it is assumed to be exactly
zero.  This is imposed for phenomenological reasons, although from a
mathematical stand-point, there is no obstacle to assuming a uniform
charge density for the background.  In the following material it will
be apparent that gauge charge conservation has nontrivial implications
to the dynamics.  One final aspect to the gauge
charge~\myref{gaugefixing-scalarkinetictermb} to be noted is that when
all phases are all frozen, $\der_0 \gs_i$, the charge is automatically
zero.  In our later numerical simulations, the initial charge will be
made zero by this method.

         \section{U(1) Model With a Single Flat Direction} \label{sec-gaugefixing-model1} 
We now perform the above gauge fixing for the simple U(1) example
introduced in Section~\ref{sec-susyflat-derivemonomials} consisting of
two oppositely charged scalar fields $\phi_1$~and~$\phi_2$, and one
complex degree of freedom in the flat direction VEV.
The potential for the model includes mass terms and the D-term,
\begin{equation}
V(\phi_1,\phi_2) = m_1^2|\phi_1|^2 + m_2^2|\phi_2|^2 + \frac18 {e}^2(|\phi_1|^2-|\phi_2|^2)^2
\label{gaugefixing-model1potential}
\end{equation}
where it is assumed both masses are of order the SUSY breaking scale,
$\msusy$.  One also reads from the D-term that $\phi_1 = f_1
e^{i\gs_1}$ and $\phi_2=f_2 e^{i\gs_2}$ are charged $+1/2$ and $-1/2$
respectively.  The phase $\ga=(\gs_1-\gs_2)$ is then the gauge variant
goldstone mode while the combination $(\gs_1+\gs_2)$ is a physical
gauge invariant phase. We transform the vector as specified above
in~\myref{gaugefixing-gaugeinvariantvector} and we write the
Lagrangian in terms of the gauge invariant "physical'' fields
$\tB_\mu,\;f_1,\;f_2$,~and~$(\gs_1+\gs_2)$,
\begin{eqnarray}
&&(\der_\mu f_1)^2 + (\der_\mu f_2)^2 + \left( \frac{f_1^2+f_2^2}{4} \right) [\der_\mu(\gs_1+\gs_2)]^2 - V \left( f_1 ,\, f_2 \right) \nonumber \\
&& - \frac{1}{4} F \left( {\tilde B} \right)^2
+ \frac14{e}^2 \left( f_1^2 + f_2^2 \right) {\tilde B}_\mu {\tilde B}^\mu
- \frac12 e {\tilde B}^\mu \left( f_1^2 - f_2^2 \right) \der_\mu(\gs_1+\gs_2)
\label{gaugefixing-model1L}
\end{eqnarray}
which is the unitary gauge action in which the goldstone mode $\ga$
has necessarily dropped out.  The effective vector mass as well as the
gauge current are easily read from the above Lagrangian,
\begin{equation}
\cM^2 = \frac12 e^2 (f_1^2 + f_2^2)
\;\;\;,\;\;\;
J_\mu = \frac12 e(f_1^2-f_2^2)\der_\mu(\gs_1+\gs_2)
\label{gaugefixing-currentmodel1}
\end{equation}
We make the following demands on the classical solutions: (i) they are
only time dependent, (ii) they result in $\langle J_0 \rangle=0$ and
(iii) the phase $\langle \gs_1+\gs_2\rangle$ is dynamic so that the
fields $\langle\phi_i\rangle$ may rotate in the complex plane (see
Figure~\ref{fig-phievolution}).  These assumptions and the above form
of the gauge charge density~\myref{gaugefixing-currentmodel1} restrict
the solutions to the case $\langle f_1\rangle = \langle f_2
\rangle\equiv F$ which puts the fields precisely on the flat
direction.  However, one can show that the classical equations of
motion will force $F=0$ given our assumptions.  To show this, one
writes the equations for $\langle f_1\rangle $ and $\langle
f_2\rangle$, and takes the difference of the two equations to
obtain,
\begin{equation}
(m_1^2-m_2^2) F = 0
\end{equation}
Thus, in order to obtain a solution with nonzero $\langle f_i\rangle$,
we must relax one of our assumptions.  A simple course is to take the
special case $m_1=m_2$, but before doing this, we consider
alternatives.  We must retain a nonzero VEV $F$ as well as a dynamic
phase $\langle\gs_1+\gs_2 \rangle$ as these define precisely the case
we wish to study.  It is also difficult to explain a case with a
uniform nonzero gauge charge density from a phenomenological
standpoint, so we do not relax this assumption.\footnote{Assuming the
  charge is one of the standard model gauge charges, it would have to
  be extremely small or its coupling extremely weak in order to have
  not been observed.}  We could relax our assumption of only time
dependence on the fields -- it may be that our model becomes
consistent only when classical spatial fluctuations to the fields
(which would be small after inflation but present) are included and
local current conservation $\der_\mu J^\mu=0$ is enforced.  This case
seems interesting, and it implies a classical instability (and thus
decay) of the flat direction, but it is not explored
here.\footnote{One final possibility is that our model is simply
  ``sick'' as a physical theory due to our assumptions on the gauge
  symmetry and the field content having been too narrow.  However, one
  should not conclude this until the other cases have been explored.}
We take the special case $m_1=m_2$, and study this theory because it
is well defined, not completely trivial and also helpful for
understanding the multi-flat direction case considered in the
following subsection.  Although there is a fine tuning involved, one
expects that the model will still be predictive in the case of
slightly different masses for a time duration of approximately
$\frac1{m_1-m_2}$ which is the time required for the field
oscillations to acquire a significant phase difference.  As there are
arguments for some of the MSSM scalar fields being nearly degenerate
in mass (see Section~\ref{sec-susyflat-susybreaking}), this model
should not be ruled out on phenomenological grounds.

We continue to determine the classical background for the case
$m_1=m_2$. The vacuum expectation values of the fields are then
specified,
\begin{equation}
\langle\tilde B_\mu\rangle = \cB_\mu
\;\;\;,\;\;\;
\langle \gs_1 + \gs_2 \rangle = \gS
\;\;\;,\;\;\;
\langle f_1\rangle = \frac12F
\;\;\;,\;\;\;
\langle f_2 \rangle = \frac12F
\label{gaugefixing-model1vevs}
\end{equation}
With regard to the vector field, because the background vector is
assumed to be only time dependent, one can show the non-dynamical
component satisfies $\cB_0=\frac{\langle J_0\rangle}{\cM}=0$ (see also
Appendix~\ref{calc-u1-lagrangian}).  Additionally, at the end of
inflation it is assumed $\vec \cB=0$.  The spatial component of the
vector $\vec \cB$ can then develop a large nonzero value only from the
effect of parametric resonance.  Since $\cM' \ll \cM$, it is not
immediately clear that a parametric resonance is possible.  We take
the background solution to be $\vec \cB=0$, and the presence of a
parametric resonance for the vector will be checked by studying the
Lagrangian for the perturbations in the following discussion.
The background equations of motion then involve only the scalar fields
$F$ and $\gS$ from the above Lagrangian, and have the form,
\begin{eqnarray}
& F'' + \left(m^2 R^2 - \frac14 {\gS'}^2 - \frac{R''}{R}\right)F = 0 \nonumber \\
& \frac14\der_0\left[F^2 \gS' \right] = 0 \nonumber \\
& \frac{R''}{R}+ \left(\frac{R'}{R}\right)^2 = 4\pi G R^2 (2 V(\Phi) + \rho_\chi)
\label{gaugefixing-backgroundeqs}
\end{eqnarray}
where the dependence on the scale factor has been restored and the
energy density of the inflaton $\rho_\chi$ has been incorporated.
Note that when written in the complex representation, our VEVs are
defined
$\langle\phi_1\rangle=\langle\phi_2\rangle\equiv\Phi\equiv\frac{F}{R}e^{i\gS}$
and using physical time, the above equations of motion are simply the
form~\myref{inflation-flatdireq} of
Section~\ref{sec-inflation-vevevolution} where the solutions to the
equations have also been discussed.

The decay of the flat direction VEV may be quantified by studying the
perturbations to the background, and these are parametrized as
follows,
\begin{eqnarray}
\tilde B_\mu &=& \cB_\mu + b_\mu \nonumber \\
(\gs_1-\gs_2) &=& \gS + \frac{\gs}{F} \nonumber \\
f_1&=& \frac{F + f + \gd}{2} \nonumber \\
f_2&=& \frac{F +f - \gd}{2} 
\label{gaugefixing-model1decomposition}
\end{eqnarray}
Note that while the background quantities have only time dependence,
the perturbations $b_\mu,\;f,\;\gs,$~and~$\gd$ possess the full
space-time dependence.  Substituting this decomposition into the
Lagrangian one obtains the quadratic Lagrangian for the perturbations,
\begin{eqnarray}
&& \frac12\left[(\der_\mu f)^2 + (\der_\mu\gs)^2 + (\der_\mu\gd)^2 \right] + \frac12\gS'\left(f\gs'- f'\gs \right) \nonumber \\
&& 
-\frac12 \left( m^2 - \frac14 \Sigma'^2 \right) f^2 
-\frac12 \left(m^2 - \frac14 \Sigma'^2 \right) \sigma ^2 
-\frac12 \left(\frac14{e}^2 F^2 + m^2 - \frac14 \Sigma'^2\right)\delta ^2
\nonumber \\
&& -\frac14 [F_{\mu\nu}(b)]^2 +\frac12 \pfrac{{e}^2F^2}{4} b_\mu b^\mu - \left(\frac12 e F \gd\gS' \right) b_0
\label{gaugefixing-model1Lquadratic}
\end{eqnarray}
Let us examine the couplings above.  Notice the vector perturbation
$\vec b$ has already decoupled at quadratic order since the
current~\myref{gaugefixing-currentmodel1} does not contain a part
linear in the perturbations. The elimination of the non-dynamical
component of the vector $b_0$ contributes a mass term for $\gd$ but
otherwise does not effect the quadratic Lagrangian (see
Appendix~\ref{calc-u1-lagrangian}).  The excitation $\gd$ is the Heavy
Higgs which to leading order has mass $\cM^2$ and also decouples.  The
perturbations $f$ and $\gs$ are apparently mixed, but these are simply
the flat direction perturbations and they are degenerate with mass $m$
to leading order.\footnote{To quickly see that $f$ and $\gs$ are the
  flat direction perturbations, look at the definition of the
  perturbations~\myref{gaugefixing-model1decomposition}, and see that
  they correspond to a shift on the VEV $F$ and $\gS$.}  The
degeneracy of the eigenvalues means the corresponding $I,J$ matrix
elements are zero to leading order and thus will not lead to a
parametric resonance.  To conclude, the eigenstates and eigenvalues in
this model are stationary and the model is incapable of leading to
parametric resonance.  This conclusion is expected as there are no
states in the above model lighter than $m$ for the flat direction
quanta to decay into.  This shortcoming is corrected in the multi-flat
direction model presented next.\footnote{In principle a decay may
  occur in the single flat direction model with $m_1\ne m_2$, though
  as discussed above there are subtleties to this case. These
  subtleties are mentioned again in the Conclusions.}

         \section{U(1) Model With Multiple Flat Directions} \label{sec-gaugefixing-model2} 
One suspects that introduction of more degrees of freedom to our U(1)
model may change the outcome.  We thus perform the same analysis on
the slightly more complicated model involving four complex scalar
fields specified by the potential,
\begin{eqnarray}
V(\phi_1,\phi_2,\phi_3,\phi_4) &=& 
m^2\left(|\phi_1|^2 + |\phi_2|^2\right)  
+ \tm^2\left(|\phi_3|^2 + |\phi_4|^2\right) \nonumber \\
&\mbox{}& +\; \frac18 {e}^2(|\phi_1|^2-|\phi_2|^2 + |\phi_3|^2-|\phi_4|^2)^2
\label{gaugefixing-model2potential}
\end{eqnarray}
where the charge assignments can be read from the D-term and where the
mass choices will be discussed shortly.  Again the fields are
decomposed $\phi_i=f_ie^{i\gs_i}$. The physical phases and the
goldstone mode are,
\begin{eqnarray*}
&& \theta_1= (\gs_1-\gs_2-\gs_3+\gs_4)/2
\;\;\;,\;\;\;
\theta_2= (\gs_1+\gs_2)/2
\;\;\;\;\;\;
\theta_3 = (\gs_3+\gs_4)/2 \\
&& \ga = (\gs_1-\gs_2 + \gs_3-\gs_4)/2
\end{eqnarray*}
The Lagrangian in the unitary gauge is then,
\begin{eqnarray}
&&\sum_{i=1}^4 (\der_\mu f_i)^2
+ \frac14( f_1^2 + f_2^2 + f_3^2+f_4^2) (\der_\mu\theta_1)^2
+ (f_1^2 + f_2^2) (\der_\mu\theta_2)^2 \nonumber \\
&& +\; (f_3^2 + f_4^2) (\der_\mu\theta_3)^2 
+ (f_1^2-f_2^2) (\der_\mu\theta_1)(\der_\mu\theta_2)
- (f_3^2 - f_4^2) (\der_\mu\theta_1)(\der_\mu\theta_3)  \nonumber \\
&& -\; V \left( f_1 ,\, f_2 \right)
- \frac{1}{4} F \left( {\tilde B} \right)^2
+ \frac14{e}^2 \left( f_1^2 + f_2^2 + f_3^2+f_4^2 \right) {\tilde B}_\mu {\tilde B}^\mu \nonumber \\
&& -\; e {\tilde B}^\mu \left[ \frac12(f_1^2+f_2^2-f_3^2-f_4^2)(\der_\mu\theta_1) + (f_1^2-f_2^2)(\der_\mu\theta_2) + (f_3^2-f_4^2)(\der_\mu\theta_3) \right] \nonumber \\
&\mbox{}& 
\label{gaugefixing-model2L}
\end{eqnarray}
where the gauge current $J_\mu$ is the coefficient of the term linear
in $\tB_\mu$ and it is equivalently written in terms of the $\gs_i$
fields as,
\begin{equation}
J_\mu = e\left[ f_1^2 \der_\mu\gs_1 - f_2^2\der_\mu\gs_2 + f_3^2\der_\mu\gs_3 - f_4^2\der_\mu\gs_4 \right]
\end{equation}
where,
\begin{equation}
\gs_1 = \gth_2 + \frac12\theta_1 \;\;\;,\;\;\;
\gs_2 = \gth_2 - \frac12\theta_1 \;\;\;,\;\;\;
\gs_3 = \gth_3 - \frac12\theta_1 \;\;\;,\;\;\;
\gs_4 = \gth_3 + \frac12\theta_1 
\end{equation}
Next we determine the background.  The background for the vector is
again zero.  This leaves seven scalar fields
$f_1,f_2,f_3,f_4,\theta_1,\theta_2,\theta_3$ participating in the flat
direction background dynamics.  We wish to select solutions to the
corresponding equations of motion for which the fields will satisfy
the gauge currrent conservation $\langle J_0 \rangle=0$.  By
examination of the current term in~\myref{gaugefixing-model2L}, one
finds a convenient solution in which $\langle \theta_1\rangle=
const.$, $f_1=f_2$ and $f_3=f_4$.  This solution is convenient because
it evolves exactly on the flat direction as the previous model did.
The equations of motion are thus two copies of the
system~\myref{gaugefixing-backgroundeqs} each copy evolving
independently except for weak coupling through the Friedmann
equations.  The VEV's are defined as follows,
\begin{eqnarray*}
&&\langle f_1 \rangle = \langle f_2 \rangle \equiv F
\;\;\;,\;\;\;
\langle \theta_2 \rangle \equiv \gS 
\\[0.4em]
&&\langle f_3 \rangle = \langle f_4 \rangle \equiv G
\;\;\;,\;\;\;
\langle \theta_3 \rangle \equiv \tgS 
\\[0.4em]
&&\langle \theta_1 \rangle = 0
\end{eqnarray*}
or in the complex representation, this is written
\begin{eqnarray}
\langle\phi_1\rangle &=& \langle\phi_2\rangle \equiv \frac{\Phi}{\sqrt2} = Fe^{i\gS} \\
\langle\phi_3\rangle &=& \langle\phi_4\rangle \equiv \frac{\tilde\Phi}{\sqrt2} = Ge^{i\tgS}
\label{gaugefixing-model2background}
\end{eqnarray}
where $\Phi$ and $\tilde\Phi$ have been defined with a factor $\sqrt2$
to make the Lagrangian for these complex VEVs canonical.  The
background and perturbations are then specified,
\begin{equation}
\begin{array}{l}
f_1 \equiv F + (f + \gd)/2    \\[0.4em]
f_2 \equiv F + (f - \gd)/2    \\[0.4em]   
f_3 \equiv G + (g + \tgd)/2   \\[0.4em]   
f_4 \equiv G + (g - \tgd)/2   
\end{array}
\hspace{3em}
\begin{array}{l}
\theta_1 \equiv \theta \equiv \frac{\sqrt{F^2+G^2}}{2FG}\gth_c + \cO\pfrac{m^3}{M^3}\\[0.4em] 
\theta_2 \equiv \gS + \frac{\gs}{2F} \\[0.4em]              
\theta_3 \equiv \tgS + \frac{\tgs}{2G} \\[0.4em]            
\tilde B_\mu \equiv b_\mu
\end{array}
\label{gaugefixing-model2decomposition}
\end{equation}
where $b_\mu$ is the vector field perturbation, and where the
definitions have been constructed again to make the fields canonical.
The elimination of the time-like component of the vector $b_0$ is
performed again in Appendix~\ref{calc-u1-lagrangian}.  The flat
direction perturbations $\{f,\gs\}$ and $\{g,\tgs\}$ decouple from
eachother and from the remaining fields, and they are degenerate with
masses $m$ and $\tm$ respectively.  The transverse and longitudinal
vector degrees of freedom decouple to leading order in the expansion
$\frac{m}{M}$.  The remaining three degrees of freedom $\gd,\tgd$ and
$\theta_c$ are strongly mixed.  Specifically, forming the vector $X =
(\gd,\tgd,\theta_c)$, the quadratic Lagrangian is written in the
canonical form~\myref{heiseqs-canonicalL} with matrices $(\tgO^2 +
K^TK)$ and $K$ which are
\begin{eqnarray}
&&(\tgO^2+K^TK)_{11} = k^2 + m^2 R^2 -\frac{R''}{R} + \left[  {e}^2  + \frac{3\Sigma'^2}{F^2+G^2}\,\right]F^2 \nonumber \\
&&(\tgO^2+K^TK)_{22} = k^2 + {\tilde m}^2 R^2 -\frac{R''}{R} + \left[  {e}^2 + \frac{3\tilde \Sigma'^2}{F^2+G^2}\,\right]G^2 \nonumber \\
&&(\tgO^2+K^TK)_{33} = k^2 + \frac{(m^2 G^2+{\tilde m}^2 F^2)R^2}{F^2+G^2} - \frac{R''}{R} 
+\frac{3\,\left( F\,G'-G\,F'\right)^2}{\left(F^2 +G^2 \right)^2} \nonumber \\
&&(\tgO^2+K^TK)_{12} =  F\,G\, \left( {e}^2 + \frac{3 \Sigma'\,\tilde{\Sigma}'}{F^2+G^2}\right) \nonumber \\
&&(\tgO^2+K^TK)_{13} = \frac{F}{\sqrt{F^2+G^2}} \,\left[ \frac{3\,\Sigma'\,\left(F\,G'-G \, F'\right) }{ F^2+G^2}\right] \nonumber \\
&&(\tgO^2+K^TK)_{23} = \frac{G}{\sqrt{F^2+G^2}} \,\left[ \frac{3\,\tilde{\Sigma}'\,\left(F\,G'-G \, F'\right) }{ F^2+G^2}\right]
\label{gaugefixing-model2tM} \\
&& K_{12} = 0
\;\;\;,\;\;\;
K_{13}= \frac{-G\,\Sigma'}{\sqrt{F^2 + G^2}}
\;\;\;,\;\;\;
K_{23} = \frac{F\,\tilde{\Sigma}'}{\sqrt{F^2 + G^2}}
\label{gaugefixing-model2K}
\end{eqnarray}
and with $\go^2_i$ and $\Gamma_{ij}$ which are,
\begin{eqnarray}
\go^2_1 &=& k^2 - \frac{R''}{R} + \frac{{e}^2}{2}R^2(|\Phi|^2+|\tilde\Phi|^2) \nonumber \\
&\mbox{}& + \;\left[ 
           \frac{R^2 \left(m^2 |\Phi|^2 + \tilde{m}^2 |\tilde\Phi|^2\right)}{|\Phi|^2 +|\tilde\Phi|^2}
 + \frac{3  \left(\mbox{Im}\left[\frac{\Phi'}{\Phi}|\Phi|^2  + \frac{\tilde\Phi'}{\tilde\Phi}|\tilde\Phi|^2 \right]\right)^2}{(|\Phi|^2 +|\tilde\Phi|^2)^2} 
   \right] + \cO \pfrac{m^2}{M^2}  \nonumber\\
\go^2_2 &=& k^2 -\frac{R''}{R} + \left[\frac{R^2 \left({\tilde m}^2 |\Phi|^2 + m^2 |\tilde\Phi|^2\right)}{|\Phi|^2 +|\tilde\Phi|^2} 
        + \frac{3|\Phi|^2 |\tilde\Phi|^2}{(|\Phi|^2+|\tilde\Phi|^2)^2}\left|\frac{\Phi'}{\Phi}-\frac{\tilde\Phi'}{\tilde\Phi} \right|^2
\right] \nonumber \\
&\mbox{}& +\; \cO \pfrac{m^2}{M^2}  \nonumber\\
\go^2_3 &=& k^2 -\frac{R''}{R} + \left[ \frac{R^2 \left({\tilde m}^2 |\Phi|^2 + m^2 |\tilde\Phi|^2\right)}{|\Phi|^2 +|\tilde\Phi|^2} \right] + \cO \pfrac{m^2}{M^2}   
\nonumber \\[1em]
\Gamma_{12} &=& 0   \nonumber\\
\Gamma_{13} &=& - \frac{|\Phi||\tilde\Phi|}{|\Phi|^2+|\tilde\Phi|^2} \left|\frac{\Phi'}{\Phi}-\frac{\tilde\Phi'}{\tilde\Phi} \right| \nonumber \\
\Gamma_{23} &=& R^2 \left(m^2 - \tilde{m}^2\right)\frac{\mbox{Im}\left[\frac{\Phi'}{\Phi}-\frac{\tilde\Phi'}{\tilde\Phi}\right] }{\left|\frac{\Phi'}{\Phi}-\frac{\tilde\Phi'}{\tilde\Phi} \right|^2} 
- \frac{\mbox{Im}\left[\frac{\Phi'}{\Phi}|\Phi|^2  + \frac{\tilde\Phi'}{\tilde\Phi}|\tilde\Phi|^2 \right]}{|\Phi|^2 +|\tilde\Phi|^2} \,.
\label{gaugefixing-model2eigenvalues}
\end{eqnarray}
Where the expressions for $\go_i$ and $\Gamma_i$ have been written in
the complex representation of the fields.  The mixing between the
three fields is a strong mixing that can not be removed by any small
rotation.  The system is diagonalized by an $\cO(1)$ time dependent
rotation, and is expected to lead to parametric resonance.
From the eigenvalues, one sees the system includes a heavy Higgs
excitation as well as two light excitations.\footnote{The light
  degrees of freedom are the third flat direction of this model.  The
  gauge current conservation on the background and our mass
  assignments however prevented this flat direction from obtaining a
  VEV} These results~\myref{gaugefixing-model2eigenvalues} will be
applied in the following section to verify the parametric resonance
and to compute the decay of this flat direction model.

\newpage \chapter{Results} \label{sec-results}  
Here we perform the numerical implementation of the four field U(1)
model presented in section~\ref{sec-gaugefixing-model2}.  The model is
converted into a form appropriate for numerical solution, the initial
conditions of the system are discussed, and some useful measures of
the decay are also presented.  Using arguments based on the scaling of
the Heisenberg equations (or equivalently equipartition of energy), it
is then shown that of the four parameters $m$, $\tm$, $|\gF_0|$,
$|\tgF_0|$ that could be varied, only the ratios $\tm/m$ and
$|\tgF_0|/|\gF_0|$ need be varied with the remaining cases being
obtained through rescaling.  The production rate of quanta is reported
for a few values of $\tm/m$ and for a continuous range of values
$|\tgF_0|/|\gF_0|$ in which the particle production is exponential.
Additionally, the decay time of the toy flat direction model is
determined in this resonance band with all data having been rescaled
to phenomenologically relevant $TeV$ scale.

         \section{Numerical Setup of the U(1) Model With Multiple Flat Directions} \label{sec-results-numericalsetup} 
We study the nonperturbative decay of the multi-flat direction model
presented in section~\ref{sec-gaugefixing-model2}.
The background equations are of the form specified
in~\myref{gaugefixing-backgroundeqs} for the two separately evolving
flat directions.
The Heisenberg Equations of motion~\myref{heiseqs-bogequations} will
be solved with the driving terms determined
in~\myref{gaugefixing-model2eigenvalues}.  It is necessary to give the
flat directions initial charges, and rather than impose the charge by
hand, the charge is introduced dynamically according to the susy
breaking potential terms motivated
in~\myref{susyflat-lambdacouplings}.  The effect on the solutions is
to give the phase of the flat direction an initial kick so as to
induce a rotation in the complex plane (see
Figure~\ref{fig-phievolution}).  These quartic terms become
sub-dominant to the mass term after a short period of time,
so their modification to the eigenvalues and Gamma matrix
elements~\myref{gaugefixing-model2eigenvalues} may be neglected after
this time.  We additionally introduce a matter fluid, to represent the
inflaton energy density which dominates throughout.  The background
fields defined in~\myref{gaugefixing-model2background} are again,
\begin{eqnarray}
\langle\phi_1\rangle &=& \langle\phi_2\rangle \equiv \frac{\Phi}{\sqrt2} = \frac{F}{R}e^{i\gS} \\
\langle\phi_3\rangle &=& \langle\phi_4\rangle \equiv \frac{\tilde\Phi}{\sqrt2} = \frac{G}{R}e^{i\tgS}
\end{eqnarray}
where the scale factor has been restored.  The scale $M = 10^{-2}M_P$
is introduced which represents the approximate scale of the VEVs, and
the following dimensionless quantities which will be used for numerics
are then defined,
\begin{eqnarray}
&& \eta_* \equiv e M \eta \nonumber\\
&& m_* \equiv \frac{m}{e M} \;\;\;,\;\;\;
{\tilde m}_* \equiv \frac{{\tilde m}}{e M} \;\;\;,\;\;\;
k_* \equiv \frac{k}{e M} 
\;\;\;,\;\;\; \lambda_* \equiv \frac{\lambda}{e^2} \;\;\;,\;\;\; {\tilde \lambda}_* \equiv \frac{\tilde \lambda}{e^2}
\nonumber\\
&& F_* \equiv \frac{F}{M} \;\;\;,\;\;\; G_* \equiv \frac{G}{M} \;\;\;,\;\;\;
\rho_{\chi*} \equiv \frac{\rho_\chi}{e^2 \, M^2 M_p^2}
\label{results-computerunits}
\end{eqnarray}
With these definitions, the coupling $e$ becomes an overall constant
in the Lagrangian which will not enter the equations of motion.  Also
the scale $M$ will now appear only in the equation for the scale
factor.  Our choice of quartic couplings 
\begin{equation}
\lambda = \frac{m^2}{10 \, \vert \Phi_0\vert^2} \quad\,,\quad \tilde{\lambda} = \frac{\tilde{m}^2}{10 \, \vert \tilde{\Phi}_0\vert^2}\,.
\label{results-lambdacouplings}
\end{equation}
which are of the form described in~\myref{susyflat-lambdacouplings}.
The above the couplings $\gl$ and $\tgl$ are assumed real with complex
phases having been absorbed by a redefinition of the fields (and thus
appearing in the initial values of the phases).
The additional factor of $\frac1{10}$ above was chosen to be as large
as possible without introducing instability in the background
equations.
Using our dimensionless representation, the complete set of background
equations are,
\begin{eqnarray}
&&F_*''+\left(m_*^2 R^2- \frac{R''}{R}-\Sigma'^2 \right) F_* + \frac{\lambda_*}{2} \,F_*^3 \cos\left(4\,\Sigma\right) = 0\nonumber \\
&&\left(F_*^2\,\Sigma'\right)' -\frac{\lambda_*}{2} \,F_*^4 \sin\left(4\,\Sigma\right)=0 \nonumber \\
&&G_*''+\left({\tilde m}_*^2 R^2- \frac{R''}{R}-{\tilde \Sigma}'^2 \right) G_* + \frac{{\tilde \lambda}_*}{2} \,G_*^3 \cos\left(4\,{\tilde \Sigma}\right) = 0\nonumber \\
&&\left(G_*^2\,{\tilde \Sigma}'\right)' -\frac{{\tilde \lambda}_*}{2} \,G_*^4 \sin\left(4\,{\tilde \Sigma}\right)=0 \nonumber \\
&& \frac{R''}{R} + \frac{R'^2}{R^2} = 4 \pi \left\{ \frac{M^2}{M_p^2} \left[
\begin{array}{l}
m_*^2 \, F_*^2 + \frac{\lambda_* \, F_*^4}{4 \, R^2} \cos \left( 4 \Sigma \right) \\
+ \; {\tilde m}_*^2 \, G_*^2 + \frac{{\tilde \lambda_*} \, G_*^4}{4 \, R^2} \cos \left( 4 {\tilde \Sigma} \right) 
\end{array}\right] + R^2 \, \rho_{\chi*} \right\}
\label{results-backgroundeqs}
\end{eqnarray}
where prime now denotes derivatives with respect to $\eta_* \,$, and
where the equation for the scale factor was expressed in a way so that
the time derivatives of the fields $\gF$ and $\tgF$ do not appear.  In
the above background equations, the evolution of the inflaton energy
density will simply be that of a dominating matter fluid
$\rho_{\chi*}=\rho_{\chi*}(0)\pfrac{R(0)}{R}^3$. In the dimensionless
units, the initial conditions of the background equations are,
\begin{eqnarray}
R(0) = 1
\;\;\;,\;\;\;
\rho_{\chi*}(0) = 1 
\\
F_*(0) = F_0
\;\;\;,\;\;\;
\gS(0) = 0.25
\;\;\;,\;\;\;
G_*(0) = G_0
\;\;\;,\;\;\;
\tgS(0) = 0.25 
\\
\pfrac{F_*e^{i\gS}}{R}' = 0
\;\;\;,\;\;\;
\pfrac{G_*e^{i\tgS}}{R}' = 0
\end{eqnarray}
where the initial condition on the scale factor is chosen $R=1$ for
convenience.  The initial condition for $R'$ is then set by the
Friedmann Equation.  The phases $\gS$ and $\tgS$ must be nonzero in
order for an initial charge to develop through the $\gl$ and $\tgl$
terms.  Also in the above, we have chosen $\rho_{\chi*}$ large enough
initially so that $H \mg m,\tm$ and thus it is consistent to set the
fields $\Phi$ and $\tgF$ as frozen, thus fixing the initial conditions
for $F_*',G_*',\gS',\tgS'$.  

The Heisenberg Equations in the dimensionless computer units are,
\begin{eqnarray}
& \alpha' = \left(-i \omega_* -I_*\right)\alpha + \left(\frac{\omega_*'}{2\omega_*}-J_*\right)\beta \nonumber \\
& \beta' = \left(i \omega_*-I_*\right) \beta + \left(\frac{\omega_*'}{2\omega_*} -J_*\right)\alpha
\label{results-bogequations}
\end{eqnarray}
where we have introduced the dimensionless diagonal frequency matrix
and dimensionless $\Gamma,I,J$ matrices
\begin{eqnarray}
\omega_* &\equiv& \frac{\go}{eM}  \;\;\;,\;\;\; \Gamma_* \equiv \frac{\Gamma}{e \, M} \\
I_*,J_* &\equiv& \frac12 \left(\sqrt{\omega_*} \Gamma_* \frac{1}{\sqrt{\omega_*}} \pm \frac{1}{\sqrt{\omega_*}}\Gamma_* \sqrt{\omega_*} \right)
\end{eqnarray}
Recall that $\go = \rm{diag}(\go_1,\go_2,\go_3)$.  The reader is also
reminded of the adiabaticity matrix elements defined
in~\myref{heiseqs-adiabaticitymatrix} which again are,
\begin{eqnarray}
A &=& \frac{\omega'}{\omega^2} - \left( \Gamma \frac{1}{\omega} - \frac{1}{\omega}\Gamma \right) \nonumber \\
A_{ij}  &=& \left\{
\begin{array}{ll}
\frac{\omega'_i}{\omega_i^2} & i=j \\
\Gamma_{ij}\left(\frac{1}{\omega_i}-\frac{1}{\omega_j} \right) &i\ne j
\end{array}\right\}
\;\; \mbox{no summation}
\label{results-adiabaticitymatrix}
\end{eqnarray}
where one notices these quantities are explicitly dimensionless and invariant
under the redefinitions~\myref{results-computerunits}.  

The initial conditions for the Bogolyubov coefficients again are
$\alpha_0 = \mathbf{1}$, $\beta_0 = 0$.  The evolution of the
Heisenberg Equations is begun at the time when the smallest eigenvalue
of~\myref{gaugefixing-model2eigenvalues} is positive and equal to
twice the $R''/R$ term of these eigenvalues.  This ensures that any
significant production of quanta will be due to the effects of the
evolving flat direction VEVs and not due to the gravitational
background which appears through the term $R''/R$.

The occupations numbers are computed with $n_i\equiv
(\gb^*\gb^T)_{ii}$, and these along with the corresponding
eigenfrequencies are used to compute the energy density of the
produced quanta.  In particular, the ratio of energy density of the
produced quanta to the energy density of the background flat
directions is,
\begin{eqnarray}
r_{prod} &\equiv& \frac{\rho_{prod}}{\rho_{flat}} = \pfrac{16\,\pi\,e^2}{T +\tT} \left(\sum_{i=1} ^3 \int dk_*\,k_*^2 \omega_{i*} (k_*) n_i (k_*)\right) \label{results-energydensityratio} \\
&\mbox{}& \nonumber \\
T  &\equiv& F_*^2 \left[ \left(\frac{F_*'}{F_*} -\frac{R'}{R} \right)^2 +\Sigma'^2 \right]         + R^2 F_*^2 \left[ m_*^2 + \frac{\lambda_*}{4} \, \frac{F_*^2}{R^2} \cos(4 \,\Sigma) \right] \nonumber\\
\tT&\equiv& G_*^2 \left[ \left(\frac{G_*'}{G_*} -\frac{R'}{R} \right)^2 +\tilde{\Sigma}'^2 \right] + R^2 G_*^2 \left[ \tilde{m_*} ^2 + \frac{\tilde{\lambda_*}}{4} \, \frac{G_*^2}{R^2} \cos(4 \,\tilde{\Sigma}) \right] \nonumber 
\end{eqnarray}
This ratio is zero at the start of the evolution, and we evolve the
equations until the ratio becomes unity at which point, we consider
the flat directions to have decayed.  Once $r_{prod}=1$, the present
analysis based on the quadratic Lagrangian is also invalid since we
are ignoring the back-reaction of the produced quanta on the evolution
of the flat directions.  For example, we do not account for the
decrease in the amplitudes of the flat directions due to the particle
production.  However the instant $r_{prod}$ is a useful measure of the
decay of the flat direction.  As a measure of the time for the decay,
we use the number of rotations of the first flat direction, $N \equiv
\left( \Sigma - \Sigma_0 \right) / 2 \pi$ as this is somewhat easier
to visualize than the physical (or conformal) time and in particular
it is independent of the scale $m$.  We thus denote by $N_{decay}$ the
value of $N$ at which the ratio $r_{prod}$ equals to one.

For numerical solution, the background
equations~\myref{results-backgroundeqs} which are five second order
equations are easily rewritten as a ten dimensional system of the form
$X' = f(X)$ by defining the state vector
$X =(F,F',\gS,\gs',G,G',\tgS,\tgS',R,R')$.
The numerical solutions are easily obtained using a typical desktop
computer (1 Gigahertz Processor, 1 Gigabyte memory) in approximately a
minute or less.  The Heisenberg equations~\myref{results-bogequations}
may also be written in this generic form with a state vector of length
36 composed of the two $3\times 3$ complex matrices $\ga,\gb$.
However, because these equations require the background solutions for
the computation of the driving terms, it is convenient to solve both
the background equations and the Heisenberg equations as a single
system in a vector $X$ of length 46.  Numeric solution of this system
is demanding due to the increased number of dependent variables, and
also due to the periods of non-adiabatic evolution resulting in a
numerically stiff differential equation for these periods.  The
solutions for a single value of momentum and for approximately five to
twenty complete rotations of the flat directions can require up to few
hours of computer time using a standard differential equation solver
such as ODEPACK.

There is one further aspect to the numerical solution which requires
attention and this concerns the series expansion in the small
parameter $\frac{m}{M}$ which was used in obtaining the driving
terms~\myref{gaugefixing-model2eigenvalues}.  The series expansion is
in fact more properly an expansion based on the following hierarchy,
\begin{equation}
\left\{ 
m^2\,, 
{\tilde m}^2 \,,
\left|\frac{\dot\Phi}{\Phi}\right|^2 \,, \left|\frac{\dot{\tilde\Phi}}{\tilde\Phi}\right|^2 
\right\} 
\ll \left(|\Phi|^2+|\tilde\Phi|^2\right)\,.
\label{results-seriesapproximation}
\end{equation}
This hierarchy is actually more restrictive than the condition
$\frac{m}{M} \ll 1$ because the non-adiabatic evolution can make terms
such as $\left|\frac{\dot\Phi}{\Phi}\right|$ sometimes larger than the
masses $m,\tm$.  However, if $m,\tm$ are chosen small enough, the
above hierarchy and the series expansion are valid for the whole
numerical evolution.  In the numerics, this has been arranged to be
the case.

To summarize, once solutions to the background and Heisenberg
equations have been obtained through the above setup, the scaling
arguments (discussed next) may be safely applied, and the numeric
results obtained.

         \section{Scaling of the Numeric Solutions} \label{sec-results-scaling} 
In this section the scaling arguments of
Section~\ref{sec-heiseqs-scaling} are applied to show that of the four
parameters $\Phi_0$, $\tgF_0$, $m$, and $\tm$, only the ratios
$\frac{m}{\tm}$ and $\frac{|\gF_0|}{|\tgF_0|}$ need be varied in the
numerical simulations.  The solutions of the remaining cases may be
obtained through the rescaling shown in Table~\ref{table-rescaling}.
The heavy scale in our model is the scale of the VEVs and specifically
it is taken to be $M=\sqrt{\Phi_0\tgF_0}$.  The light scale is the
scale of the masses and we use $m$ for this scale.  Recall the scaling
is reliable only when both the known solution and the rescaled
solution have sufficiently small ratios $\frac{m}{M}$.  However, in
practice, all four parameters $\Phi_0$, $\tgF_0$, $m_*$ and $\tm_*$
are varied in the numerical simulations, and the expected scaling will
be verified in the following section (see Figure~\ref{fig-scaling}).

\begin{table}[!bthp]
\begin{center}
\begin{tabular}{|c|l|l|}
\hline 
& \textbf{known sol.} & \textbf{rescaled sol.} \\
\hline 
&&\\
\textbf{Scales} & $m$ , $\tm$ , $\Phi_0$ , $\tgF_0$ & $\mu m$ , $\mu \tm$ , $\gg \Phi_0$ , $\gg\tgF_0$ \\
&&\\
\hline
&&\\
\textbf{Background Solutions} & $\Phi(t)$ , $\tgF(t)$  & $ \gamma\Phi\pfrac{t}{\mu}$ , $\gamma\tgF\pfrac{t}{\mu}$ \\
&&\\
\hline
\end{tabular}
\caption[Scaling of the Background
Solutions]{\label{table-rescalingbackground} With known solutions to
  the background equations for $\Phi$ and $\tgF$ of the
  form~\myref{results-backgroundscaling} at the scales $m$ and
  $M=(|\Phi_0||\tgF_0|)^{1/2}$, one may infer rescaled solutions at the
  scales $\mu m$ and $\gamma M$.  The table shows how the solutions
  and their functional dependence change under the rescaling.}
\end{center}
\end{table}
The scaling of the background solutions was simply assumed in
Section~\ref{sec-heiseqs-scaling}.  We now briefly justify how the
background rescales.  We use the complex representation of the fields
and physical time for which the equation for a flat direction is,
\begin{equation}
\ddot\Phi + 3H\dot\Phi + m^2\Phi +\pfrac{m^2}{10|\Phi_0|^2}\Phi^3 = 0
\label{results-backgroundscaling}
\end{equation}
and where the assumed form for $\gl$
in~\myref{results-lambdacouplings} has been substituted.  Here the
Hubble parameter is driven externally by the dominating inflaton so we
may take $H=\frac{2}{3t}$ which is the time evolution for a matter
dominated universe.  With these two substitutions, the rescaling of
the solutions is straightforward to show and the results are outlined
in Table~\ref{table-rescalingbackground}.  This Table may be compared
to Table~\ref{table-rescaling} constructed earlier, and note in
particular, the time dependence has rescaled in the same way in both
tables.  What the above rescaling means is the trajectory of the
background retains its shape under changes in $m$ and $M$, with the
only changes being the overall scale of the trajectory and how fast
the trajectory is traversed.
Also, from Table~\ref{table-rescaling}, one reads the occupation numbers
will rescale as follows,
\begin{equation}
\left\{ n_1 \;,\; n_2 \;,\; n_3 \right\} \rightarrow \left\{ \frac{\mu}{\gamma} \, n_1 \;,\; n_2 \;,\; n_3 \right\}
\label{results-rescalingn}
\end{equation}
and applying this along with the rescaling of the background and the
rescaling of the frequencies $\go_i$, it is possible to show that
$r_{prod}$ defined in~\myref{results-energydensityratio} will rescale
as,
\begin{equation}
r_{prod} \rightarrow 
\frac{\mu^2}{\gamma^2}\,r_{prod}
\label{results-scalingrprod}
\end{equation}
which is the main result of this section.  This formula implies that
$r_{prod}$ may be expressed,
\begin{equation}
r_{prod} \simeq \frac{\tilde{m} \,  m }{ |\tilde{\Phi}_0| \, |\Phi_0|} \times f \left( \frac{\tilde m}{m} ,\, 
\frac{\vert {\tilde \Phi}_0 \vert}{\vert \Phi_0 \vert} ,\, N \right)
\label{results-functionf}
\end{equation}
where the the dependence on the overall scales of the masses and the
VEVs is made explicit. The function $f$ may be obtained numerically,
and in general it exhibits an exponential growth with the number of
rotations $N$ as well a strong dependence on its arguments
$|\tgF_0|/|\Phi_0|$ and $\tilde m / m$.  However both the exponential
growth and the scaling will only be obtained within a band of
parameter values for $|{\tilde\Phi}_0|/|\Phi_0|$, and outside this
band, neither will be obtained.  Within the resonance
band~\myref{results-functionf} can be cast in the form
\begin{equation}
r_{prod}  \simeq C \,
\frac{\tilde{m} \,  m }{ |\tilde{\Phi}_0| \, |\Phi_0|} \, 10^{\,\sigma 
\, N} \,,
\label{results-growthexp}
\end{equation}
where $C$ and $\sigma$ are two time--independent quantities which are
functions of the ratios $\tilde{m}/m$ and $| \tilde{\Phi}_0|/|
\Phi_0|$.  One may also invert this expression to obtain,
\begin{equation}
N_{decay} \simeq \frac{1}{\sigma} \, {\rm log}_{10} \left(\frac{| 
\tilde{\Phi}_0| \, | \Phi_0|}{ C\,\tilde{m}\,m} 
\right)\,.
\label{results-ndecay}
\end{equation}
And thus, so long as the parameters are within the resonance band, the
decay time $N_{decay}$ is driven mostly by the the production rate
$\gs$ with other effects contributing small logarithmic corrections.

         \section{Numerical Results} \label{sec-results-results} 
The numerical model was presented in the first subsection along with
some useful physical measures such as the adiabaticity matrix $A_{ij}$
in~\myref{results-adiabaticitymatrix}, and the ratio of the energy
densities $r_{prod}$ in~\myref{results-energydensityratio}.  The
adiabaticity matrix elements are a quick method for identifying a
resonance band, but the ratio of the energy densities will provide a
more precise method, and the number of rotations $N_{decay}$ at which
the ratio becomes unity will be our primary measure of the decay time.
In the previous section it was determined that only the ratios $m/\tm$
and $|\tgF_0|/|\gF_0|$ have any physical relevance in our numerics, so
in the following, the results will be shown in terms of these two
ratios.

We begin by showing the background evolution in
Figure~\ref{fig-background} for a sample set of initial conditions and
parameters.  As discussed in the previous section, the background
trajectories retain their shape under a rescaling of the masses $m$
and the VEVs.  Thus, the trajectory shapes shown are relevant to any
mass scale including the Electroweak scale.  The dimensionless charge
asymmetry~\myref{inflation-chargeasymmetry} may also be determined,
and for instance the flat direction $\Phi$ shown in
Figure~\ref{fig-background} is computed to have a charge asymmetry
$\ge_{B-L}\approx 0.26$.  In the following, the relative size of the
trajectories is varied simply by adjusting the initial value of the
ratio $\frac{|\tgF_0|}{|\gF_0|}$.

\begin{fullpagefigure}
\begin{center}

\hspace{-1em}\includegraphics[scale=0.4]{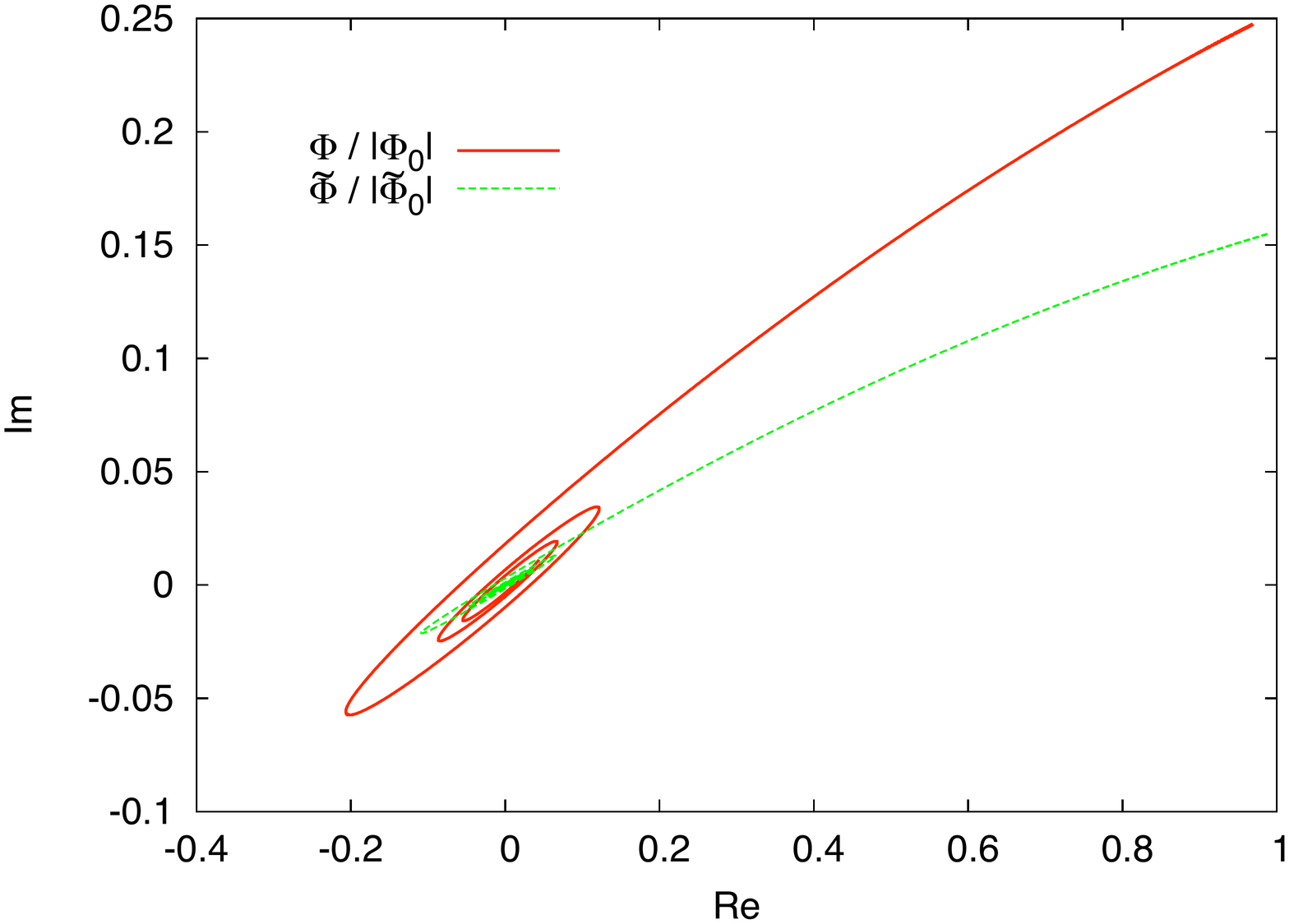}

\vspace{-3em}
\hspace{-1em}\includegraphics[scale=0.4]{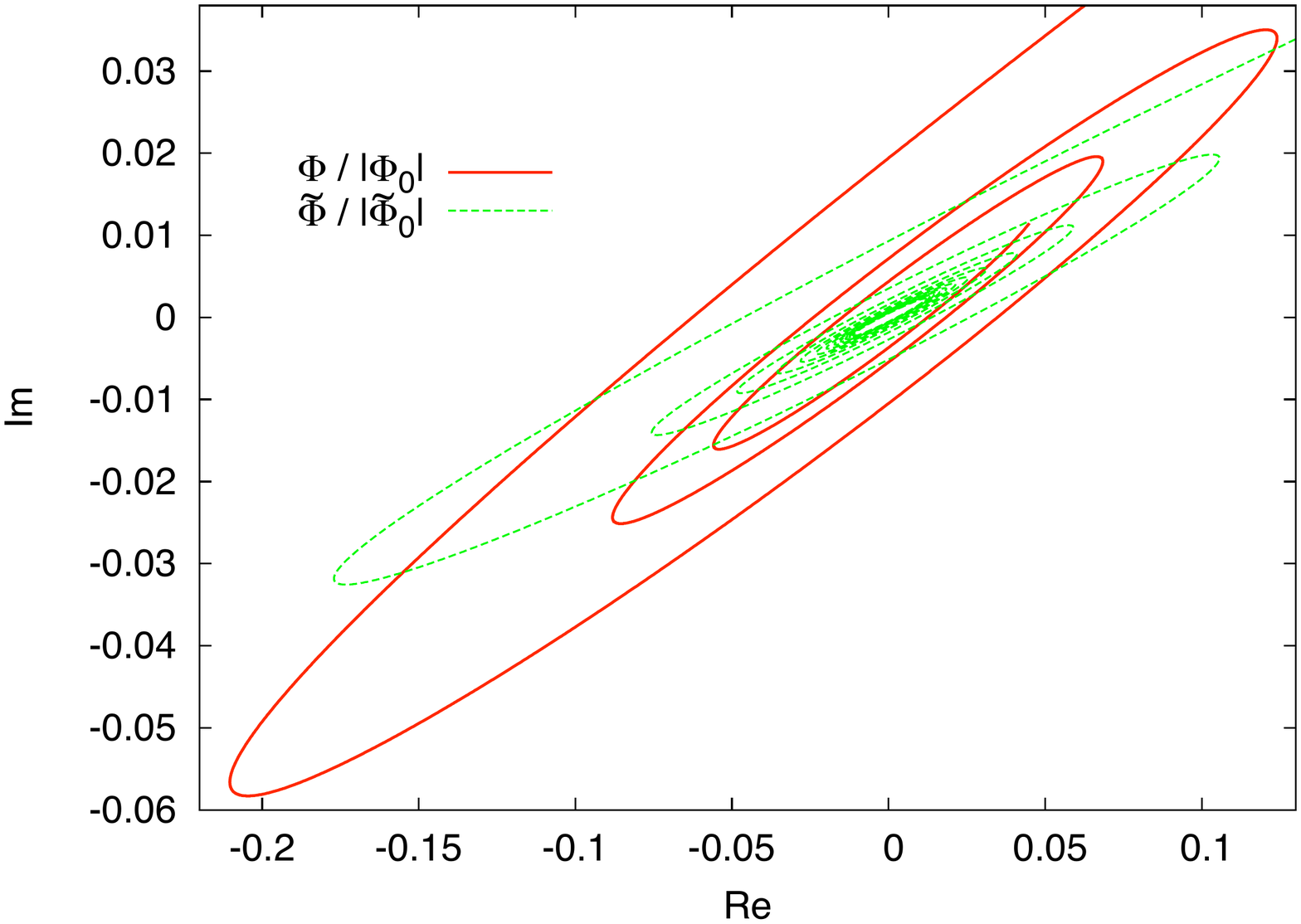}

\vspace{-2em}
\caption[Evolution of the two Flat Directions Used in Numerical
Simulations]{\label{fig-background} The evolution of the flat
  directions for the choice of masses $m_* = 10^{-6} ,\; \tilde{m}/m =
  3.72$ and initial phases $\Sigma_0 = 0.25$ , $\tilde{\Sigma}_0 =
  0.156$. The initial values of the VEVs are chosen such that
  $|\tilde\Phi_0| / |{\Phi}_0| = 15$ and $|{\Phi}_0| = 10^{-2} M_p$.
  We show the real and imaginary parts of the flat directions for the
  first three rotations of $|\Phi_0|$ (solid red) and
  $\vert\tilde\Phi\vert$ (solid green); the right panel shows the same
  evolution as the left panel, but magnified.  The trajectory shown
  for the field $\Phi$ is chosen to match that of
  Figure~\ref{fig-phievolution}, hence the charge asymmetry is
  approximately the same $\ge_{B-L}\approx 0.26$.}
\end{center}
\end{fullpagefigure}

\begin{figure}[!thbp]
\begin{center}
\mbox{}\hspace{-3em}\includegraphics[scale=0.55]{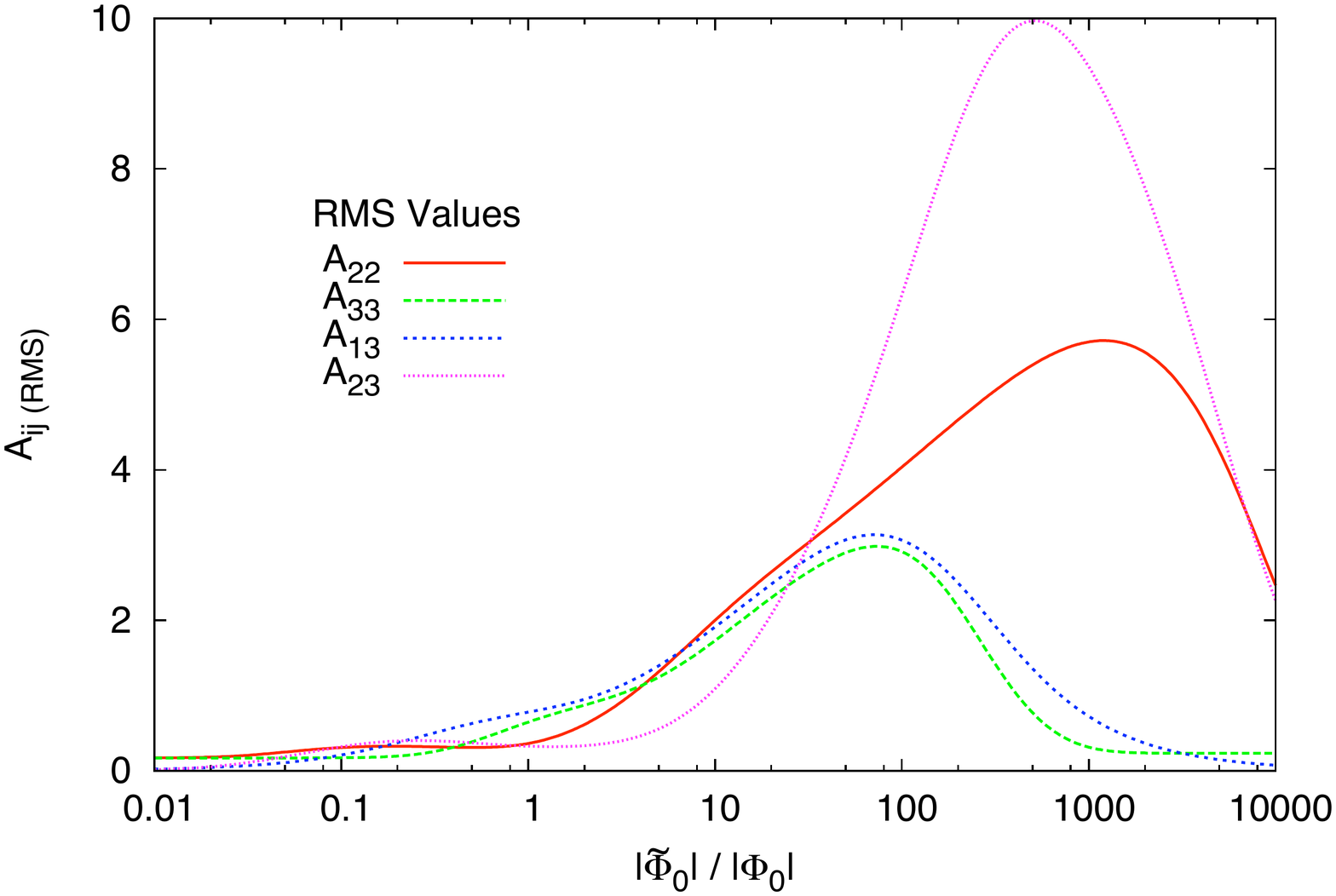}

\vspace{-3em}
\caption[Root-Mean-Square Leading Adiabatic Matrix
Elements]{\label{fig-adiabatic} The root mean square (over physical
  time) of the four leading adiabatic matrix elements
  $A_{22},A_{33},A_{13}$ and $A_{23}$ computed according
  to~\myref{results-adiabaticitymatrix}. The computation is performed
  for the modes with momentum $k=0$, for five rotations of the flat
  direction $\Phi$.  While the RMS values are not directly correlated
  with the growth exponent $\sigma$ shown in
  Figure~\ref{fig-growthexp}, they indicate the parameter space where
  particle production may be possible.}
\end{center}
\end{figure}

Now with just information of the above background evolution, one may
compute the eigenvalues~\myref{gaugefixing-model2eigenvalues}, Gamma
matrix elements~\myref{gaugefixing-model2eigenvalues} and also the
adiabaticity matrix elements~\myref{results-adiabaticitymatrix}.  The
adiabaticity matrix elements are a good predictor for when particle
production may occur.  In Figure~\ref{fig-adiabatic}, the
Root-mean-square leading adiabaticity matrix elements are computed
over the course of five rotations of the flat direction $\Phi$ while
varying the ratio of the initial VEVs.
The RMS values of $A_{ij}$ suggest there is indeed a resonance band.
One notices that the predicted band is weighted more towards larger
values of $|\tgF_0|/|\gF_0|$ in the figure.  This may be explained by
the fact that the heavier field $\tgF$ becomes unfrozen at a later
value of the scale factor than the lighter field $\gF$ and this
results in an effectively smaller ratio $|\tgF|/|\gF|$ once both
fields are oscillating.  For example, although the ratio is
approximately $15$ for the trajectory shown in
Figure~\ref{fig-background}, the effective ratio when both fields are
unfrozen and oscillating is clearly closer to unity.

\begin{figure}[!thbp]
\begin{center}
\mbox{}\hspace{-3em}\includegraphics[scale=0.55]{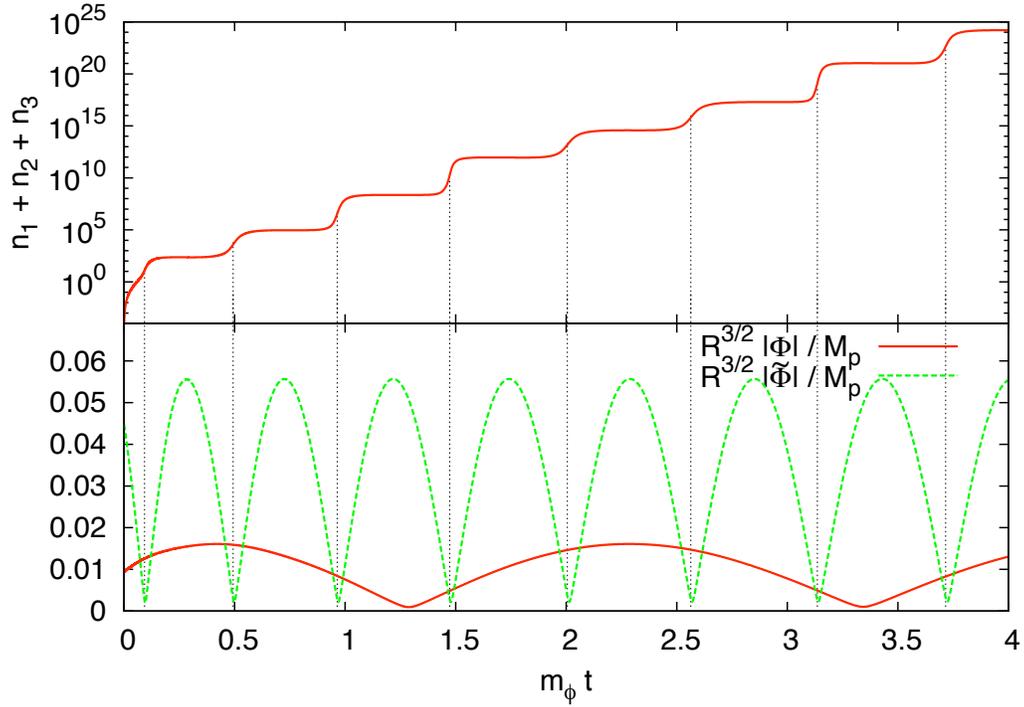}

\vspace{-3em}
\caption[Correlation of Quanta Production to Background
Evolution]{\label{fig-correlation1} Upper panel: Number density of the
  produced quanta for a given momentum $k_* = 10^{-7} \,$. Lower
  panel: amplitudes of the two flat directions. We note that particle
  production occurs whenever the two amplitudes are comparable, as
  discussed in the main text. The parameters chosen for this evolution
  are $m_* = 10^{-6} ,\, F_* = 2 ,\, {\tilde m} / m = 3.72 ,\, \vert
  {\tilde \Phi}_0 \vert / \vert \Phi_0 \vert = 15$}
\end{center}
\end{figure}

One may now solve the Heisenberg Equations of motion for the same
range of parameters to confirm the existence of a resonance band.  To
obtain some further intuition for the particle production, the total
occupation number $n_{tot}=n_1+n_2+n_3$ is shown in
Figure~\ref{fig-correlation1} compared to the above background
evolution.  Notice particle production occurs for short periods of
time when the fields have comparable magnitude.  One may see this
feature in the expressions for the elements of the $\Gamma$
matrix~\myref{gaugefixing-model2eigenvalues}.  The appearance of
factors such as $|\Phi|/\sqrt{|\Phi|^2+|\tilde\Phi|^2}$ and
$|\tilde\Phi|/\sqrt{|\Phi|^2+|\tilde\Phi|^2}$ in these
expressions indicate a suppression when the ratio $\tgF/\gF$ is very
large or very small, but the $\Gamma$ matrix, and thus particle
production, is unsuppressed for those instants when $\tgF\sim\gF$ is
obtained.  Finally, we should mention that the particle production
does not always occur in brief bursts.  Specifically, when the fields
are continually of a comparable magnitude, the production can then be
a more continuous process.

\begin{figure}[!thbp]
\begin{center}
\mbox{}\hspace{-3em}\includegraphics[scale=0.55]{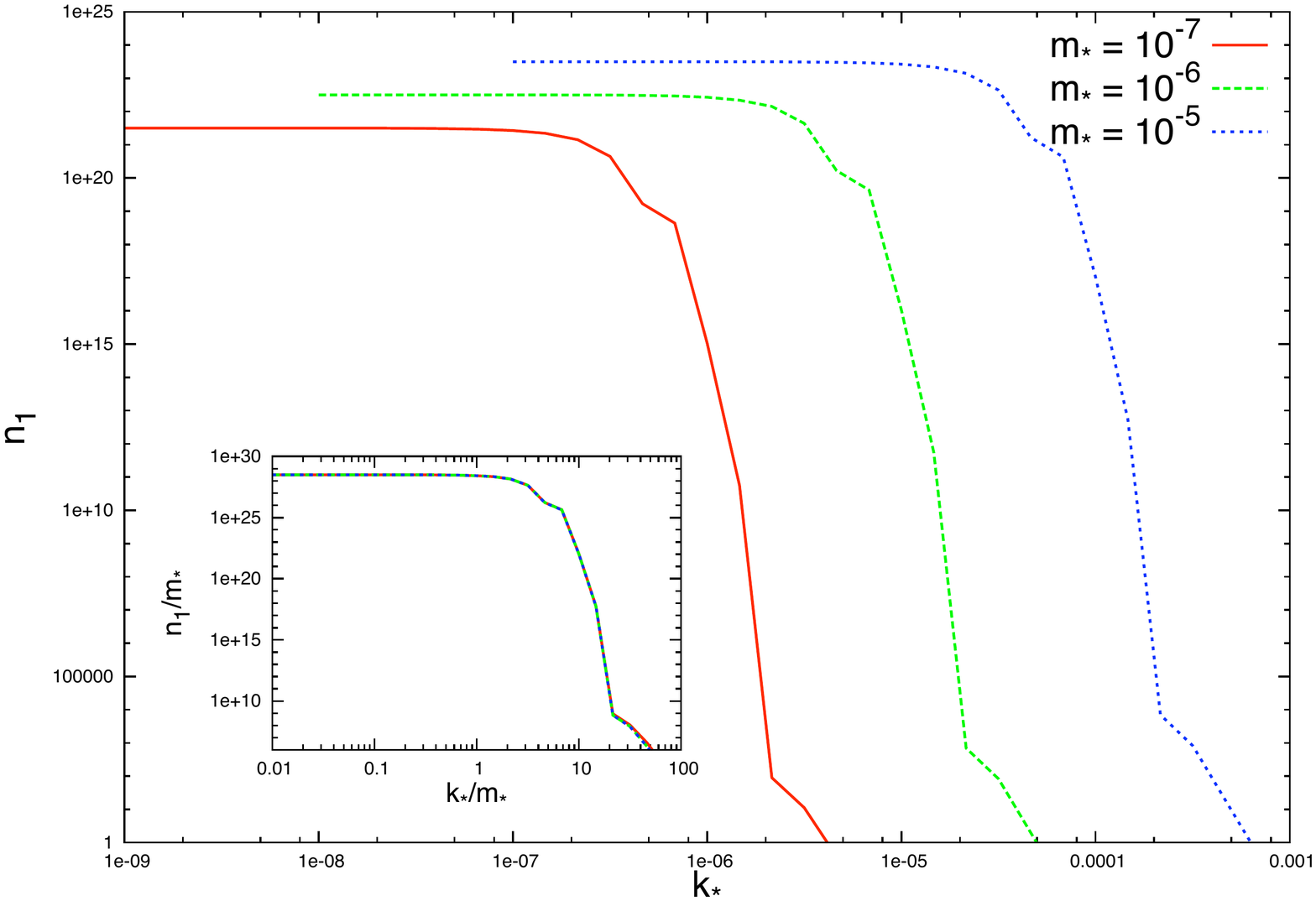}

\vspace{-3em}
\caption[Verification of the Scaling of Occupation
Numbers]{\label{fig-scaling} Occupation number for the first (heavy)
  eigenstate after three rotations of the first direction, for a
  specific set of parameters: $\vert \Phi_0 \vert = 10^{-3} \, M_p$,
  $\vert {\tilde \Phi}_0 \vert / \vert \Phi_0 \vert = 4$, ${\tilde m}
  / m = 3.72$, $\Sigma_0 = 0.25$, ${\tilde \Sigma}_0 = 0.156$. The
  larger figure shows the occupation number as a function of the
  momentum for three specific choices of the flat direction mass
  $m_*$. The insert shows the same result, but plotting $n_1/ m_*$
  vs. $k_* / m_*$. In terms of these variables, the three curves
  overlap. Since $m_* \propto \frac{m}{M}$, this confirms the
  scaling~\myref{results-rescalingn} for the occupation number, and
  the fact that the momenta of the quanta produced also scale as
  $\frac{m}{M}$.}
\end{center}
\end{figure}

Next we check the rescaling discussed in the previous section.  The
Heisenberg equations are solved for different values of the parameters
$m_*$,$\tm_*$,$\Phi_0$, and $\tgF_0$, and the occupation numbers are
rescaled according to~\myref{results-rescalingn} in order to compare
the spectra at a given time.  This procedure is illustrated in
Figure~\ref{fig-scaling} where we show the spectrum of the first
(heavy) eigenstate for three distinct cases in which the ratio
$\frac{m}{M}$ is varied over the range $10^{-7} < \frac{m}{M} <
10^{-5}$. The three spectra coincide once $n_1$ and $k$ are rescaled
according to
Table~\ref{table-rescaling}. Although we do not show it here, we also
verified that the occupation numbers for the light eigenstates do not
change with $\frac{m}{M}$.

\begin{figure}[!thbp]
\begin{center}
\mbox{}\hspace{-3em}\includegraphics[scale=0.55]{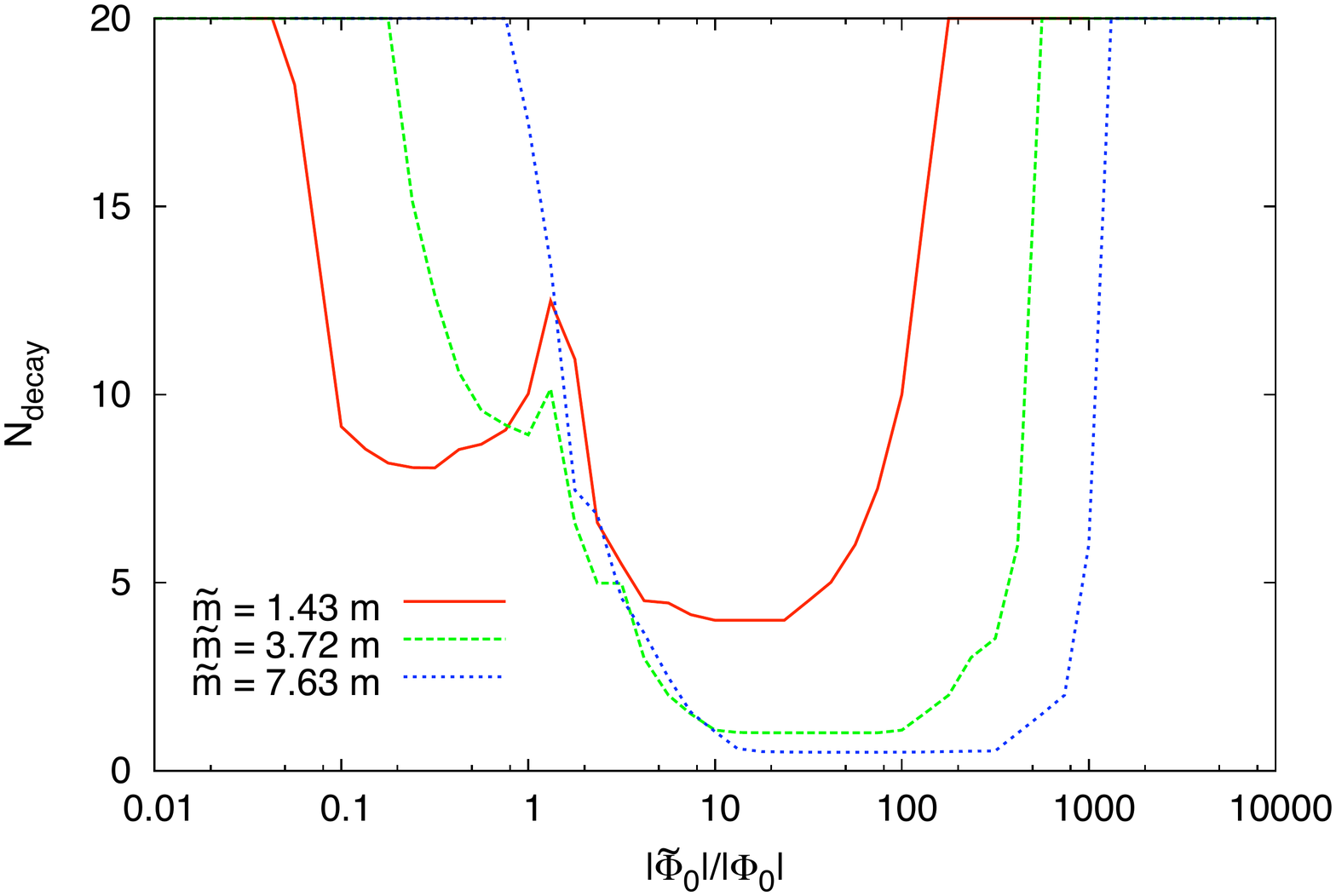}

\vspace{-3em}
\caption[Numerical Determination of the Decay Rate of the Flat
Directions]{\label{fig-production} Number of rotations of the
  first VEV where the production criterion $r_{prod} = 1$ is satisfied
  and its dependence on the initial ratio of the VEVs $|
  \tilde{\Phi}_0|/| \Phi_0|$, for three different mass ratios.  The
  numerical analysis was made for the first twenty rotations of the
  first flat direction and the results were rescaled to $m/e=
  10^4\,{GeV}$ and $ \sqrt{| \tilde{\Phi}_0| \, | \Phi_0|} = 10^{-2}
  M_p $.}
\end{center}
\end{figure}

Next the time evolution of $r_{prod}$ is computed, and this is done
for three values of the ratio $m/\tm$ and a sampling of the values
$|\Phi_0|/|\tgF_0|$ over the resonance bands.  The scaling
relation~\myref{results-scalingrprod} is applied to determine the
number of rotations at which $r_{prod}=1$ is satisfied with
$m=\msusy$ and $M=10^{-2}M_P$, and these values are recorded.  The
results are shown in Figure~\ref{fig-production}.

\begin{figure}[!thbp]
\begin{center}
\mbox{}\hspace{-3em}\includegraphics[scale=0.55]{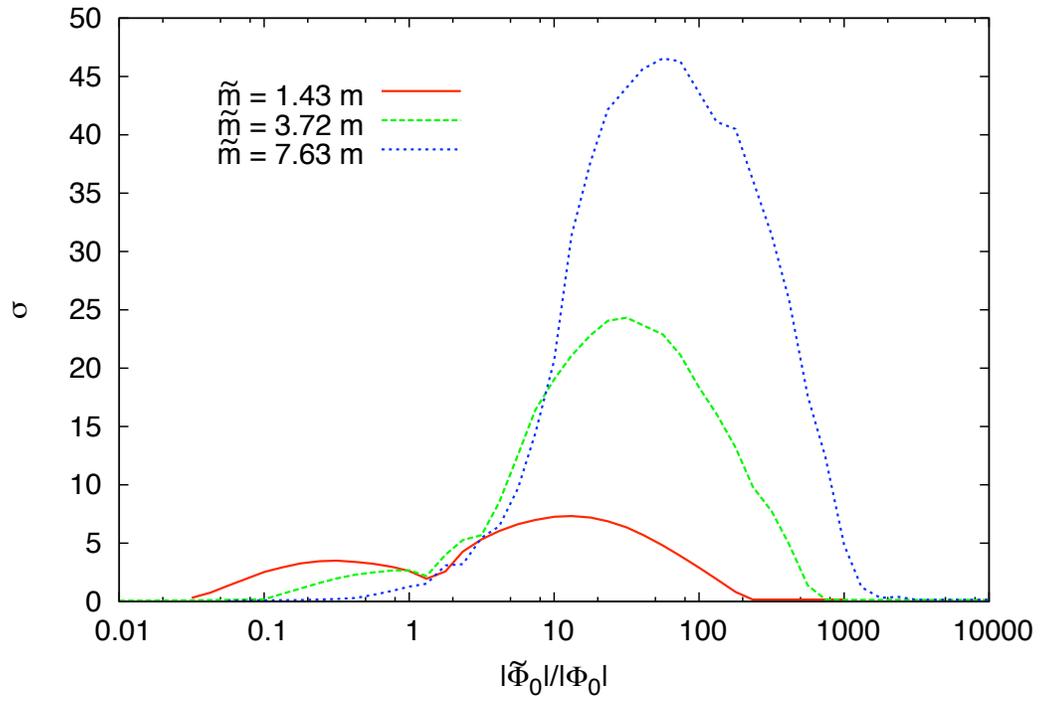}

\vspace{-3em}
\caption[Growth Exponents of the Decay]{\label{fig-growthexp} The
  dependence of the growth exponent $\sigma$ defined
  in~\myref{results-growthexp} on the initial ratio of the VEVs $|
  \tilde{\Phi}_0|/| \Phi_0|$, for three different mass ratios.}
\end{center}
\end{figure}

The data of Figure~\ref{fig-production} exhibits a resonance band for
each of the three selected mass ratios.
Note also that different values of $\frac{m}{M}$ were used to obtain
this data.  Thus the figure is a second verification of the scaling
relation.  Before rescaling there was noticeable scatter in
neighboring data points, and after rescaling, the data collapsed to
the smooth curves shown in the figure.
The growth exponent defined in~\myref{results-growthexp} was also
computed for the same parameter set and these are shown in
figure~\ref{fig-growthexp}

The data shown in the
Figures~\ref{fig-production}~and~\ref{fig-growthexp} are the main
quantitative results of this thesis.  They explicitly show the decay
of our toy model flat direction for realistic values of the parameter
space.  The degree to which these results may carry over to MSSM flat
directions is discussed in the Conclusions which follow.

\clearpage 
\chapter{Conclusions} \label{sec-conclusions}  
Inflation provides a strong motivation for the study of the dynamics
of coherent fields in the early universe.  The inflaton itself is a
coherent field, and as we have shown, other fields such as
supersymmetric flat directions may be made coherent through the
process of inflation.  The motivation for this thesis and the attempt
was to determine the evolution and decay of flat direction VEVs in the
early universe.  As discussed, the VEVs will break at least some of
the standard model gauge symmetries and establish an effective vacuum
into which the inflaton would decay and in which the inflaton decay
products would thermalize.  However, evidence was presented that flat
direction fields may decay quickly and nonperturbatively, and thus
restore the vacuum to a gauge symmetric state perhaps even before the
inflaton has decayed.  The final thermalization of the inflaton decay
products was not studied, but once the gauge symmetric vacuum is
restored, existing techniques such as those of
\cite{davidson-sarkar-2000} would be applicable.

Our study began with a very brief introduction to supersymmetric
theories and the scalar potential of these theories.  It was shown
that the scalar potential -- which is a functional of all the scalar
fields of ones model -- can possess special nonzero combinations of
fields for which the potential is exactly zero.  These flat directions
are the light scalar fields described above.  We thus studied the
growth of the fields during inflation, and established reasonable
upper bounds on their VEVs as shown in
Figure~\ref{fig-vevbound}. After the conclusion of inflation while the
universe is behaving as a matter dominated universe, a special time
was noted when the Hubble parameter became comparable to the mass of
the light scalar.  At this time, the complex scalar would begin to
execute well defined oscillation.  The motion is generally an
elliptic-spiral like trajectory in the complex plane as shown in
Figure~\ref{fig-phievolution}~or~\ref{fig-background} with the
frequency of rotation being simply the mass of the field.  The
orientation of rotation and the ellipticity of the trajectory will
additionally quantify a conserved charge for the field which in the
case of the MSSM flat directions is typically the difference of baryon
number and lepton number or $B-L$.  During this time both the inflaton
and the flat directions both oscillate and reduce in amplitude as
matter constituents and as shown in Figure~\ref{fig-rhoevolution}.
Both will eventually decay however, and the order of their decays is
somewhat important.  If the inflaton decays first, and into a
relativistic species, then the flat direction may overtake the
inflaton's decay products later.  However if there are multiple flat
directions which have obtained obtained large expectation values, then
the flat direction VEVs may decay first through a rapid
nonperturbative process.

An example of the nonperturbative decay was shown here.  It is shown
in a toy $U(1)$ gauge model that the decay can occur in approximately
ten or twenty oscillations of the flat direction fields for a range of
parameter choices.  This is in contrast to the perturbative estimates
which would put the flat direction decay to be long after the decay of
the inflaton.  The
Figures~\ref{fig-production}~and~\ref{fig-growthexp} are thus the main
quantitative results of this thesis.

In the following we examine some notable aspects of our models.  To
begin, we note our study \cite{sexton-etall-2008} was the first
concrete calculation of a nonperturbative decay in a gauge model.  One
consequence of the gauge symmetry on our results has been the gauge
current constraint.  For instance, in the model of a single flat
direction, the gauge current
constraint~\myref{gaugefixing-currentmodel1} essentially would not
admit a time dependent solution except in the circumstance when the
masses of the two fields involved were degenerate.  We were thus led
to select this case in order to consistently apply our formalism.  The
general case appears to only admit a space-time dependent solution,
and to pursue this model apparently would require lattice simulation.
Similarly, in the model of multiple flat directions of
Section~\ref{sec-gaugefixing-model2} which possesses three flat
directions, the gauge current constraint forced one of three flat
directions from obtaining a VEV in the solution.  A simple lesson from
our study is that the current conservation constraint can often have a
non-trivial effect on the dynamics.  We note that similar conclusions
were drawn in the work of \cite{enqvist-jokinen-mazumdar-2004}.  

In the above, the conservation of current was considered on just the
background.  It should be possible to consider gauge current
conservation at the level of the perturbations as well.  Specifically,
the conservation should appear in the Heisenberg Equations of
motion. These equations do not conserve current on their own however,
so it must be that the driving terms which are derived from the $U(1)$
symmetric Lagrangian will contain this information.  Verifying this
conservation on the perturbations is another direction which may be
pursued.

Having studied the $U(1)$ models, a logical next step would be to
pursue the same analysis on an MSSM flat direction.  As an example
consider again the work of \cite{enqvist-jokinen-mazumdar-2004} in
which the background evolution of the $H_u L$ multi-flat direction was
solved.  These authors obtained only time-dependent solutions to this
model.  However, using the methods presented in this thesis, the
spatial stability may in principle be studied.\footnote{The classical
  development of spatial instability in an Affleck-Dine condensate has
  been studied for example in \cite{kusenko-shaposhnikov-1998} where
  the connection is made to non-topological solitons in gauge theories
  known as Q-balls \cite{qballs} (see also \cite{dine-kusenko-2003}).
  Due to limitations of time, the relation of this thesis work to the
  work of \cite{kusenko-shaposhnikov-1998} has not been explored.} In
particular, for the $H_u L$ directions or some other choice of MSSM
directions, obtaining the driving terms to the Heisenberg equations
would be more involved due to the added gauge symmetry, but the series
expansion in the hierarchy $\frac{m}{M}$ for example would greatly
facilitate these calculations.  In general, the techniques learned by
studying the $U(1)$ models here can make these more complicated
problems feasible.

Putting aside the above modelling considerations, we consider again
the connection between flat direction decay and the equipartition
principle which was made in Section~\ref{sec-heiseqs-scaling} and in
the Results.  The most significant consequence of equipartition is the
production of heavy quanta.  Our analysis and numerics both revealed
that heavy quanta may be produced in the flat direction decay.  The
production of heavy quanta is not uncommon in preheating scenarios as
was discussed for a toy model of inflaton preheating in
Section~\ref{sec-reheating} and has also been shown in more detailed
studies (for example
\cite{kofman-linde-starobinsky-1996,guidice-etall-1999}).  One might
speculate that equipartition is a property that all preheating models
which involve heavy states might share in common.

However, it is unsettling that states with nearly Planck mass could be
produced in such large quantities.
One must look to the expected decay time of the heavy quanta to
resolve the difficulty.  A preliminary estimate of the decay time was
made at the conclusion of
Section~\ref{sec-reheating-nonperturbativeflat} and the estimate
indicates a decay time which may be comparable to the mass of the
heavy quanta.  In order to treat these states consistently as
particles, we had to assume the decay time was somewhat longer than
one period of oscillation of the heavy state.  This time would still
be very short compared to the period of a flat direction state, and in
particular the ratio of the two periods is simply $\mbyM$.  This
suggests the heavy state occupation numbers would be depleted quickly
on the time scale of our simulations.  Equivalently, this means that
the inclusion of interaction terms would be significant for the heavy
states.  Further study in this area is warranted.  However, we note
that the exponential production on our model did not greatly rely on
the presence of large occupation numbers for the heavy state. This was
checked by artificially setting to zero the mixing between the states
via $I=J=0$ so that the only source of particle production could be
from the evolving eigenfrequencies and not the evolving eigenvectors.
After this change, an exponential particle production was still
obtained for the light sector of the model.  

The above situation in which the final states are short-lived is an
interesting one and to our knowledge has not been addressed in the
literature.  However, the effects of interactions and the
back-reaction of produced particles on the condensate have been
studied for example in \cite{kofman-linde-starobinsky-1996}, so
analysis of a similar nature may be possible.  Alternatively, the
issue may be resolved through classical lattice simulation.
In general, classical lattice simulations may be useful to compare
against the generic results obtained here.  For instance, classical
stability analysis such as the determination of lyapunov exponents may
be applied in combination with a lattice simulation \cite{lyapunov}.
In principle, methods such as these may provide further insight.


Finally, we present some discussion of observables.  Our model suffers
in this respect in the same way as other models of preheating. The
difficulty lies in the process of thermalization itself which erases
most all information about what happened during the reheating process
aside from gross thermodynamic quantities such as the baryon asymmetry
and the reheating temperature.
However other observables may survive the reheating process.  These
include such effects as modulation of the cosmological perturbations,
gravitational waves, the creation of dark matter, primordial black
holes and topological defects \cite{kofman-reheating-1996}.
The most likely place one may look for a signature of the existence of
a flat direction VEV may be in the statistics of the cosmological
perturbations and in particular the imprint left on the Cosmic
Microwave Background.  In some specific models known as D-term
inflation in which the inflaton is supersymmetric (see
\cite{lyth-riotto-review-1998} for an introduction), it has been noted
the flat direction VEVs may contribute to the cosmological
perturbations \cite{enqvist-mcdonald-2000,enqvist-mazumdar-2003}.  For
instance, bounds may be placed on the magnitude of the flat direction
VEV (based on the COBE normalization) which are in the vicinity of
$10^{-3}M_P$ to $10^{-2}M_P$ and which correspond to a lowest
non-vanishing terms in the superpotential $n=4$ and $n=6$
\cite{enqvist-mcdonald-2000}.  For comparison, see
Figure~\ref{fig-vevbound}.  More recently, it has been suggested that
the effect of parametric resonance of a field coupled to the flat
direction may generically amplify the perturbations of this field so
they may be observable in the CMB as curvature perturbations
\cite{racine-brandenberger-2008}.  Again, in the absence of a direct
signature, one may constrain proposed models based on non-observation
of such a signal \cite{racine-brandenberger-2008}.  We note this last
work appears to have been somewhat motivated by our own work on which
this thesis is based \cite{sexton-etall-2008}.

This completes the main body of this thesis on the dynamics of
supersymmetric flat directions.  While much has been determined in
this study, we have emphasized in the above that there remains more
work that can be done on this interesting subject.

\appendix
\chapter{Notation} \label{calc-notation} 
The notation used in the thesis is here summarized for the convenience
of both specialist and non-specialist readers.

The notation for fields is as follows: Fundamental complex scalar
fields (specifically the fields of which the flat directions are
composed) are denoted by $\phi_i=f_ie^{\gs_i}$ with $i$ running over
the whole set of fields.  However, the inflaton field is denoted by
$\chi$ (the author apologizes as this is an unusual convention).  The
U(1) vector field is denoted by $B_\mu$.  The SU(2) and SU(3) vector
fields are $A^a_\mu$, where $a$ is a group index, and it should be
clear from the context which symmetry is being considered.  Fermions
are denoted by $\psi$.

The fundamental fields are typically decomposed into a classical
background plus quantized fluctuations away from this background. The
classical backgrounds are written in upper case and the perturbations
in lower case. In the case of the vector, the classical background is
written in calligraphic letters, for instance $B_\mu = \cB_\mu +
b_\mu$ where $b_\mu$ are the quantized fluctuations.  Note the
background value of the vector $\cB_\mu$ is always zero in the context
of this thesis (See Section~\ref{sec-gaugefixing}).  Finally, the
notation for derivatives is,
$\frac{d\chi}{dx^{\mu}} \equiv (\der_\mu \chi)^2 \equiv (\chi,_\mu)^2$

When discussing the classical background of the scalar fields
$\Phi_i$, the terms Vacuum Expectation Value (VEV) or scalar
condensate are sometimes used as these are equivalent descriptions.
When the term VEV is used, it is implied that it is the VEV determined
by a local observer in some causally connected patch of the universe
at some time, $\langle\phi_i \rangle_{\mathsubscript{patch}}
\equiv\Phi_i$, and this is not to be confused with the VEV determined
over ``all'' space which is zero $\langle \phi_i
\rangle_{\mathsubscript{all\;space}}=0$. We sometimes use the term
``Flat Direction'' when referring to the VEV, although strictly
speaking the flat direction is a property of the scalar potential
while the VEV is a realization of the field configuration on the
potential (see Section~\ref{sec-susyflat}).  Effective vacuum and Bose
Einstein condensate are also used to describe the classical
background.

In describing the ``decay'' of the flat direction condensate, we
sometimes refer to this as ``decoherence'' or ``the development of
spatial instability'' as these are all equivalent statements for the
decay in the context of our models.

The notation for energy scales is as follows: The mass of the scalar
fields which comprise the flat directions will be $m$ or $\tm$ and
these will always be of order $\msusy\sim\tev=1000\,\gev$ which is the
assumed scale of the supersymmetric particle spectrum.  The $\tev$
energy scale (approximately 1000 times the mass of the proton) is
heavy by the standards of nuclear and particle physics, but it is
relatively light compared to other scales in Cosmology such as the
Grand Unified Scale, $M_{\mathsubscript{GUT}} \sim 10^{16}\,\gev$ or
the Planck scale, $M_P\sim 10^{19}\,\gev$.  Note the Planck mass is
defined by $G_N = \frac{1}{M_P^2}$.  This is not to be confused with
the reduced Planck mass $M_P^* = M_P/\sqrt{8\pi} \approx 2.4\times
10^{18}\,\gev$ which is not used in the thesis.  An arbitrary heavy
mass scale will be written as $M$, and in the results section $M$ is
defined specifically as the scale of the VEVs $|\Phi_i|\sim
10^{-2}M_P$.  The ratio of the $\tev$ scale states to one of these
heavy scales is then generically written $\mbyM$.  A quantity is
unsuppressed when it is zeroth order in the expansion in $\mbyM$.

The gauge couplings are denoted by $"g"$ in Section~\ref{sec-susyflat}
where supersymmetry is discussed, but in our model and results
sections~\ref{sec-gaugefixing}~and~\ref{sec-results}, we instead use
$e$.  This is not the electron's charge as it would be in Quantum
Electrodynamics, but the notation is convenient.

The terms parametric resonance, non-adiabatic evolution and
nonperturbative dynamics are often used interchangeably.

Finally, the general metric tensor is $g_{\mu\nu}$ and the Minkowski
metric tensor is $\eta_{\mu\nu}$.  The cosmological scale factor is
$R$.  The conformal time is $\eta$ with time derivatives written as a
prime (such as $R'$), and physical time is $t$ with time derivatives
written with a dot (such as $\dot R$).

\chapter{Details Regarding U(1) Gauge Symmetric Actions}\label{calc-u1-lagrangian}
The action for a $U(1)$ gauge field, and $n$ complex scalars charged
under this gauge group is presented in the main
text~\myref{gaugefixing-u1action}.  The procedure for gauge fixing
such actions to the unitary gauge was also presented there.  Some
details which were not explained in Section~\ref{sec-gaugefixing} are
explained here.  In particular, it is shown here (i) how to
incorporate the expansion of the universe, (ii) how to decompose the
vector field Lagrangian and eliminate the nondynamical time-like
component $B_0$ and (iii) how to diagonalize the Lagrangians for the
systems of two fields and four fields in the expansion parameter~$\mbyM$.

\textbf{Incorporating the Expansion of the Universe:} With a change of
spatial coordinates and a rescaling of fields one may put the
Lagrangian~\myref{gaugefixing-u1action} in a form which is more easily
quantized.  The background metric is taken to be the
Friedmann-Robertson-Walker metric with the spatial curvature constant
set to zero.  The line element for this metric is $ds^2=dt^2 - R(t)^2
d{\vec x}^2$, where $R(t)$ is the scale factor.  It is convenient to
use the conformal time coordinate $\eta$ which is related to the
physical time by $R d\eta = dt$.  The line element in these
coordinates is,
\begin{equation*}
g_{\mu\nu}dx^\mu dx^\nu = R(t)^2 \eta_{\mu\nu}dx^\mu dx^\nu = R(t)^2(dt^2 - d{\vec x}^2)
\end{equation*}
where $\eta_{\mu\nu}$ is the Minkowski metric.  After choosing the
conformal time, the action~\myref{gaugefixing-u1action} is written in
a way that the raising and lowering of the space-time indeces is
simply done with the minkowski metric as follows,
\begin{equation*}
S=\int d^4x R^4 \left\{-\frac{1}{4R^4}\eta^{\mu\mu'}\eta^{\nu\nu'}F_{\mu\nu}F_{\mu'\nu'} +  \frac{1}{R^2}\eta^{\mu\nu}(D_\mu\phi_i)(D_\nu\phi_i)^*  - V(\phi_1,...\phi_n) \right\}
\end{equation*}
Additionally, if we redefine the scalar fields $\phi=\frac{\gvf}{R}$,
the action rewrites,
\begin{equation*}
S=\int d^4x \left\{-F_{\mu\nu}F^{\mu\nu} + (D_\mu\gvf_i)(D^\mu\gvf_i)^*  + \frac{R''}{R}\gvf_i^2 -  R^4V\left(\frac{\gvf_1}{R},...\frac{\gvf_n}{R}\right) \right\}
\end{equation*}
where primes above denote $\frac{\der}{\der \eta}$ where the scale
factor now appears through the effective mass term and also may appear
in the potential, though the potentials we consider such as that from
the D-term are quartic in the scalar fields, and so $R^4\phi^4 =
\gvf^4$ and the scale factor drops from these terms as well.  To
summarize, for the Lagrangians~\myref{gaugefixing-u1action}, the
prescription for incorporating the scale factor is to replace the mass
terms as follows
\begin{equation*}
m_i^2\phi^2 \rightarrow \left(R^2 m^2_i -\frac{R''}{R}\right)\gvf_i^2
\end{equation*}
and when later computing physical quantities to convert back to
comoving coordinates via $R d \eta = dt$ and to physical fields via
$\phi_i = \frac{\gvf}{R}$.
%

\textbf{Decomposing the Vector Field and Eliminating the Nondynamical
  Time-like Component:} We now decompose the vector field in our U(1)
actions~\myref{gaugefixing-u1action}.  We note that $B_\mu$ is coupled
to the scalars through the current $J^\mu$ as well as through the
effective mass term $\cM^2$ both defined in
(\ref{gaugefixing-scalarkineticterm}).  The part of the lagrangian
which contains the vector is thus,
\begin{equation*}
-\frac14F(B)^2 + \frac12 \cM^2 B_\mu B^\mu - B_\mu J^\mu
\end{equation*}
where the field strength tensor is $F_{\mu\nu} = \der_\mu B_\nu -
\der_\nu B_\mu$.  We split $B_\mu$ into it's space and time components
$B_0$ and $B_i$ as follows
\begin{eqnarray*}
\mathcal{L}_{massless}
&=& -\frac{1}{4} F_{\mu\nu}F^{\mu\nu} \\
&=& -\frac12 F_{0i}F^{0i} - \frac14 F_{ij}F^{ij} \\
&=& \frac12 F_{0i}F_{0i} - \frac14 F_{ij}F_{ij} \\
&=&\frac{1}{2}\left( B_i,_0 - B_0,_i\right)^2 -\frac14 \left( B_j,_i -  B_i,_j\right)^2  \\
 &=&\frac12 B_i'B_i'+ \frac12 B_0,_i B_0,_i - B_i' B_0,_i -\frac12 B_j,_i B_j,_i +\frac12 B_j,_i B_i,_j \\
 &=&\frac12  B^{i\prime} B^{i\prime} + \frac12 B^0,_i B^0,_i + B^{i\prime} B^0,_i -\frac12 B^j,_iB^j,_i +\frac12 B^j,_i B^i,_j \\
 &=& \frac12 {\vec B}' \cdot {\vec B}' + \frac12\nabla B_0 \cdot \nabla B_0 + {\vec B}'\cdot\nabla B_0 - \frac12|\nabla \times \vec B|^2 
\end{eqnarray*}
where the identity 
$\ge_{ijk}\ge_{ilm} = (\gd_{jl}\gd_{km} -\gd_{jm}\gd_{kl})$ 
has been applied.  The Lagrangian for the massive vector field is then
\begin{equation}
\cL= 
\frac12 \left\{ \begin{array}{l}
|{\vec B}'|^2 + |\nabla B_0|^2 + 2{\vec B}'\cdot\nabla B_0 - |\nabla \times \vec B|^2  \\[0.4em]
+\;\cM^2(B_0^2 - |\vec{B}|^2) - 2 J_0 B_0 + 2 \vec J \cdot \vec B \end{array}
\right\}
\label{u1-lagrangian-vectorL}
\end{equation}
We split the vector into a transverse and longitudinal part $\vec B =
{\vec B}_T + \nabla L$ where $\nabla \cdot {\vec B}_T=0$.
The transverse part of the vector decouples from the Lagrangian since
it transforms under the Lorentz group as a spin-1 field, while the
scalar longitudinal mode of the vector as well as the fundamental
scalar fields transform as Lorentz scalars.  Hence the transverse mode
will decouple at quadratic order in the perturbations.  We thus only
consider the longitudinal mode $\nabla L$ as well as $B_0$ which is
non dynamical and will be replaced by substitution of its equation of
motion. The equations of motion for $B_\mu$ are
\begin{eqnarray*} 
F_{\mu\nu},^\mu + \cM^2 B_\nu  &=& j_\mu \\
\Box B_\nu - (B_\mu,^\mu),_\nu + \cM^2 B_\nu  &=& j_\mu 
\end{eqnarray*}
and from these, the equations of motion for $B_0$ and $B_i$ are,
\begin{eqnarray} 
- \nabla^2 B_0 - \der_t(\nabla\cdot\vec B) + \cM^2 B_0 &=& J_0   \label{b0eq} \\
-\Box \vec B - \nabla (\der_0B_0 - \nabla\cdot\vec B) - \cM^2 \vec B  &=& \vec J
\end{eqnarray}
We solve for $B_0$ by inverting the equation,
\begin{eqnarray*}
(- \nabla^2 + \cM^2) B_0 &=& J_0+ (\nabla^2 L')
\label{calc-u1-lagrangian-b0eom}
\end{eqnarray*}
The part of the Lagrangian~\myref{u1-lagrangian-vectorL} corresponding
to the longitudinal mode is then determined to be,
\begin{eqnarray}
\frac12 \left\{|\nabla L'|^2 - B_0\left[ J_0+ \nabla^2 L' \right] - \cM^2|\nabla L|^2 + 2 \vec J\cdot\nabla L \right\}
\label{calc-u1-lagrangian-Llong}
\end{eqnarray}
The remaining aspects of the calculation
will be done in momentum space for which the expression for $B_0$ is,
\begin{equation}
B_0=\frac{J_0 - k^2 L'}{k^2+\cM^2}
\end{equation}
where $k=|\vec k|$, 
The longitudinal part of the
Lagrangian~\myref{calc-u1-lagrangian-Llong} in momentum space is,
\begin{eqnarray}
\frac12 \left\{
\pfrac{k^2 \cM^2}{k^2+\cM^2} {L^*}'L' 
+ \frac{k^2(J_0^* L'+ J_0 {L^*}')}{k^2+\cM^2}
- \frac{J^*_0J_0}{k^2+\cM^2} \right. \nonumber \\
\left. - \; \cM^2k^2 L^*L + ({\vec J}^* L  + {\vec J} L^*)\cdot\vec k 
\right\} 
\label{calc-u1-lagrangian-Lk}
\end{eqnarray}
The longitudinal mode can then be made canonical with the field
redefinition,
\begin{equation}
L = \pfrac{\sqrt{k^2+\cM^2}}{k\cM}\hat L
\label{calc-u1-lagrangian-norm1}
\end{equation}
Notice in the simple case where the current is exactly zero, the
Lagrangian for the longitudinal mode of the vector after applying the
field redefinition is,
\begin{equation}
\frac12 \left\{ 
{\hat L}^{*\prime}{\hat L}' 
- (\cM^2 +k^2) {\hat L}^*{\hat L}
\right\} 
\end{equation}
which is the expected result for a freely propagating longitudinal
mode of the massive vector field.  The above
normalization~\myref{calc-u1-lagrangian-norm1} will be performed on
the specific examples which follow in the case of nonzero current
$J_0$.  The longitudinal mode should still obtain the above dispersion
relation $\go^2 = k^2 +\cM^2$ when the fields are evolving
adiabatically.

To this point the Lagrangian has been written in the fundamental field
description.  However for the purpose of computing observables, the
Lagrangians are always expressed in background fields plus
perturbations.  The models we are interested in have backgrounds which
satisfy $\langle B_\mu \rangle=0$ and also $\langle J_0\rangle=0$, so
we must only consider the linear perturbations in these quantities,
$b_\mu$ and $j_\mu$ in order to obtain a quadratic Lagrangian from the
above result~\myref{calc-u1-lagrangian-Lk}.  Additionally, we will
employ the expansion parameter $\mbyM$ which was also applied in
Section~\ref{sec-heiseqs-scaling}. Our objective is to obtain the
quantities $\Gamma_{ij}$ and $\go_i$ to leading order in this
expansion.

\textbf{Diagonalization of the Two field Model:} The Lagrangian for
the perturbations is~\myref{gaugefixing-model1Lquadratic} from which
we read the current density and effective mass of the vector to be,
\begin{eqnarray*}
&j_0 = \frac12 e F \gS'\gd 
\;\;\;,\;\;\;
\vec j = 0 \\[0.5em]
&\cM = \frac12 eF
\end{eqnarray*}
The Lagrangian~\myref{gaugefixing-model1Lquadratic} after the
elimination of $b_0$ is,
\begin{eqnarray}
&& \frac12\left[(\der_\mu f)^2 + (\der_\mu\gs)^2 + (\der_\mu\gd)^2 \right] + \frac12\gS'\left(f\gs'- f'\gs \right) \nonumber \\
&& 
-\;\frac12 \left( m_+^2 - \frac14 \Sigma'^2 \right) f^2 
-\frac12 \left(m_+^2 - \frac14 \Sigma'^2 \right) \sigma ^2 \nonumber \\ 
&&
-\;\frac12 \left(\frac14{e}^2 F^2 + m_+^2 + \frac34 \Sigma'^2\right)\delta ^2 
\end{eqnarray}
where the vector field is not shown since it has been decoupled.
The remaining coupled system involves $f$ and $\gs$ and these fields
are obtained to be degenerate in mass $m$.  They are simply the flat
direction perturbations.

\textbf{Diagonalization of the Four Field Model:} In the four field
model, the current density of \myref{gaugefixing-model2L} and the
effective vector mass are obtained in terms of the
perturbations~\myref{gaugefixing-model2decomposition} to be,
\begin{eqnarray*}
&j_0 = e \left[
(F^2-G^2)\theta'
+ 2F\gS'\gd + 2G\tgS' \tgd
\right] \\[0.5em]
&\vec j = e (F^2-G^2)\nabla \theta \\[0.5em]
&\cM = e\sqrt{F^2+G^2}
\end{eqnarray*}
Substituting these into~\myref{calc-u1-lagrangian-Lk}, one notices
there will be coupling between $L$ and $\theta$ in their kinetic terms
at leading order.  Our procedure will be to first find the
transformation which makes these kinetic terms canonical (temporarily
discarding mass terms and mixed terms).  We then apply this
transformation to obtain the complete Lagrangian in the
form~\myref{heiseqs-canonicalL}.  The kinetic terms are,
\begin{eqnarray*}
&\mbox{}& \frac12 \left\{ 
{\hat L}^{*\prime}{\hat L}' 
+ \frac{k}{\sqrt{k^2+\cM^2}} \left(\frac{j_0^*}{\cM}{\hat L}'+ \frac{j_0}{\cM} {\hat L}^{*\prime}\right)
- \frac{\pfrac{|j_0|}{\cM}^2}{1+\pfrac{k}{\cM}^2} + (F^2+G^2) {\theta^*}'\theta' \right\} \\
&=& \frac12\left\{
{\hat L}^{*\prime}{\hat L}'
+ {\hat\theta}^{*\prime}{\hat\theta}'
+ \frac{k}{\cM}\frac{F^2-G^2}{\sqrt{4F^2 G^2 + (F^2+G^2)^2\pfrac{k}{\cM}^2}}
({\hat\theta}^{*\prime}{\hat L}' + {\hat\theta}'{\hat L}^{*\prime})
\right\}
\end{eqnarray*}
where $j_0$ has been substituted and the kinetic term for $\gth$ has
been normalized.  This last expression is compactly written in matrix
notation in terms of the fields $L$ and $\theta$ as follows,
\begin{eqnarray*}
&\mbox{}&
\frac12 (L^* , \theta^*)'
\left[ \begin{array}{ll}n_L & 0 \\ 0 & n_\gth\end{array}\right] 
\left[ \begin{array}{ll}1 & Q \\ Q & 1\end{array}\right] 
\left[ \begin{array}{ll}n_L & 0 \\ 0 & n_\gth\end{array}\right] 
\left(\begin{array}{l}L \\ \theta \end{array}\right)'
\end{eqnarray*}
where
\begin{eqnarray*}
Q &=& \frac{k}{\cM}\frac{F^2-G^2}{\sqrt{4F^2 G^2 + (F^2+G^2)^2\pfrac{k}{\cM}^2}} \\
n_\theta &=& \sqrt{\frac{4F^2 G^2 + (F^2+G^2)^2\pfrac{k}{\cM}^2}{(F^2+G^2)(1+\pfrac{k}{\cM}^2)}} \\
n_L &=& \frac{k\cM}{\sqrt{k^2+\cM^2}}
\end{eqnarray*}
One may apply the following additional rotations and rescalings,
\begin{equation}
\left(\begin{array}{l}L_c \\ \theta_c \end{array}\right)
 =
R_1^TR_0^T
\left[ \begin{array}{ll}\sqrt{1+Q} & 0 \\ 0 & \sqrt{1-Q}\end{array}\right]
R_0
\left[ \begin{array}{ll}n_L & 0 \\ 0 & n_\gth\end{array}\right] 
\left(\begin{array}{l}L \\ \theta \end{array}\right)
\label{u1-lagrangian-Ltheta}
\end{equation}
where $R_0$ is a time independent rotation and $R_1$ is a small time dependent rotation,
\begin{eqnarray*}
R_0 &=& \frac{1}{\sqrt2}\left(\begin{array}{ll}1 & 1 \\ -1 & 1\end{array}\right) \\
R_1 &=& \left[ \begin{array}{ll}1 - \frac{k^2 q^2}{2} & k q \\ -k q & 1 - \frac{k^2 q^2}{2}\end{array}\right] \\
q &=& \frac{-F^2+G^2}{4e FG \sqrt{F^2+G^2}}
\end{eqnarray*}
and these rotations have been chosen specifically to decouple the
longitudinal mode of the vector from the other scalar degrees of
freedom.  The expression~\myref{u1-lagrangian-Ltheta} may be inverted
and expanded finally to obtain,
\begin{eqnarray}
\gth &=& \frac{\sqrt{F^2+G^2}}{2FG}\gth_c + \cO\pfrac{m^3}{M^3}\\
L &=& \frac{L_c}{k} - \frac{(F^2-G^2)}{2e F G \sqrt{F^2+G^2}}\gth_c + \frac{k}{2 {e}^2(F^2+G^2)} L_c + \cO\pfrac{m^2}{M^2}
\end{eqnarray}
This final result is substituted into the
Lagrangian~\myref{calc-u1-lagrangian-Lk}.  The remaining mixed scalar
degrees of freedom are arranged in a vector $X=(\gd,\tgd,\gth_c)$, and
the Lagrangian for these has the desired
form~\myref{heiseqs-precanonicalL}.  Specifically, the $K$ matrix, the
matrix $(\tgO^2 + K^TK)$, the eigenvalues $\go_i$ and the matrix
$\Gamma$ are all shown in the main
text~(\ref{gaugefixing-model2tM}-\ref{gaugefixing-model2eigenvalues}).

\chapter{Calculating the Variance of a Scalar Test Field During Inflation}\label{calc-variance}
We determine the variance of test scalar field $\phi$ during an
asymptotically de~Sitter expansion.  The sampling is done inside a
patch of space of width $L$.  We are not interested in wavelengths
smaller than $L$, so we smear out these fast fluctuations by use of a
window function $W_L(x)$,
\begin{equation}
\phi_L(t) = \int d^3x  W_L(x)  \phi(x,t) 
\;\;\;\;,\;\;\;\;
W_{L}(x) = \frac{1}{(2\pi)^{3/2}L^3}e^{-\frac{x^2}{2L^2}}
\label{inflation-windowfunction}
\end{equation}
in which $W_L(x)$ is any reasonable normalized distribution of order 1
inside the patch, and dropping to zero quickly outside of the patch.
In the above, it is taken to be a normalized Gaussian.  The quantity
$\phi_L$ is then the quantity which we must determine the variance of.
The calculation is performed using conformal time $dt = R d\eta$, and
a redefined scalar field $\phi=\frac{\gvf}{R}$.  The fields are
decomposed as in~\myref{inflation-modeexpansion}, and the
variance is thus obtained to be,
\begin{eqnarray}
\langle\gvf_L^2 \rangle &=& \langle 0 | \left[\int d^3x \; W_L(x) \gvf \right]^2|0 \rangle \\
&=& \frac12 \int \frac{d^3k}{(2\pi)^3} |h_k|^2 |w(kL)|^2 \\
&\approx& \frac12 \int_0^{L^{-1}} \frac{d^3k}{(2\pi)^3} |h_k|^2
\label{calc-inflation-variance}
\end{eqnarray}
where $w(kL)$ is the Fourier Transform of the window function, and
we've used the fact that $w(kL)$ is order 1 for $|k|<L^{-1}$ and drops
to zero quickly for $|k|>L^{-1}$.  The effect of the window function
is simply to impose a momentum cutoff.  In the de~Sitter metric, the
equations for the modes is listed in the main
text~\myref{inflation-modeequation2} which again is
\begin{equation}
h_k'' +\left[k^2 + \left(\frac{m^2}{H_I^2}  - 2\right)\frac1{\eta^2}\right] h_k =0
\label{calc-inflation-modeequation2}
\end{equation}
and may be put in the form of Bessel's equation with the definitions
$x\equiv-k\eta$, and $h_k(\eta) \equiv \sqrt{x}f(x)$,
\begin{equation}
x^2\frac{d^2f(x)}{dx^2} + x \frac{df(x)}{dx} + (x^2-n^2)f(x) = 0
\;\;\;,\;\;\;
n^2= \frac94 - \frac{m^2}{H^2}
\end{equation}
the solutions are, $f(x)=A_k J_n(x) + B_kY_n(x)$, and so the mode
functions take the form,
\begin{equation}
h_k(\eta) = \sqrt{-k\eta}\left(A_k J_n(-k\eta) + B_k Y_n(-k\eta)\right)
\;\;\;,\;\;\;
n^2= \left(\frac94 - \frac{m^2}{H^2}\right)
\label{calc-inflation-modesolution}
\end{equation}
The constants $A_k$ and $B_k$ must now be determined from the initial
conditions for the modes which are fixed by the quantum fluctuations.
Technically this means we must canonically normalize the modes $h_k$
at early times (or large k).  The
equation~(\ref{calc-inflation-modeequation2}) for the modes at sub-horizon scales,
asympotes to $h_k''+k^2h_k=0$ with solutions $e^{\pm ik\eta}$.
Referring to~(\ref{inflation-modeexpansion}), we want $h_k(\eta) =
\frac{1}{\sqrt{2k}}e^{-ik\eta}$ at early times for our fields to be
canonically normalized.  Our solutions~\myref{calc-inflation-modesolution}
are then normalized as follows,
\begin{equation}
h_k(\eta) = \sqrt{\frac{-\eta\pi}{4}}e^{i\left(\frac{n\pi}{2}+\frac{\pi}{4} \right)} \left(J_n(-k\eta) + i Y_n(-k\eta)\right) 
\label{variance-hk}
\end{equation}
where one may apply the asymptotic form of the Bessell functions which
is,
\begin{equation}
\left(J_n(-k\eta) + i Y_n(-k\eta)\right) \sim \sqrt{\frac{2}{\pi (-k\eta)}}e^{i\left(-k\eta-\frac{n\pi}{2}-\frac{\pi}{4}\right)}
\;\;\;,\;\;\;
\mbox{for large }(-k\eta)
\end{equation}
from which it is quick to verify that the
solutions~\myref{variance-hk} have the correct asymptotic form far
within the horizon.  We consider these solutions in the limit of small
mass $\frac{m}{H_I}<1$ where $n^2>0$, and large mass $\frac{m}{H_I}>1$
where $n^2<0$, and we note the Bessell solutions will still hold in
the latter case when $n$ is pure imaginary.  We are interested in
calculating the variance, so we substitute in our general solution,
and we approximate our solutions in the long wavelength limit as our
integration will only be over scales larger than the horizon.  The
steps of the calculation are shown first for the case $n^2>0$,
\begin{eqnarray*}
\langle\gvf_L(\eta)^2 \rangle_{n^2>0}  &\approx& \frac12 \int_0^{L^{-1}} \frac{d^3k}{(2\pi)^3} |h_k|^2  \\
&\approx& \frac12 \int_0^{L^{-1}} \frac{dk}{(2\pi)^2} k^2\frac{(-\eta\pi)}{4}\left|J_n(-k\eta) + i Y_n(-k\eta)\right|^2  \\
\langle\phi^2(t) \rangle_{n^2>0} &\approx& \frac12\int_0^{L_{phys}^{-1}} \frac{dk_{phys}}{(2\pi)^2} \frac{k_{phys}^2}{4H_I}\left|J_n\left(\frac{k_{phys}}{H_I} \right) +i Y_n\left(\frac{k_{phys}}{H_I}\right)\right|^2  \\
&\approx& \frac12 H_I^2\int_0^{(H_I L_{phys})^{-1}} \frac{dx}{(2\pi)^2} \frac{x^2}{4}\left|J_n(x) +i Y_n(x)\right|^2  \\
&\approx& \frac12 H_I^2\int_0^{(H_I L_{phys})^{-1}} \frac{dx}{(2\pi)^2} \frac{x^2}{4} \left|-\frac{2^n\gG(n)}{\pi} (x)^{-n}\right|^2  \;\;\;\mbox{for }x\ll 1 \\
&\approx& H_I^2 \pfrac{2^{(2n-5)}\gG(n)^2}{\pi^4}\int_0^{(H_I L_{phys})^{-1}} \frac{dx}{x} x^{(3-2n)}
\end{eqnarray*}
where we have put the expression in terms of the physical momenta, $k=
k_{phys}R$, and original fields $\phi=\gvf/R$, then put the
integration over the dimensionless variable $x=\frac{k_{phys}}{H_I}$.
We may similarly calculate the case $n^2<0$,
\begin{eqnarray*}
\langle\gvf_L(\eta)^2 \rangle_{n^2<0}  &\approx& \frac12 \int_0^{L^{-1}} \frac{d^3k}{(2\pi)^3} |h_k|^2  \\
&\approx& \frac12\int_0^{L^{-1}} \frac{dk}{(2\pi)^2} k^2\frac{(-\eta\pi)}{4}e^{-|n|\pi} \left|J_{i|n|}(-k\eta) + i Y_{i|n|}(-k\eta)\right|^2  \\
\langle\phi^2(t) \rangle_{n^2<0} &\approx& \frac12\int_0^{L_{phys}^{-1}} \frac{dk_{phys}}{(2\pi)^2} \frac{k_{phys}^2}{4H_I}e^{-|n|\pi}\left|J_{i|n|}\left(\frac{k_{phys}}{H_I} \right) +i Y_{i|n|}\left(\frac{k_{phys}}{H_I}\right)\right|^2  \\
&\approx& \frac12 H_I^2\int_0^{(H_I L_{phys})^{-1}} \frac{dx}{(2\pi)^2} \frac{x^2}{4}e^{-|n|\pi}\left|J_{i|n|}(x) +i Y_{i|n|}(x)\right|^2  \\
&\approx& H_I^2\pfrac{1}{16|n|\pi^3}\int_0^{(H_I L_{phys})^{-1}} \frac{dx}{x} x^3 \;\;\;\mbox{for }n\approx\frac{m}{H_I} \mg 1
\end{eqnarray*}

\chapter{Inflaton Oscillations After Inflation}\label{calc-inflatonoscillation}
Here, the approximate evolution of the inflaton field at the
conclusion of inflation is determined in the case of a quadratic
inflaton potential.  Note, the result has been used
in~\myref{inflation-inflatonoscillations} of the main text. Also note
that the approximate evolution of the flat direction
fields~\myref{inflation-flatdiroscillations} used in the main text may
be determined through a similar method as that shown below for the
inflaton (see also~\cite{affleck-dine-1985}).

To begin, the Friedmann equation and the equation of motion for the
dominating isotropic and homogeneous inflaton field are,
\begin{eqnarray}
& H^2 = \pfrac{\dot s}{s}^2 = \frac{8\pi G}{3} \left( \frac12 {\dot\chi}^2 + V(\chi) \right) \label{e1} \\
& \ddot \chi + 3H \dot\chi + \frac{dV}{d\chi} = 0 \hspace{2em}\mbox{ where }\hspace{2em}V = \frac12 m^2 \chi^2 \label{e2}
\end{eqnarray}
At late times, after inflation as occurred, this field is oscillating
around the minimum of its potential $(V=0,\chi=0)$.  The motion is
determined by the above equations, and one assumes a solution of
damped oscillations,
\begin{equation} \chi(\tau) = X(\tau) \sin [m\tau] \hspace{4em}X(\tau) = \frac{A_1}{\tau} + \frac{A_3}{\tau^3} + ... \label{trial} \end{equation}
where we've defined the time parameter $\tau=(t-t_0)$ with $t_0$ being
the onset of the inflaton oscillations.  Specifically,
$\dot\chi(\tau=0) =0 $ and this constraint requires $X(\tau)$ must
not contain even powered terms.  We may solve for the series $A_1,
A_3,...$ by an iterative procedure.  Substitute the trial solution
into \myref{e1}, and set $A_1=A$.  Also set $\sin mt = s$ and $\cos mt
= c$ to simplify the notation,
\begin{eqnarray}
\dot X &=& -At^{-2} + O(t^{-4})  \nonumber \\
\ddot X &=& 2At^{-3} + O (t^{-5}) \nonumber \\
\dot\chi &=& {\dot X} s + m X c \nonumber \\
&=&  As t^{-2}  + mcA t^{-1}  \label{phidot} \\
\ddot \chi &=&  \ddot X s + 2 m \dot X c - m^2 X s  \nonumber \\
&=&  - 2 m c A t^{-2} -m^2 A s t^{-1} \label{phiddot}
\end{eqnarray}
where only the lowest order terms have been retained. Then, from
\myref{e1},
\begin{eqnarray}
H^2 &=&   \frac{4\pi G}{3}\left( ({\dot X} s + m X c)^2 + m^2  X^2 s^2 \right) 	 \nonumber \\
&=&   \frac{4\pi G}{3}\left( m^2  X^2 (s^2 +c^2) + 2 m{\dot X}  X s c\right) \nonumber \\
&=&   \frac{4\pi G}{3} m^2  X^2 \left(1 + 2 s c \frac{\dot X}{m  X}\right)  \nonumber \\
&=&   \frac{4\pi G}{3} m^2 \frac{A^2}{t^2} \left(1 - \frac{2 s c}{mt} \right) \label{hexpansion}
\end{eqnarray}
Now substitute for $\dot\chi$, $\ddot \chi$ and $H$ into \myref{e2}
and keep track of the order $t^{-n}$.
\begin{eqnarray*}
0 &=&\ddot \chi + 3H \dot\chi + m^2 \chi \\
0 &=& (- 2 m c t^{-2} -m^2 A s t^{-1} + O(t^{-3})) \\
&\mbox{}& +\; 3 \left(\left[\frac43\pi G m^2\right]^{1/2}At^{-1} + O(t^{-3/2})\right) (  As t^{-2}  + mcA t^{-1}) \\
&\mbox{}& +\; m^2 s (A t^{-1} + O(t^{-3})) \\
0 &=&( -m^2 A s + m^2 A s )t^{-1} + \left(-2mcA + 3 mc A^2\left[\frac43\pi G m^2\right]^{1/2}  \right) t^{-2} + O(t^3) 
\end{eqnarray*}
the first term vanishes, indicating that the guess for the solution
was correct, and the condition for the second term to vanish fixes the
constant $A$,
\begin{equation} A =A_1= \frac1{\sqrt{3\pi G m^2}} \label{a} \end{equation}
Considering higher order terms would yield the coefficients
$A_3,A_5,...$, which will not be done.  Substituting \myref{a} into
\myref{hexpansion} yields the Hubble parameter to lowest order,
\begin{equation} H \approx \frac2{3t} \end{equation}
which is the Hubble parameter of a perfect fluid of nonrelativistic
and pressureless matter.

\chapter{Thermodynamics of a Relativistic Plasma}\label{calc-thermodynamics}
Here we present some expressions used in the main text for a universe
dominated by an isotropic and homogenous relativistic fluid in thermal
equilibrium.  First, the total energy density of the fluid is related
to the temperature through the Stefan-Boltzman Law,
\begin{equation*}
\rho\approx \rho_{rad}  = \frac{\pi^2}{30} g_* T_{rad}^4
\;\;\;,\;\;\;
g_*(T) = g_b + \frac78 g_f
\end{equation*}
where $g_*$ is the effective number of degrees of freedom in the fluid
and $g_b$ and $g_f$ are the numbers of bosonic and fermionic degrees
of freedom respectively.  Using the Stefan-Boltzman law in combination
with the Friedmann equation, one determines the temperature in terms
of the Hubble parameter,
\begin{eqnarray}
\rho_{rad}  &=& \frac{3M_p^2}{8\pi} H^2 \nonumber \\
\frac{\pi^2}{30} g_* T^4  &=& \frac{3M_p^2}{8\pi} H^2 \nonumber \\
T &=& \pfrac{45M_p^2}{4g_* \pi^3}^{1/4} H^{1/2} \nonumber \\
  &=& \pfrac{45}{4g_* \pi^3}^{1/4} \sqrt{H M_P} 
\label{calc-thermodynamics-relateTH}
\end{eqnarray}
The entropy density for this relativistic fluid may be determined to
be, 
\begin{equation*}
s = \frac{2\pi^2}{45} g_* T^3 
\end{equation*}
The entropy density may then be written in terms of the energy density
using the Stefan Boltzman Law,
\begin{eqnarray}
s &=& \frac{2\pi^2}{45} g_* \pfrac{30\rho_{rad}}{g_*\pi^2}^{3/4} \nonumber \\
&=& \frac{2\pi^{1/2}}{45} (30)^{3/4}g_*^{1/4} \rho_{rad}^{3/4}  \nonumber \\
&\approx& \rho_{rad}^{3/4}
\label{thermodynamics-relatesrho}
\end{eqnarray}
Finally, the number density of a component relativistic species $X$
is,
\begin{equation*}
n_X = \frac{\zeta(3)}{\pi^2}g_X T^3 \hspace{4em} \zeta(3) \approx 1.2
\end{equation*}
where $\zeta$ is the Riemann zeta function.  

\chapter{Determination of Various Scale Factors and Events}\label{calc-rhoevolution}
We determine some expressions used in the main text which involve the
scale factor at different events during the evolution of the inflaton
and flat direction.  The reader is referred to
Figure~\ref{fig-rhoevolution} as well as to the following table which
describes the succession of events with $R$ increasing as one moves
down the table from the top.
\begin{center}
\begin{tabular}{|l|l|c|}
\hline
\textbf{Event} & $\mathbf{R}$ & \textbf{Eq. of state}\\
\hline
inflaton begins oscillations & $R_0$ & \\
&& matter dom \\
flat direction begins to oscillate & $R_m$& \\
&& \\
inflaton decays& $R_{d\chi}$ &    ------ \\
&& radiation dom\\
flat direction overtakes inflaton & $R_{eq}$ & ------ \\
&& matter dom \\
\hline
\end{tabular}
\end{center}
Other scale factors not shown in the table are $R_{d\Phi}$, the scale
factor at the instant of flat direction decay and $R_{RH}$, the scale
factor at the instant of thermalization.  Only the ratios of scale
factors are physically meaningful and to determine any ratio, we can
apply the following relations,
\begin{eqnarray*}
R \propto t^{2/3} \;\;\;\rightarrow\;\;\; 
t\propto H^{-1} \;\;\;\rightarrow\;\;\; 
&R\propto H^{-2/3}& \;\;\;\mbox{during matter domination}\\
R \propto t^{1/2} \;\;\;\rightarrow\;\;\; 
t\propto H^{-1} \;\;\;\rightarrow\;\;\; 
&R\propto H^{-1/2}& \;\;\;\mbox{during radiation domination}
\end{eqnarray*}
and we will also need to know the energy densities of the
flat direction to the inflaton at some reference times which we will
take to be the onset of their respective oscillations.
\begin{equation*}
r_\chi \equiv \frac{\rho_{\chi_0}}{m_\chi^2 M_P^2} 
\;\;\;,\;\;\;
r_\Phi \equiv \frac{\rho_{\Phi_m}}{m^2 \Phi_0^2}
\;\;\;,\;\;\;
r_m \equiv \pfrac{\rho_\Phi}{\rho_\chi}_m = \frac{r_\Phi}{r_\chi}\pfrac{\Phi_0}{M_P}^2
\end{equation*}
where the last expression is the ratio of the energy densities at the
onset of flat direction oscillations.\footnote{For the inflaton, the
  energy density at the end of slow-roll inflation is
  $\rho_\chi\approx\frac12 m_\chi^2 \chi^2$ where
  $\chi=\frac{MP}{\sqrt{3\pi}}$, thus $r_\chi=\frac{1}{6\pi}$.  For
  the flat direction the energy density is simply $m^2|\Phi_m|^2$
  where $|\Phi_m|$ depends on the evolution of this field during
  inflation, thus $r_\Phi=1$.  In order to trace the dependence on
  these quantities in the resulting expressions, the substitution are
  not made here.}  For example, using the above tools,
the energy density of the inflaton at the instant of inflaton decay is
obtained,
\begin{equation*}
\rho_{\Gamma_\chi} \sim r_\chi (\Gamma_\chi M_P)^2
\end{equation*}
Similarly, the scale factor when the flat direction overtakes the
inflaton $R_{eq}$ is obtained,
\begin{eqnarray}
\frac{\rho_\Phi}{\rho_\chi} &\sim& 1 \nonumber \\
\frac{\rho^m_\Phi\pfrac{R_m}{R_{eq}}^3}{\rho_\chi^m\pfrac{R_m}{R_{d\chi}}^3\pfrac{R_{d\chi}}{R_{eq}}^4} &\sim& 1 \nonumber \\
\frac{R_{eq}}{R_{d\chi}} &\sim& \frac{1}{r_m} 
\label{rhoevolution-Req}
\end{eqnarray}
which is valid for $r_m < 1$.  The condition for which the flat
direction dominates is,
\begin{eqnarray}
\frac{R_{eq}}{R_{d\Phi}} & < & 1 \nonumber \\
\frac{R_{eq}}{R_{d\chi}}\frac{R_{d\chi}}{R_{d\Phi}} &<& 1 \nonumber \\
\frac{\Gamma_\Phi }{\Gamma_\chi} &<& r_m^2
\label{rhoevolution-flatdirdominates}
\end{eqnarray}
After substitution for $r_m$, this expression defines a boundary in
the plane $(\Phi_0,\Gamma_\phi)$ separating regions in which the flat
direction dominates to regions where the inflaton dominates which is
shown in Figure~\ref{fig-flatdirdominates}.

Next the reheat temperature is determined assuming the thermalization
rate is controlled by $2\rightarrow 3$ interactions (with no
assumptions on $\Gamma_\chi$ and $\Gamma_\Phi$ ). From
Section~\ref{sec-reheating-perturbative} the approximate rate is,
\begin{eqnarray*}
& \Gamma_{2\rightarrow 3} \sim \ga \pfrac{m_\chi}{T_{RH}}^2 \Gamma_{2\rightarrow 2} \\
& \Gamma_{2\rightarrow 2} \sim n_\chi \gs
\;\;\;,\;\;\;
\gs = \pfrac{\ga}{m_\chi}^2\pfrac{R}{R_{d\chi}}^2
\end{eqnarray*}
and putting these expressions together,
\begin{eqnarray*}
\Gamma_{2\rightarrow 3} &\sim& \ga \pfrac{m_\chi}{T_{RH}}^2\frac{\rho_{d\chi}}{m_\chi}\pfrac{\ga}{m_\chi}^2\pfrac{R}{R_{d\chi}}^2 \\
&\sim& \ga \pfrac{m_\chi}{T_{RH}}^2\frac{r_\chi (\Gamma_\chi M_P)^2}{m_\chi}\pfrac{\ga}{m_\chi}^2\pfrac{R}{R_{d\chi}}^2
\end{eqnarray*}
Now considering the case with no flat directions, and assuming
$\Gamma_{therm}>\Gamma_\chi$ so reheating happens instantly at
$R_{d\chi}$ we obtain the following consistency requirement on $\ga$,
\begin{eqnarray}
\Gamma_{2\rightarrow 3} &\simg& \Gamma_\chi \nonumber \\
\ga \pfrac{m_\chi}{0.1\sqrt{\Gamma_\chi M_P}}^2\frac{r_\chi(\Gamma_\chi M_P)^2}{m_\chi}\pfrac{\ga}{m_\chi}^2  &\simg& \Gamma_\chi \nonumber \\
\ga^3  &\simg& \frac{1}{100 r_\chi} \frac{m_\chi}{M_P} 
\label{rhoevolution-alphanoflat}
\end{eqnarray}
Next considering the case involving flat directions and supressed
reaction rates, we simply replace $\pfrac{\ga}{m_\chi}\rightarrow
\pfrac{\ga}{\Phi}$ in the expression for $\gs$ above.  We consider
three situations, $\Gamma_{therm}\sim \{\Gamma_\chi,
\Gamma_{2\rightarrow 3}, \Gamma_\Phi\}$ which can each be realized by
tuning $\Phi_0$, $\ga$ and/or $\Gamma_\Phi$.  Considering the case
where reheating is controlled by $\Gamma_\chi$, one obtains the
consistency condition,
\begin{eqnarray}
\Gamma_{2\rightarrow 3} &\simg& \Gamma_\chi \nonumber \\
\ga \pfrac{m_\chi}{T_{RH}}^2\frac{\rho_{d\chi}}{m_\chi}\pfrac{\ga}{\Phi}^2\pfrac{R}{R_{d\chi}}^2 &\simg& \Gamma_\chi \nonumber \\
\ga \pfrac{m_\chi}{T_{RH}}^2\frac{\rho_{d\chi}}{m_\chi}\frac{\ga^2}{\Phi_0^2\pfrac{R_{m}}{R_{d\chi}}^3} &\simg& \Gamma_\chi \nonumber \\
\ga \pfrac{m_\chi}{0.1\sqrt{\Gamma_\chi M_P}}^2\frac{r_\chi\Gamma_\chi^2 M_P^2}{m_\chi}\frac{\ga^2}{\Phi_0^2}\pfrac{m}{\Gamma_\chi}^2 &\simg& \Gamma_\chi \nonumber \\
100r_\chi\ga^3 \pfrac{m_\chi}{M_P}\pfrac{m}{M_P}^2\pfrac{M_P}{\Gamma_\chi}^2 &\simg& \pfrac{\Phi_0}{M_P}^2
\label{rhoevolution-alphawithflat}
\end{eqnarray}
%
Next considering the case where reheating is controlled by
$\Gamma_{2\rightarrow 3}$ which happens when $\Gamma_\Phi <
\Gamma_{2\rightarrow 3} < \Gamma_\chi$.  One determines $T_{RH}$
self-consistently using the expresion for $\Gamma_{2\rightarrow 3}$,
\begin{eqnarray}
\Gamma_{2\rightarrow 3} &\sim& \ga \pfrac{m_\chi}{T_{RH}}^2\frac{\rho_{d\chi}}{m_\chi}\pfrac{\ga}{\Phi}^2\pfrac{R_{RH}}{R_{d\chi}}^2 \nonumber \\
\Gamma_{2\rightarrow 3} &\sim& \ga \pfrac{m_\chi}{T_{RH}}^2\frac{r_\chi\Gamma_\chi^2 M_P^2}{m_\chi}\pfrac{\ga^2}{\Phi_0^2\pfrac{R_{m}}{R_{RH}}^3}\pfrac{R_{RH}}{R_{d\chi}}^2 \nonumber \\
\Gamma_{2\rightarrow 3} &\sim& \ga^3 \pfrac{m_\chi}{T_{RH}}^2\frac{r_\chi\Gamma_\chi^2}{m_\chi}\pfrac{M_P}{\Phi_0}^2\pfrac{R_{d\chi}}{R_{m}}^3\pfrac{R_{RH}}{R_{d\chi}}^5 \nonumber \\
\Gamma_{2\rightarrow 3} &\sim& \ga^3 \pfrac{m_\chi}{T_{RH}}^2\frac{r_\chi\Gamma_\chi^2}{m_\chi}\pfrac{M_P}{\Phi_0}^2\pfrac{m}{\Gamma_\chi}^2\pfrac{\Gamma_\chi}{\Gamma_{2\rightarrow 3}}^{5/2} \nonumber \\
10^7\pfrac{T_{RH}}{M_P}^9 &\sim& r_\chi\ga^3 \pfrac{m_\chi}{M_P}\pfrac{M_P}{\Phi_0}^2\pfrac{m}{M_P}^2\pfrac{\Gamma_\chi}{M_P}^{5/2} \nonumber \\
\pfrac{T_{RH}}{M_P} &\sim& \left(r_\chi\ga^3 10^{-7}\right)^{1/9} \pfrac{m_\chi}{M_P}^{1/9}\pfrac{M_P}{\Phi_0}^{2/9}\pfrac{m}{M_P}^{2/9}\pfrac{\Gamma_\chi}{M_P}^{5/18} \nonumber \\
\mbox{}\label{rhoevolution-trh1}
\end{eqnarray}
This result is also continuous and consistent with the previous
result.  In the final case, reheating is controlled by $\Gamma_\Phi$,
which occurrs when $\Gamma_{2\rightarrow 3} <\Gamma_\Phi$.

Now specifing to the case when the flat direction decays
perturbatively, we determine the expressions for $\Gamma_\Phi$ first
assuming the decay happens during matter domination before inflaton
decay.  Note that $\Gamma^{pert}_\Phi$ indicates the instant of decay
of the whole flat direction condensate which is the single particle
decay rate at a specific value of the scale factor defined as
$R_{d\Phi}$,
\begin{eqnarray*}
\frac{\Gamma_\Phi^{pert}}{m}  &\sim& \frac{\ga^2 m^2}{\Phi^2} \nonumber \\
\frac{\Gamma_\Phi^{pert}}{m}  &\sim& \frac{\ga^2 m^2}{\Phi_0^2}\pfrac{R_{d\Phi}}{R_m}^3 \nonumber \\
\pfrac{\Gamma_\Phi^{pert}}{m}^3  &\sim& \frac{\ga^2 m^2}{\Phi_0^2} \nonumber \\
\pfrac{\Gamma_\Phi^{pert}}{m}  &\sim& \pfrac{\ga m}{\Phi_0}^{2/3}
\end{eqnarray*}
Next the case when the flat direction decays during radiation domination and for which the above
calculation gets modified on the third step which now becomes
\begin{eqnarray}
\frac{\Gamma_\Phi^{pert}}{m}  &\sim& \frac{\ga^2 m^2}{\Phi_0^2} \pfrac{R_{d\Phi}}{R_{d\chi}}^3\pfrac{R_{d\chi}}{R_m}^3 \nonumber \\
\frac{\Gamma_\Phi^{pert}}{m}  &\sim& \frac{\ga^2 m^2}{\Phi_0^2} \pfrac{\Gamma_\chi}{\Gamma^{pert}_\Phi}^{3/2}\pfrac{m}{\Gamma_\chi}^2 \nonumber \\
\pfrac{\Gamma_\Phi^{pert}}{m}^{5/2}  &\sim& \frac{\ga^2 m^2}{\Phi_0^2} \pfrac{m}{\Gamma_\chi}^{1/2} \nonumber \\
\frac{\Gamma_\Phi^{pert}}{m}  &\sim& \pfrac{\ga m}{\Phi_0}^{4/5} \pfrac{m}{\Gamma_\chi}^{1/5}
\label{rhoevolution-GammaPhipert}
\end{eqnarray}
From this last result, we can determine the condition on $\Phi_0$ for
flat direction domination,
\begin{eqnarray}
\frac{R_{eq}}{R_{d\Phi}^{pert}} &<& 1 \nonumber \\
\frac{\Gamma_\Phi^{pert}}{m} \frac{m}{\Gamma_\chi} &<& r_m^2 \nonumber \\
\pfrac{\ga m}{\Phi_0}^{4/5} \pfrac{m}{\Gamma_\chi}^{1/5} \frac{m}{\Gamma_\chi} &<& \pfrac{r_\Phi}{r_\chi}^2\pfrac{\Phi_0}{M_P}^4 \nonumber \\
\ga^{1/6}\pfrac{m}{M_P}^{5/12}\pfrac{M_P}{m_\chi}^{3/4}\pfrac{r_\chi}{r_\phi}^{5/12} &<& \pfrac{\Phi_0}{M_P}
\label{rhoevolution-flatdirdominates2}
\end{eqnarray}
where we have substituted for $\Gamma_\chi$.  We may determine the
flat direction decay rate in the case the flat direction
dominates by,
\begin{eqnarray*}
\frac{\Gamma_\Phi^{pert}}{m}  &\sim& \frac{\ga^2 m^2}{\Phi^2} \nonumber \\
\frac{\Gamma_\Phi^{pert}}{m}  &\sim& \frac{\ga^2 m^2}{\Phi_0^2}\pfrac{R^{pert}_{d\Phi}}{R_m}^3 \nonumber \\
\pfrac{\Gamma_\Phi^{pert}}{m}  &\sim& \frac{\ga^2 m^2}{\Phi_0^2}\pfrac{R^{pert}_{d\Phi}}{R_{eq}}^3\pfrac{R_{eq}}{R_{d\chi}}^3\pfrac{R_{d\chi}}{R_m}^3 \nonumber \\
\pfrac{\Gamma_\Phi^{pert}}{m}  &\sim& \frac{\ga^2 m^2}{\Phi_0^2}\pfrac{H_{eq}}{\Gamma_\Phi^{pert}}^2\pfrac{\Gamma_\chi}{H_{eq}}^{3/2}\pfrac{m}{\Gamma_\chi}^2 \nonumber \\
\pfrac{\Gamma_\Phi^{pert}}{m}^3  &\sim& \frac{\ga^2 m^2}{\Phi_0^2}\frac{H_{eq}^{1/2}}{\Gamma_\chi^{1/2}} \nonumber \\
\pfrac{\Gamma_\Phi^{pert}}{m}^3  &\sim& \pfrac{\ga m}{M_P}^2\pfrac{r_\Phi}{r_\chi} \nonumber \\
\pfrac{\Gamma_\Phi^{pert}}{m}  &\sim& \pfrac{\ga m}{M_P}^{2/3}\pfrac{r_\Phi}{r_\chi}^{1/3}
\end{eqnarray*}
Note this result is also obtained by
substituting~\myref{rhoevolution-flatdirdominates2}
into~\myref{rhoevolution-GammaPhipert}.  Also in the above we have
computed $H_{eq}$ from~\myref{rhoevolution-Req} as follows
\begin{eqnarray*}
\frac{R_{eq}}{R_{d\chi}} &\sim& \frac{1}{r_m} \\
\pfrac{\Gamma_\chi}{H_{eq}}^{1/2} &\sim& \frac{1}{r_m} \\
\Gamma_\chi r_m^2 &\sim& H_{eq} \\
\Gamma_\chi \pfrac{r_\Phi}{r_\chi}^2\pfrac{\Phi_0}{M_P}^4 &\sim& H_{eq}
\end{eqnarray*}

\chapter{Bogolyubov Transformations for Multiple Scalar Fields}\label{calc-bogtransform-general}
The material of this appendix is supplementary to that of
Section~\ref{sec-heiseqs-formalism}.  Some calculations which were not
performed in this section are performed here, and the derivation of the
equations~\myref{heiseqs-bogequations} is here made from an
alternative perspective.  As a reminder, the results of
Section~\ref{sec-heiseqs-formalism} and this appendix were first
obtained in~\cite{nilles-peloso-sorbo-2001}, and the notation and
conventions of this paper are used in the following.

The starting point is a set of creation and annihilation operators
$a_i$, $a_i^\dagger$ which we wish to transform to a different set
$b_i$, $b_i^\dagger$ through a Bogolyubov transformation,
\begin{eqnarray}
a_i &=& \alpha_{ij} b_j + \beta^*_{ij} b^\dagger_j \\
a^\dagger_i &=& \alpha^*_{ij}b^\dagger_j + \beta_{ij}b_j
\end{eqnarray}
where $\alpha_{ij}$ and $\beta_{ij}$ are complex matrices.  We require
canonical commutation relations on both sets of operators, and this
gives constraints on the $\alpha_{ij}$ and $\beta_{ij}$ coefficients.
First calculate the commutator $[a_i,a^\dagger_j]$
\begin{eqnarray}
[a_i, a^\dagger_j] &=& \gd_{ij} \nonumber \\
\;[ \alpha_{ik} b_k + \beta^*_{ik} b^\dagger_k , \alpha^*_{jm}b^\dagger_m + \beta_{jm}b_m] &=& \gd_{ij}\nonumber \\
\ga_{ik}\ga^*_{jm}[b_k,b_m^\dagger] + \gb^*_{ik}\gb_{jm}[b^\dagger_k,b_m] + \ga_{ik}\gb_{jm}[b_k,b_m] + \gb^*_{ik}\ga^*_{jm}[b_k^\dagger,b_m^\dagger] &=& \gd_{ij} \nonumber \\
\ga_{ik}\ga^*_{jk} - \gb^*_{ik}\gb_{jk} &=& \gd_{ij} \nonumber \\
\mbox{} \label{bog-c1}
\end{eqnarray}
and then the commutator $[a_i,a_j]$,
\begin{eqnarray}
[a_i, a_j] &=& 0 \nonumber \\
\;[ \alpha_{ik} b_k + \beta^*_{ik} b^\dagger_k , \alpha_{jm}b_m + \beta^*_{jm}b^\dagger_m] &=& 0\nonumber \\
\ga_{ik}\ga_{jm}[b_k,b_m] + \gb^*_{ik}\gb^*_{jm}[b^\dagger_k,b^\dagger_m] + \ga_{ik}\gb^*_{jm}[b_k,b^\dagger_m] + \gb^*_{ik}\ga_{jm}[b_k^\dagger,b_m] &=& 0 \nonumber \\
\ga_{ik}\gb^*_{jm}\gd_{km} - \gb^*_{ik}\ga_{jm}\gd_{km} &=& 0 \nonumber \\
\ga_{ik}\gb^*_{jk} - \gb^*_{ik}\ga_{jk} &=& 0 \nonumber \\
\mbox{}\label{bog-c2}
\end{eqnarray}
so that by requiring both sets of operators to satisfy canonical
commutation relations, we have determined two constraints
(\ref{bog-c1}) and (\ref{bog-c2}), which in matrix notation are,
\begin{eqnarray} 
\ga \ga^\dagger - \gb^*\gb^T &=& 1 \label{bog-mc1}\\
\ga \gb^\dagger - \gb^*\ga^T &=& 0 \label{bog-mc2}
\end{eqnarray}
Note that in the one-dimensional case where $\ga$ and $\gb$ are
complex numbers, the second constraint is satisfied trivially, and the
remaining constraint is $|\alpha|^2-|\beta|^2 = 1$.  

Now, we wish to apply the above Bogolyubov transformations to a system
of $N$ evolving real scalar fields $\hat{\phi}_i(x,t)$ described by a
Lagrangian with canonical kinetic terms and a time dependent and
non-diagonal mass matrix $M^2$
\begin{equation}
S=  \int d^4x \frac12 \left(\phi_i,_\mu \phi_i,_\mu - \phi_iM^2_{ij}\phi_j \right) \label{bog-L}
\end{equation}
Note that here the physical time $t$ will be used instead of the
conformal time as is used in the main text.  We define a basis of time
dependent Heisenberg operators, ${\hat a}_{i}(k,t)$ and ${\hat
  a}_{i}^\dagger(k,t)$ as well as a basis of fixed time Schroedinger
operators $a_{i}(k)={\hat a}_{i}(k,0)$ and $a_{i}^\dagger(k) = {\hat
  a}^\dagger_{i}(k,0)$ with initial time $t=0$.  The operators are
indexed $i=1,...N$ with $N$ being the number of fields.  The
transformation between the two bases is,
\begin{eqnarray}
{\hat a}_{i}(k,t) &=& \ga_{ij}(k,t) a_{j}(k) + \gb_{ij}^*(k,t) a_{j}^\dagger(k) \\
{\hat a}_{i}^\dagger(k,t) &=&  \ga^*_{ij}(k,t) a_{j}^\dagger(k)  +  \gb_{ij}(k,t) a_{j}(k) 
\end{eqnarray}
where now $\ga_{ij}(k,t)$ and $\gb_{ij}(k,t)$ are momentum and time
dependent.  We will suppress the momentum dependence and time
dependence below where possible in order to simplify notation.  The
real scalar fields are written in terms of our creation and
annihilation operators as follows,
\begin{eqnarray}
\hat{\phi}_i &=& \int d^3k \left[ e^{ikx}\frac{1}{\sqrt{2\go_i}}{\hat a}_i(k,t) + e^{-ikx}\frac{1}{\sqrt{2\go_i}}{\hat a}_i^\dagger(k,t) \right] \label{bog-definephi}  \\
&=& \int d^3k \left[ e^{ikx}\frac{1}{\sqrt{2\go_i}}(\ga_{ij} a_{j} + \gb_{ij}^* a_{j}^\dagger) + e^{-ikx}\frac{1}{\sqrt{2\go_i}}(\ga^*_{ij} a_{j}^\dagger  +  \gb_{ij} a_j ) \right] \nonumber \\
&=& \int d^3k \left[ e^{ikx}\frac{1}{\sqrt{2\go_i}}(\ga_{ij} +\gb_{ij})a_j + e^{-ikx}\frac{1}{\sqrt{2\go_i}}(\ga_{ij}^* + \gb_{ij}^*)a_j^\dagger \right] \nonumber \\
&=& \int d^3k \left[ e^{ikx}h_{ij}(t) a_j + e^{-ikx}h_{ij}^*(t) a_j^\dagger \right] \nonumber \\
\hat{\Pi} &=& \int d^3k \left[ e^{ikx} {\tilde h}_{ij}(t) a_j + e^{-ikx}{\tilde h}_{ij}^*(t)a_j^\dagger \right] 
\label{bog-definepi}
\end{eqnarray}
where it is implied in the first line above that our set of Heisenberg
operators diagonalize the Hamiltonian at the time $t$.  We've also
defined $h_{ij}(t)\equiv\frac{\ga_{ij}+\gb_{ij}}{\sqrt{2\go_i}}$ and
parametrized the conjugate momentum $\hat{\Pi}_i$ via the function
${\tilde h}_{ij}$ which must be determined.  The commutation relation for the
fields will fix the form of $\hat\Pi$ and thus~${\tilde h}_{ij}$,
\begin{eqnarray*}
\;[ \gvf_i(x), \Pi_m(y)] &=& i\gd^3(x-y)\gd_{im}  \\
\int d^3k d^3k' \left(\begin{array}{l}
e^{ikx+ik'y}h_{ij} {\tilde h}_{mn} [a_j(k), a_n(k')]  \\
+\;e^{ikx-ik'y}h_{ij} {\tilde h}^*_{mn} [a_j(k),a_n^\dagger(k')]  \\
+\;e^{-ikx+ik'y}h^*_{ij}{\tilde h}_{mn} [a_j^\dagger(k), a_n(k')]  \\
+\;e^{ikx-ik'y} h^*_{ij}{\tilde h}^*_{mn} [a_j^\dagger(k), a_n^\dagger(k')]
\end{array}\right) &=& i\gd^3(x-y)\gd_{im}  \\
\int d^3k \left(e^{ik(x-y)} h_{ij}{\tilde h}^*_{mj} - e^{-ik(x-y)}h^*_{ij} {\tilde h}_{mj} \right) &=& i\gd^3(x-y)\gd_{im}  \\
\left(h_{ij}{\tilde h}^*_{mj} - h^*_{ij} {\tilde h}_{mj} \right) &=& i\gd_{im} 
\end{eqnarray*}
This is a nonlinear algebraic constraint from which we can solve for ${\tilde h}$ 
In matrix notation, this constraint is written,
\begin{equation}
h{\tilde h}^\dagger - h^*{\tilde h}^T = \mathbf{1}i
\label{bog-hconstraint}
\end{equation}
The solution may be guessed, and it is (in matrix notation),
\begin{eqnarray}
h &=& \frac{1}{\sqrt{2\go}}(\ga+\gb) \\
{\tilde h} &=& \frac{i\go}{\sqrt{2\go}}(-\ga +\gb)
\end{eqnarray}
where $\go$ is the diagonal matrix of eigenfrequencies, and these
solutions are verified on (\ref{bog-hconstraint}) by
application of the constraints~(\ref{bog-mc1}-\ref{bog-mc2}).
Finally, the equations of motion will determine the evolution of the
$\ga_{ij}$'s and $\gb_{ij}$'s if we are given their initial values.
With this goal in mind, we assume the Lagrangian introduced at the
beginning (\ref{bog-L}).  The mass matrix of this Lagrangian is
diagonalized with a time dependent rotation matrix $C$,
\begin{equation}
C^T(t) M^2(t) C(t) = m_d^2(t)
\end{equation}
where $m_d^2$ is diagonal.  The conjugate momenta are $\Pi_i
\equiv \frac{\der\cL}{\der\dot{\phi}_i} = \dot\phi$, and the
Hamiltonian written in momentum space is,
\begin{equation}
H = \int d^3x \frac12 \left(\Pi_i \Pi_i +\phi_i \gO^2_{ij}\phi_j\right) 
\;\;\;,\;\;\;
\gO^2_{ij} = (M^2_{ij} + k^2\gd_{ij})
\end{equation}
Hamilton's Equations of motion are
\begin{eqnarray}
\dot\phi &=& \Pi = \frac{\der\mathcal{L}}{\der\dot\phi} \nonumber \\
\dot\Pi &=& -\gO^2 \phi
\end{eqnarray}
However, Hamilton's equations are written in fields $\phi_i$ and
$\Pi_i$ which do not yield a diagonal Hamiltonian.  From our
Lagrangian (\ref{bog-L}), we see that the fields for the diagonal
basis will instead be,
\begin{equation}
\hat{\phi}_i = C^T\phi_i \;\;\;,\;\;\; \hat{\Pi}_i = C^T \Pi_i  \label{bog-relatephi}
\end{equation}
and we thus identify $\hat\phi_i$ with the fields defined above in
(\ref{bog-definephi}) which we assumed to be a diagonal basis.  Now we
may use (\ref{bog-relatephi}) and
(\ref{bog-definephi}-\ref{bog-definepi}) to write Hamilton's equations
(Heisenberg's equations of motion) in terms of the $\alpha_{ij}$'s and
$\beta_{ij}$'s,
\begin{eqnarray}
& \dot\alpha = \left(-i \omega -I\right)\alpha + \left(\frac{\dot\omega}{2\omega}-J\right)\beta \nonumber \\
& \dot\beta = \left(i \omega-I\right) \beta + \left(\frac{\dot\omega}{2\omega} -J\right)\alpha
\label{bogequations}
\end{eqnarray}
where matrix notation is used again, and the antisymmetric $\Gamma$
and $I$ matrices and the symmetric $J$ matrix are defined,
\begin{equation} I,J = \frac12 \left(\sqrt{\omega} \Gamma \frac{1}{\sqrt{\omega}} \pm \frac{1}{\sqrt{\omega}}\Gamma \sqrt{\omega} \right) \;\;\;,\;\;\; \Gamma = C^T \dot C
\label{defineIJGamma}
\end{equation}

\end{document}

%% file: texdefs.tex
\newcommand{\siml}{\raisebox{-.1ex}{\renewcommand{\arraystretch}{0.3}
$\begin{array}{@{}c} \scriptstyle < \\ \scriptstyle \sim \end{array}$}}

\newcommand{\simg}{\raisebox{-.1ex}{\renewcommand{\arraystretch}{0.3}
$\begin{array}{@{}c} \scriptstyle > \\ \scriptstyle \sim \end{array}$}}

\DeclareMathSymbol{\mg}{\mathrel}{symbols}{"1D}
\newcommand{\ml}{\ll}

\newcommand{\ga}{\alpha}
\newcommand{\gb}{\beta}
\renewcommand{\gg}{\gamma}
\newcommand{\gd}{\delta}
\renewcommand{\ge}{\epsilon}

\newcommand{\gvf}{\varphi}

\newcommand{\gl}{\lambda}

\newcommand{\gth}{\theta}

\newcommand{\gs}{\sigma}

\newcommand{\go}{\omega}

\newcommand{\gG}{\Gamma}
\newcommand{\gD}{\Delta}
\newcommand{\gF}{\Phi}

\newcommand{\gL}{\Lambda}
\newcommand{\gS}{\Sigma}

\newcommand{\gO}{\Omega}

\newcommand{\cB}{{\cal B}}

\newcommand{\cL}{{\cal L}}
\newcommand{\cM}{{\cal M}}

\newcommand{\cO}{{\cal O}}

\newcommand{\cR}{{\cal R}}

\newcommand{\tg}{{\tilde g}}

\newcommand{\tm}{{\tilde m}}

\newcommand{\tB}{{\tilde B}}

\newcommand{\tT}{{\tilde T}}

\newcommand{\der}{\partial}

\newcommand{\beq}{\begin{equation}}
\newcommand{\eeq}{\end{equation}}
\newcommand{\barr}{\begin{array}}
\newcommand{\earr}{\end{array}}

\newcounter{oldcounter}

\newcommand{\bd}{{\bar d}}
\newcommand{\bee}{{\bar e}}

\newcommand{\bk}{{\bar k}}

\newcommand{\bq}{{\bar q}}

\newcommand{\bga}{{\bar \alpha}}
\newcommand{\bgb}{{\bar\beta}}

\newcommand{\tgd}{{\tilde \delta}}

\newcommand{\tgl}{{\tilde \lambda}}

\newcommand{\tgs}{{\tilde \sigma}}

\newcommand{\tgS}{{\tilde \Sigma}}

\newcommand{\tgF}{{\tilde \Phi}}

\newcommand{\tgO}{{\tilde \Omega}}

\newcommand{\ba}[2]{\[\begin{array}{#2}\label{#1}}
\newcommand{\ea}{\end{array}\]}
\newcommand{\be}{\begin{equation}}
\newcommand{\ee}{\end{equation}}
\newcommand{\bea}{\begin{eqnarray}}
\newcommand{\eea}{\end{eqnarray}}

%% file: thes-all.bbl
\begin{thebibliography}{99}
\bibitem{inflation-evidence} Inflation resolves the horizon, flatness
  and monopole problems and successfully generates the cosmological
  perturbations (reviewed in Section~\ref{sec-inflation}).  A
  ``smoking gun'' signature of inflation would come from measurement
  of a primordial gravitational wave signature in the cosmic microwave
  background \cite{inflation-smokinggun}.  This signature may be very
  weak, and has not yet been detected.
\bibitem{equipartition} The study of equipartition in classical
   systems of coupled oscillators began over 50 years ago.  Note that
   many well-motivated systems may not strictly obtain an
   equipartition of energy.  There is still active research on these
   problems.  For instance see \cite{equipartition2}
\end{thebibliography}
